\documentclass[12pt]{article}
\usepackage{graphicx,amsmath,setspace,amssymb,mathtools}
\usepackage{units}
\usepackage{color}
\usepackage{cite}
\usepackage{hyperref}
\usepackage{multirow}
\usepackage{pbox}
\usepackage{array}
\usepackage{float}
\usepackage{lineno}
\usepackage{relsize}
\usepackage{hyperref}
\usepackage[caption=true]{subfig}
\usepackage[font=small,labelfont=bf]{caption}
\numberwithin{equation}{section}
\usepackage[normalem]{ulem}

\newlength{\dinwidth}
\newlength{\dinmargin}
\DeclarePairedDelimiter\abs{\lvert}{\rvert}

\def\lapproxeq{\lower .7ex\hbox{$\;\stackrel{\textstyle                                                    
<}{\sim}\;$}}                                                    
\def\gapproxeq{\lower .7ex\hbox{$\;\stackrel{\textstyle                                                    
>}{\sim}\;$}}                                                    
\def\be{\begin{equation}}                                                    
\def\ee{\end{equation}}                                                    
\def\bea{\begin{eqnarray}}                      
\def\eea{\end{eqnarray}}

\def\sh{\hat s}
\def\sh2{{\hat s}^2}

\begin{document}


\newcolumntype{P}[1]{>{\centering\arraybackslash}p{#1}}

\vspace*{0.5cm}
\begin{center}
{\Large \bf Approximate N$^{3}$LO Parton Distribution Functions  \\ \vspace*{0.2cm} with Theoretical Uncertainties:}\\ 
\vspace*{0.5cm}{\Large \bf MSHT20aN$^3$LO PDFs }\\

\vspace*{1cm}
J. McGowan$^a$, T. Cridge$^a$, L. A. Harland-Lang$^{b}$, 
and R.S. Thorne$^a$\\                                               
\vspace*{0.5cm}                                                    

$^a$ Department of Physics and Astronomy, University College London, London, WC1E 6BT, UK \\ 
$^b$ Rudolf Peierls Centre, Beecroft Building, Parks Road, Oxford, OX1 3PU, UK   \\  

\begin{abstract}
    \noindent We present the first global analysis of parton distribution functions (PDFs) at approximate N$^{3}$LO in the strong coupling constant $\alpha_{s}$, extending beyond the current highest NNLO achieved in PDF fits. To achieve this, we present a general formalism for the inclusion of theoretical uncertainties associated with the perturbative expansion in the strong coupling. We demonstrate how using the currently available knowledge surrounding the next highest order (N$^{3}$LO) in $\alpha_{s}$ can provide consistent, justifiable and explainable approximate N$^{3}$LO (aN$^{3}$LO) PDFs. This includes estimates for uncertainties due the the currently unknown N$^{3}$LO ingredients, but also implicitly some missing higher order uncertainties (MHOUs) beyond these. Specifically, we approximate the splitting functions, transition matrix elements, coefficient functions and $K$-factors for multiple processes to N$^{3}$LO. Crucially, these are constrained to be consistent with the wide range of already available information about N$^{3}$LO to match the complete result at this order as accurately as possible. Using this approach we perform a fully consistent approximate N$^{3}$LO global fit within the MSHT framework. This relies on an expansion of the Hessian procedure used in previous MSHT fits to allow for sources of theoretical uncertainties. These are included as nuisance parameters in a global fit, controlled by knowledge and intuition based prior distributions. We analyse the differences between our aN$^{3}$LO PDFs and the standard NNLO PDF set, and study the impact of using aN$^{3}$LO PDFs on the LHC production of a Higgs boson at this order. Finally, we provide guidelines on how these PDFs should be be used in phenomenological investigations.
\end{abstract}
\vspace*{0.5cm}  

\end{center}

\begin{spacing}{0.8}
\clearpage

\pagebreak
\tableofcontents
\clearpage
\end{spacing}

\section{Introduction}\label{sec: intro}

In recent years, the level of precision achieved at the LHC has reached far beyond what was once thought possible. This has initiated a new era of high precision phenomenology that has pushed the need for a robust understanding of theoretical uncertainty to new levels. Due to the perturbative nature of calculations in Quantum Chromodynamics (QCD), with respect to the strong coupling constant $\alpha_{s}$, a leading theoretical uncertainty arises from the truncation of perturbative expansions~\cite{Mojaza:scaleamb,Czakon:dynamicscale}. The current state of the art for parton distribution functions (PDFs) is next-to-next-to leading order (NNLO)~\cite{Thorne:MSHT20,Hou:2019efy,NNPDF3.1,NNPDF:2021njg,NNPDF:2021uiq,PDF4LHC22,ABMP16,ATLAS:2021vod}. However, these PDF sets do not generally include theoretical uncertainties arising from the truncation of perturbative calculations that enter the fit. The consideration of these so-called Missing Higher Order Uncertainties (MHOUs), and how to estimate them, is the topic of much discussion among groups involved in fitting PDFs~\cite{NNPDFscales,consistency,Lucian,Ball:corr2021}.

More recently, a method of utilising a scale variation approach to estimating these uncertainties has been included in an NLO PDF fit~\cite{NNPDFscales}. This approach is based upon the fact that to all orders, a physical calculation must not depend on any unphysical scales introduced into calculations. Therefore varying the factorisation and renormalisation scales is, in principle, a first attempt at estimating the level of theory uncertainty from missing higher orders (MHOs). Motivated by the renormalisation group invariance of physical observables, this method is theoretically grounded to all orders. However, the method of scale variations has been shown to be less than ideal in practice~\cite{consistency,Bonvini:mhous}. An obvious difficulty is the arbitrary nature in the chosen range of the scale variation, as well as the choice of central scale. Expanding on this further, even if a universal treatment of scale variations was agreed upon, these variations are unable to predict the effect of various classes of logarithms (e.g.  small-$x$, mass threshold and leading large-$x$ contributions) present at higher orders. 
As an example, studies of fits including small-$x$ resummation have recently been done~\cite{NNPDFsx,xFitterDevelopersTeam:2018hym}, showing 
significant PDF changes. 
Since it is these type of contributions that are often the most dominant at higher orders, this is an especially concerning pitfall in the use of scale variations to estimate MHOUs. Rather more subtle are the challenges encountered when considering and accounting for correlations between fit and predictions of PDFs~\cite{consistency,Ball:corr2021}. 
An alternative method to the above is to parameterise the missing higher orders with a set of nuisance parameters, using the available (albeit incomplete) current knowledge~\cite{Tackmann:2019,Tackmann:review}.

In this paper we present the first study of an approximate ${\rm N}^3{\rm LO}$ (aN$^{3}$LO) PDF fit. In particular, we first consider approximations to the N$^{3}$LO structure functions and DGLAP evolution of the PDFs, including the relevant heavy flavour transition matrix elements. We make use of all available knowledge to constrain an approximate parameterisation of the N$^{3}$LO theory, including the calculated Mellin moments, low-$x$ logarithmic behaviour and the full results where they exist. Then for the case of hadronic observables (where less N$^{3}$LO information is available), we include approximate N$^{3}$LO $K$-factors which are guided by the size of known NLO and NNLO corrections. Based on the uncertainty in our knowledge of each N$^3$LO function, we obtain a theoretical confidence level (C.L.) constrained by a prior. The corresponding theoretical uncertainties are therefore regulated by our theoretical understanding or lack thereof. Applying the above procedure, we have performed a full global fit at approximate N$^{3}$LO, with a corresponding theoretical uncertainty included within a nuisance parameter framework. As we will show, adopting this procedure allows the correlations and sources of uncertainties to be easily controlled. The preferred form of the aN$^{3}$LO corrections is determined from the fit quality to data, subject to theoretical constraints from the known information about higher orders.

We note that the source of the above uncertainty is due the (currently unknown) missing ingredients at N$^{3}$LO, and hence to be precise this corresponds to a `missing N$^{3}$LO' uncertainty. However under the common assumption that the dominant uncertainty from missing higher orders (MHOs) is due this uncertainty at the next not fully known N$^{3}$LO order, one can also expect this to provide a reasonable estimate of MHOs in the fit. Indeed, by allowing the unknown theory parameters to be determined by the fit to data, sensitivity to orders beyond N$^{3}$LO is explicitly introduced. As we will see, this is particularly transparent in the case of the hadronic $K$--factors, which are more directly interpreted as giving a full MHO uncertainty, although a similar sensitivity to higher orders (in particular at low $x$) is observed in the DGLAP evolution of the PDFs. Therefore, while we assume that the majority of this uncertainty is due to the missing information at N$^{3}$LO, it is the case that some is associated with orders even beyond this, most obviously further effects due to small-$x$ logarithms. Nonetheless, there is in general a distinction between the missing N$^{3}$LO uncertainty we explicitly include and the uncertainty from MHOs beyond this and hence we will take care throughout this paper to distinguish the two where appropriate, even if the separation is not always clear cut.

The outline of this paper is as follows. In Section~\ref{sec: theo_framework} we present the theoretical framework, describing the method and conventions used for the rest of the paper. Section~\ref{sec: structure} describes the structure functions and their role in QCD calculations. In Sections~\ref{sec: n3lo_split}, \ref{sec: n3lo_OME} and \ref{sec: n3lo_coeff} we present our approximations for the N$^{3}$LO DIS theory functions, while in Section~\ref{sec: n3lo_K} we present the $K$-factors at aN$^{3}$LO. In Section~\ref{sec: results} we present the MSHT aN$^{3}$LO PDFs with theoretical uncertainties and analyse the implications of the approximations in terms of a full MSHT global fit. Section~\ref{sec: predictions} contains examples of using these aN$^{3}$LO PDFs in predictions up to N$^{3}$LO. Finally in Sections~\ref{sec: availability} and \ref{sec: conclusion} we present recommendations for how to best utilise these PDFs and summarise our results.
\section{Theoretical Procedures}\label{sec: theo_framework}

In this section we describe the mathematical procedures used to implement N$^{3}$LO approximations into the MSHT PDF framework. These procedures are discussed in terms of the Hessian minimisation method employed by the MSHT fit and extended by theoretically grounded arguments to accommodate theoretical uncertainties.

The above will be achieved by adapting the underlying theory description of the data from NNLO to N$^{3}$LO (a formal description of how this will be done for the $F_{2}$ structure function is discussed in Section~\ref{sec: structure}). 
Not all the ingredients necessary for full N$^{3}$LO theory predictions are known, where there is missing information the N$^{3}$LO theory predictions will therefore include additional theoretical nuisance parameters, allowing their variation via an additional degree of freedom in specific theoretical pieces. These theoretical nuisance parameters will be constrained via an additional $\chi^{2}$ penalty in the global fit and will accommodate a level of uncertainty for each added approximate N$^{3}$LO ingredient (more information on how these prior variations are decided is included in Section's~\ref{subsec: genframe}, \ref{subsec: genframe_continuous} and \ref{subsec: kN3LO}). From this point, the fitting procedure remains similar to previous MSHT fits with a number of extra theory nuisance parameters which are treated in the same manner as experimental nuisance parameters inherent in PDF fits i.e. they can be fit to the data via an expanded Hessian matrix.

\subsection{Hessian Method with Nuisance Parameters}\label{subsec: hessian_method}

Following the notation and description from~\cite{Ball:corr2021}, in the Hessian prescription, the Bayesian probability can be written as
\begin{equation}\label{eq: bayes_PTD_orig}
    P(T|D) \propto \exp \left( -\frac{1}{2}(T - D)^{T} H_{0} (T - D)\right)
\end{equation}
where $H_{0}$ is the Hessian matrix and $T = \{T_{i}\}$ is the set of theoretical predictions fit to $N$ experimental data points $D = \{D_{i}\}$ with $i=1,\ldots, N$. In this section we explicitly show the adaptation of this equation to accommodate extra theoretical parameters (with penalties) into the total $\chi^{2}$ and Hessian matrices.

To adapt this equation to include a single extra theory parameter, we can make the transformation $T \rightarrow T + t u = T^{\prime}$, where $t$ is the chosen central value of the theory parameter considered and $u$ is some non-zero vector such that $u u^{T}$ is the theory covariance matrix for $t$. In defining this new theoretical prescription $T^{\prime}$, we are making the general assumption that the underlying theory is now not necessarily identical to our initial NNLO theory\footnote{For the aN$^{3}$LO prescription defined in this paper this is indeed the case, although for any extra theory parameters that do not inherently change the theory from $T$ (for example where there is no known N$^{3}$LO information to be included), this transformation still holds in the case that $t = 0$.} $T$.

We now seek to include a nuisance parameter $\theta$, centered around $t$, to allow the fit to control this extra theory addition. We demand that when $\theta = t$, $T^{\prime}$ remains unaffected with the theory addition unaltered from its central value $t$. This leads us to the expression,
\begin{equation}
T^{\prime} + (\theta - t)u = T + t u + (\theta - t)u. 
\end{equation}
Redefining the nuisance parameter as the shift from its central value $t$ ($\theta^{\prime} = \theta - t$) we define $\theta^{\prime}$ centered around 0. 
To constrain $\theta^{\prime}$ within the fitting procedure, we must also define a prior probability distribution $P(\theta^{\prime})$ centered around zero and characterised by some standard deviation $\sigma_{\theta^{\prime}}$,
\begin{equation}\label{eq: prior_theta_1}
    P(\theta^{\prime}) = \frac{1}{\sqrt{2\pi}\sigma_{\theta^{\prime}}}\exp(-\theta^{\prime\ 2} / 2\sigma_{\theta^{\prime}}^{2}).
\end{equation} 
Throughout this paper, we refer to the chosen variation of theory predictions in the language of the standard deviation $\sigma_{\theta^{\prime}}$ presented here. A caveat to this however is that technically speaking, this standard deviation is chosen with a level of arbitrariness based on general assumptions and known information about the theory (we will show how this is done in more detail in Section's~\ref{subsec: genframe}, \ref{subsec: genframe_continuous} and \ref{subsec: kN3LO}). Although this definition of $\sigma_{\theta^{\prime}}$ lacks the full extent of statistical meaning of a true standard deviation, the same is also true for scale variations as well as various experimental systematic uncertainties, which are often not strictly Gaussian. Furthermore, a more robust statistical meaning is recovered for the constraints on various theoretical parameters after a fit is performed, where we become less sensitive to a prior.
Using this information and making the redefinition $u \rightarrow u / \sigma_{\theta^{\prime}}$ (in order to normalise the covariance matrix), we can update Equation~\eqref{eq: bayes_PTD_orig} to be
\begin{align}
    P(T|D\theta) &\propto \exp \left( -\frac{1}{2}(T^{\prime} + \frac{(\theta - t)}{\sigma_{\theta^{\prime}}} u - D)^{T} H_{0} (T^{\prime} + \frac{(\theta - t)}{\sigma_{\theta^{\prime}}} u - D)\right) \\
    P(T^{\prime}|D\theta^{\prime}) &\propto \exp \left( -\frac{1}{2}(T^{\prime} + \frac{\theta^{\prime}}{\sigma_{\theta^{\prime}}} u - D)^{T} H_{0} (T^{\prime} + \frac{\theta^{\prime}}{\sigma_{\theta^{\prime}}} u - D)\right)\label{eq: bayes_PTDt}
\end{align}
From here, Bayes theorem tells us
\begin{equation}
    P(T^{\prime}|D\theta^{\prime})P(\theta^{\prime}|D) = P(\theta^{\prime}|T^{\prime}D)P(T^{\prime}|D)
\end{equation}
where our nuisance parameter $\theta^{\prime}$ is assumed to be independent of the data i.e. $P(\theta^{\prime}|D) = P(\theta^{\prime})$. Integrating over $\theta^{\prime}$ gives
\begin{equation}\label{eq: bayes_PTD}
    P(T^{\prime}|D) = \underbrace{\int d\theta^{\prime} P(\theta^{\prime}|T^{\prime}D)}_{=1} P(T^{\prime}|D) = \int d\theta^{\prime} P(T^{\prime}|D\theta^{\prime}) P(\theta^{\prime}).
\end{equation}
Combining Equations~\eqref{eq: prior_theta_1},~\eqref{eq: bayes_PTDt} and~\eqref{eq: bayes_PTD} it is possible to show that,
\begin{equation}\label{eq: bayes_PTD2}
    P(T^{\prime}|D) \propto \int d\theta \exp\left(-\frac{1}{2}\left[(T^{\prime} + \frac{\theta^{\prime}}{\sigma_{\theta^{\prime}}} u - D)^{T} H_{0} (T^{\prime} + \frac{\theta^{\prime}}{\sigma_{\theta^{\prime}}} u - D) + \theta^{\prime\ 2}/\sigma_{\theta^{\prime}}^{2}\right]\right).
\end{equation}
To make progress with this equation we consider the exponent and refactor terms in powers of $\theta^{\prime}$,
\begin{equation}
     \left(u^{T}H_{0} u + 1\right)\frac{\theta^{\prime\ 2}}{\sigma_{\theta^{\prime}}^{2}} + 2u^{T} H_{0} (T^{\prime}-D) \frac{\theta^{\prime}}{\sigma_{\theta^{\prime}}} + (T^{\prime}-D)^{T}H_{0}(T^{\prime}-D).
\end{equation}
Defining $M^{-1} = \frac{1}{\sigma_{\theta^{\prime}}^{2}}\left(u^{T}H_{0}u + 1\right) $ and completing the square gives,
\begin{multline}\label{eq: square_comp}
    M^{-1}\left[\theta^{\prime} + \frac{1}{\sigma_{\theta^{\prime}}}M u^{T} H_{0} (T^{\prime}-D) \right]^{2} - \frac{1}{\sigma_{\theta^{\prime}}^{2}}M\left(u^{T} H_{0} (T^{\prime}-D)\right)^{2} \\+ (T^{\prime}-D)^{T}H_{0}(T^{\prime}-D).
\end{multline}
In Equation~\eqref{eq: square_comp}, we are able to simplify the first term by defining,
\begin{equation}
\overline{\theta}^{\prime}(T,D) = \frac{1}{\sigma_{\theta^{\prime}}} M u^{T} H_{0} (D-T^{\prime}).
\end{equation}
Expanding the second term leaves us with,
\begin{equation}
  \left(u^{T} H_{0} (T^{\prime}-D)\right)^{2} = (T^{\prime}-D)^{T} H_{0} u u^{T} H_{0} (T^{\prime}-D)
\end{equation}
The second and third term in Equation~\eqref{eq: square_comp} can then be combined to give,
\begin{equation}
(T^{\prime}-D)^{T}\left(H_{0}-\frac{1}{\sigma_{\theta^{\prime}}^{2}}M H_{0} u u^{T} H_{0}\right)(T^{\prime}-D).
\end{equation}
Further to this we note that the following is true:
\begin{multline}\label{eq: relation_hessian}
    (H_{0}^{-1} + u u^{T})\left(H_{0}-\frac{1}{\sigma_{\theta^{\prime}}^{2}}M H_{0} u u^{T} H_{0}\right) = 1 + uu^{T}H_{0} - \frac{1}{\sigma_{\theta^{\prime}}^{2}}Muu^{T}H_{0} - \frac{1}{\sigma_{\theta^{\prime}}^{2}}Mu u^{T}H_{0}u u^{T} H_{0} \\= 1 + uu^{T}H_{0} - \frac{1}{\sigma_{\theta^{\prime}}^{2}}Muu^{T}H_{0} - \frac{1}{\sigma_{\theta^{\prime}}^{2}}Mu (\sigma_{\theta^{\prime}}^{2}M^{-1} - 1) u^{T} H_{0} = 1.
\end{multline}
Using Equation~\eqref{eq: relation_hessian} we are finally able to rewrite Equation~\eqref{eq: bayes_PTD2} as,
\begin{equation}\label{eq: single_nuisance_bayes}
    P(T^{\prime}|D) \propto \int d\theta^{\prime} \exp \left(-\frac{1}{2} M^{-1} (\theta^{\prime} - \overline{\theta}^{\prime})^{2} - \frac{1}{2} (T^{\prime}-D)^{T}(H_{0}^{-1} + u u^{T})^{-1} (T^{\prime}-D)\right).
\end{equation}
At this point we can make a choice whether to redefine our Hessian matrix as $H = (H_{0}^{-1} + u u^{T})^{-1}$, or keep the contributions completely separate. By redefining the Hessian we can include correlations between the standard set of MSHT parameters included in $H_{0}$ and the new theoretical parameter $\theta^{\prime}$ contained within $u u^{T}$. However, by doing so we  lose information about the specific contributions to the total uncertainty i.e. we cannot then decorrelate the theoretical and standard PDF uncertainties a posteriori. Whereas for the decorrelated choice, although we sacrifice knowledge related to the correlations between the separate sources of uncertainty, we are able to treat the sources completely separably. Interpreting Equation~\eqref{eq: single_nuisance_bayes} as in Equation~\eqref{eq: bayes_PTD_orig} we can write down the two $\chi^{2}$ contributions,
\begin{align}
\chi^{2}_{1} &= (T^{\prime} - D)^{T} (H_{0}^{-1} + u u^{T})^{-1} (T^{\prime}-D) = (T^{\prime} - D)^{T} H (T^{\prime}-D), \\
\chi^{2}_{2} &= M^{-1} (\theta^{\prime} - \overline{\theta}^{\prime})^{2}.
\label{eq: penalty_example}
\end{align}
Where $\chi^{2}_{1}$ is the contribution from the fitting procedure, $\chi^{2}_{2}$ is the posterior penalty contribution applied when the theory addition strays too far from its fitted central value and $M$ is the posterior error matrix for this contribution. This will be discussed further in following sections.

\subsection{Multiple Theory Parameters}\label{subsec: mult_params}

In the case of multiple $N_{\theta^{\prime}}$ theory parameters, Equation~\eqref{eq: bayes_PTDt} becomes
\begin{equation}
	P(T^{\prime}|D\theta^{\prime}) \propto \exp \left( -\frac{1}{2}\sum_{i,j}^{N_{\mathrm{pts}}}\bigg(T^{\prime}_{i} + \sum_{\alpha = 1}^{N_{\theta^{\prime}}}\frac{\theta^{\prime}_{\alpha}}{\sigma_{\theta^{\prime}_{\alpha}}} u_{\alpha, i} - D_{i}\bigg) H^{0}_{ij} \bigg(T^{\prime}_{j} + \sum_{\beta = 1}^{N_{\theta^{\prime}}}\frac{\theta^{\prime}_{\beta}}{\sigma_{\theta^{\prime}_{\beta}}} u_{\beta,j} - D_{j}\bigg)\right)
\end{equation}
where we have explicitly included the sum over the number of data points $N_{\mathrm{pts}}$ in the matrix calculation for completeness.

The prior probability for all N$^{3}$LO nuisance parameters also becomes
\begin{equation}\label{eq: prior_theta}
    P(\theta^{\prime}) =\prod_{\alpha = 1}^{N_{\theta^{\prime}}} \frac{1}{\sqrt{2\pi}\sigma_{\theta_{\alpha}^{\prime}}}\exp(-\theta_{\alpha}^{\prime\ 2} / 2\sigma_{\theta_{\alpha}^{\prime}}^{2}).
\end{equation}
Constructing $P(T^{\prime}|D)$ using Bayes theorem as before, results in the expression,
\begin{multline}
    P(T^{\prime}|D) \propto \int d^{N_{\theta^{\prime}}}\theta^{\prime} \exp \Bigg( -\frac{1}{2}\Bigg[\sum_{i,j}^{N_{\mathrm{pts}}}\bigg(T^{\prime}_{i} + \sum_{\alpha = 1}^{N_{\theta^{\prime}}}\frac{\theta^{\prime}_{\alpha}}{\sigma_{\theta^{\prime}_{\alpha}}} u_{\alpha, i} - D_{i}\bigg) H^{0}_{ij} \times \\ \times \bigg(T^{\prime}_{j} + \sum_{\beta = 1}^{N_{\theta^{\prime}}}\frac{\theta^{\prime}_{\beta}}{\sigma_{\theta^{\prime}_{\beta}}} u_{\beta,j} - D_{j}\bigg) + \sum_{\alpha, \beta}^{N_{\theta^{\prime}}}\frac{\theta^{\prime}_{\alpha}}{\sigma_{\theta^{\prime}_{\alpha}}}\frac{\theta^{\prime}_{\beta}}{\sigma_{\theta^{\prime}_{\beta}}}\delta_{\alpha \beta}\Bigg]\Bigg).
\end{multline}
Following the same procedure as laid out in the previous section, defining $M_{\alpha\beta}^{-1} = (\delta_{\alpha\beta} + u_{\alpha, i}H^{0}_{ij}u_{\beta, j}) / \sigma_{\theta_{\alpha}^{\prime}}\sigma_{\theta_{\beta}^{\prime}}$ and completing the square leaves us with,
\begin{multline}
     (T^{\prime}_{i} - D^{\prime}_{i}) H^{0}_{ij} (T^{\prime}_{j} - D^{\prime}_{j}) + \sum_{\alpha,\beta}^{N_{\theta^{\prime}}}M_{\alpha\beta}^{-1}\left[\left(\theta^{\prime}_{\alpha} + \sum_{i,j}^{N_{\mathrm{pts}}}\sum_{\delta=1}^{N_{\theta^{\prime}}}\frac{1}{\sigma_{\theta_{\alpha}^{\prime}}}M_{\alpha \delta}u_{\delta,i}H^{0}_{ij} (T^{\prime}_{j} - D_{j}) \right)^{2} \right.\\ \left.- \left(\sum_{i,j}^{N_{\mathrm{pts}}}\sum_{\delta=1}^{N_{\theta^{\prime}}}\frac{1}{\sigma_{\theta_{\alpha}^{\prime}}}M_{\alpha \delta}u_{\delta,i}H^{0}_{ij} (T^{\prime}_{j} - D_{j})\right)^{2}\right],
\end{multline}
where the summation over the $\beta$ index in $M^{-1}_{\alpha\beta}$ is implicit in the squared terms of the squared bracket expressions.

As in the previous section for a single parameter, we can define,
\begin{align}
    \overline{\theta}_{\alpha}^{\prime}(T^{\prime},D) &= \sum_{i,j}^{N_{\mathrm{pts}}}\sum_{\delta = 1}^{N_{\theta^{\prime}}}\frac{1}{\sigma_{\theta_{\alpha}^{\prime}}}M_{\alpha\delta}u_{\delta,i}H^{0}_{ij}(D_{j} - T^{\prime}_{j}) \\
    H_{ij} &= \left( \left(H^{0}_{ij}\right)^{-1} + \sum^{N_{\theta^{\prime}}}_{\alpha = 1} u_{\alpha,i}u_{\alpha, j}\right)^{-1}
    \label{eq: new_Hessian}
\end{align}
which leads to the final expression for $P(T|D)$,
\begin{multline}
    P(T^{\prime}|D) \propto \int d^{N_{\theta^{\prime}}}\theta^{\prime} \exp \left( -\frac{1}{2}\left[\sum_{\alpha, \beta}^{N_{\theta^{\prime}}} \left(\theta^{\prime}_{\alpha} - \overline{\theta}_{\alpha}^{\prime}\right)M_{\alpha\beta}^{-1}\big(\theta^{\prime}_{\beta} - \overline{\theta}_{\beta}^{\prime}\big) \right. \right. \\ \left. \left. + \sum_{i,j}^{N_{\mathrm{pts}}}\left(T^{\prime}_{i} - D_{i}\right) H_{ij} \big(T^{\prime}_{j} - D_{j}\big)\right]\right).
\end{multline}
which can be interpreted analogously to the single parameter case in \eqref{eq: single_nuisance_bayes}.

\subsection{Decorrelated parameters}\label{subsec: decorr_params}

In the treatment above we investigated the case of correlated parameters whereby the Hessian matrix was redefined in Equation~\eqref{eq: new_Hessian}. In performing this redefinition we sacrifice the information contained within $u_{\alpha,i}u_{\alpha, j}$ in order to gain information about the correlations between the original PDF parameters making up $H^{0}_{ij}$ and any new N$^{3}$LO nuisance parameters. As stated earlier, in this case, we can perform a fit to find $H_{ij}$ but one is unable to separate this Hessian matrix into individual contributions.

As will be discussed in later sections, the $K$-factors we include in the N$^{3}$LO additions are somewhat more separate from other N$^{3}$LO parameters considered. The reason for this is that not only are they concerned with the cross section data directly, they are also included for processes separate from inclusive DIS\footnote{It is true that we may still expect some indirect correlation with the parameters controlling the N$^{3}$LO splitting functions, which are universal across all processes. However, as we will show, these correlations are small and can be ignored.}.

Hence, we have some justification to include the aN$^{3}$LO $K$-factors' nuisance parameters as completely decorrelated from other PDF parameters (including other N$^{3}$LO theory parameters). 
To do this we rewrite Equation~\eqref{eq: new_Hessian} as,
\begin{equation}
    \left( \left(H^{0}_{ij}\right)^{-1} + \sum^{N_{\theta^{\prime}}}_{\alpha = 1} u_{\alpha,i}u_{\alpha, j} + \sum_{p = 1}^{N_{p}}\sum^{N_{\theta_{K}}}_{\delta = 1} u_{\delta,i}^{p}u_{\delta, j}^{p}\right)^{-1} = \left(H_{ij}^{-1} + \sum_{p = 1}^{N_{p}}K_{ij,p}^{-1} \right)^{-1} = H_{ij}^{\prime}
\end{equation}
where $N_{\theta^{\prime}} \rightarrow N_{\theta^{\prime}} + N_{\theta_{K}}$, $K_{ij,p}$ defines the extra decorrelated contributions from the N$^{3}$LO $K$-factor's parameters, stemming from $N_{p}$ processes; $H_{ij}$ is the Hessian matrix including correlations with parameters associated with N$^{3}$LO structure function theory; and $H_{ij}^{\prime}$ is the fully correlated Hessian matrix. It is therefore possible to construct these matrices separately and perform the normal Hessian eigenvector analysis (described in Section~\ref{subsec: eigenvector_results}) on each matrix in turn. In doing this, we maintain a high level of flexibility in our description by assuming the sets of parameters (contained in $H_{ij}^{-1}$ and $K_{ij,p}$) to be suitably orthogonal.

\section{Structure Functions at \texorpdfstring{N$^{3}$LO}{N3LO}}\label{sec: structure}

The general form of a structure function $F(x,Q^{2})$ is a convolution between the PDFs $f_{i}(x, Q^{2})$ and some defined process dependent coefficient function $C(x,\alpha_{s}(Q^{2}))$,
\begin{equation}\label{eq: structure_form}
    F(x,Q^{2}) = \sum_{i=q,\bar{q},g}\left[C_{i}(\alpha_{s}(Q^{2}))\otimes f_{i}(Q^{2})\right](x)
\end{equation}
where we have the sum over all partons $i$ and implicitly set the factorisation and renormalisation scales as $\mu_{f}^{2} = \mu_{r}^{2} = Q^{2}$, a choice that will be used throughout this paper for DIS scales. We also note that the relevant charge weightings are implicit in the definition of the coefficient function for each parton.

In Equation~\eqref{eq: structure_form}, the perturbative and non-perturbative regimes are separated out into coefficient functions $C_{i}$ and PDFs $f_{i}$ respectively.
Since these coefficient functions are perturbative quantities, they are an important aspect to consider when transitioning to N$^{3}$LO.

The PDFs $f_{i}(x, Q^{2})$ in Equation~\eqref{eq: structure_form} are non-perturbative quantities. However, their evolution in $Q^{2}$ is perturbatively calculable. In a PDF fit, the PDFs are parameterised at a chosen starting scale $Q^{2}_{0}$, which is in general different to the scale $Q^{2}$ at which an observable (such as $F(x,Q^{2})$) is calculated. It is therefore important that we are able to accurately evolve the PDFs from $Q_{0}^{2}$ to the required $Q^{2}$ to ensure a fully consistent and physical calculation. To permit this evolution, we introduce the standard factorisation scale $\mu_{f}$.

The flavour singlet distribution is defined as,
\begin{equation}\label{eq:sing}
\Sigma(x,\mu_{f}^{2})=\sum^{n_{f}}_{i=1}\left[q_{i}(x,\mu_{f}^{2})+\overline{q}_{i}(x,\mu_{f}^{2})\right],
\end{equation}
where $q_{i}(x,\mu_{f}^{2})$ and $\overline{q}_{i}(x,\mu_{f}^{2})$ are the quark and anti-quark distributions respectively, as a function of Bjorken $x$ and the factorisation scale $\mu_{f}^{2}$. The summation in Equation~\eqref{eq:sing} runs over all flavours of (anti-)quarks $i$ up to the number of available flavours $n_{f}$.

This singlet distribution is inherently coupled to the gluon density. 
Because of this, we must consider the gluon carefully when describing the evolution of the flavour singlet distribution with the energy scale $\mu_{f}$. The Dokshitzer-Gribov-Lipatov-Altarelli-Parisi (DGLAP)~\cite{DGLAP} equations that govern this evolution are:
\begin{equation}
\frac{d \boldsymbol{f}}{d \ln \mu_{f}^{2}} \equiv \frac{d}{d \ln \mu_{f}^{2}}\left(\begin{array}{c}{\Sigma} \\ {g}\end{array}\right)=\left(\begin{array}{cc}{P_{q q}} & {\ n_{f}P_{q g}} \\ {P_{g q}} & {P_{g g}}\end{array}\right) \otimes\left(\begin{array}{c}{\Sigma} \\ {g}\end{array}\right) \equiv \boldsymbol{P} \otimes \boldsymbol{f}
\label{eq: DGLAP}	
\end{equation}
where $P_{ij}: i,j \in q,g$ are the splitting functions and the factorisation scale $\mu_{f}$ is allowing the required evolution up to the physical scale $Q^{2}$. The matrix of splitting functions $\boldsymbol{P}$ appropriately couples the singlet and gluon distribution by means of a convolution in the momentum fraction $x$. We note here that $P_{qq}\equiv P_{q \rightarrow gq}$ is decomposed into non-singlet (NS) and a pure-singlet (PS) parts defined by,
\begin{equation}
P_{qq}(x)=P^{+}_{\mathrm{NS}}(x)+P_{\mathrm{PS}}(x),
\label{eq: NStoPS}
\end{equation}
where the $P^{+}_{NS}$ is a non-singlet distribution splitting function which has been calculated approximately to four loops in~\cite{4loopNS}\footnote{In this discussion, we only consider the $P^{+}_{NS}$ non-singlet distribution as this is the distribution which contributes to the singlet evolution. Other non-singlet distributions are briefly discussed in Section~\ref{sec: n3lo_split}.}. The non-singlet part of $P_{qq}$ dominates at large-$x$ but as $x\rightarrow 0$, this contribution is highly suppressed due to the relevant QCD sum rules. On the other hand, due to the involvement of the gluon in the pure-singlet splitting function (as described above), this contribution grows towards small-$x$ and therefore begins to dominate.

Turning to the splitting function matrix, each element can be expanded perturbatively as a function of $\alpha_s$ up to N$^{3}$LO as,
\begin{equation}
	\boldsymbol{P}(x,\alpha_{s})=\alpha_{s}\boldsymbol{P}^{(0)}(x)+\alpha_{s}^{2}\boldsymbol{P}^{(1)}(x)+\alpha_{s}^{3}\boldsymbol{P}^{(2)}(x)+\alpha_{s}^{4}\boldsymbol{P}^{(3)}(x)+\dots\ ,
	\label{eq: splittingExpand}
\end{equation}
where we have omitted the scale argument of $\alpha_{s}(\mu_{r}^{2}=\mu_{f}^{2})\equiv\alpha_{s}$ for brevity and $\boldsymbol{P}^{(0)}$, $\boldsymbol{P}^{(1)}$, $\boldsymbol{P}^{(2)}$ are known~\cite{DGLAP, NLO1, NLO2, NLO3, NLO4, Moch:2004pa, Vogt:2004mw}. $\boldsymbol{P}^{(3)}$ are the four-loop quantities which we approximate in Section~\ref{sec: n3lo_split} using information from~\cite{Catani:1994sq,Lipatov:1976zz,Kuraev:1977fs,Balitsky:1978ic,Fadin:1998py,Ciafaloni:1998gs,Marzani:smallx, S4loopMoments, S4loopMomentsNew, 4loopNS}. 

Considering Equation~\eqref{eq: structure_form}, $\Sigma(Q^{2})$ and $g(Q^{2})$ are the singlet and gluon PDFs respectively, evolved to the required $Q^{2}$ energy of the process via Equation~\eqref{eq: DGLAP}. For more information on the relevant formulae used in this convolution, the reader is referred to~\cite{PinkBook}.

Thus far, we have limited our discussion to only light quark flavours. However, as we move through the full range of $Q^{2}$ values, the number of partons which are kinematically accessible increases. More specifically, as we pass over the charm and bottom mass thresholds (where $Q^{2} = m_{c, b}^{2}$) we must account for the heavy quark PDFs and their corresponding contributions.

To deal with the heavy quark contributions to the total structure function, whilst remaining consistent with the light quark picture described above, we consider
\begin{equation}
    f_{\alpha}^{n_{f} + 1}(x, Q^{2}) = \left[A_{\alpha i}(Q^{2}/m_{h}^{2}) \otimes f_{i}^{n_{f}}(Q^{2})\right](x),
\end{equation}
where we have an implied summation over partons $i$ and $A_{\alpha i}$ are the heavy flavour transition matrix elements~\cite{Buza:OMENLO, Buza:OMENNLO} which explicitly depend on the heavy flavour mass threshold $m_{h}$, where these contributions are activated\footnote{The indices here run as $\alpha \in \{H,q,g\}$ and $i \in \{q,g\}$, since $n_{f}$ is the number of light flavours.}. We also denote the PDFs as $f_{i}^{n_{f}}$ and $f_{i}^{n_{f} + 1}$ to indicate whether the PDF has been evolved with only light flavours ($n_{f}$) or also with heavy flavours ($n_{f} + 1$). In this work we only consider contributions at heavy flavour threshold i.e. where $Q^{2} = m_{h}^{2}$. We then define the PDFs:
\begin{subequations}
\begin{equation}\label{eq: OME_fq}
    f_{q}^{n_{f} + 1}(x, Q^{2}) = \left[A_{qq,H}(Q^{2}/m_{h}^{2}) \otimes f_{q}^{n_{f}}(Q^{2}) + A_{qg,H}(Q^{2}/m_{h}^{2}) \otimes f_{g}^{n_{f}}(Q^{2})\right](x)
\end{equation}
\begin{equation}\label{eq: OME_fg}
    f_{g}^{n_{f} + 1}(x, Q^{2}) = \left[A_{gq,H}(Q^{2}/m_{h}^{2}) \otimes f_{q}^{n_{f}}(Q^{2}) + A_{gg,H}(Q^{2}/m_{h}^{2}) \otimes f_{g}^{n_{f}}(Q^{2})\right](x)
\end{equation}
\begin{equation}\label{eq: OME_fh}
    f_{H}^{n_{f} + 1}(x, Q^{2}) = \left[A_{Hq}(Q^{2}/m_{h}^{2}) \otimes f_{q}^{n_{f}}(Q^{2}) + A_{Hg}(Q^{2}/m_{h}^{2}) \otimes f_{g}^{n_{f}}(Q^{2})\right](x)
\end{equation}
\end{subequations}
where we have an implicit summation over light flavours of $q$ and a generalised theoretical description to involve heavy flavour contributions\footnote{Note that the notation $A_{\alpha i, H}$ is exactly equivalent to $A_{\alpha i}$. When $H$ is not present in the final state of matrix element interactions, we opt for the $A_{\alpha i,H}$ notation. This is to remind the reader that these elements are considering only those interactions involving a heavy quark.}. Equation~\eqref{eq: OME_fq} and Equation~\eqref{eq: OME_fg} are the light flavour quark and gluon PDFs defined earlier, modified to include contributions mediated by heavy flavour loops. Whereas in Equation~\eqref{eq: OME_fh} we describe the heavy flavour PDF, perturbatively calculated from the light quark and gluon PDFs.

By considering the number of vertices (and hence orders of $\alpha_{s}$) required for each of these transition matrix elements to contribute to their relevant `output' partons, we are immediately able to show:

\begin{minipage}{0.47\linewidth}
\begin{align}
A_{qq,H} &= \delta(1-x)\ +\ \mathcal{O}(\alpha_{s}^{2}) \nonumber \\
A_{qg,H} &= \mathcal{O}(\alpha_{s}^{2}) \nonumber \\
A_{gq,H} &= \mathcal{O}(\alpha_{s}^{2}) \nonumber
\end{align}
\end{minipage}
\begin{minipage}{0.47\linewidth}
\begin{align}
A_{gg,H} &= \delta(1-x)\ +\ \mathcal{O}(\alpha_{s}) \nonumber \\
A_{Hq} &= \mathcal{O}(\alpha_{s}^{2}) \nonumber \\
A_{Hg} &= \mathcal{O}(\alpha_{s}) \label{eq: OME_orders}
\end{align}
\end{minipage}
\vspace{0.3cm}

\noindent
where $A_{qq,H}$ and $A_{gg,H}$ include LO $\delta$-functions to ensure this description is consistent with the light quark picture discussed earlier. It is therefore the $A_{Hg}$ transition matrix element which provides our lowest order contribution to the heavy flavour sector (i.e. $g \rightarrow H \overline{H}$). 

The insertion of scale independent contributions to $A_{\alpha i}$ introduce unwanted discontinuities at NNLO into the PDF evolution.
In order to ensure the required smoothness and validity of the structure functions across $(x, Q^{2})$, these discontinuities must be accounted for elsewhere in the structure function picture.
Equating the coefficient functions above the mass threshold $m_{h}^{2}$ (describing the total number of flavours including heavy flavour quarks) and those below this threshold, discontinuities are able to be absorbed by a suitable redefinition of the coefficient functions. This procedure provides the foundation for the description of different flavour number schemes.

There are two number schemes which are preferred at different points in the $Q^2$ range. Towards $Q^2 \leq m_{h}^2$ we adopt the Fixed Flavour Number Scheme (FFNS). Towards $\frac{Q^2}{m_{h}^2} \rightarrow \infty$, the heavy contributions can be considered massless and therefore the Zero Mass Variable Flavour Number Scheme (ZM-VFNS) is assumed.
In order to join the FFNS and ZM-VFNS schemes seamlessly together, we ultimately wish to describe the General Mass Variable Number Scheme (GM-VFNS)~\cite{Aivazis:gmvf} (which is valid across all $Q^2$). This scheme can then account for discontinuities from transition matrix elements and re-establish a smooth description of the structure functions.

In~\cite{Thorne:GMVF} an ambiguity in the definition of the GM-VFNS scheme was pointed out (namely the freedom to swap $\mathcal{O}(m_{h}^{2}/Q^{2})$ terms without violating the definition of the GM-VFNS). We note here that since~\cite{Thorne:MSTW09}, MSHT PDFs have employed the TR scheme to define the distribution of $\mathcal{O}(m_{h}^{2}/Q^{2})$ terms, the specific details of which are found in~\cite{Thorne:GMVF,Thorne:GMVFNNLO,thorne2012}. The general method to relate the FFNS and GM-VFNS number schemes is to compare the prediction for a result e.g. the $F_2$ structure function in the FFNS scheme:
\begin{align}\label{eq: FFexpanse}
F_{2}(x,Q^2) &= F_{2,q}(x, Q^{2})\ +\ F_{2, H}(x, Q^{2}) \nonumber\\
&= C_{q,i}^{\mathrm{FF},\ n_{f}}\otimes f_{i}^{n_{f}}(Q^2) + C_{H,i}^{\mathrm{FF},\ n_{f}}\otimes f_{k}^{n_{f}}(Q^2) \nonumber\\
&= C_{q,q}^{\mathrm{FF},\ n_{f}}\otimes f_{q}^{n_{f}}(Q^2) + C_{q,g}^{\mathrm{FF},\ n_{f}}\otimes f_{g}^{n_{f}}(Q^2)\nonumber \\
&+ C_{H,q}^{\mathrm{FF},\ n_{f}}\otimes f_{q}^{n_{f}}(Q^2) + C_{H,g}^{\mathrm{FF},\ n_{f}}\otimes f_{g}^{n_{f}}(Q^2)
\end{align}
and the GM-VFNS scheme,
\begin{multline}\label{eq: GM-VFNS_gen}
     F_{2}(x,Q^2) = \mathlarger{\mathlarger{\sum}}_{\alpha \in \{H,q,g\}}\left(C_{q,\alpha}^{\mathrm{VF},\ n_{f}+1}\otimes A_{\alpha i}(Q^2/m_{h}^2)\otimes f_{i}^{n_{f}}(Q^2) \right.\\\left.+ C_{H,\alpha}^{\mathrm{VF},\ n_{f}+1}\otimes A_{\alpha i}(Q^2/m_{h}^2)\otimes f_{i}^{n_{f}}(Q^2)\right),
\end{multline}
where $F_{2,q}$ and $F_{2,H}$ are the light and heavy flavour structure functions respectively\footnote{The extra contribution from $F_{2,H}$ allows for the possibility of final state heavy flavours.}. $C^{\mathrm{FF}, n_{f}}$ and $C^{\mathrm{VF}, n_{f}+1}$ are the FFNS (known up to NLO~\cite{FFNLO1,FFNLO2} with some information at NNLO~\cite{Catani:FFN3LO1,Laenen:FFN3LO2,Vogt:FFN3LO3} including high-$Q^{2}$ transition matrix elements at $\mathcal{O}(\alpha_{s}^{3})$~\cite{bierenbaum:OMEmellin,ablinger:3loopNS,Blumlein:2021enk,ablinger:3loopPS,ablinger:agq,Blumlein:AQg,Vogt:FFN3LO3}) and GM-VFNS coefficient functions respectively, and $A_{\alpha i}(Q^2/m_{h}^2)$ are the transition matrix elements. We note that the above also applies to other structure functions and for clarity, in the following we consider the light and heavy structure functions separately.

\subsection*{$F_{2,q}$}

Expanding the first term in Equation~\eqref{eq: GM-VFNS_gen} in terms of the transition matrix elements results in,
\begin{align}\label{eq: expanse_q}
F_{2,q}(x,Q^2) &= C_{q, H}^{\mathrm{VF},\  n_{f}+1}\otimes\bigg[A_{Hq}(Q^2/m_{h}^2)\otimes f_{q}^{n_{f}}(Q^2) + A_{Hg}(Q^2/m_{h}^2)\otimes f_{g}^{n_{f}}(Q^2)\bigg] \nonumber \\ & + C_{q, q}^{\mathrm{VF},\ n_{f}+1}\otimes\bigg[A_{qq, H}(Q^2/m_{h}^2)\otimes f_{q}^{n_{f}}(Q^2) + A_{qg, H}(Q^2/m_{h}^2)\otimes f_{g}^{n_{f}}(Q^2)\bigg] \nonumber \\ & + C_{q, g}^{\mathrm{VF},\ n_{f}+1}\otimes\bigg[A_{gq, H}(Q^2/m_{h}^2)\otimes f_{q}^{n_{f}}(Q^2) + A_{gg, H}(Q^2/m_{h}^2)\otimes f_{g}^{n_{f}}(Q^2)\bigg],
\end{align}
which is valid at all orders. The first term in Equation~\eqref{eq: expanse_q} is the contribution to the light quark structure function from heavy quark PDFs (since the term contained within square brackets is exactly our definition in Equation~\eqref{eq: OME_fh}). Due to this, the coefficient function $C_{q,H}$ describes the transition of a heavy quark to a light quark via a gluon and is therefore forbidden to exist below NNLO. The second and third terms here are the purely light quark and gluon contributions, with extra corrections from heavy quark at higher orders.

Using the definitions in Equation~\eqref{eq: OME_orders} we can obtain an equation for $F_{2,q}(x, Q^{2})$ up to $\mathcal{O}(\alpha_{s}^{3})$ as,
\begin{multline}\label{eq: fullN3LO_q}
	F_{2,q}(x, Q^{2}) = C_{q,q}^{\mathrm{VF},\ (0)} \otimes f_{q}(Q^{2}) + \frac{\alpha_s}{4\pi}\ \bigg\{C_{q,q,\ n_{f}+1}^{\mathrm{VF},\ (1)}\otimes f_{q}(Q^{2}) + C_{q,g,\ n_{f}+1}^{\mathrm{VF},\ (1)}\otimes f_{g}(Q^{2})\bigg\} \\+ \left(\frac{\alpha_s}{4\pi}\right)^{2}\ \bigg\{\bigg[C_{q,q,\ n_{f}+1}^{\mathrm{VF},\ (2)} + C_{q,q}^{\mathrm{VF},\ (0)} \otimes A_{qq,H}^{(2)}\bigg]\otimes f_{q}(Q^{2}) + \bigg[C_{q,g,\ n_{f}+1}^{\mathrm{VF},\ (2)} \\+ C_{q,g,\ n_{f}+1}^{\mathrm{VF},\ (1)}\otimes A_{gg, H}^{(1)} + C_{q,q}^{\mathrm{VF},\ (0)} \otimes A_{qg,H}^{(2)}\bigg]\otimes f_{g}(Q^{2})\bigg\} \\+ \left(\frac{\alpha_s}{4\pi}\right)^{3}\ \bigg\{\bigg[C_{q,q,\ n_{f}+1}^{\mathrm{VF},\ (3)} + C_{q,q,\ n_{f}+1}^{\mathrm{VF},\ (1)}\otimes A_{qq, H}^{(2)} + C_{q,g,\ n_{f}+1}^{\mathrm{VF},\ (1)}\otimes A_{gq, H}^{(2)} \\+ C_{q,q}^{\mathrm{VF},\ (0)} \otimes A_{qq,H}^{(3)}\bigg]\otimes f_{q}(Q^{2}) \\+ \bigg[C_{q,g,\ n_{f}+1}^{\mathrm{VF},\ (3)} + C_{q,g,\ n_{f}+1}^{\mathrm{VF},\ (1)}\otimes A_{gg, H}^{(2)} + C_{q,q,\ n_{f}+1}^{\mathrm{VF},\ (1)}\otimes A_{qg, H}^{(2)} \\+ C_{q,g,\ n_{f}+1}^{\mathrm{VF},\ (2)}\otimes A_{gg, H}^{(1)}+ C_{q,q}^{\mathrm{VF},\ (0)} \otimes A_{qg,H}^{(3)}\bigg]\otimes f_{g}(Q^{2}) \\+ C_{q,H}^{\mathrm{VF},\ (2)}\otimes A_{Hg}^{(1)}\otimes f_{g}(Q^{2})\bigg\} + \mathcal{O}(\alpha_{s}^{4})
\end{multline}
where $C_{q,q}^{\mathrm{VF},\ (0)} = \delta(1-x)$ up to charge weighting. Equation~\eqref{eq: fullN3LO_q} defines the light quark structure function to N$^{3}$LO including heavy flavour corrections\footnote{We also note that $\alpha_{s}^{n_{f}+1} \ne \alpha_{s}^{n_{f}}$ and account for this, but omit in expressions such as Equation~\eqref{eq: fullN3LO_q} for simplicity.}.

\subsection*{$F_{2,H}$}

Moving to the heavy quark structure function in Equation~\eqref{eq: FFexpanse}, as above the second term in Equation~\eqref{eq: GM-VFNS_gen} can be expanded in terms of the transition matrix elements to obtain,
\begin{align}\label{eq: expanse_H}
F_{2,H}(x,Q^2) &= C_{H, H}^{\mathrm{VF},\ n_{f}+1}\otimes\bigg[A_{Hq}(Q^2/m_{h}^2)\otimes f_{q}^{n_{f}}(Q^2) + A_{Hg}(Q^2/m_{h}^2)\otimes f_{g}^{n_{f}}(Q^2)\bigg] \nonumber \\ & + C_{H, q}^{\mathrm{VF},\ n_{f}+1}\otimes\bigg[A_{qq, H}(Q^2/m_{h}^2)\otimes f_{q}^{n_{f}}(Q^2) + A_{qg, H}(Q^2/m_{h}^2)\otimes f_{g}^{n_{f}}(Q^2)\bigg] \nonumber \\ & + C_{H, g}^{\mathrm{VF},\ n_{f}+1}\otimes\bigg[A_{gq, H}(Q^2/m_{h}^2)\otimes f_{q}^{n_{f}}(Q^2) + A_{gg, H}(Q^2/m_{h}^2)\otimes f_{g}^{n_{f}}(Q^2)\bigg],
\end{align}
which is valid at all orders. Similar to Equation~\eqref{eq: expanse_q}, we have a contribution from the heavy flavour quarks, the light quarks and the gluon respectively. However in this case, due to the required gluon intermediary, the coefficient functions associated with the light quark flavours and gluon are forbidden to exist below NNLO. Considering the $C_{H,H}$ function, we are able to choose this to be identically the ZM-VFNS light quark coefficient function $C_{q,q}$ up to kinematical suppression factors, since at $Q^{2} \rightarrow \infty$ these functions must be equivalent~\cite{Aivazis:gmvf,thorne2012,Aivazis:1993kh}.

The full heavy flavour structure function then reads as,
\begin{multline}
F_{2,H}(x, Q^{2}) = \frac{\alpha_s}{4\pi}\bigg[C_{H,g}^{\mathrm{VF},\ (1)} + C_{H,H}^{\mathrm{VF},\ (0)}\otimes A_{Hg}^{(1)}\bigg]\otimes f_{g}(Q^{2}) \\
+ \bigg(\frac{\alpha_s}{4\pi}\bigg)^{2} \bigg\{\bigg[C_{H,q}^{\mathrm{VF},\ (2)} + C_{H,H}^{\mathrm{VF},\ (0)}\otimes A_{Hq}^{(2)}\bigg]\otimes f_{q}(Q^{2}) \\
+ \bigg[C_{H,g}^{\mathrm{VF},\ (2)} + 
C_{H,g}^{\mathrm{VF},\ (1)}\otimes A_{gg,H}^{(1)} + C_{H,H}^{\mathrm{VF},\ (1)}\otimes A_{Hg}^{(1)} + C_{H,H}^{\mathrm{VF},\ (0)}\otimes A_{Hg}^{(2)}\bigg]\otimes f_{g}(Q^{2})\bigg\} \\
+ \bigg(\frac{\alpha_s}{4\pi}\bigg)^{3} \bigg\{\bigg[C_{H,q}^{\mathrm{VF},\ (3)} + C_{H,g}^{\mathrm{VF},\ (1)}\otimes A_{gq,H}^{(2)} \\
+ C_{H,H}^{\mathrm{VF},\ (1)}\otimes A_{Hq}^{(2)} + C_{H,H}^{\mathrm{VF},\ (0)}\otimes A_{Hq}^{(3)}\bigg]\otimes f_{q}(Q^{2}) \\
+ \bigg[C_{H,g}^{\mathrm{VF},\ (3)} + 
C_{H,g}^{\mathrm{VF},\ (2)}\otimes A_{gg,H}^{(1)} + C_{H,g}^{\mathrm{VF},\ (1)}\otimes A_{gg,H}^{(2)} + C_{H,H}^{\mathrm{VF},\ (2)}\otimes A_{Hg}^{(1)} \\ + C_{H,H}^{\mathrm{VF},\ (1)}\otimes A_{Hg}^{(2)} + C_{H,H}^{\mathrm{VF},\ (0)}\otimes A_{Hg}^{(3)}\bigg]\otimes f_{g}(Q^{2})\bigg\} \\
\label{eq: fullN3LO_H}
\end{multline}
where combining Equation~\eqref{eq: fullN3LO_q} and Equation~\eqref{eq: fullN3LO_H}, one can obtain the full structure function $F_{2}(x, Q^{2})$.
Equating the FFNS expansion from Equation~\eqref{eq: FFexpanse} to the above expressions in the GM-VFNS setting, one can find relationships between the two pictures. In Section~\ref{sec: n3lo_coeff} we use this equivalence to enable the derivation of the GM-VFNS functions at N$^{3}$LO.

To summarise, we have identified the leading theoretical ingredients entering the structure functions and detailed how these affect the PDFs. As we will discuss further, when pushing these equations to N$^{3}$LO, there is already some knowledge available. For example, the N$^{3}$LO ZM-VFNS coefficient functions are known precisely for $n_{f}=3$ from~\cite{Vermaseren:2005qc}, as are a handful of Mellin moments~\cite{S4loopMoments,S4loopMomentsNew,4loopNS,bierenbaum:OMEmellin} and leading small and large-$x$ terms~\cite{Catani:1994sq,Lipatov:1976zz,Kuraev:1977fs,Balitsky:1978ic,Fadin:1998py,Ciafaloni:1998gs,Marzani:smallx,ablinger:3loopNS,Blumlein:2021enk,ablinger:3loopPS,ablinger:agq,Vogt:FFN3LO3} associated with the splitting functions and transition matrix elements at N$^{3}$LO. Using this information, we approximate these functions to N$^{3}$LO and incorporate the results into the first approximate N$^{3}$LO global PDF fit.
\section{\texorpdfstring{N$^{3}$LO}{N3LO} Splitting Functions}\label{sec: n3lo_split}

Splitting functions at N$^{3}$LO allow us to more accurately describe the evolution of the PDFs. These functions are estimated here and the resulting approximations are included within the framework described in Section~\ref{sec: theo_framework} and below in Section~\ref{subsec: genframe}. In all singlet cases we set $n_{f} = 4$ before constructing our approximations and ignore any corrections to this from any further change in the number of flavours\footnote{An exception to this are the cases of $P_{qg}$ and $P_{qq}^{\mathrm{PS}}$ where we have already defined $P_{qg} \equiv n_{f}P_{qg}$ and $P_{qq}^{\mathrm{PS}} \equiv n_{f}P_{qq}^{\mathrm{PS}}$. Therefore the leading $n_{f}$ dependence is already taken into account.}. In the non-singlet case, we calculate the approximate parts of $P_{qq}^{NS\ (3)}$ with $n_{f}=4$ however, there is a relatively large amount of information about the $n_{f}$-dependence included from~\cite{4loopNS}. Therefore in the final result we choose to allow the full $n_{f}$-dependence to remain for the non-singlet splitting function.

\subsection{Approximation Framework: Discrete Moments}\label{subsec: genframe}

In order to estimate the missing N$^3$LO uncertainty in the splitting functions (also transition matrix elements considered in the following Section~\ref{sec: n3lo_OME}), and ultimately include these into the framework described in Section~\ref{subsec: mult_params}, one must acquire some approximation at N$^{3}$LO. Here we discuss using available sets of discrete Mellin moments for each function, along with any exact leading terms already calculated, to obtain N$^{3}$LO estimations. To perform the parameterisation of the unknown N$^{3}$LO quantities, we follow a similar estimation procedure as in~\cite{vogt:NNLOns,vogt:NNLOs} following the form,
\begin{equation}
	F(x)=\sum_{i=1}^{N_{m}}A_{i}f_{i}(x) + f_{e}(x).
	\label{eq: splitAnsatz}
\end{equation}
In Equation~\eqref{eq: splitAnsatz}, $N_{m}$ is the number of available moments, $A_{i}$ are calculable coefficients, $f_i(x)$ are functions chosen based on our intuition and theoretical understanding of the full function, and $f_{e}(x)$ encapsulates all the currently known leading exact contributions at either large or small-$x$. To describe this, consider a toy situation where we are given four data points described by some unknown degree 9 polynomial. Along with this information, we are told the dominant term at small-$x$ is described by $3x$. In this case, one may wish to attempt to approximate this function by means of a set of 4 simultaneous equations formed from Equation~\eqref{eq: splitAnsatz} equated to each of the four data points (or constraints). The result of this is then a unique solution for each chosen set of functions $\{f_{i}(x)\}$. However, a byproduct of this is that for each $\{f_{i}(x)\}$, one lacks any means to control the uncertainty in these approximate solutions. In order to allow a controllable level of uncertainty into this approximation, one must introduce an extra degree of freedom. This degree of freedom will be introduced through an unknown coefficient $a \equiv A_{N_{m} + 1}$, which for convenience, will be absorbed into the definition of $f_{e}(x) \rightarrow f_{e}(x, a)$. In this toy example one is then able to choose to define the functions $f_{i}(x)$ as,
\begin{alignat}{5}\label{eq: example_split_ansatz}
    f_{1}(x) \quad &= \quad x^{3} \quad \text{or} \quad x^{4}, \nonumber \\
    f_{2}(x) \quad &= \quad x^{5}, \quad \text{or} \quad x^{6} \nonumber \\
    f_{3}(x) \quad &= \quad x^{7} \quad \text{or} \quad x^{8}, \nonumber \\
    f_{4}(x) \quad &= \quad x^{9}, \nonumber \\
    f_{e}(x, a) \quad &= \quad 3x + ax^{2},
\end{alignat}
where we have prioritised approximating the small-$x$ behaviour more precisely than the large-$x$ behaviour. This could easily be adapted and even reversed depending on which region of $x$ we are most sensitive to, however in this paper we will be more focused on small-$x$. There is also an inherent functional uncertainty from the ambiguity in the choice of functions for $f_{1,2,3}(x)$ in this toy example, in principle the number of functions in the functional variation can be larger than demonstrated here and indeed a larger choice of functions will be used for all $f_i(x)$ when we apply this in practice in subsequent sections. Using these functions, one is then able to assemble a set of potential approximations to the overall polynomial, each uniquely defined by a set of functions and corresponding coefficients $\{A_{i}, f_{i}\}$  for each value of $a$.

As mentioned, for the N$^{3}$LO additions considered in this framework we use the available calculated moments as constraints for the corresponding simultaneous equations. 
A summary of all the known and used ingredients for all N$^{3}$LO approximations is provided in Appendix~\ref{app: n3lo_known}. The details of these known quantities will be discussed in detail in Section~\ref{subsec: 4loop_split} and Section~\ref{subsec: 3loop_OME}. We also mention here that towards the small-$x$ regime, the leading terms present in the splitting functions and transition matrix elements exhibit the relations,
\begin{subequations}
\begin{align}
F_{gg}(x\rightarrow 0) \simeq & \frac{C_{A}}{C_{F}}F_{gq}(x\rightarrow 0), \label{eq: PggPgqRelation}\\
F_{qq}(x\rightarrow 0) \simeq & \frac{C_{F}}{C_{A}}F_{qg}(x\rightarrow 0),
\end{align}
\label{eq: Relations}
\end{subequations}
where $F_{ij} \in \{P_{ij}, A_{ij,H}\}$ and $C_{A}, C_{F}$ are the usual QCD constants. Although Equation~\eqref{eq: Relations} are exact at leading order, it is known that as we expand to higher orders, these will break down due to the effect of large sub-leading logarithms. Due to this, we do not demand this relation as a constraint in our approximations. Instead we discuss the validity of Equation~\eqref{eq: Relations} in comparison with the aN$^{3}$LO functions.

Following from~\cite{vogt:NNLOns,vogt:NNLOs}, we must choose a set of candidate functions for each $f_{i}(x)$. Our convention is to assign these functions such that at small-$x$, $f_{1}(x)$ is dominant, while at large-$x$, $f_{N_{m}}(x)$ is dominant. With $f_{i}(x)\ \forall i \in \{2,\dots, N_{m} - 1\}$, dominating in the region between. 
The sets of functions assigned to each $f_{i}(x)$ are determined for each N$^{3}$LO function based on knowledge from lower orders and our intuition about what to expect at N$^{3}$LO.

Analogous to our toy polynomial example, we allow the inclusion of an unknown next-to-leading small-$x$ logarithm (NLL) term (NNLL in the $P_{gg}$ case) into the $f_{e}$ function of our parameterisation. The coefficient of this NLL (NNLL) term is then controlled by a variational parameter $a$. This parameter uniquely defines the solution to the sets of simultaneous equations considered i.e. for each set of functions $f_{i}(x)$ there exists a unique solution for every possible choice of $a$. The final step to consider in this approximation is how to choose the prior allowed variation of $a$ in a sensible way for each N$^{3}$LO approximation. To do this, we consider the criteria outlined below:

\vspace{5mm}

\begin{tabular}{ p{3cm} p{11.5cm} }
\textbf{Criterion 1}: & At sufficiently small-$x$ ($x < 10^{-5}$), for a fixed value of $a$, we require $f_{e}(x, a)$ to be contained within the range of variation for $F(x)$ predicted from the combinations of functions in \eqref{eq: example_split_ansatz}. For example, after fixing $a$, $f_{e}(x, a)$ it should lie within the variation predicted for $F(x)$ from the entire set of potential approximations defined in \eqref{eq: example_split_ansatz}. In practice this means that we require the small-$x$ behaviour to not be in large tension with the large-$x$ description.\\
[0.5cm]
\textbf{Criterion 2}: & At large-$x$ ($x > 10^{-2}$) the N$^3$LO contribution should have relatively little effect. More specifically, we do not expect as large of a divergence as we do at small-$x$. Due to this, we require that the trend of the N$^{3}$LO approximation follow the general trend of the NNLO function at large-$x$.\\
\end{tabular}

\vspace{5mm}

The allowed variation in $a$ gives us an uncertainty which, at its foundations, is chosen via a conservative estimate based on all the available prior knowledge about the function and lower orders being considered. We note that given we are including known information about the higher order, it is not guaranteed that a value of $a=0$ will satisfy either criterion 1 or 2. Indeed, typically the NLL coefficient in the splitting functions is the opposite sign to and larger than the LL contribution, for example in the NNLO splitting functions and the known NLL term in the N$^3$LO
splitting function $P_{gg}$. To determine a full predicted uncertainty for the function and allow for a computationally efficient fixed functional form, the variation of $a$ can absorb the uncertainty from the ambiguity in the choice of functions $f_{i}(x)$ (essentially expanding the allowed range of $a$ -- as will be shown in the following sections). Since the functions are approximations themselves, increasing the allowed variation of $a$ to encapsulate the total uncertainty predicted by the initial treatment described above is a valid simplification.

A worked example following this procedure is provided for the $P_{qg}^{(3)}$ and $A_{Hg}^{(3)}$ functions in Section's~\ref{subsec: 4loop_split} and~\ref{subsec: 3loop_OME} respectively.

\subsection{4-loop Approximations}\label{subsec: 4loop_split}

\subsection*{$P_{qg}^{(3)}$}

We begin by considering the four-loop quark-gluon splitting function. Here we provide a more detailed explanation of the method described in Section~\ref{subsec: genframe} which will then be applied to the remaining splitting functions considered in this section. Four even-integer moments are known for $P_{qg}^{(3)}(n_{f} = 4)$ from~\cite{S4loopMoments,S4loopMomentsNew}, along with the LL small-$x$ term from \cite{Catani:1994sq}. 

The functions made available for the $P_{qg}$ analysis are,
\begin{alignat}{5}\label{eq: PqgComb}
    f_{1}(x) \quad &= \quad \frac{1}{x} \quad &&\text{or} \quad \ln^{4}x \quad &&\text{or} \quad \ln^{3} x \quad &&\text{or} \quad \ln^{2} x,\nonumber \\
    f_{2}(x) \quad &= \quad \ln x, \nonumber \\
    f_{3}(x) \quad &= \quad 1 \quad &&\text{or} \quad x \quad &&\text{or} \quad x^{2},\nonumber \\
    f_{4}(x) \quad &= \quad \ln^{4}(1-x) \quad &&\text{or} \quad \ln^{3}(1-x) \quad &&\text{or} \quad \ln^{2}(1-x) \quad &&\text{or} \quad \ln(1-x),\nonumber \\
    f_{e}(x,\rho_{qg}) \quad &= \quad \frac{C_{A}^{3}}{3\pi^{4}}\bigg(\frac{82}{81}\ +&&2\zeta_{3}\bigg)\frac{1}{2}\frac{\ln^{2}1/x}{x}\ +\ &&\rho_{qg}\ \frac{\ln 1/x}{x},
\end{alignat}
where $\rho_{qg}$ is the variational parameter. This is then varied between $-2.5 < \rho_{qg} < -0.9$, which has been chosen to satisfy the criteria described in Section~\ref{subsec: genframe}. The set of functions in Equation~\eqref{eq: PqgComb} is chosen from the analysis of lower orders. Specifically, following the pattern of functions from lower orders, it can be shown that at this order we expect the most dominant large-$x$ term to be $\ln^{4}(1-x)$ and $\ln^{4}x$ to be the highest power of $\ln x$ at small-$x$.

\begin{figure}
\centering
\includegraphics[width=0.49\textwidth]{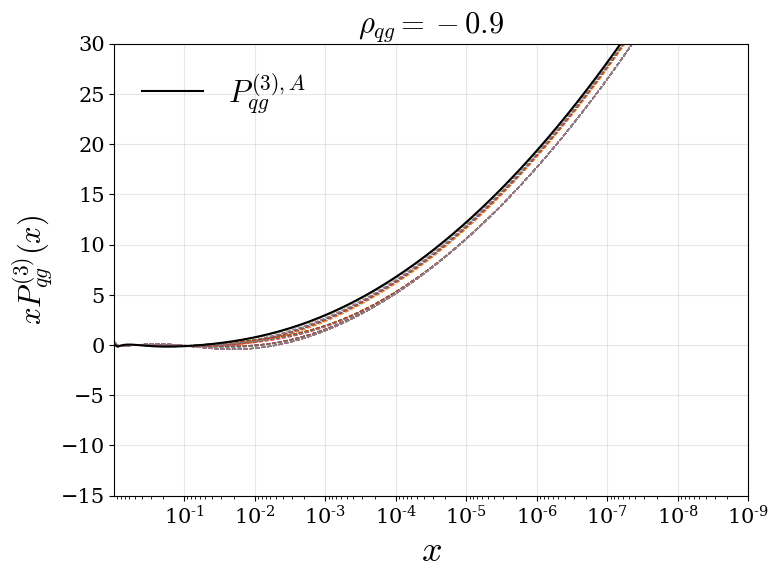}
\includegraphics[width=0.49\textwidth]{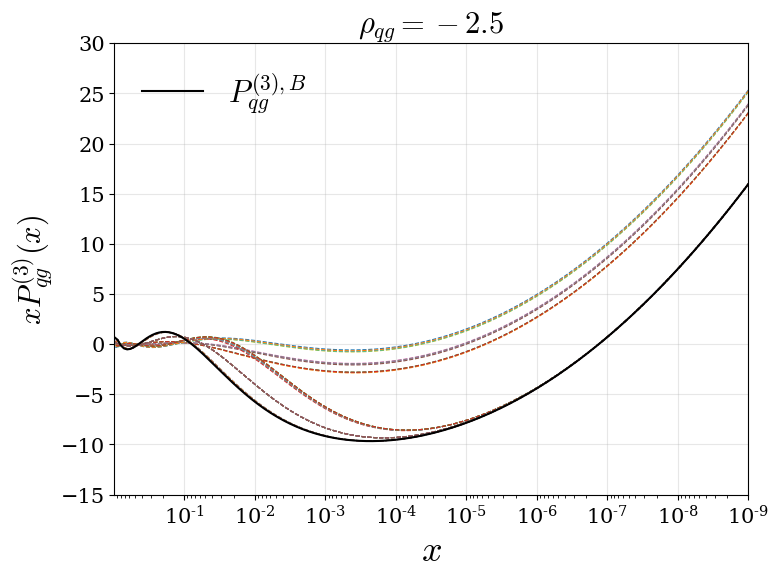}
\caption{Combinations of functions with an added variational factor ($\rho_{qg}$) controlling the
NLL term. Combinations of functions at the upper (left) and lower (right) bounds of the variation are shown. The solid lines indicate the upper and lower bounds for this function chosen from the relevant criteria.}
\label{fig: PqgRangeofa}
\end{figure}
Fig.~\ref{fig: PqgRangeofa} displays an example of the variation found from the different choices of functions that encapsulate the chosen range of $\rho_{qg}$. We also show the upper (A) and lower (B) bounds (at small-$x$) for the entire uncertainty (solid line) combining the variation in the functions and in the variation of $\rho_{qg}$. The upper ($P_{qg}^{(3),A}$) and lower ($P_{qg}^{(3), B}$) bounds are given by,
\begin{multline}
P_{qg}^{(3), A} = 1.6699\ \frac{1}{x} + 2.4167\ \ln x - 2.2011\ x^{2} + 0.0024228\ \ln^{4}(1-x) \\+ \frac{C_{A}^{3}}{3\pi^{4}}\bigg(\frac{82}{81}\ + 2\zeta_{3}\bigg)\frac{1}{2}\frac{\ln^{2}1/x}{x}\ -0.9\ \frac{\ln 1/x}{x},
\end{multline}
\begin{multline}
P_{qg}^{(3), B} = 12.582\ \ln^{2} x + 5.3065\ \ln x + 1.7957\ x^{2} - 0.0041296\ \ln^{4}(1-x) \\+ \frac{C_{A}^{3}}{3\pi^{4}}\bigg(\frac{82}{81}\ + 2\zeta_{3}\bigg)\frac{1}{2}\frac{\ln^{2}1/x}{x}\ -2.5\ \frac{\ln 1/x}{x}.
\end{multline}
Using this information, a fixed functional form is chosen to be,
\begin{multline}\label{eq: pqg_fff}
P_{qg}^{(3)} = A_{1}\ \ln^{2} x + A_{2}\ \ln x +  A_{3}\ x^{2} + A_{4}\ \ln^{4}(1-x) \\+ \frac{C_{A}^{3}}{3\pi^{4}}\bigg(\frac{82}{81}\ + 2\zeta_{3}\bigg)\frac{1}{2}\frac{\ln^{2}1/x}{x}\ + \rho_{qg}\ \frac{\ln 1/x}{x}
\end{multline}
and $\rho_{qg}$ is allowed to vary as $-2.5 < \rho_{qg} < -0.8$. This fixed functional form identically matches with the lower bound $P_{qg}^{(3),B}$ and the expansion of the variation of $\rho_{qg}$ enables (to within $\sim 1\%$) the absorption of the small-$x$ upper bound uncertainty (predicted from $P_{qg}^{(3),A}$) into the variation\footnote{Explicitly, the range is expanded from $-2.5 < \rho_{qg} < -0.9$ to $-2.5 < \rho_{qg} < -0.8$, in order to absorb the functional variation lost by moving to a fixed functional form (for implementation purposes).}. In other areas of $x$ there are larger deviations from the upper bound ($\sim 10\%$) when using this convenient fixed functional form. However, in these regions the function is already relatively small, therefore any larger percentage deviations are negligible. 
Also since the heuristic choice of variation found earlier is intended as a guide, we are not bound by any solid constraints to precisely reconstruct it with our subsequent choice of fixed functional form. Therefore it is entirely justified to be able to slightly adapt the shape of the variation in less dominant regions.

\subsection*{$P_{qq}^{\mathrm{NS},\ (3)}$}

As discussed in Section~\ref{sec: structure}, 
the quark-quark splitting function 
is comprised of a pure-singlet and non-singlet contribution. We approximate each part independently, although the final quark-quark singlet function will be almost completely dominated by the pure-singlet, except at very high-$x$.

The four-loop non-singlet splitting function has been the subject of relatively extensive research and is known exactly for a number of regimes. For example in~\cite{4loopNS}, some important exact contributions to the four-loop non-singlet splitting functions are presented, along with 8 even-integer moments for each of the $+$ and $-$ distributions~\cite{4loopNS}. In this discussion we are exclusively approximating the non-singlet $+$-distribution, as this is the part that contributes to the full singlet quark-quark splitting function. The other relevant non-singlet distributions $P_{\mathrm{NS}}^{(3),\ -}$ and $P_{\mathrm{NS}}^{(3),\ \mathrm{sea}}$ (described in \cite{Moch:2004pa}), are set to the central values predicted from \cite{4loopNS} since any variation in these functions are negligible. All presently known information is used in this approximation, with results similar to that seen in~\cite{4loopNS} but with our own choice of functions.
\begin{alignat}{5}\label{eq: PnsComb}
    f_{1}(x) \ &= \ \frac{1}{(1-x)_{+}}, \qquad \quad f_{2}(x) \ = \ (1-x)\ \ln(1-x), \qquad f_{3}(x)\ =\ (1-x)\ \ln^{2}(1-x), \nonumber \\
    f_{4}(x) \ &= \ (1-x)\ \ln^{3}(1-x), \quad f_{5}(x) \ = \ 1, \quad f_{6}(x) \ = \ x, \quad
    f_{7}(x) \ = \ x^{2}, \quad f_{8}(x) \ =\ \ln^{2} x, \nonumber
\end{alignat}
\begin{multline}
    f_{e}(x, \rho_{qq}^{\mathrm{NS}}) \ = \ C_{F} n_{c}^{3} P_{\mathrm{L}, 0}^{(3)}(x)+C_{F} n_{c}^{2} n_{f} P_{\mathrm{L}, 1}^{(3)}(x)+P_{L n_{f}}^{(3)+}(x) + \rho_{qq}^{\mathrm{NS}}\ \ln^{3} x \\ -55.876\ \ln^{4}x-2.8313\ \ln^{5}x-0.14883\ \ln^{6}x-2601.7-2118.9\ \ln (1-x)  \\ + n_{f} \left(4.6584\ \ln^{4}x+0.2798\ \ln^{5}x+312.16+337.93\ \ln(1-x)\right)
\end{multline}
where the functions $C_{F} n_{c}^{3} P_{\mathrm{L}, 0}^{(3)}(x)+C_{F} n_{c}^{2} n_{f} P_{\mathrm{L}, 1}^{(3)}(x)$ and $P_{L n_{f}}^{(3)+}(x)$ can be found in Equation (4.11) and Equation (4.14) respectively within~\cite{4loopNS}, and $\rho_{qq}^{\mathrm{NS}}$ is our variational parameter. Note that the ansatz from Equation~\eqref{eq: splitAnsatz} has been extended to include 8 pairs of functions and coefficients, to accommodate 8 known moments. Within the $f_{e}(x, \rho_{qq}^{\mathrm{NS}})$ part of Equation~\eqref{eq: PnsComb}, we have chosen to vary the coefficient of the most divergent unknown small-$x$ term ($\ln^{3}x$) with the variation across $0 < \rho_{qq}^{\mathrm{NS}} < 0.014$. Due to the high level of information and larger number of functions allowed to be included, we ignore any functional uncertainty and explicitly define each function. Therefore the only variation needed to be considered as an uncertainty stems from the variation of $\rho_{qq}^{\mathrm{NS}}$. 

The resulting approximation is then,
\begin{multline}\label{eq: pns_fff}
P_{\mathrm{NS}}^{(3),\ +} = A_{1}\ \frac{1}{(1-x)_{+}} + A_{2}\ (1-x)\ \ln(1-x) + A_{3}\ (1-x)\ \ln^{2}(1-x) \\ + A_{4}\ (1-x)\ \ln^{3}(1-x) + A_{5} + A_{6}\ x + A_{7}\ x^{2} + A_{8}\ \ln^{2}x + f_{e}(x,\rho_{qq}^{\mathrm{NS}}),
\end{multline}
where no alterations are made to the allowed range of $0 < \rho_{qq}^{\mathrm{NS}} < 0.014$. 

\subsection*{$P_{qq}^{\mathrm{PS},\ (3)}$}

We now restrict our analysis to focus on approximating the pure-singlet part of $P_{qq}^{(3)}$, thereby providing a more accurate set of functions with a focus on the small-$x$ regime. To ensure the $P_{qq}^{\mathrm{PS}\ (3)}$ function does not interfere with the large-$x$ regime (where the non-singlet description dominates) the ansatz from Equation~\eqref{eq: splitAnsatz} is adapted to be:
\begin{equation}
	P_{ij}^{(3)}(x)=\bigg\{A_{1}f_{1}(x)+A_{2}f_{2}(x)+A_{3}f_{3}(x)+A_{4}f_{4}(x)\bigg\}(1-x)+f_{e}(x, \rho_{qq}^{\mathrm{PS}}).
	\label{eq: splitAnsatzPqq}
\end{equation}
This modified parameterisation guarantees that any instabilities in the pure singlet approximation will not wash out the non-singlet behaviour at large-$x$.

Using four available even-integer moments for $n_{f} = 4$~\cite{S4loopMoments,S4loopMomentsNew} and the exact small-$x$ information~\cite{Catani:1994sq}, the chosen set of functions for this approximation is,
\begin{alignat}{5}\label{eq: PqqComb}
    f_{1}(x) \quad &= \quad \frac{1}{x} \quad &&\text{or} \quad \ln^{4} x,\nonumber \\
    f_{2}(x) \quad &= \quad \ln^{3} x \quad &&\text{or} \quad \ln^{2} x \quad &&\text{or} \quad \ln x,\nonumber \\
    f_{3}(x) \quad &= \quad 1 \quad &&\text{or} \quad x \quad &&\text{or} \quad x^{2},\nonumber \\
    f_{4}(x) \quad &= \quad \ln^{4}(1-x) \quad &&\text{or} \quad \ln^{3}(1-x) \quad &&\text{or} \quad \ln^{2}(1-x) \quad &&\text{or} \quad \ln(1-x),\nonumber \\
    f_{e}(x, \rho_{qq}^{\mathrm{PS}}) \quad &= \quad \frac{C_{A}^{2}C_{F}}{3\pi^{4}}\bigg(\frac{82}{81}&&+2\zeta_{3}\bigg)\frac{1}{2}\frac{\ln^{2}1/x}{x}\ +\ &&\rho_{qq}^{\mathrm{PS}}\ \frac{\ln 1/x}{x},
\end{alignat}
where $\rho_{qq}^{\mathrm{PS}}$ is varied as $-0.7 < \rho_{qq}^{\mathrm{PS}} < 0$. For the variation produced from stable combinations of these functions, we coincidentally end up with the same functional form for both the upper $P_{\mathrm{PS}}^{(3),\ A}$ and lower $P_{\mathrm{PS}}^{(3),\ B}$ bounds. Therefore trivially, the fixed functional form is defined as:
\begin{multline}
P_{\mathrm{PS}}^{(3)} = \bigg\{A_{1}\ \frac{1}{x} + A_{2}\ \ln^{2}x + A_{3}\ x^{2} + A_{4}\ \ln^{2}(1-x)\bigg\}(1-x)\ + \\\frac{C_{A}^{2}C_{F}}{3\pi^{4}}\bigg(\frac{82}{81}+2\zeta_{3}\bigg)\frac{1}{2}\frac{\ln^{2}1/x}{x}\ +\ \rho_{qq}^{\mathrm{PS}}\ \frac{\ln 1/x}{x}(1-x)
\end{multline}
where the variation of $\rho_{qq}^{\mathrm{PS}}$ is unchanged and the entire predicted variation is encapsulated in this form.

\subsection*{$P_{gq}^{(3)}$}

As with the previous singlet splitting functions, four even-integer moments for $n_{f} = 4$ are known~\cite{S4loopMoments,S4loopMomentsNew} along with the LL small-$x$ information~\cite{Lipatov:1976zz,Kuraev:1977fs,Balitsky:1978ic}. The set of functions made available for the combinations in our approximation are stated as,
\begin{alignat}{5}\label{eq: PgqComb}
    f_{1}(x) \quad &= \quad \frac{\ln 1/x}{x} \quad &&\text{or} \quad \frac{1}{x},\nonumber \\
    f_{2}(x) \quad &= \quad \ln^{3}x, \nonumber \\
    f_{3}(x) \quad &= \quad x \quad &&\text{or} \quad x^{2},\nonumber \\
    f_{4}(x) \quad &= \quad \ln^{4}(1-x) \quad &&\text{or} \quad \ln^{3}(1-x) \quad &&\text{or} \quad \ln^{2}(1-x) \quad &&\text{or} \quad \ln(1-x),\nonumber \\
    f_{e}(x, \rho_{gq}) \quad &= \quad \frac{C_{A}^{3}C_{F}}{3\pi^{4}}\zeta_{3}\frac{\ln^{3}1/x}{x}\ +\ &&\rho_{gq}\ \frac{\ln^{2} 1/x}{x},
\end{alignat}
where $\rho_{gq}$ is set as $\rho_{gq} = -1.8$. In this case, the variation from the choice of functions is large enough to satisfy the criteria in Section~\ref{subsec: genframe} and encapsulate a sensible $\pm 1\sigma$ variation without including any further variation in $\rho_{gq}$. 
Similarly to previous approximations, for stable variations we estimate this variation with the fixed functional form, 
\begin{equation}
P_{gq}^{(3)} = A_{1}\ \frac{\ln 1/x}{x} + A_{2}\ \ln^{3} x + A_{3}\ x + A_{4}\ \ln(1-x) + \frac{C_{A}^{3}C_{F}}{3\pi^{4}}\zeta_{3}\frac{\ln^{3}1/x}{x}\ +\ \rho_{gq}\ \frac{\ln^{2} 1/x}{x}
\end{equation}
where the allowed range of $\rho_{gq}$ is expanded to $-1.8 < \rho_{gq} < -1.5$ to approximate the variation from the choice of functions. As with the $P_{qg}^{(3)}$ fixed functional form, this new range recovers a variation which is within $\sim 1\%$ of the original, in the dominant areas of $x$.

\subsection*{$P_{gg}^{(3)}$}

Finally we move to the approximation of the gluon-gluon splitting function, where four available even-integer moments for $P_{gg}^{(3)}(n_{f} = 4)$ are known from \cite{S4loopMoments, S4loopMomentsNew}. The list of functions (including the known small-$x$ LL and NLL terms from \cite{Lipatov:1976zz,Kuraev:1977fs,Balitsky:1978ic,Fadin:1998py,Ciafaloni:1998gs}) used for the approximation is,
\begin{alignat}{6}\label{eq: PggComb}
    f_{1}(x) \quad &= \quad \frac{1}{x} \quad &&\text{or} \quad \ln^{3} x \quad &&\text{or} \quad \ln^{2} x,\nonumber \\
    f_{2}(x) \quad & = \quad \ln x,\nonumber \\
    f_{3}(x) \quad &= \quad 1 \quad &&\text{or} \quad x \quad &&\text{or} \quad x^{2},\nonumber \\
    f_{4}(x) \quad &= \quad \frac{1}{(1-x)}_{+} \quad &&\text{or} \quad \ln^{2}(1-x) \quad &&\text{or} \quad \ln(1-x),\nonumber
\end{alignat}
\begin{multline}
    f_{e}(x, \rho_{gg}) \quad = \quad \frac{C_{A}^{4}}{3\pi^{4}}\zeta_{3}\frac{\ln^{3}1/x}{x} +\frac{1}{\pi^{4}}\bigg[C_{A}^{4}\bigg(-\frac{1205}{162}+\frac{67}{36} \zeta_{2}+\frac{1}{4} \zeta_{2}^{2}-\frac{11}{2} \zeta_{3}\bigg)  \\+n_{f} C_{A}^{3}\bigg(-\frac{233}{162}+\frac{13}{36} \zeta_{2}-\frac{1}{3}\zeta_{3}\bigg) \\+n_{f} C_{A}^{2} C_{F}\bigg(\frac{617}{243} -\frac{13}{18} \zeta_{2}+\frac{2}{3} \zeta_{3}\bigg)\bigg]\frac{1}{2}\frac{\ln^{2}1/x}{x}\ +\ \rho_{gg}\ \frac{\ln 1/x}{x},
\end{multline}
where $\rho_{gg}$ is varied as $-5 < \rho_{gg} < 15$ and $n_{f} = 4$. The fixed functional form is then chosen to be,
\begin{multline}
P_{gg}^{(3)} = A_{1}\ \ln^{2} x + A_{2}\ \ln x + A_{3}\ x^{2} + A_{4}\ \ln^{2}(1-x) + \frac{C_{A}^{4}}{3\pi^{4}}\zeta_{3}\frac{\ln^{3}1/x}{x} \\ +\frac{1}{\pi^{4}}\bigg[C_{A}^{4}\bigg(-\frac{1205}{162}+\frac{67}{36} \zeta_{2}+\frac{1}{4} \zeta_{2}^{2}-\frac{11}{2} \zeta_{3}\bigg)  +n_{f} C_{A}^{3}\bigg(-\frac{233}{162}+\frac{13}{36} \zeta_{2}-\frac{1}{3}\zeta_{3}\bigg) \\+n_{f} C_{A}^{2} C_{F}\bigg(\frac{617}{243} -\frac{13}{18} \zeta_{2}+\frac{2}{3} \zeta_{3}\bigg)\bigg]\frac{1}{2}\frac{\ln^{2}1/x}{x}\ +\ \rho_{gg}\ \frac{\ln 1/x}{x}, \quad (n_{f} = 4)
\end{multline}
where we maintain the variation of $\rho_{gg}$ from above, as the fixed functional form manages to encapsulate the variation predicted, without any extra allowed $\rho_{gg}$ variation.

\subsection{Predicted \texorpdfstring{aN$^{3}$LO}{aN3LO} Splitting Functions}\label{subsec: expansion_split}

\begin{figure}
\begin{center}
\includegraphics[width=0.97\textwidth]{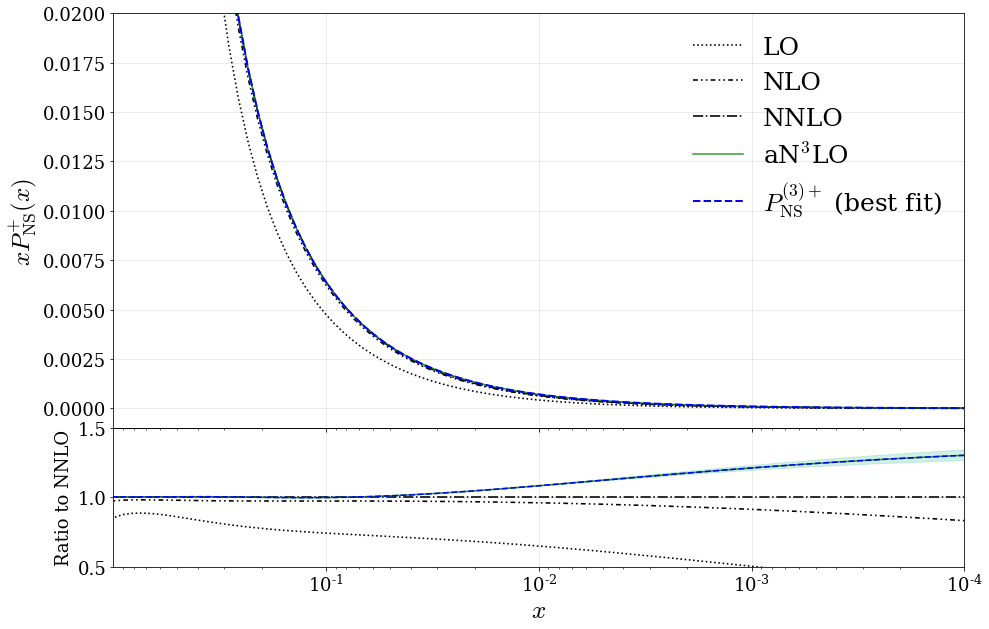}
\end{center}
\caption{\label{fig: split_variation_NS}Perturbative expansion up to aN$^{3}$LO for the non-singlet splitting function $P_{qq}^{\mathrm{NS},\ +}$ including any corresponding allowed $\pm 1\sigma$ variation (shaded green region). The best fit value (blue dashed line) displays the prediction for this function determined from a global PDF fit.}
\end{figure}
\begin{figure}
\begin{center}
\includegraphics[width=0.97\textwidth]{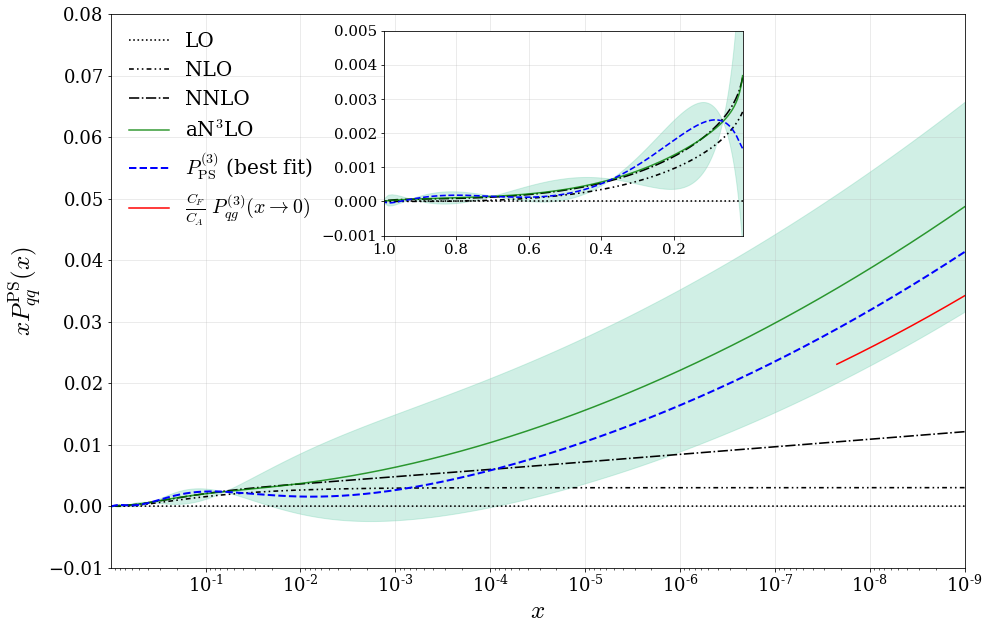}
\includegraphics[width=0.97\textwidth]{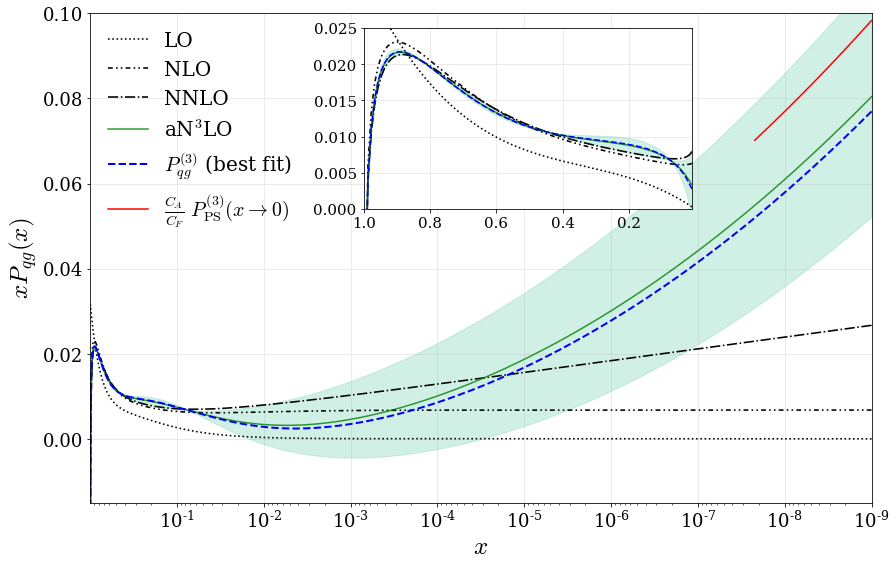}
\end{center}
\caption{\label{fig: split_variation_sing_q}Perturbative expansions up to aN$^{3}$LO for the quark singlet splitting functions $P_{qq}^{\mathrm{PS}}$ (top) and $P_{qg}$ (bottom) including any corresponding allowed $\pm 1\sigma$ variation (shaded green region). The best fit values (blue dashed line) display the predictions for each function determined from a global PDF fit.}
\end{figure}
\begin{figure}
\begin{center}
\includegraphics[width=0.97\textwidth]{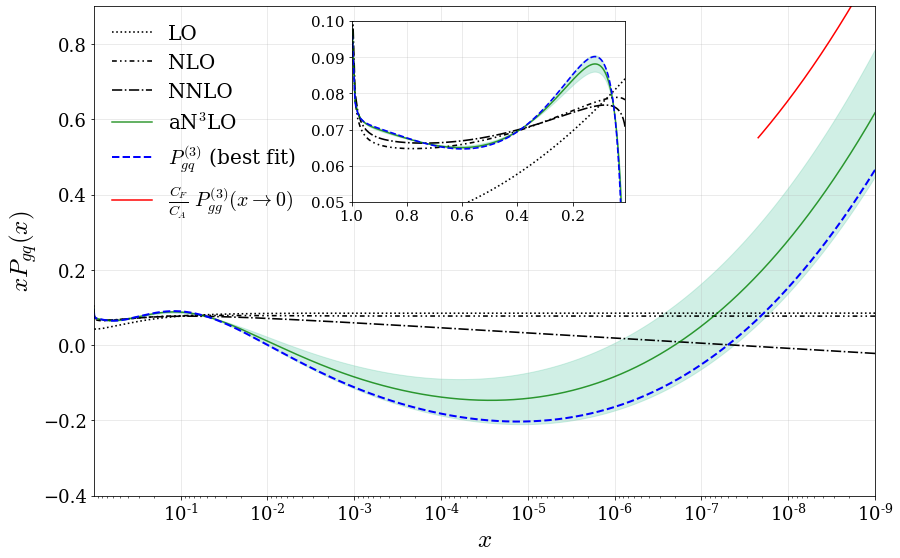}
\includegraphics[width=0.97\textwidth]{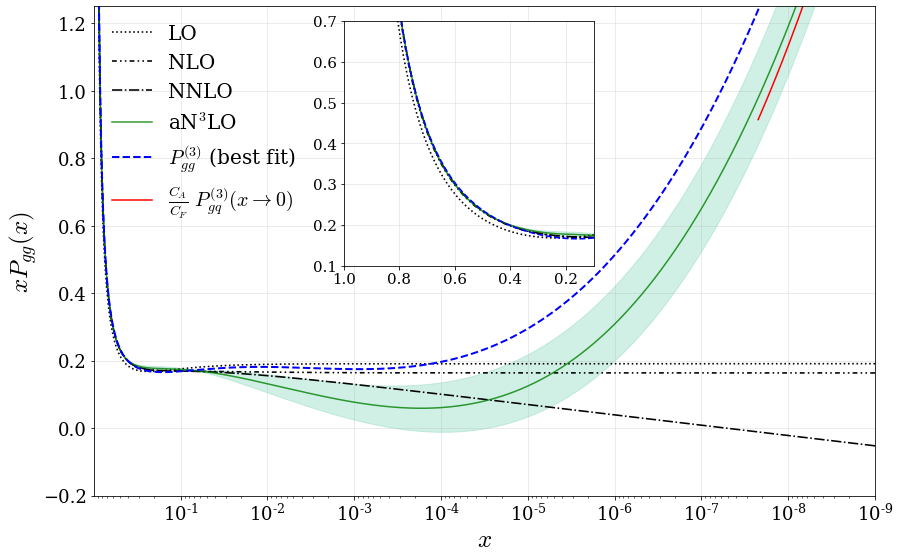}
\end{center}
\caption{\label{fig: split_variation_sing_g}Perturbative expansions for the gluon splitting functions $P_{gq}$ (top) and $P_{gg}$ (bottom) including any corresponding allowed $\pm 1\sigma$ variation (shaded green region). The best fit value (blue dashed line) displays the prediction for this function determined from a global PDF fit.}
\end{figure}
Fig.'s~\ref{fig: split_variation_NS}, \ref{fig: split_variation_sing_q} and \ref{fig: split_variation_sing_g} show the perturbative expansions for each splitting function up to approximate N$^{3}$LO. Included with these expansions are the predicted variations ($\pm 1 \sigma$) from Section~\ref{subsec: 4loop_split} (shown in green) and the aN$^{3}$LO best fits (shown in blue -- discussed further in Section~\ref{sec: results}). As a general feature, we observe that the singlet N$^3$LO approximations are much more divergent than lower orders due to the presence of higher order logarithms at small-$x$, further highlighting the need for an understanding of MHOUs beyond the default NNLO considered in current PDF sets in a way that is not reliant on the NNLO central value.

Considering the non-singlet case shown in Fig.~\ref{fig: split_variation_NS}, we see a very close agreement at large-$x$ between $P_{qq}^{\mathrm{NS}}$ expanded to NNLO and aN$^{3}$LO. This is a general feature of the non-singlet distribution, since by design, this distribution is largely unaffected by small-$x$ contributions. The ratio plot in Fig.~\ref{fig: split_variation_NS} provides clearer evidence for this, since it is only towards small-$x$ (where the non-singlet distribution tends towards 0) that any noticeable difference between NNLO and aN$^{3}$LO can be seen.

The contributions to $P_{qq}^{\mathrm{PS}}$, $P_{qg}$, $P_{gq}$ and $P_{gg}$ shown in Fig.'s~\ref{fig: split_variation_sing_q} and \ref{fig: split_variation_sing_g} respectively, display a much richer description at aN$^{3}$LO. In all cases, the divergent terms (with $x \rightarrow 0$) present in the approximations have a large effect from intermediate-$x$ ($\sim 10^{-2}$) down to very small-$x$ values. The asymptotic relationships (red line) Equation~\eqref{eq: Relations} defined using the best fit values of the aN$^{3}$LO expansions (i.e. comparable to the blue dashed line) are also shown in Fig.'s~\ref{fig: split_variation_sing_q} and \ref{fig: split_variation_sing_g}. As discussed earlier, these relations are violated by large sub-leading small-$x$ terms and are therefore provided here as a qualitative comparison. Furthermore, we also observe a close resemblance to the N$^{3}$LO asymptotic results in Fig.~4 of \cite{Marzani:smallx}. Specifically for quark evolution, we show that the data prefers a similar form ($P_{qq}^{\mathrm{PS}}$ and $P_{qg}$) to the resummed splitting function results in \cite{Marzani:smallx} whereas for gluon evolution, this agreement is less prominent.

Superimposed onto these variations in Fig.'s \ref{fig: split_variation_sing_q} and \ref{fig: split_variation_sing_g} are the best fit values for the splitting functions, as predicted from a global fit of the full MSHT approximate N$^{3}$LO PDFs. The full fit results will be discussed in more detail in Section~\ref{sec: results}, however we note here that the fit produces relatively good agreement with the prior allowed variations for each of the splitting functions. For all functions except for $P_{gg}$, the best fit results lie within their $\pm 1 \sigma$ variation range. This result implies that constraints from the data included in the global fit are in good agreement with the penalties describing quark evolution (i.e. $P_{\mathrm{PS}}$ and $P_{qg}$ in Fig.~\ref{fig: split_variation_sing_q}). For the gluon evolution in Fig.~\ref{fig: split_variation_sing_g} we observe a small level of tension with the data pushing towards a slightly harder small-$x$ gluon than preferred by the penalty constraints for $P_{gg}$. An important caveat to these best fit results is that the data included in the fit is sensitive to all orders in $\alpha_{s}$. Therefore by proxy, the best fit predictions are also sensitive to corrections at all orders. This will certainly be a driving factor for any violations away from the expected N$^{3}$LO behaviour. However, since the ultimate goal of this investigation is to provide a theoretical uncertainty, the violation from higher orders is manifested into the defined penalties and therefore accounted for in the fit as a source of MHOU.

Finally, an important feature that can be seen across all these splitting function plots are points of zero aN$^{3}$LO uncertainty in the high-$x$ regions. The regions where these points occur are where the moments are constraining the chosen fixed functional forms very tightly. In particular, for $N_{m}$ moments (constraints) in Equation~\eqref{eq: splitAnsatz}, we are left with $N_{m}-1$ points of zero uncertainty predicted from our approximations. As stated, these points are dependent on the choice of our fixed functional form and are therefore regions where the uncertainty has been underestimated when compared to the functional uncertainty which the fixed form approximates. To provide a more complete estimate of the uncertainty in these areas, it would be necessary to smooth the uncertainty band out across these regions (or take into account several fixed functional forms). However, this shortcoming only occurs towards large-$x$, where the uncertainty is naturally smaller across these functions. Therefore if the uncertainty was smoothed, the effect would be negligible for the theoretical uncertainty this work aims to include in a PDF global fit. Further to this, these functions are ingredients in the DGLAP convolution where any smaller details are washed out by more dominant features inside convolutions with PDFs. For these reasons, we opt for computational efficiency and leave these points as shown.

\subsubsection{Moment Analysis}

\begin{table}
\centerline{
\begin{tabular}{c|c|cccc}
\hline
 & Moment & LO & NLO & NNLO & N$^{3}$LO \\
\hline
\multirow{4}{*}{$P^{\mathrm{PS}}_{qq}$} 
 & $N=2$ & $-0.056588$ & $-0.06362642$ & $-0.06395712$ & $-0.06412109$\\
 & $N=4$ & $-0.11104$ & $-0.1261481$ & $-0.12804822$ & $-0.12835549$\\
 & $N=6$ & $-0.14329$ & $-0.16188618$ & $-0.16433013$ & $-0.16470246$\\
 & $N=8$ & $-0.166448$ & $-0.18751366$ & $-0.19033329$ & $-0.19074888$\\
\hline
\multirow{4}{*}{$P_{qg}$} 
 & $N=2$ & $0.042442$ & $0.05008496$ & $0.04991043$ & $0.04983007$\\
 & $N=4$ & $0.023342$ & $0.02203438$ & $0.02110201$ & $0.02112623$\\
 & $N=6$ & $0.016674$ & $0.01387744$ & $0.01311037$ & $0.01316929$\\
 & $N=8$ & $0.013086$ & $0.00979920$ & $0.00919186$ & $0.00927006$\\
\hline
\multirow{4}{*}{$P_{gq}$} 
 & $N=2$ & $0.056588$ & $0.06362642$ & $0.06395712$ & $0.06412109$\\
 & $N=4$ & $0.015562$ & $0.01903295$ & $0.0195455$ & $0.01965547$\\
 & $N=6$ & $0.008892$ & $0.0112073$ & $0.01158133$ & $0.0116615$\\
 & $N=8$ & $0.006232$ & $0.00801547$ & $0.00831037$ & $0.0083761$\\
\hline
\multirow{4}{*}{$P_{gg}$} 
 & $N=2$ & $-0.042442$ & $-0.05008496$ & $-0.04991043$ &   $-0.04983007$\\
 & $N=4$ & $-0.242978$ & $-0.26161441$ & $-0.26280015$ & $-0.26326763$\\
 & $N=6$ & $-0.32551$ & $-0.35114066$ & $-0.35335022$ & $-0.35384552$\\
 & $N=8$ & $-0.38091$ & $-0.41151668$ & $-0.41447721$ & $-0.41495604$\\
\hline
\end{tabular}
}
\caption{\label{tab: split_moments}Numerical moments of singlet and gluon splitting function moments up to N$^{3}$LO for $\alpha_{s}=0.2$ and $n_{f}=4$.}
\end{table}
\begin{figure}[t]
\begin{center}
\includegraphics[width=\textwidth]{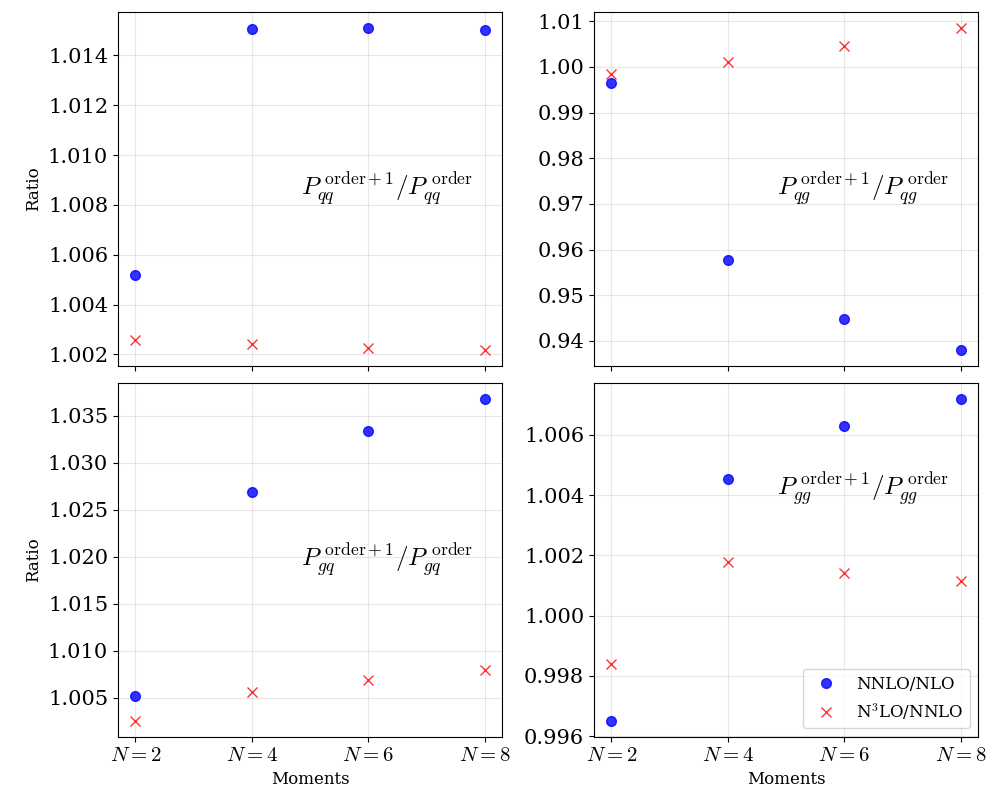}
\end{center}
\caption{\label{fig: split_moment_ratio}The low-integer numerical Mellin moments of relevant singlet splitting functions (excluding $P_{qq}^{\mathrm{NS},\ +}$) as a ratio between orders. In all cases the expected perturbative convergence is demonstrated.}
\end{figure}
Tracking back to the moments found for the splitting functions~\cite{S4loopMoments, S4loopMomentsNew} (shown in Table~\ref{tab: split_moments} and as a ratio in Fig.~\ref{fig: split_moment_ratio}), we are able to identify the expected convergence in the perturbative expansions up to N$^{3}$LO. Fig.~\ref{fig: split_moment_ratio} illustrates the relative size of the NNLO and N$^{3}$LO contributions to the low even-integer moments. 

Until recently (at the time of writing), there were only 3 moments available for the functions $P_{gq}$ and $P_{gg}$ approximated here. However, in \cite{S4loopMomentsNew} an extra moment was published for these two gluon splitting functions. 
This extra information led to our predictions at small-$x$ being more in line with the resummation results in \cite{Marzani:smallx} mentioned earlier. 
This is an example of how extra information can be added as and when it is available to update any approximations and utilise our full knowledge of the next highest order. By adopting this procedure, we immediately benefit from a slightly increased precision (with a relevant theoretical uncertainty) instead of having to delay the inclusion of higher order theory (for potentially decades) until a complete analytical calculation of the next order in $\alpha_{s}$ is known.

\subsection{Numerical Results}\label{subsec: num_res_split}

We now consider the DGLAP evolution equations for the singlet and gluon shown in Equation~\eqref{eq: DGLAP}. We expand this equation to $\alpha^{4}_{s}$ and investigate the effects of the variation in the N$^{3}$LO contributions. 

For the purposes of this analysis, the approximate functions~\eqref{eq: singGluApprox}, taken from \cite{Vogt:2004mw}, are used as sample distributions at an energy scale of $\mu_{f}^{2} \simeq 30\ \mathrm{GeV}^{2}$, a scale chosen due to its relevance to DIS processes included in the MSHT global fit.
\begin{subequations}\label{eq: singGluApprox}
\begin{align}
x\Sigma(x,\mu_{f}^{2}=30\ \mathrm{GeV}^{2})\ =&\quad 0.6\ x^{-0.3}(1-x)^{3.5}(1+5x^{0.8}) \label{eq: singApprox} \\ xg(x,\mu_{f}^{2}=30\ \mathrm{GeV}^{2})\ =& \quad 1.6\ x^{-0.3}(1-x)^{4.5}(1-0.6x^{0.3}) \label{eq: gluApprox}
\end{align}
\end{subequations}
The expressions above are order independent and so provide a robust means to isolate the effects arising from higher orders in the splitting functions. For convenience we also assume
\begin{align}
\alpha_{s}(\mu_{r}^{2}=\mu_{f}^{2}=30\ \mathrm{GeV}^{2})\simeq 0.2.
\end{align}
where $\mu_{r}$ and $\mu_{f}$ are the renormalisation and factorisation scales respectively.

\subsubsection*{Singlet Evolution}

\begin{figure}[t]
\centering
\includegraphics[width=0.49\textwidth]{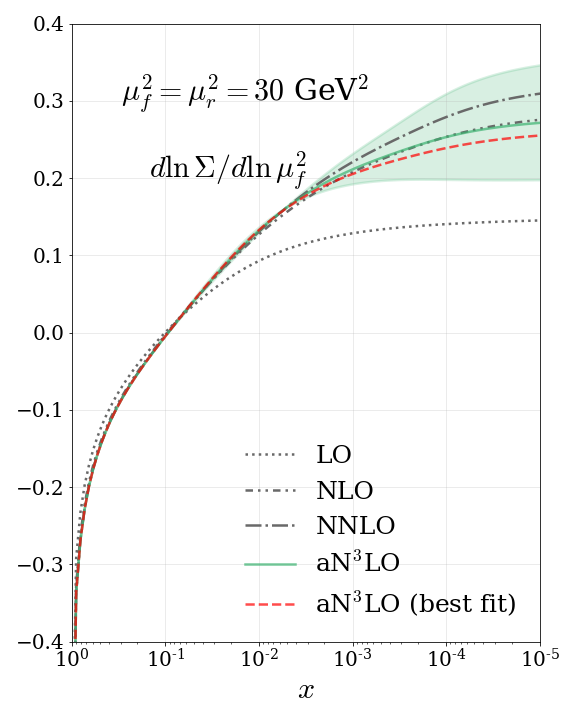}
\includegraphics[width=0.49\textwidth]{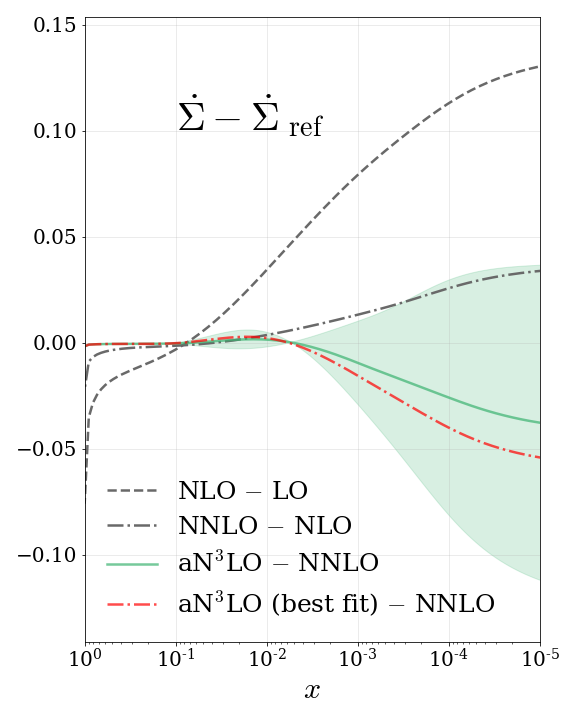}
\caption{\label{fig: SingEvolutionApprox} The flavour singlet quark distribution evolution equation Equation~\eqref{eq: DGLAP} shown for orders up to the approximate N$^3$LO (left). The relative shift between subsequent orders of the flavour singlet evolution (right) where $\dot{\Sigma} = d \ln \Sigma/d \ln \mu_{f}^{2}$. }
\end{figure}
Fig.~\ref{fig: SingEvolutionApprox} demonstrates the result of including the respective N$^3$LO expansions from Section~\ref{subsec: 4loop_split} in an analysis of the evolution equation. Towards small-$x$ this variation increases due to the larger uncertainty in the $P_{qq}^{\mathrm{PS}}$ and $P_{qg}$ splitting functions at aN$^{3}$LO. On the right of Fig.~\ref{fig: SingEvolutionApprox}, the difference plot displays the respective shifts from the previous order and demonstrates how this shift changes up to N$^3$LO. These results predict a reduction in the evolution of the singlet towards small-$x$ from NNLO. Inspecting Fig.~\ref{fig: split_variation_sing_q}, we can see that this reduction is stemming from the contribution of the gluon with the $P_{qg}$ function at 4-loops, which is the dominant contribution to the evolution. Towards larger $x$ values ($10^{-2} < x < 10^{-1}$) we see a fractional increase in the quark evolution, also following the shape of the $P_{qg}$ function. These results can therefore give some indication as to how we expect our gluon PDF to behave at N$^{3}$LO; since the structure functions are directly related to the quarks (through LO), the singlet evolution should remain fairly constant. Therefore we can expect that the fit will prefer a slightly harder gluon at small-$x$ and a softer gluon between $10^{-2} < x < 10^{-1}$ relative to NNLO.

Fig.~\ref{fig: SingEvolutionApprox} displays a good level of agreement between the allowed N$^{3}$LO shift and the evolution at NLO and NNLO (within $\pm1\sigma$ variation bands from theoretical uncertainties). Also shown in Fig.~\ref{fig: SingEvolutionApprox} is the evolution prediction using the best fit results for $P_{qq}^{(3)}$ and $P_{qg}^{(3)}$ (red dashed). This prediction tends to follow slightly below the center of the $1\sigma$ uncertainty band, where the data has balanced the two variations and is more in line with the NLO evolution than NNLO due to a negative contribution below $10^{-2}$.  Considering the magnitude of shifts from each order, the predicted shift from NNLO to aN$^{3}$LO is slightly larger than that from NLO to NNLO, contradicting what may be expected from perturbation theory. However, we remind the reader that these best fit results are, to some degree, sensitive to all orders in perturbation theory through the data constraint. Due to this, the resultant best fit can be thought of as an approximate asymptote to all orders. Interpreting the approximation in this way, restores our faith in perturbation theory and becomes an entirely plausible estimation of the missing higher orders.

Fig.~\ref{fig: SingEvolutionApprox} also exhibits an example of how the points of zero uncertainty (discussed in Section \ref{subsec: 4loop_split}) can affect the evolution predictions. We can see that at most the uncertainty is being underestimated by $<1\%$ and therefore, for the reasons discussed earlier, we do not consider these regions further here.

\subsubsection*{Gluon Evolution}

\begin{figure}[t]
\centering
\includegraphics[width=0.49\textwidth]{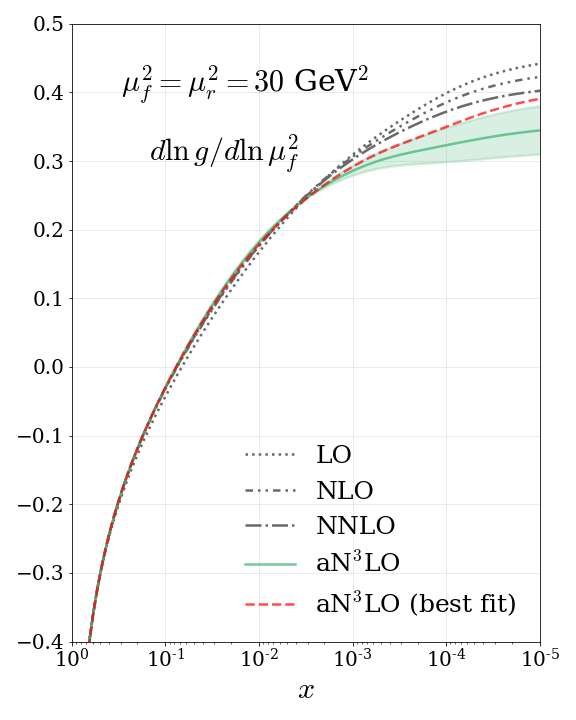}
\includegraphics[width=0.49\textwidth]{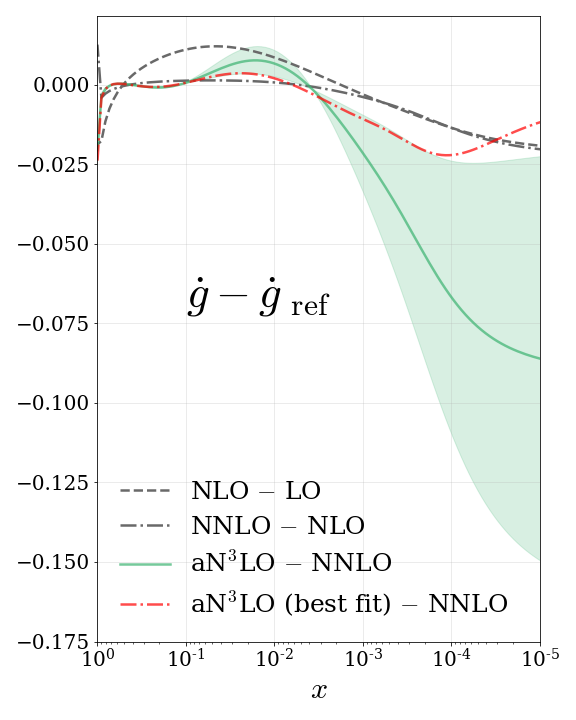}
\caption{\label{fig: GluonEvolutionApprox} The gluon distribution evolution equation Equation~\eqref{eq: DGLAP} shown for orders up to the approximate N$^3$LO (left). The relative shift between subsequent orders of the gluon evolution (right) where $\dot{g} = d \ln g/d \ln \mu_{f}^{2}$.}
\end{figure}
Fig.~\ref{fig: GluonEvolutionApprox} displays the result of including the aN$^{3}$LO splitting function contributions into the gluon evolution equation. As with the singlet evolution case, this extra contribution is currently inducing a notable variation at N$^3$LO. The general trend at small-$x$ is a reduction in the value of the evolution equation due to the N$^3$LO prediction for $P_{gg}$. On the right hand side of Fig.~\ref{fig: GluonEvolutionApprox} we observe the respective shifts from lower orders and how this shift changes up to N$^3$LO. 

In the gluon evolution, there is a large variation coming from the uncertainty in the $P_{gg}^{(3)}$ function. Therefore when $P_{gg}^{(3)}$ is convoluted with the gluon PDF at small-$x$, one could expect a potentially large shift from NNLO. The best fit gluon evolution prediction in Fig.~\ref{fig: GluonEvolutionApprox} is produced by utilising the best fit results for $P_{gq}^{(3)}$ and $P_{gg}^{(3)}$ functions (red dashed). In this prediction we see that the fit prefers a reduction in the evolution from NNLO, which is contained within the $\pm 1 \sigma$ band until around $x \lesssim 10^{-4}$. Since at low-$Q^{2}$, the quark and gluon are comparable at small-$x$, this reduction is likely driven from the form of $P_{gq}$ in Fig.~\ref{fig: split_variation_sing_g}. Combining this with the smaller gluon PDF at low-$Q^{2}$ therefore acts to slow the gluon evolution despite $P_{gg}$ increasing. Furthermore, the best fit is seemingly more in line with the perturbative expectation of the evolution than the chosen variation\footnote{Due to the presence of more divergent higher order logarithms at this level, it is not certain or by any means guaranteed that the shift at N$^{3}$LO will follow the same trend outlined from lower orders.}. Since this variation is chosen from the known information about the perturbative expansions, this is a manifestation of how the framework we present here can capture the relevant sources of theoretical uncertainty (and account for these via a penalty in a PDF fit). This is encouraging, as even with the large amount of freedom for this gluon evolution, it seems that the data is constraining and balancing the two contributions from the splitting functions in a sensible fashion. As discussed in the singlet evolution case, the relative shift from NNLO to N$^{3}$LO is slightly larger than one might hope for when dealing with a perturbative expansion. However, since this best fit is impacted to all orders from the experimental data (up to the leading logarithms at N$^{3}$LO i.e. even higher orders involve more divergent logarithms which are missed in this theoretical description), we can interpret this shift as an approximate all order shift and once again restore its validity in perturbation theory.

As with the singlet case above, negligible points of non-zero uncertainty are displayed in Fig.~\ref{fig: GluonEvolutionApprox}. For the reasons discussed in the singlet case and in Section~\ref{subsec: 4loop_split}, these are not an area of concern at the current level of desired uncertainty and are therefore not considered further.
\section{\texorpdfstring{N$^{3}$LO}{N3LO} Transition Matrix Elements}\label{sec: n3lo_OME}

Heavy flavour transition matrix elements, $A_{ij}$, as described in Section~\ref{sec: structure}, are exact quantities that describe the transition of all PDFs with $n_{f}$ active flavours into a scheme with $n_{f} + 1$ active flavours. Due to discontinuous nature of $A_{ij}$ at the heavy flavour mass thresholds, they are also present in the coefficient functions to ensure an exact cancellation of this discontinuity in physical quantities. This combination then preserves the smooth nature of the structure function, as demanded by the renormalisation group flows.

The general expansion of the heavy-quark transition matrix elements in powers of $\alpha_s$ reads,
\begin{equation}
A_{i j}=\delta_{i j}+\sum_{\ell=1}^{\infty} \alpha_{\mathrm{s}}^{\ell} A_{i j}^{(\ell)}=\delta_{i j}+\sum_{\ell=1}^{\infty} \alpha_{\mathrm{s}}^{\ell} \sum_{k=0}^{\ell} L_{\mu}^{k} a_{i j}^{(\ell, k)},
\end{equation}
where at each order the terms proportional to powers of $\mathrm{L}_{\mu} = \ln (m_{h}^{2} / \mu^{2})$ are determined by lower order transition matrix elements and splitting functions. Therefore the focus only needs to be on the $a_{ij}^{(\ell,0)}$ expressions, as the rest are not only known~\cite{Buza:OMENLO,Buza:OMENNLO}, but are guaranteed not to contribute at mass thresholds due to the presence of $L_{\mu}$. These $\mu$-independent terms can be decomposed in powers of $n_{f}$ as
\begin{equation}
a_{ij}^{(3,0)} =a_{ij}^{(3,0),\ 0}+n_{f} a_{ij}^{(3,0),\ 1},
\end{equation}
where a number of the $n_f$-dependent and independent terms are known exactly. The $n_{f}$ parts are however sub-leading and so as a first approximation, are set to zero in this work. In keeping with the framework set out in Section~\ref{subsec: genframe} for the N$^{3}$LO splitting functions, we will make use of the available known information (even-integer Mellin moments~\cite{bierenbaum:OMEmellin} and leading small and large-$x$ behaviour~\cite{ablinger:3loopNS,Blumlein:2021enk,ablinger:3loopPS,ablinger:agq,Blumlein:AQg,Vogt:FFN3LO3}) about the heavy flavour transition matrix elements to approximate the $\mu$-independent contributions $a_{ij}^{(3,0)}$.
As discussed above, we make the choice to completely ignore any terms that do not contribute at mass threshold since not only are these sub-leading but can also be ignored by explicitly setting $\mu^{2} = m_{h}^{2}$.

\subsection{3-loop Approximations}\label{subsec: 3loop_OME}

\subsection*{$A_{Hg}$}\label{subsec: AHg}

The $A_{Hg}^{(3)}$ function is still under calculation at the time of writing. Currently the first five even-integer moments are known for the $\overline{\mathrm{MS}}$ scheme $A_{Hg}^{(3)}$~\cite{bierenbaum:OMEmellin}, along with the leading small-$x$ terms~\cite{Vogt:FFN3LO3}.

The $n_{f}$-dependent contribution to the 3-loop unrenormalised $A_{Hg}$ transition matrix element has also been approximated in \cite{Vogt:FFN3LO3}, while all other contributions to $A_{Hg}^{(3)}(n_{f} = 0)$ were already known. For this approximation we work in the $\overline{\mathrm{MS}}$ scheme using the framework set out in Section~\ref{subsec: genframe}. We then approximate the function using the set of functions,
\begin{alignat}{5}\label{eq: aQgComb}
    f_{1,2}(x) \quad &= \quad \ln^{5}(1-x) \quad &&\text{or} \quad \ln^{4}(1-x) \quad &&\text{or} \quad \ln^{3}(1-x) \quad &&\text{or} \quad \ln^{2}(1-x) \nonumber\\ & &&\text{or} \quad \ln(1-x), \nonumber \\
    f_{3,4}(x) \quad &= \quad 2 - x \quad &&\text{or} \quad 1 \quad &&\text{or} \quad x \quad &&\text{or} \quad x^{2}, \nonumber \\
    f_{5}(x) \quad &= \quad \ln x \quad &&\text{or} \quad \ln^{2} x,\nonumber \\
    f_{e}(x, a_{Hg}) \quad &= \quad \bigg(224\ \zeta_{3} - \frac{41984}{27} - 160\ &&\frac{\pi^{2}}{6} \bigg)  \frac{\ln 1/x}{x} + a_{Hg}\ &&\frac{1}{x}
\end{alignat}
where $a_{Hg}$ is varied as $6000 < a_{Hg} < 13000$. This variation is chosen from the criteria outlined in Section~\ref{subsec: genframe} and is comparable to that chosen in \cite{Vogt:FFN3LO3}.
\begin{figure}
\includegraphics[width=0.49\textwidth]{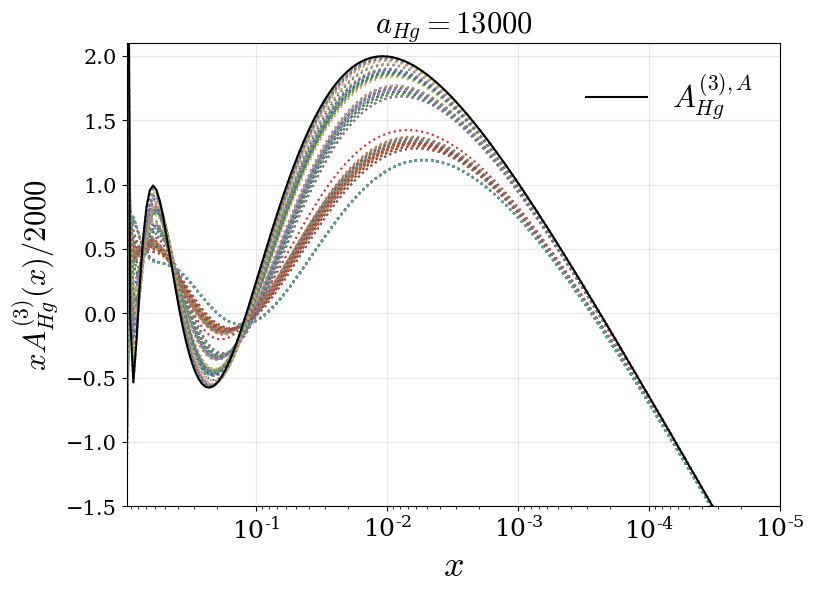}
\includegraphics[width=0.49\textwidth]{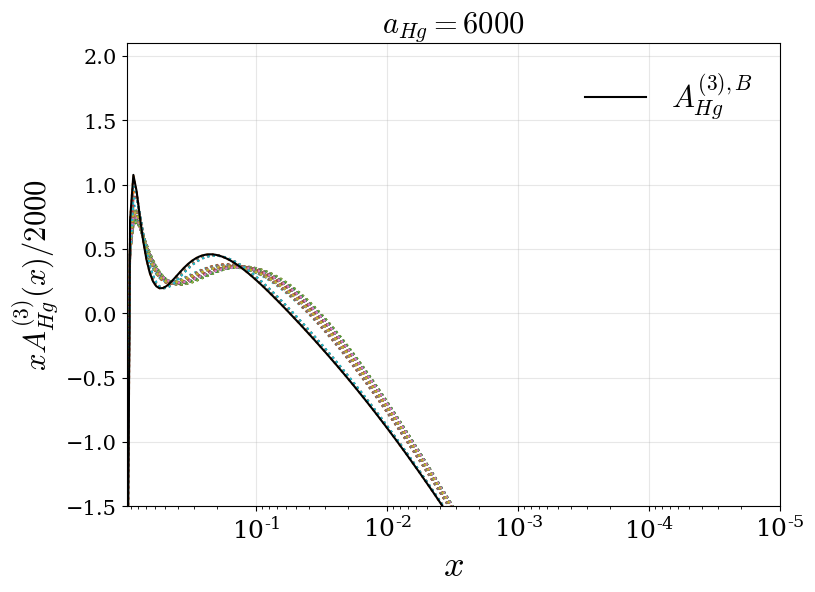}
\caption{\label{fig: AQg} Combinations of functions with an added variational factor ($a_{Hg}$) controlling the
NLL term. Combinations of functions at the upper (left) and lower (right) bounds of the variation are shown. The solid lines indicate the upper and lower bounds for this function chosen from the relevant criteria.}
\end{figure}
Fig.~\ref{fig: AQg} displays the approximation of the $\overline{\mathrm{MS}}$ $A_{Hg}^{(3)}$ with the variation from different combinations of functions in Equation~\eqref{eq: aQgComb} at the chosen limits of $a_{Hg}$. Comparing with Fig.~3 in \cite{Vogt:FFN3LO3}, we see a slightly larger range of allowed variation. A small proportion of this difference can be accounted for by the difference in renormalisation schemes, with the majority of this change being from the differences in the criteria from Section~\ref{subsec: genframe}. The upper ($A_{Hg}^{(3), A}$) and lower ($A_{Hg}^{(3), B}$) bounds in the small-$x$ region (shown in Fig.~\ref{fig: AQg}) are given by,
\begin{multline}
    A_{Hg}^{(3), A} = 44.1703\ \ln^{5}(1-x) + 268.024\ \ln^{4}(1-x) + 45271.0\ x - 68401.4\ x^{2} \\+ 36029.8\ \ln x + \bigg(224\ \zeta_{3} - \frac{41984}{27} - 160\ \frac{\pi^{2}}{6}\bigg)\ \frac{\ln 1/x}{x} + 12000\ \frac{1}{x}
\end{multline}
\begin{multline}
    A_{Hg}^{(3), B} = -18.9493\ \ln^{5}(1-x) - 138.763\ \ln^{4}(1-x)  - 31692.1\ x + 33282.3\ x^{2} \\- 3088.75\ \ln^{2} x + \bigg(224\ \zeta_{3} - \frac{41984}{27} - 160\ \frac{\pi^{2}}{6}\bigg)\ \frac{\ln 1/x}{x} + 6000\ \frac{1}{x}
\end{multline}
Using this information, we then choose the fixed functional form,
\begin{multline}
    A_{Hg}^{(3)} = A_{1}\ \ln^{5}(1-x) + A_{2}\ \ln^{4}(1-x) + A_{3}\ x + A_{4}\ x^{2} + A_{5}\ \ln x \\+ \bigg(224\ \zeta_{3} - \frac{41984}{27} - 160\ \frac{\pi^{2}}{6}\bigg)\ \frac{\ln 1/x}{x} + a_{Hg}\ \frac{1}{x}
\end{multline}
where the variation of $a_{Hg}$ remains unchanged as it already encapsulates the predicted variation to within the $\sim 1\%$ level.

\subsection*{$A_{Hq}^{\mathrm{PS}}$}\label{subsec: Aps}

The $A_{Hq}^{\mathrm{PS}}$ transition matrix element has been calculated exactly in \cite{ablinger:3loopPS}. Here we attempt to qualitatively reproduce this result via an efficient parameterisation to an appropriate precision. 

Using the expressions for the small and large-$x$ limits~\cite{ablinger:3loopPS} and the known first six even-integer moments converted into $\overline{\mathrm{MS}}$~\cite{bierenbaum:OMEmellin}, we provide a user-friendly approximation as,
\begin{multline}\label{eq: ApsApprox}
A_{Hq}^{\mathrm{PS}, (3)} =\ (1-x)^{2} \bigg\{-152.523\ \ln^{3}(1-x) -107.241\ \ln^{2}(1-x)\bigg\} \\ - 4986.09\ x + 582.421\ x^{2} - 1393.50\ x\ln^{2}x -4609.79\ x\ln x\\ - 688.396\ \frac{\ln 1/x}{x} + (1-x)\ 3812.90\ \frac{1}{x} + 1.6\ \ln^{5}x - 20.3457\ \ln^{4} x \\ + 165.115\ \ln^{3}x - 604.636\ \ln^{2}x + 3525.00\ \ln x \\  + (1 - x) \bigg\{0.246914 \ln^{4}(1-x) - 4.44444 \ln^{3}(1-x) - 2.28231 \ln^{2}(1-x) \\ - 357.427 \ln(1-x) + 116.478\bigg\}
\end{multline}
where the first two lines have been approximated and the last four lines are the exact leading small and large-$x$ terms. We note here that the approximated part of this parameterisation is in a much less important region of $x$ than the exact parts, therefore any small differences in the approximated part from the exact function are unimportant.

\subsection*{$A_{qq, H}^{\mathrm{NS}}$}\label{subsec: AqqHns}

Moving to the non-singlet $A_{qq, H}^{\mathrm{NS}}$ function, we attempt to parameterise the work from \cite{ablinger:3loopNS,Blumlein:2021enk}. Specifically, we make use of the known even integer moments up to $N=14$~\cite{bierenbaum:OMEmellin}, converted into the $\overline{\mathrm{MS}}$ scheme, with the even moments corresponding to the ($+$) non-singlet distribution.

As for $A_{Hg}^{(3)}$, the approximation is performed using
the set of functions,
\begin{alignat}{5}\label{eq: ansComb}
    f_{1}(x) \quad &= \quad \ln x, \qquad f_{2}(x) \quad = \quad \ln^{2}x, \nonumber \\
    f_{3,4}(x) \quad &= \quad 1 \quad \text{or} \quad x \quad \text{or} \quad x^{2} \quad \text{or} \quad \ln(1-x),\nonumber \\
    f_{5}(x) \quad &= \quad 1/x, \quad f_{6}(x) \quad = \quad \ln^{3}(1-x), \quad f_{7}(x) \quad = \quad \ln^{2}(1-x), \nonumber \\
    f_{e}(x, a_{qq,H}^{\mathrm{NS}}) \quad &= \quad a_{qq,H}^{\mathrm{NS}}\ \ln^{3}x
\end{alignat}
where $a_{qq,H}^{\mathrm{NS}}$ is varied as $-90 < a_{qq,H}^{\mathrm{NS}} < -37$. 
To contain this variation in a fixed functional form we employ:
\begin{multline}
    A_{qq,H}^{\mathrm{NS},\ (3)\ +} = A_{1}\ \frac{1}{(1-x)_{+}} + A_{2}\ \ln^{3}(1-x) + A_{3}\ \ln^{2}(1-x) + A_{4}\ \ln(1-x) + A_{5} \\+ A_{6}\ x + A_{7}\ \ln^{2} x + a_{qq,H}^{\mathrm{NS}}\ \ln^{3} x
\end{multline}
where the variation of $a_{qq,H}^{\mathrm{NS}}$ is unchanged.

\subsection*{$A_{gq, H}$}\label{subsec: AgqH}

The 3-loop $A_{gq, H}$ function has been calculated exactly in \cite{ablinger:agq}. As with the $A_{Hq}^{\mathrm{PS}}$ function above, we attempt to provide a simple and computationally efficient approximation to this exact form. To do this, we use the known even-integer moments (converted to the $\overline{\mathrm{MS}}$ scheme) and small and large-$x$ information from \cite{ablinger:agq, bierenbaum:OMEmellin}. Gathering a fixed set of functions $f_{i}(x)$ and omitting any variational parameter $a_{gq,H}$, due to the higher amount of information available, the resulting approximation to the $\overline{\mathrm{MS}}$ $A_{gq, H}^{(3)}$ is:
\begin{multline}
A_{gq, H}^{(3)} = -237.172\ \ln^{3}(1-x) - 201.497\ \ln^{2}(1-x) + 7247.70\ \ln(1-x) + 39967.3\ x^{2} \\- 22017.7 - 28459.1\ \ln x - 14511.5\ \ln^{2} x \\+ 341.543\ \frac{\ln 1 / x}{x} + 1814.73\ \frac{1}{x} - \frac{580}{243}\ \ln^{4}(1 - x) - \frac{17624}{729}\ \ln^{3}(1 - x) \\- 135.699\ \ln^{2}(1 - x)
\end{multline}
where the first two lines have been approximated and the last two lines are the exact small and large-$x$ limits.

\subsection*{$A_{gg, H}$}\label{subsec: AggH}

Work is ongoing for the 3-loop contribution to $A_{gg, H}$~\cite{gluonContrib, gluonContrib2}. Due to this, the entire approximation of $A_{gg, H}^{(3)}$ presented here is based on the first 5 even-integer Mellin moments~\cite{bierenbaum:OMEmellin}. To reduce the wild behaviour of this approximation from only using the Mellin moment information (converted into the $\overline{\mathrm{MS}}$ scheme), we introduce a second mild constraint in the form of the relations in Equation~\eqref{eq: Relations}. These relations are closely followed by the gluon-gluon functions up to NNLO, but there is no guarantee that this behaviour will continue at N$^{3}$LO. This constraint is given as,
\begin{equation}\label{eq: agg_smallx}
A_{gg,H}(x\rightarrow0) \simeq \frac{C_{A}}{C_{F}}A_{gq, H}(x\rightarrow 0).
\end{equation}
It can be expected that even though this relation may not be followed exactly, it should not stray too far from this general `rule of thumb'. Due to this a generous contingency of $\pm 50\%$ is allowed when using this rule. Furthermore, to ensure this relation is only used as a guide, we allow the variation to move beyond this rule as long as the criteria in Section~\ref{subsec: genframe} are still satisfied. As a result of this change in prescription and because the allowed variation is now on a much larger scale than that of any functional uncertainty, we choose a fixed functional form from the start and use the criteria described above to guide our choice of variation.
\begin{multline}
    A_{gg,H}^{(3)} = A_{1}\ \ln^{2}(1-x) + A_{2}\ \ln(1-x) + A_{3}\ x^{2} + A_{4}\ \ln x + A_{5}\ x + a_{gg,H}\ \frac{\ln x}{x}
\end{multline}
where $-2000 < a_{gg,H} < -700$.

\subsection{Predicted \texorpdfstring{aN$^{3}$LO}{aN3LO} Transition Matrix Elements}\label{subsec: expansion_OME}

\begin{figure}[t]
\begin{center}
\includegraphics[width=0.97\textwidth]{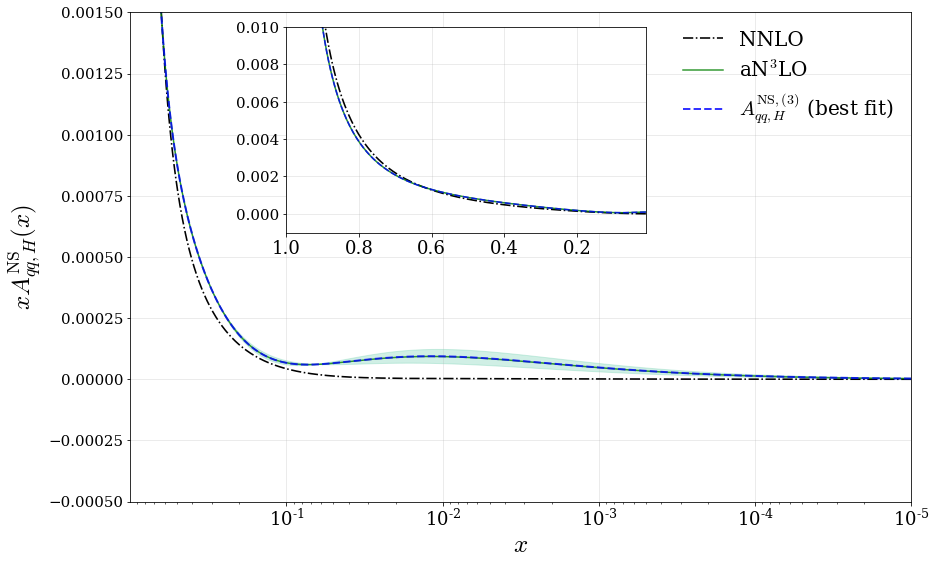}
\end{center}
\caption{\label{fig: OME_variation_ns}Perturbative expansions for the transition matrix element $A^{\mathrm{NS}}_{qq,\ H}$ including any corresponding allowed $\pm 1\sigma$ variation (shaded green region). This function is shown at the mass threshold value of $\mu = m_{h}$. The best fit value (blue dashed line) displays the prediction for this function determined from a global PDF fit.}
\end{figure}
\begin{figure}
\begin{center}
\includegraphics[width=0.97\textwidth]{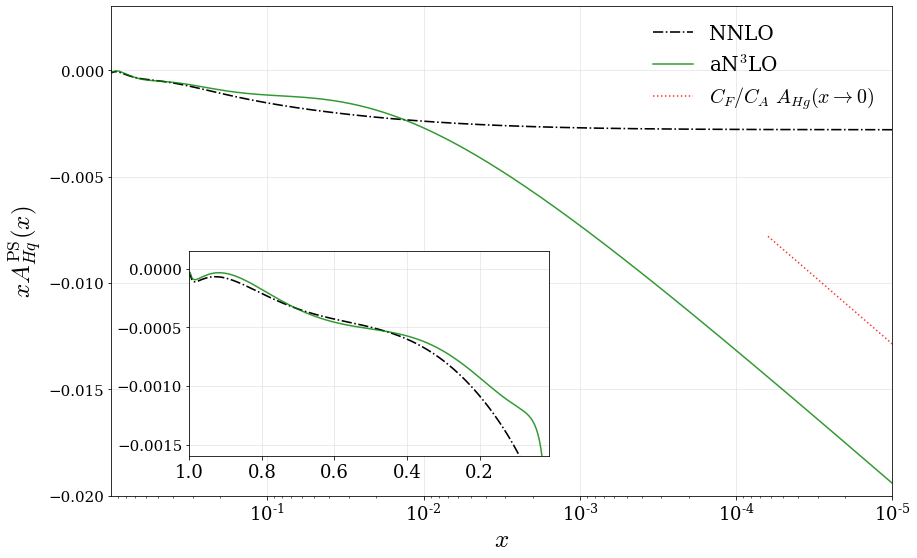}
\includegraphics[width=0.97\textwidth]{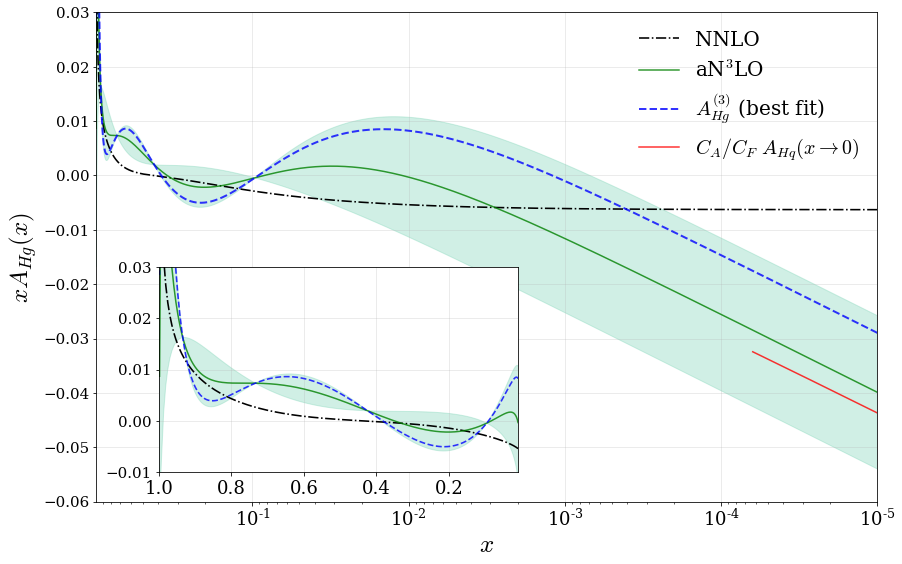}
\end{center}
\caption{\label{fig: OME_variation_H}Perturbative expansions for the transition matrix elements $A^{\mathrm{PS}}_{Hq}$ and $A_{Hg}$ including any corresponding allowed $\pm 1\sigma$ variation (shaded green region). These functions are shown at the mass threshold value of $\mu = m_{h}$. The best fit values (blue dashed line) display the predictions for these functions determined from a global PDF fit.}
\end{figure}
\begin{figure}
\begin{center}
\includegraphics[width=0.97\textwidth]{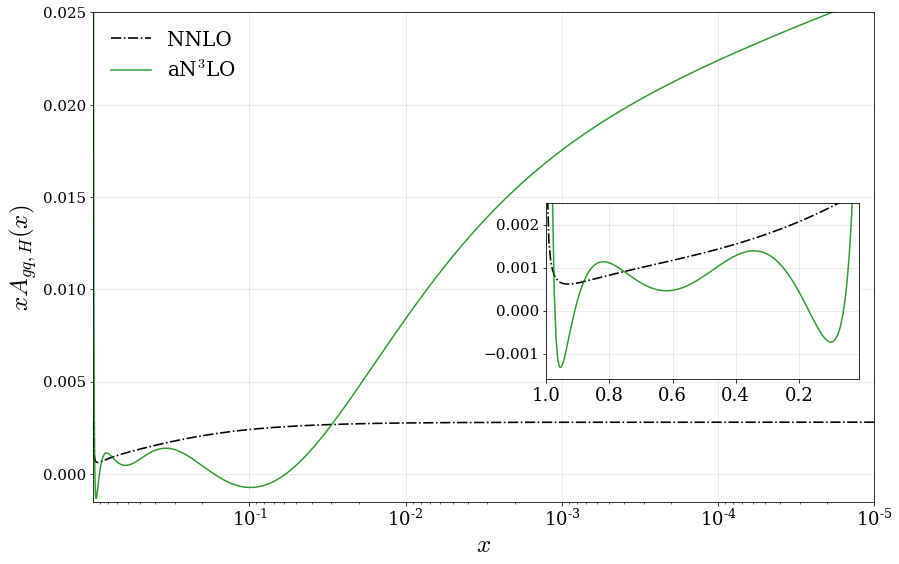}
\includegraphics[width=0.97\textwidth]{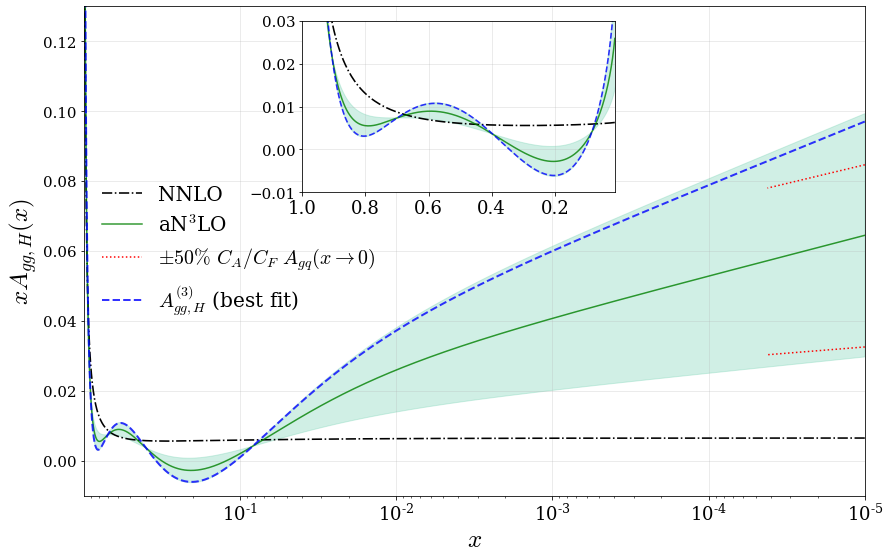}
\end{center}
\caption{\label{fig: OME_variation_gluon}Perturbative expansions for the transition matrix elements $A_{gq,H}$ and $A_{gg,H}$ including any corresponding allowed $\pm 1\sigma$ variation (shaded green region). These functions are shown at the mass threshold value of $\mu = m_{h}$. The best fit values (blue dashed line) display the predictions for these functions determined from a global PDF fit.}
\end{figure}
Fig.'s~\ref{fig: OME_variation_ns}, \ref{fig: OME_variation_H} and \ref{fig: OME_variation_gluon} show the perturbative expansions for each of the $n_{f}$-independent contributions to the transition matrix elements at the mass threshold value of $\mu = m_{h}$. Included with these expansions are the predicted variations ($\pm 1 \sigma$) from Section~\ref{subsec: 3loop_OME} (shown in green) and the approximate N$^{3}$LO best fits (shown in blue - discussed further in Section~\ref{sec: results}).

$A_{qq,H}^{\mathrm{NS}}$ in Fig.~\ref{fig: OME_variation_ns} behaves as expected with little variation from NNLO until the magnitude of this function is very small. The approximations for the more dominant $A^{\mathrm{PS}}_{Hq}$ and $A_{Hg}$ functions in Fig.~\ref{fig: OME_variation_H} exhibit some slight sporadic behaviour towards large-$x$ due to the increased logarithmic influence. However, since this is in a region where the magnitude of these functions become small, any instabilities will have a minimal effect on the overall result. The major feature prevalent across both these functions is the large deviation away from the NNLO behaviour, especially at small-$x$ (and also mid-$x$ for $A_{Hg}$).

Similarly for $A_{gq,H}$ in Fig.~\ref{fig: OME_variation_gluon} (upper), we see some irregular behaviour towards large-$x$. As with $A_{Hq}^{\mathrm{PS}}$ and $A_{Hg}$, this behaviour is in a region where the magnitude of $A_{gq,H}$ is small. As discussed in Section~\ref{subsec: 3loop_OME}, $A_{gq, H}^{(3)}$ is approximated without any variation due to the range of available information being large\footnote{Although an exact expression has been calculated for $A_{gq, H}^{(3)}$~\cite{ablinger:agq}, this function is not yet available in a computationally efficient format i.e. numerical grids.}. Due to this, and the fact that the region of potential instability (large-$x$) is highly suppressed, we can accept this function with negligible effect on any results. As more information becomes available about all these functions, it will be interesting to observe how the behaviour across $x$ changes.

The $A_{gg,H}$ function shown in Fig.~\ref{fig: OME_variation_gluon} (lower) displays the $\pm 50\%$ bounds of violation we allow for the relation Equation~\eqref{eq: Relations}. It follows that the allowed variation is conservative enough to include a generous violation of Equation~\eqref{eq: Relations} at N$^{3}$LO, with the prediction that the function is positive at small-$x$. This is an area where small-$x$ information would clearly be very beneficial. With this information currently in progress, it will be very interesting to compare how well this variation captures the true small-$x$ $A_{gg,H}$ behaviour.

The final best fit values shown in Fig.'s~\ref{fig: OME_variation_ns}, \ref{fig: OME_variation_H} and \ref{fig: OME_variation_gluon} are determined from a global PDF fit with various datasets seen to be constraining these functions within the $\pm\ 1 \sigma$ variations. As observed, we are able to show good agreement between the allowed variations and the best fit predictions. The perturbative expansion predicted for $A_{gg, H}$ is the least well constrained while also violating its expected relation with $A_{gq,H}$ more than one may originally expect. Since the small-$x$ region in all cases changes dramatically at N$^{3}$LO, one potential explanation is that this function is compensating for an inaccuracy in another area of the theory. However, when comparing with the relationship between $A_{Hg}$ and $A_{Hq}^{\mathrm{PS}}$, Equation~\eqref{eq: Relations} also exhibits a significant violation at this order. This could suggest that for the N$^{3}$LO transition matrix elements, this relation may not be the best indicator of precision or consistency. Finally, we remember that the best fit in this case may be feeling a larger effect from higher orders, especially due to these functions only existing from NNLO. For example, in Section~\ref{subsec: expansion_split} we observed a high level of divergence introduced at 4-loops in the splitting functions. The best fit results shown here may therefore be sensitive to a similar level of divergence further along in their corresponding perturbative expansions.

As previously discussed, this lack of knowledge is contained within our choice of the predicted variations of these functions. Therefore this treatment only seeks to add to the predicted level of theoretical uncertainty from missing N$^{3}$LO contributions, as one expects.

\subsection{Numerical Results}\label{subsec: num_res_OME}

\begin{figure}
\begin{center}
\includegraphics[width=0.49\textwidth]{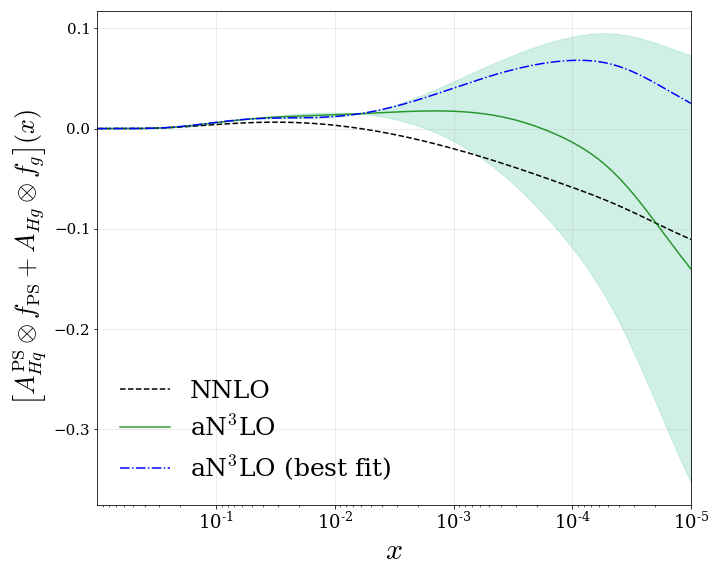}
\includegraphics[width=0.49\textwidth]{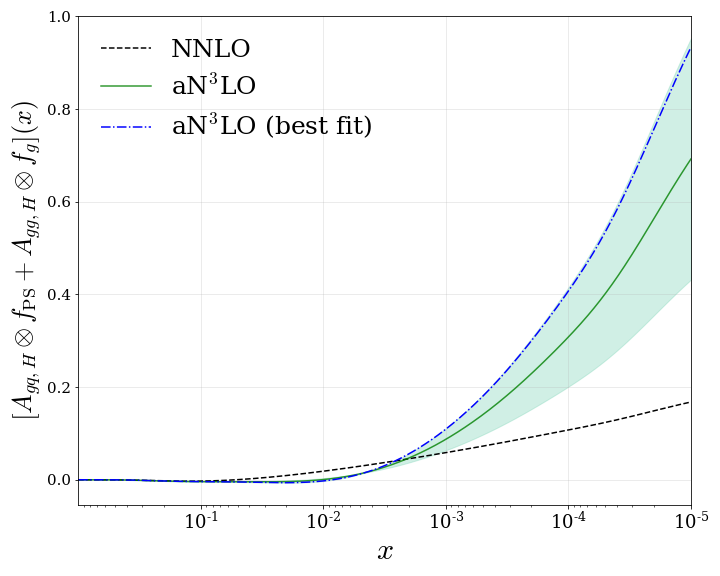}
\end{center}
\caption{\label{fig: step_results}Heavy flavour evolution contributions to the heavy quark ($H + \overline{H}$ (left)) and gluon (right) PDFs provided at $\mu \simeq 30\ \mathrm{GeV}^{2}$. These results include the $\mu = m_{h}$ contributions from $A^{\mathrm{PS}}_{Hq}$, $A_{Hg}$, $A_{gq,H}$ and $A_{gg,H}$ transition matrix elements up to aN$^{3}$LO.}
\end{figure}
For these results, the same toy PDFs presented in Section~\ref{subsec: num_res_split} are employed which approximate the general order-independent PDF features at $Q^{2} \simeq 30\ \mathrm{GeV}^{2}$. Note that due to the higher $Q^{2}$, these results are more representative of the b-quark. The left plot in Fig.~\ref{fig: step_results} shows the result of including the N$^{3}$LO transition matrix element approximations we have determined into Equation~\eqref{eq: OME_fh}, which is describing the heavy quark distribution $(H + \overline{H})(x, Q^{2} = m_{h}^{2})$. The right plot in Fig.~\ref{fig: step_results} is describing the heavy flavour contribution to the gluon at $(x, Q^{2} = m_{h}^{2})$ in Equation~\eqref{eq: OME_fg} where the delta function describing the leading order contribution to $A_{gg,H}$ has been subtracted out. The dominant contribution to the heavy quark (left plot) is stemming from the $A_{Hg}$ function. Whereas the dominant contribution to the gluon (right plot) is from the $A_{gg,H}$ function. As one might expect, the predictions at N$^{3}$LO are more divergent at small-$x$, however it is also true that the general trend from NNLO is being followed across most values of $x$. 

The best fit functions predicted from a global fit show the preferred aN$^{3}$LO contributions for both scenarios. The predicted behaviour from the global fit follows the results for the perturbative expansions in Section~\ref{subsec: expansion_OME}. For the $(H + \overline{H})(x, Q^{2} = m_{h}^{2})$ result (Fig.~\ref{fig: step_results} left), the aN$^{3}$LO result is positive across a much wider range of $x$. Since this is a perturbatively calculated PDF, this is an encouraging result that could potentially eliminate some of the more unphysical shortcomings at NNLO without demanding positivity of the PDF a priori.

\section{\texorpdfstring{N$^{3}$LO}{N3LO} Heavy Coefficient Functions}\label{sec: n3lo_coeff}

The final set of functions considered are the Neutral Current (NC) DIS coefficient functions which, when combined with the PDFs, form the structure functions discussed in Section~\ref{sec: structure}.\footnote{Charged current (CC) structure function data 
is limited to relatively high-$x$ values compared to NC data and is either comparatively low statistics, high-$Q^2$ proton target data from HERA or nuclear target data (again often quite low statistics) on heavy nuclear targets. In both cases the effect of N$^3$LO corrections is small compared with uncertainties, especially when considering those involved with nuclear corrections. Also, heavy flavour contributions are less well known at high orders for CC structure functions. Hence, we do not include
N$^3$LO for these processes, except dimuon data, which is particularly important for the poorly constrained strange quark, but which is a semi-inclusive DIS process, and for which we parameterise N$^3$LO corrections, as discussed in Section~\ref{sec: n3lo_K}. An improvement would be necessary for more precise proton data, from the EIC for example.} We approximate the N$^{3}$LO heavy quark coefficient functions which accompany the heavy flavour transition matrix elements from Section~\ref{sec: n3lo_OME} and also the N$^{3}$LO light quark coefficient functions. We note that our standard definition of the order of coefficient functions includes the longitudinal coefficient functions at order $\alpha_s$ at LO, at order $\alpha_s^2$ at NLO etc. This means we already include order $\alpha_s^3$ coefficient functions for the longitudinal coefficient functions at NNLO, whereas many groups only consider order $\alpha_s^2$ at NNLO. Since little is know about longitudinal coefficient functions at order $\alpha_s^4$, and the data constraints from $F_L(x,Q^2)$ are very much less precise than from $F_2(x,Q^2)$, we simply remain at the precisely known order $\alpha_s^3$ in this study.

\subsection{Approximation Framework: Continuous Information}\label{subsec: genframe_continuous}

In Sect.~\ref{subsec: genframe} we described the approximation framework employed for functions with discrete Mellin moment information, combined with any available exact information. For the N$^{3}$LO coefficient function approximations, we have access to a somewhat richer vein of information than the discrete moments discussed for the framework used in approximating the N$^{3}$LO splitting functions and transition matrix elements in Section's~\ref{sec: n3lo_split} and \ref{sec: n3lo_OME}. More specifically, approximations of the FFNS coefficient functions at $\mathcal{O}(\alpha_{s}^{3})$ are known for the heavy quark contributions to the heavy flavour structure function $F_{2,H}(x, Q^{2})$ at  $Q^{2} < m_{c,b}^{2}$~\cite{Catani:FFN3LO1,Laenen:FFN3LO2,Vogt:FFN3LO3}. These approximations include the exact LL and mass threshold contributions, with an approximated NLL term (the details of this are described in Section~\ref{subsec: NLL_coeff}). Furthermore, the N$^{3}$LO ZM-VFNS coefficient functions are known exactly~\cite{Vermaseren:2005qc}. Both of these contributions can then be combined with the transition matrix element approximations to define the GM-VFNS functions in the $Q^{2} \leq m_{c}^{2}, m_{b}^{2}$ and $Q^{2} \rightarrow \infty$ regimes. Due to this, we base our approximations for the $C_{H,\{q,g\}}^{(3)}$ functions on the known continuous information in the low and high-$Q^{2}$ regimes.

To achieve a reliable approximation for $C_{H,\{q,g\}}^{(3)}$, we first fit a regression model with a large number of functions in $(x,Q^{2})$ space made available to the model (in order to reduce the level of functional bias in the parameterisation). This produces an unstable result at the extremes of the parameterisation (large-$x$ and low-$Q^{2}$). However, it provides a basis for manually choosing a stable parameterisation to move between the two known regimes (low-$Q^{2}$ and high-$Q^{2}$).

Using the regression model predictions as a qualitative guide, we choose a stable and smooth interpolation between the two $Q^{2}$ regimes (low-$Q^{2}$ and high-$Q^{2}$) as given in Equation~\eqref{eq: ChqFF_param}. This interpolation is observed to mirror the expected behaviour observed from lower orders, the regression model qualitative prediction having been calculated independently of lower orders and the best fit quality to data. By definition, we also ensure an exact cancellation between the coefficient functions and the transition matrix elements at the mass threshold energies as demanded by the theoretical description in Section~\ref{sec: structure}.

For the contributions to the heavy flavour structure function $F_{2,H}$ the final interpolations in the FFNS regime are defined as,
\begin{multline}\label{eq: ChqFF_param}
	C_{H,\ \{q,g\}}^{\mathrm{FF},\ (3)} = 
\begin{cases}
	C_{H,\ \{q,g\},\ \text{low-$Q^{2}$}}^{\mathrm{FF},\ (3)}(x, Q^{2} = m_{h}^{2})\ e^{0.3\ (1-Q^{2}/m_{h}^{2})} \\ \qquad\qquad\qquad\qquad +\ C_{H,\ \{q,g\}}^{\mathrm{FF},\ (3)}(x, Q^{2} \rightarrow \infty)\big(1 - e^{0.3\ (1-Q^{2}/m_{h}^{2})}\big) &,\ \text{if}\ Q^{2} \geq m_{h}^{2}, \\ \\
	C_{H,\ \{q,g\},\ \text{low-$Q^{2}$}}^{\mathrm{FF},\ (3)}(x, Q^{2})&,\ \text{if}\ Q^{2} < m_{h}^{2}.
\end{cases}
\end{multline}
where $C_{H,\ \{q,g\},\ \text{low-$Q^{2}$}}^{\mathrm{FF},\ (3)}$ are the already calculated approximate heavy flavour FFNS coefficient functions at $Q^{2} \leq m_{h}^{2}$, and $C_{H,\ \{q,g\}}^{\mathrm{FF},\ (3)}(Q^{2} \rightarrow \infty)$ is the limit at high-$Q^{2}$ found from the known ZM-VFNS coefficient functions and relevant subtraction terms, themselves found from Equation~\eqref{eq: fullN3LO_H}. Both of these limits will be discussed in detail on a case-by-case basis in Section~\ref{sec: n3lo_coeff}.

For the heavy flavour contributions to $F_{2,q}$, we have no information about the low-$Q^{2}$ N$^{3}$LO FFNS coefficient functions. In this case, we use intuition from lower orders to provide a soft (lightly weighted) low-$Q^{2}$ target for our regression model in $(x,Q^{2})$.
However, since the overall contribution is very small from these functions, the exact form of these functions is not phenomenologically important at present. Further to this, our understanding from lower orders is that these functions have a weak dependence on $Q^{2}$ and so the form of the low-$Q^{2}$ description is even less important. As with the $C_{H,\ \{q,g\}}^{(3)}$ coefficient functions, the regression results provide an initial qualitative guide which exhibits instabilities in the extremes of $(x,Q^{2})$. We therefore employ a similar technique as before to ensure a smooth extrapolation across all $(x, Q^{2})$ into the unknown behaviour at low-$Q^{2}$. For these functions, the ansatz used is given as,
\begin{multline}\label{eq: CqFF_param}
	\qquad C_{q,\ \{q,g\}}^{\mathrm{FF},\ (3)} = 
\begin{cases}
	C_{q,\ q}^{\mathrm{FF},\ \mathrm{NS},\ (3)}(x, Q^{2} \rightarrow \infty)\big(1 + e^{-0.5\ (Q^{2}/m_{h}^{2}) - 3.5}\big), \qquad  \qquad \qquad \\ \\
	C_{q,\ q}^{\mathrm{FF},\ \mathrm{PS},\ (3)}(x, Q^{2} \rightarrow \infty)\big(1 - e^{-0.25\ (Q^{2}/m_{h}^{2}) - 0.3}\big),\qquad \qquad \qquad \\ \\
	C_{q,\ g}^{\mathrm{FF},\ (3)}(x, Q^{2} \rightarrow \infty)\big(1 - e^{-0.05\ (Q^{2}/m_{h}^{2}) + 0.35}\big), \qquad \qquad \qquad
\end{cases}
\end{multline}
where $C_{q,\{q,g\}}^{\mathrm{FF},\ (3)}(x, Q^{2} \rightarrow \infty)$ is the known limit at high-$Q^{2}$.

\subsection{Low-\texorpdfstring{$Q^{2}$}{Q2} \texorpdfstring{N$^{3}$LO}{N3LO} Heavy Flavour Coefficient Functions}\label{subsec: NLL_coeff}

As previously mentioned in Section~\ref{sec: structure}, the standard MSHT theoretical description of NNLO structure functions includes approximations to the low-$Q^{2}$ FFNS coefficient functions $C_{H,\{q,g\}}^{(3),\mathrm{FF}}$ from~\cite{Catani:FFN3LO1,Laenen:FFN3LO2,Vogt:FFN3LO3}. Within these functions are the precisely known LL small-$x$ terms and mass threshold information, along with an approximate NLL small-$x$ term added into the MSHT fit. In the NNLO fit these approximate NLL parameters play a very small role due to not only being sub-leading, but also only affecting the FFNS scheme below the mass thresholds. At NNLO they are therefore heuristically set to a value that is theoretically justified and suits the NNLO best fit. At N$^{3}$LO these functions begin to directly affect the form of the full GM-VFNS scheme across all $(x, Q^{2})$. For this reason, these NLL parameters need to be considered as an independent source of theoretical uncertainty. In the aN$^{3}$LO fit, the NLL parameters are left free and included into the framework set out in Section~\ref{subsec: hessian_method}.

The standard NNLO MSHT fit contains terms of the form,
\begin{equation}
	C_{H,i}^{(3),\ \mathrm{NLL}}(Q^{2} \rightarrow 0) \propto	 -4\ \frac{1}{x}\ + c_{i}^{\mathrm{LL}}\ \frac{\ln 1/x}{x}, \qquad \quad \left(c_{g}^{\mathrm{LL}} = \frac{C_{F}}{C_{A}}\ c_{q}^{\mathrm{LL}}\right),
\end{equation}
where $i = q, g$ and $c_{i}^{\mathrm{LL}}$ is the precisely known leading small-$x$ log coefficient.
In the aN$^{3}$LO fit, the NLL coefficient is allowed to vary by $\pm 50\%$ ($\pm 1\sigma$ variation). This conservative range is chosen to enable the release of tension with the variational parameters associated with the N$^{3}$LO transition matrix elements. Here we stress that this quantity is heuristically set even at NNLO, therefore our treatment is completely justified with the added benefit of now accounting for an uncertainty for this choice.

\subsection{3-loop Approximations}\label{subsec: 3loop_coeff}

\subsection*{$C_{H, q}$}\label{subsec: Chq}

In this section the $C_{H, q}$ coefficient function is investigated. As discussed in Section~\ref{sec: structure}, $C_{H, q}$ contributes to the heavy flavour structure function $F_{2, H}$. We begin by isolating this function from Equation~\eqref{eq: fullN3LO_H} and relating the FFNS and GM-VFNS schemes at all orders from Equation~\eqref{eq: FFexpanse} and Equation~\eqref{eq: expanse_H},
\begin{multline}\label{eq: general_Cq}
C_{H, q}^{\mathrm{FF}}= \left[C_{H, H}^{\mathrm{VF},\ \mathrm{N S}} +C_{H, H}^{\mathrm{VF},\ \mathrm{PS}}\right] \otimes A_{H q}^{\mathrm{P S}} \ +\ C_{H, q}^{\mathrm{VF}} \otimes \left[A_{q q, H}^{\mathrm{NS}} + A_{qq, H}^{\mathrm{PS}}\right]\ +\ C_{H, g}^{\mathrm{VF}} \otimes A_{g q, H}.
\end{multline}
Expanding this function we obtain:
\begin{align}
\mathcal{O}(\alpha_{s}): \hspace{1cm}
C_{H,q}^{\mathrm{FF},\ (1)} = 0
\label{eq: alpha1}
\end{align}
\begin{align}
\mathcal{O}(\alpha_{s}^{2}): \hspace{1cm} & 
\begin{multlined}[t]
C_{H,q}^{\mathrm{FF},\ (2)} = C_{H,H}^{\mathrm{VF},\ (0)} \otimes A_{Hq}^{PS,\ (2)} + C_{H, q}^{\mathrm{VF},\ (2)} \otimes A_{qq, H}^{\mathrm{NS},\ (0)}
\end{multlined}
\label{eq: alpha2}
\end{align}
\begin{align}
\mathcal{O}(\alpha_{s}^{3}): \hspace{1cm} & 
\begin{multlined}[t]
C_{H,q}^{\mathrm{FF},\ (3)} =  C_{H, H}^{\mathrm{VF},\ (1)}\otimes A_{Hq}^{\mathrm{PS},\ (2)} + C_{H, H}^{\mathrm{VF},\ (0)}\otimes A_{Hq}^{\mathrm{PS},\ (3)} \\+\ C_{H, q}^{\mathrm{VF},\ (3)}\otimes A_{qq,H}^{\mathrm{NS},\ (0)} + C_{H, g}^{\mathrm{VF},\ (1)} \otimes A_{gq, H}^{(2)}
\end{multlined}
\label{eq: alpha3}
\end{align}
where we recall that $A_{qq,H}^{\mathrm{NS},\ (0)} = \delta (1 - x)$.

\subsection*{NNLO}

\begin{figure}
\centering
\includegraphics[width=\textwidth]{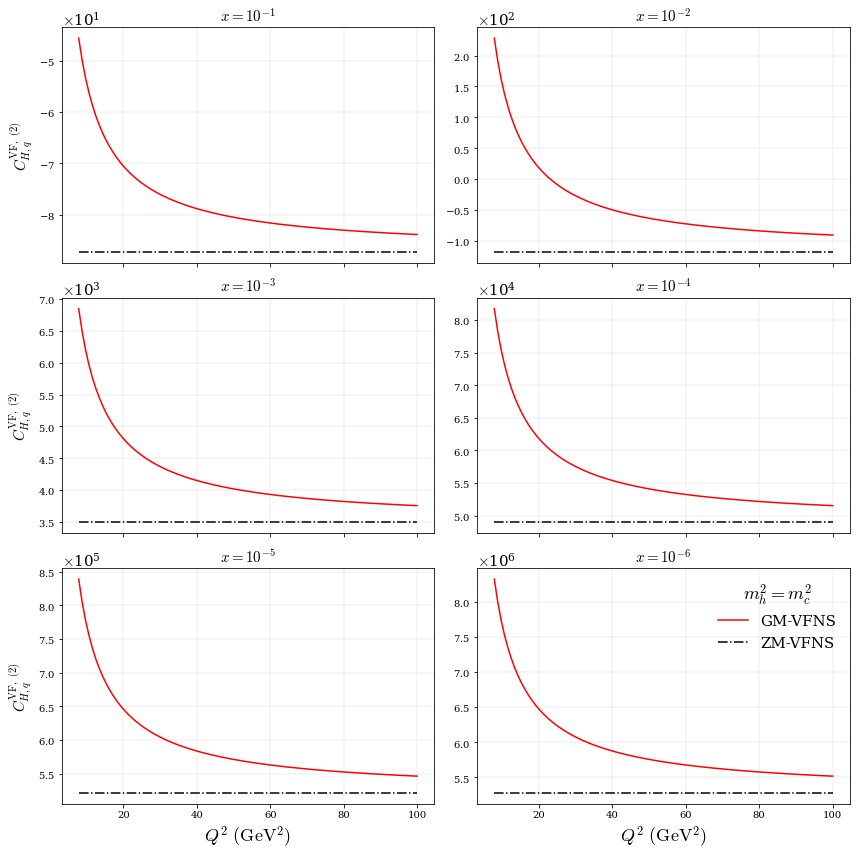}
\caption{\label{fig: NNLO_Cq_toZM} The NNLO GM-VFNS function $C_{H, q}^{\mathrm{VF},\ (2)}$ compared with the NNLO ZM-VFNS function $C_{H, q}^{\mathrm{ZM},\ (2)}$ across a variety of $x$ and $Q^{2}$ values. Mass threshold is set at the charm quark level ($m_{h}^{2} = m_{c}^{2} = 1.4\ \text{GeV}^{2}$).}
\end{figure}

The first contribution from the heavy quarks appears at the $\mathcal{O}(\alpha_{s}^{2})$ level. Fortunately there is a complete picture of this order~\cite{Thorne:GMVFNNLO} which provides some experience with the behaviour of these functions before moving into unknown territory. Fig.~\ref{fig: NNLO_Cq_toZM} shows the case for $C_{H, q}^{\mathrm{VF},\ (2)}$ converging onto $C_{H, q}^{\mathrm{ZM},\ (2)}$ at high-$Q^2$, as required by the definition of the GM-VFNS scheme outlined in Section \ref{sec: structure}.

From Fig.~\ref{fig: NNLO_Cq_toZM}, immediately some intuition can be built up surrounding the form of these functions. It can be observed that the GM-VFNS function at low-$Q^{2}$ is consistently more positive than at high-$Q^{2}$. However, the values at low and high-$Q^{2}$ are of the same order of magnitude which provides evidence that the behaviour should not be substantially different across values of $Q^{2}$ when estimating our N$^{3}$LO quantities.
Further to this, as $x \rightarrow 0$ the overall magnitude of $C_{H,q}^{(2)}$ becomes much larger, which is consistent with an inherently pure singlet quantity.

\subsection*{N$^{3}$LO}

At $\mathcal{O}(\alpha_{s}^{3})$ the N$^{3}$LO ZM-VFNS and low-$Q^{2}$ FFNS functions are known~\cite{Vermaseren:2005qc,Catani:FFN3LO1,Laenen:FFN3LO2,Vogt:FFN3LO3} and parameterisations/approximations are available (up to the level of precision discussed in Section~\ref{subsec: NLL_coeff}). Nevertheless, there is no direct information on how the full GM-VFNS function behaves at this order which is required for a full treatment of the heavy flavour coefficients. Using Equation~\eqref{eq: alpha3} to estimate the N$^{3}$LO contribution, we have
\begin{equation}\label{eq: ChqGM-VFNS}
	C_{H, q}^{\mathrm{VF},\ (3)} = C_{H,q}^{\mathrm{FF},\ (3)} - C_{H, H}^{\mathrm{VF},\ (1)}\otimes A_{Hq}^{\mathrm{PS},\ (2)} - C_{H, g}^{\mathrm{VF},\ (1)} \otimes A_{gq, H}^{(2)} - A_{Hq}^{\mathrm{PS},\ (3)}.
\end{equation}
where $A_{Hq}^{\mathrm{PS},\ (3)}$ is the N$^{3}$LO transition matrix element approximated in Section~\ref{subsec: 3loop_OME}.

It must be the case that the discontinuities introduced into the heavy flavour PDF from the transition matrix elements (at the threshold value of $Q^{2} = m_{h}^{2}$) are cancelled exactly in the structure function. The cancellation of $A_{Hq}^{\mathrm{PS},\ (3)}$ is therefore guaranteed by its inclusion into the GM-VFNS coefficient function in Equation~\eqref{eq: ChqGM-VFNS}. Since in practice the transition matrix elements are convoluted with the PDFs separately to the coefficient functions, to ensure that this statement remains the case, the parameterisation will be performed in the FFNS number scheme. By doing this, we can explicitly switch to the GM-VFNS number scheme by including the subtraction term in Equation~\eqref{eq: ChqGM-VFNS}. This procedure then ensures that $A_{Hq}^{\mathrm{PS},\ (3)}$ is subtracted off exactly with no unphysical discontinuity.

Following the methodology set out in Section~\ref{subsec: genframe_continuous}, the two regimes we wish to interpolate between are the approximate $C_{H, q}^{\mathrm{FF},\ (3)}(Q^{2} \rightarrow 0)$ limit and
\begin{multline}\label{eq: ChqFF}
	C_{H, q}^{\mathrm{FF},\ (3)}(Q^{2} \rightarrow \infty) = C_{H,q}^{\mathrm{ZM},\ (3)} + C_{H, H}^{\mathrm{VF},\ (1)}\otimes A_{Hq}^{\mathrm{PS},\ (2)} + C_{H, g}^{\mathrm{VF},\ (1)} \otimes A_{gq, H}^{(2)} + A_{Hq}^{\mathrm{PS},\ (3)},
\end{multline}
where $C_{H,q}^{\mathrm{VF},\ (3)}$ is replaced with $C_{H,q}^{\mathrm{ZM},\ (3)}$ in the high-$Q^{2}$ limit. Equation~\eqref{eq: ChqFF_param} is then stable across all $(x, Q^{2})$, exactly cancelling any discontinuity that would violate the RG flow, whilst also demanding that the known FFNS approximation (for $Q^{2} < m^{2}_{h}$) is followed\footnote{Since in practice the discontinuities from the transition matrix elements are added to PDFs regardless of what order coefficient function they are convoluted with, discontinuities of even higher order (e.g. $\alpha_{s}^{4}$ and beyond) are also present in calculations. Because the order $\alpha_{s}^{3}$ matrix elements are large these even higher order discontinuities are not insignificant. Therefore we add the same contributions to the unknown FFNS contributions below $m_{h}^{2}$ to impose continuity on structure functions. Such corrections are extremely small, except right at the transition point where they eliminate minor unphysical discontinuities.}.

\begin{figure}
\centering
\includegraphics[width=\textwidth]{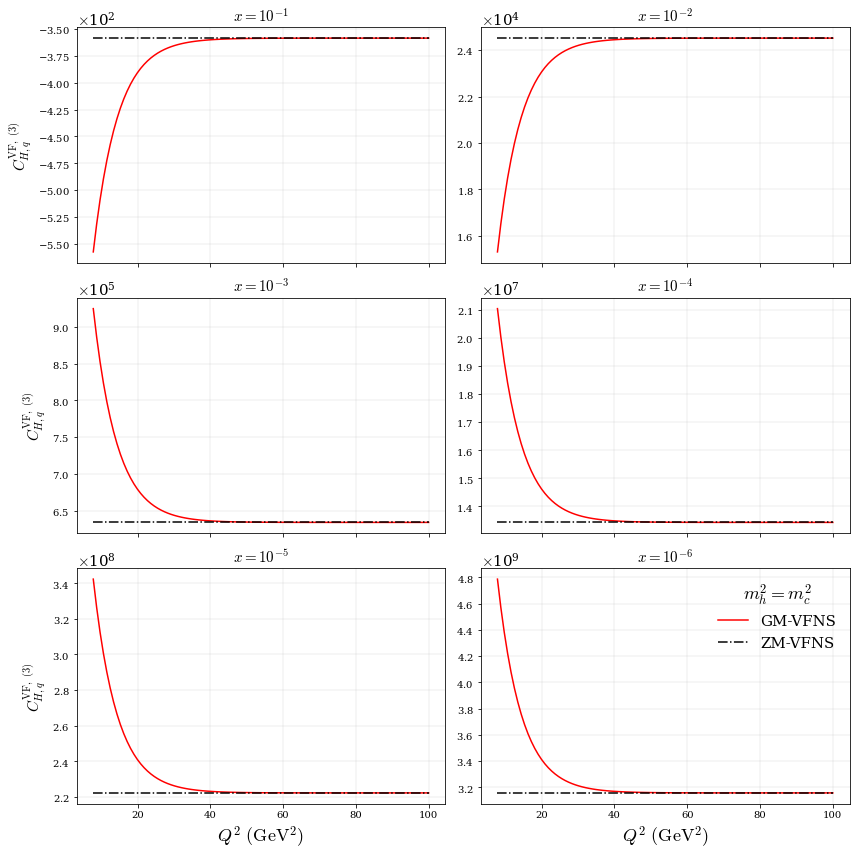}
\caption{\label{fig: GM-VFNS_estimation}N$^{3}$LO GM-VFNS function $C_{H, q}^{\mathrm{VF},\ (3)}$ compared with the N$^{3}$LO ZM-VFNS function $C_{H, q}^{\mathrm{ZM},\ (3)}$ across a variety of $x$ and $Q^{2}$ values (shown without the variation from the low-$Q^{2}$ NLL term discussed in Section~\ref{subsec: NLL_coeff}). $C_{H, q}^{\mathrm{VF},\ (3)}$ is parameterised via Equations~\eqref{eq: ChqGM-VFNS}, \eqref{eq: ChqFF} and \eqref{eq: ChqFF_param}. Mass threshold is set at the charm quark level ($m_{h}^{2} = m_{c}^{2} = 1.4\ \text{GeV}^{2}$).}
\end{figure}
Fig.~\ref{fig: GM-VFNS_estimation} shows the result of estimating $C_{H, q}^{\mathrm{VF},\ (3)}$ using the above approximation for $C_{H, q}^{\mathrm{FF},\ (3)}$ and the relevant subtraction term from Equation~\eqref{eq: ChqGM-VFNS}. Note that this plot ignores any variation from the low-$Q^{2}$ NLL term discussed in Section~\ref{subsec: NLL_coeff}, where this is fixed to its central value.

\subsection*{$C_{H, g}$}\label{subsec: Chg}

As with $C_{H,q}$, using Equation~\eqref{eq: FFexpanse} and Equation~\eqref{eq: expanse_H} to isolate $C_{H, g}$ and relate the FFNS and GM-VFNS schemes,
\begin{multline}
C_{H,g}^{\mathrm{FF}}= C_{H,g}^{\mathrm{VF}} \otimes A_{gg, H} + C_{H,q}^{\mathrm{VF},\ \mathrm{PS}} \otimes A_{qg, H} + \left[C_{H,H}^{\mathrm{VF},\ \mathrm{NS}}+C_{H,H}^{\mathrm{VF},\ \mathrm{PS}}\right] \otimes A_{H g}
\end{multline}
\begin{align}
\mathcal{O}(\alpha_{s}): \hspace{1cm}
& C_{H,g}^{\mathrm{FF},\ (1)} = C_{H,g}^{\mathrm{VF},\ (1)} + C_{H, H}^{\mathrm{VF},\ (0)}\otimes A_{Hg}^{(1)}
\label{eq: alpha1g}
\end{align}
\begin{align}
\mathcal{O}(\alpha_{s}^{2}): \hspace{1cm} & 
\begin{multlined}[t]
C_{H,g}^{\mathrm{FF},\ (2)} = C_{H,g}^{\mathrm{VF},\ (2)} + C_{H,g}^{\mathrm{VF},\ (1)}\otimes A_{gg,H}^{(1)} + C_{H,H}^{\mathrm{VF},\ (0)} \otimes A_{Hg}^{(2)} \\+ C_{H, H}^{\mathrm{VF},\ (1)}\otimes A_{Hg}^{(1)} 
\end{multlined}
\label{eq: alpha2g}
\end{align}
\begin{align}
\mathcal{O}(\alpha_{s}^{3}): \hspace{1cm} & 
\begin{multlined}[t]
C_{H,g}^{\mathrm{FF},\ (3)} = C_{H,g}^{\mathrm{VF},\ (3)} + C_{H, g}^{\mathrm{VF},\ (2)}\otimes A_{gg, H}^{(1)} + C_{H, g}^{\mathrm{VF},\ (1)}\otimes A_{gg, H}^{(2)} \\+ C_{H,H}^{\mathrm{VF},\ \mathrm{NS+PS},\ (2)} \otimes A_{Hg}^{(1)} + C_{H, H}^{\mathrm{VF},\ (1)}\otimes A_{Hg}^{(2)} \\+ C_{H, H}^{\mathrm{VF},\ (0)}\otimes A_{Hg}^{(3)}
\end{multlined}
\label{eq: alpha3g}
\end{align}
we uncover a NLO contribution to the heavy flavour structure function. This lower order contribution is a consequence of the gluon being able to directly probe the heavy flavour quarks, whereas a light quark must interact via a secondary interaction (hence the $C_{H,q}$ coefficient function beginning at NNLO).

\subsection*{NLO \& NNLO}

\begin{figure}
\centering
\includegraphics[width=\textwidth]{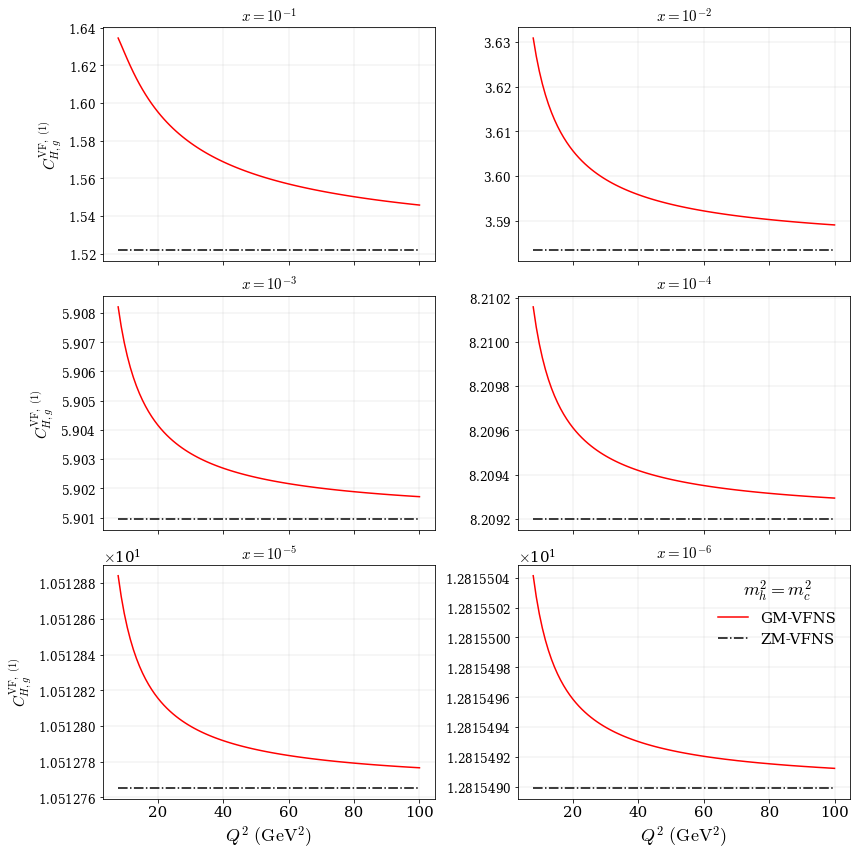}
\caption{\label{fig: NLO_Cg_toZM}The NLO GM-VFNS function $C_{H, g}^{\mathrm{VF},\ (1)}$ compared with the NLO ZM-VFNS function $C_{H, g}^{\mathrm{ZM},\ (1)}$ across a variety of $x$ and $Q^{2}$ values. Mass threshold is set at the charm quark level ($m_{h}^{2} = m_{c}^{2} = 1.4\ \text{GeV}^{2}$).}
\end{figure}

\begin{figure}
\centering
\includegraphics[width=\textwidth]{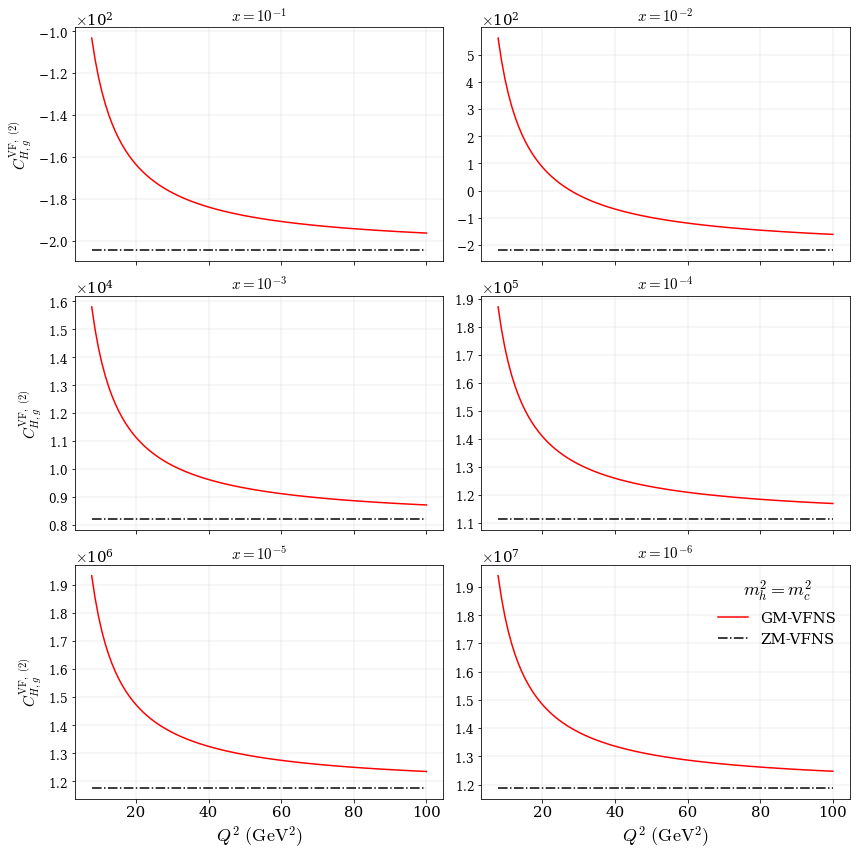}
\caption{\label{fig: NNLO_Cg_toZM}The NNLO GM-VFNS function $C_{H, g}^{\mathrm{VF},\ (2)}$ compared with the NNLO ZM-VFNS function $C_{H, g}^{\mathrm{ZM},\ (2)}$ across a variety of $x$ and $Q^{2}$ values. Mass threshold is set at the charm quark level ($m_{h}^{2} = m_{c}^{2} = 1.4\ \text{GeV}^{2}$).}
\end{figure}

The NLO and NNLO contributions to $C_{H,g}$ are known exactly~\cite{Thorne:GMVFNNLO}. To build some experience and check our understanding, we can observe how the lower order GM-VFNS functions converge onto their ZM-VFNS counterparts in Fig.~\ref{fig: NLO_Cg_toZM} and Fig.~\ref{fig: NNLO_Cg_toZM}.

At NLO and NNLO the magnitude of the functions is generally higher in the low-$Q^{2}$ limit than at high-$Q^{2}$. In both cases, the function remains at the same order of magnitude across all $Q^{2}$. However, the relative change across $Q^{2}$ is smaller at NLO, and similar to that seen for $C_{H,q}^{(2)}$ at NNLO. Due to this, we can once again expect that although more of a scaling contribution at N$^{3}$LO may be present, it should not be too substantial across the range of $Q^{2}$.

\subsection*{N$^{3}$LO}

As with the $C_{H, q}^{(3)}$ function at $\mathcal{O}(\alpha_{s}^{3})$, the FFNS result at low-$Q^{2}$ is known (up to the level of precision discussed in Section~\ref{subsec: NLL_coeff}), as well as the exact ZM-VFNS function at high-$Q^{2}$~\cite{Vermaseren:2005qc,Catani:FFN3LO1,Laenen:FFN3LO2,Vogt:FFN3LO3}. 
Considering the form of $C^{\mathrm{VF},\ (3)}_{H,g}$, there is an extra complication coming from the transition matrix element $A_{Hg}^{(3)}$. As discussed in Section~\ref{subsec: 3loop_OME}, the $A_{Hg}^{(3)}$ function is not as well known as the $A_{Hq}^{(3)}$ function considered earlier and is accompanied by the variational parameter $a_{Hg}$. 
Since it is a requirement for $C_{H, g}^{(3)}$ to exactly cancel the PDF discontinuity introduced by $A_{Hg}^{(3)}$, this variation must be compensated for and included in the description,
\begin{multline}\label{eq: ChgGM-VFNS}
C_{H,g}^{\mathrm{VF},\ (3)} = C_{H,g}^{\mathrm{FF},\ (3)} - C_{H, g}^{\mathrm{VF},\ (2)}\otimes A_{gg, H}^{(1)} - C_{H, g}^{\mathrm{VF},\ (1)}\otimes A_{gg, H}^{(2)} - C_{H,H}^{\mathrm{VF},\ \mathrm{NS+PS},\ (2)} \otimes A_{Hg}^{(1)} \\- C_{H, H}^{\mathrm{VF},\ (1)}\otimes A_{Hg}^{(2)} - A_{Hg}^{(3)}.
\end{multline}
As in Section~\ref{subsec: Chq}, transitioning to the FFNS number scheme ensures an exact cancellation via the subtraction term in Equation~\eqref{eq: ChgGM-VFNS}. Using the exact information for $C_{H,g}^{\mathrm{FF},\ (3)}(Q^{2} \rightarrow 0)$ and the known high-$Q^{2}$ limit,
\begin{multline}\label{eq: ChgFF}
C_{H,g}^{\mathrm{FF},\ (3)}(Q^{2} \rightarrow \infty) = C_{H,g}^{\mathrm{ZM},\ (3)} + C_{H, g}^{\mathrm{VF},\ (2)}\otimes A_{gg, H}^{(1)} + C_{H, g}^{\mathrm{VF},\ (1)}\otimes A_{gg, H}^{(2)} \\+ C_{H,H}^{\mathrm{VF},\ \mathrm{NS+PS},\ (2)} \otimes A_{Hg}^{(1)} + C_{H, H}^{\mathrm{VF},\ (1)}\otimes A_{Hg}^{(2)} + A_{Hg}^{(3)}
\end{multline}
where $C_{H,g}^{\mathrm{VF},\ (3)}$ is replaced with $C_{H,g}^{\mathrm{ZM},\ (3)}$ in the high-$Q^{2}$ limit. Applying the framework set out in Equation~\eqref{eq: ChqFF_param}, the resulting parameterisation is stable across all $(x,Q^{2})$. As $A_{Hg}^{(3)}$ and its variation is explicitly included in Equation~\eqref{eq: ChgGM-VFNS} this ensures the continuity of the structure function with exact cancellations of discontinuities at mass thresholds.

\begin{figure}
\centering
\includegraphics[width=\textwidth]{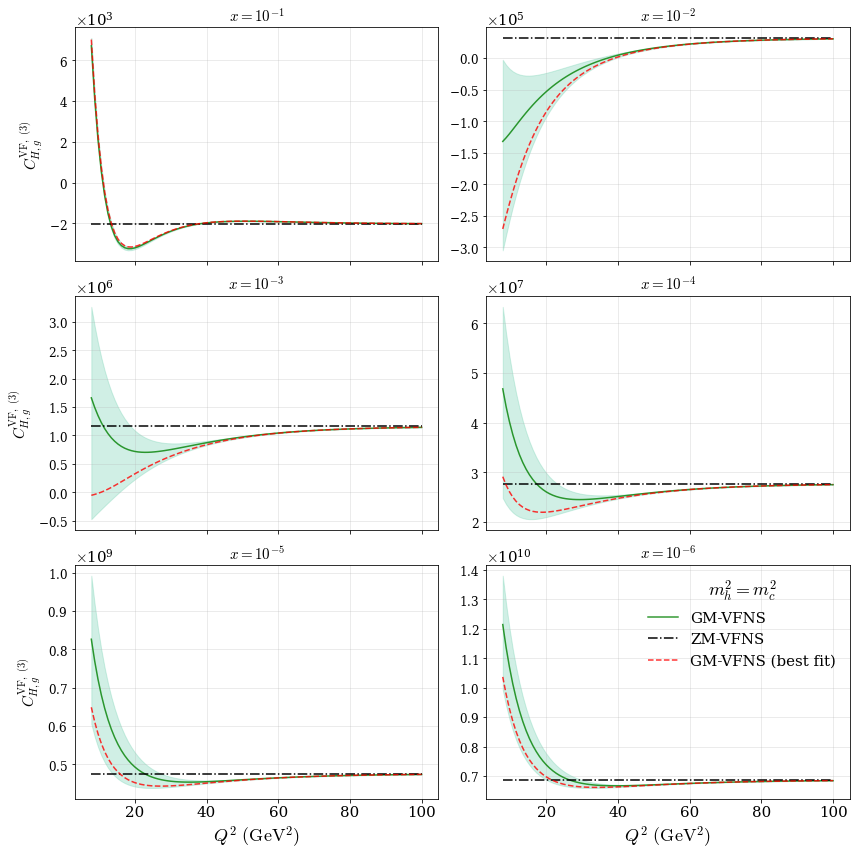}
\caption{\label{fig: GM-VFNS_estimation_g}The N$^{3}$LO GM-VFNS function $C_{H, g}^{\mathrm{VF},\ (3)}$ compared with the N$^{3}$LO ZM-VFNS function $C_{H, g}^{\mathrm{ZM},\ (3)}$ across a variety of $x$ and $Q^{2}$ values (shown without the variation from the low-$Q^{2}$ NLL term discussed in Section~\ref{subsec: NLL_coeff}). $C_{H, g}^{\mathrm{VF},\ (3)}$ is parameterised via Equations~\eqref{eq: ChgGM-VFNS}, \eqref{eq: ChgFF} and \eqref{eq: ChqFF_param}. Mass threshold is set at the charm quark level ($m_{h}^{2} = m_{c}^{2} = 1.4\ \text{GeV}^{2}$).}
\end{figure}
Fig.~\ref{fig: GM-VFNS_estimation_g} displays our approximation for the GM-VFNS coefficient function across a range of $(x, Q^{2})$ via a parameterisation for $C_{H, g}^{\mathrm{FF},\ (3)}$ and the relevant subtraction term in Equation~\eqref{eq: ChgGM-VFNS}. Fig.~\ref{fig: GM-VFNS_estimation_g} also contains the uncertainty in this approximation stemming from $A_{Hg}^{(3)}$ (see Section~\ref{subsec: AHg}). Note that Fig.~\ref{fig: GM-VFNS_estimation_g} ignores any variation from the low-$Q^{2}$ NLL term discussed in Section~\ref{subsec: NLL_coeff}, where this is fixed to its central value. The uncertainty shown in Fig.~\ref{fig: GM-VFNS_estimation_g} is suppressed as we move to high-$Q^{2}$ owing to the required convergence of the GM-VFNS onto the corresponding ZM-VFNS gluon coefficient function at N$^{3}$LO.

Included in Fig.~\ref{fig: GM-VFNS_estimation_g} is the best fit prediction for $C_{H, g}^{\mathrm{VF},\ (3)}$ (corresponding to the best fit of $A_{Hg}^{(3)}$ approximated in Section \ref{sec: n3lo_OME}). 
Overall we see the resultant shape of $C_{H,g}^{(3)}$ is within our predicted range and follows a sensible shape that matches with the known high-$Q^{2}$ FFNS behaviour. Contrasting this with NNLO, the shape across the range of $x$ values shown is less consistent. There is no guarantee that this should be the case, since we do not know how the perturbative nature of QCD will behave. However, we do maintain the relatively consistent order of magnitude across the evolution in $Q^{2}$, therefore the exact form of the shape across $Q^{2}$ will be less important in the resultant structure function picture.

\subsection*{$C_{q,q}^{\mathrm{NS}}$}\label{subsec: Cqqns}

The light quark coefficient functions involve small heavy flavour contributions at higher orders from heavy quarks produced away from the photon vertex. As discussed in Section~\ref{subsec: genframe_continuous} the low-$Q^{2}$ FFNS function in this case is unknown. However, since the heavy flavour contributions to the light quark structure function $F_{2,q}(x,Q^{2})$ are very small, any choice of sensible variation in $Q^{2}$ has a near negligible effect on the overall structure function. Further to this, as is apparent from lower order examples, it can be expected that the light quark coefficient functions remain relatively constant across $Q^{2}$.

Using Equation~\eqref{eq: FFexpanse} and Equation~\eqref{eq: expanse_q}, the non-singlet coefficient function is stated as,
\begin{equation}
C_{q,q}^{\mathrm{FF},\ \mathrm{NS}} = A_{qq,H}^{\mathrm{NS}}\otimes C_{q,q}^{\mathrm{VF}\ \mathrm{NS}},
\end{equation}
\begin{subequations}
\begin{align}
\mathcal{O}(\alpha_{s}^{0}): & \hspace{1cm}
\begin{multlined}
C_{q,q,\ \mathrm{NS}}^{\mathrm{FF},\ (0)}\ =\ C_{q,q,\ \mathrm{NS}}^{\mathrm{VF},\ (0)}
\end{multlined}
\\ \mathcal{O}(\alpha_{s}^{1}): & \hspace{1cm}
\begin{multlined}
C_{q,q,\ \mathrm{NS}}^{\mathrm{FF},\ (1)}\ =\ C_{q,q,\ \mathrm{NS}}^{\mathrm{VF},\ (1)}
\end{multlined}
\\ \mathcal{O}(\alpha_{s}^{2}): & \hspace{1cm}
\begin{multlined}\label{eq: NSlightquark - NNLO}
C_{q,q,\ \mathrm{NS}}^{\mathrm{FF},\ (2)}\ =\ C_{q,q,\ \mathrm{NS}}^{\mathrm{VF},\ (2)} + A_{qq,H}^{\mathrm{NS},\ (2)}
\end{multlined}
\\ \mathcal{O}(\alpha_{s}^{3}): & \hspace{1cm}
\begin{multlined}\label{eq: NSlightquark - N3LO}
C_{q,q,\ \mathrm{NS}}^{\mathrm{FF},\ (3)}\ =\ C_{q,q,\ \mathrm{NS}}^{\mathrm{VF},\ (3)} + A_{qq,H}^{\mathrm{NS},\ (3)} + C_{q,q,\ \mathrm{NS}}^{\mathrm{VF},\ (1)}\otimes A_{qq,H}^{\mathrm{NS},\ (2)}.
\end{multlined}
\end{align}
\label{eq: NSlightquark}
\end{subequations}
From Equation~\eqref{eq: NSlightquark} the FFNS contribution at LO and NLO is identical to the GM-VFNS and ZM-VFNS function at high-$Q^{2}$. Physically for heavy quarks to affect light quarks, a larger number of vertices than are allowed at LO and NLO must be present to enable interactions involving heavy quarks. We therefore begin our discussion at NNLO.

\subsubsection*{NNLO}

At NNLO the functions included in Equation~\eqref{eq: NSlightquark - NNLO} are known exactly~\cite{vogt:NNLOns,Buza:OMENNLO}. Assembling these together, we provide an example of how the GM-VFNS function converges to the familiar ZM-VFNS function for the light quark. By performing this exercise, expectations as to how $C_{q,q}^{\mathrm{NS}}$ will behave at N$^{3}$LO can be constructed.

\begin{figure}
\centering
\includegraphics[width=\textwidth]{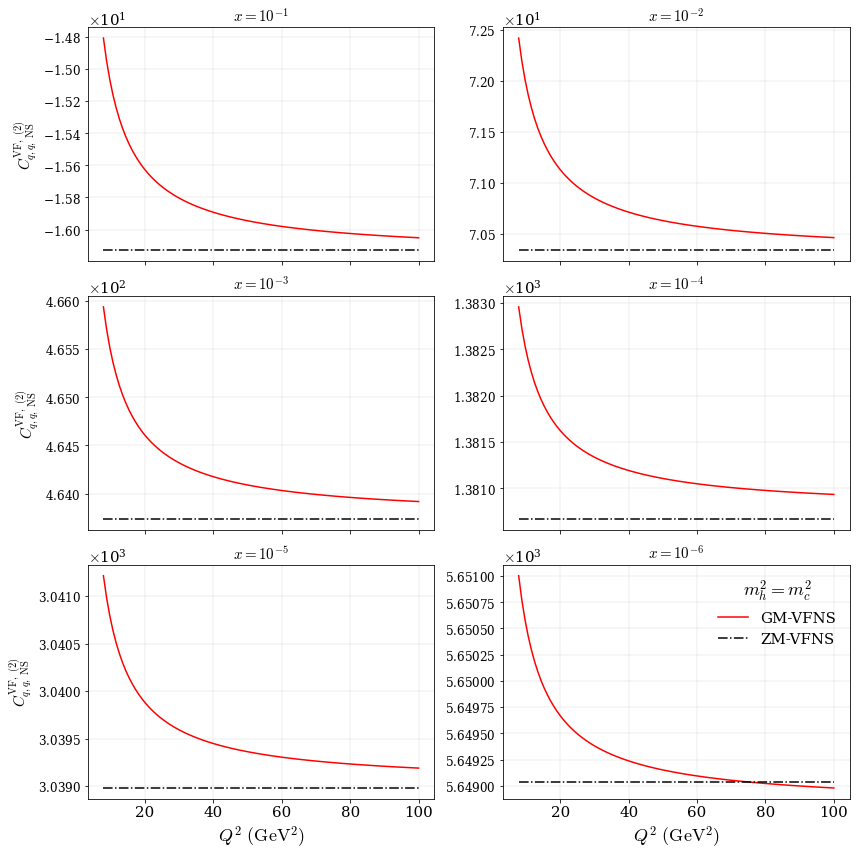}
\caption{\label{fig: Cqq_ns_NNLO} The NNLO GM-VFNS function $C_{q, q,\ \mathrm{NS}}^{\mathrm{VF},\ (2)}$ compared with the NNLO ZM-VFNS function $C_{q, q,\ \mathrm{NS}}^{\mathrm{ZM},\ (2)}$ across a variety of $x$ and $Q^{2}$ values. Mass threshold is set at the charm quark level ($m_{h}^{2} = m_{c}^{2} = 1.4\ \text{GeV}^{2}$).}
\end{figure}
From Fig.~\ref{fig: Cqq_ns_NNLO} $C_{q,q,\ \mathrm{NS}}^{\mathrm{VF},\ (2)}$ quickly converges onto the ZM-VFNS function with the difference between the low and high-$Q^{2}$ being within $10\%$ at large-$x$ and within $0.01\%$ at small-$x$. This weak scaling with $Q^{2}$ reinforces the statement that it is possible to approximate the N$^{3}$LO function relatively well without extensive low-$Q^{2}$ information.

\subsubsection*{N$^{3}$LO}

Equation~\eqref{eq: NSlightquark - N3LO} involves a mixture of functions known exactly (ZM-VFNS high-$Q^{2}$ limit~\cite{Vermaseren:2005qc}) and functions that are completely unknown ($C_{q,q,\ \mathrm{NS}}^{\mathrm{FF},\ (3)}$). This presents an issue as it is no longer possible to rely on $C_{q,q,\ \mathrm{NS}}^{FF,\ (3)}$ to constrain the low-$Q^{2}$ limit. 
Nevertheless, by utilising the experience gained from NNLO, it is feasible to choose any sensible choice for the low-$Q^{2}$ limit. In practice, due to the observed weak scaling in $Q^{2}$, the exact form at low-$Q^{2}$ will not present any noticeable differences.

A naive choice for heuristically placing the $C_{q,q,\ \mathrm{NS}}^{\mathrm{FF},\ (3)}(Q^{2}\rightarrow 0)$ function would be a constant value i.e. no scaling in $Q^{2}$. We propose to use the intuition from NNLO and the overall fit quality to give us potentially a more sensible and viable choice for the GM-VFNS approximation\footnote{The differences in fit quality for sensible choices are $<0.05\%$ compared to the overall $\chi^{2}$ for the light quark NS coefficient function.}. 
By inserting the high-$Q^{2}$ limit into the NS part of Equation~\eqref{eq: CqFF_param}, the result is a crude approximation to $C_{q,q,\ \mathrm{NS}}^{\mathrm{FF},\ (3)}(Q^{2}\rightarrow 0)$. Combining this with Equation~\eqref{eq: NSlightquark  - N3LO}, we obtain a GM-VFNS parameterisation which is relatively constant across $Q^{2}$ (similar to the NNLO behaviour) with any differences arising from the subtraction terms which are known.

\begin{figure}
\centering
\includegraphics[width=\textwidth]{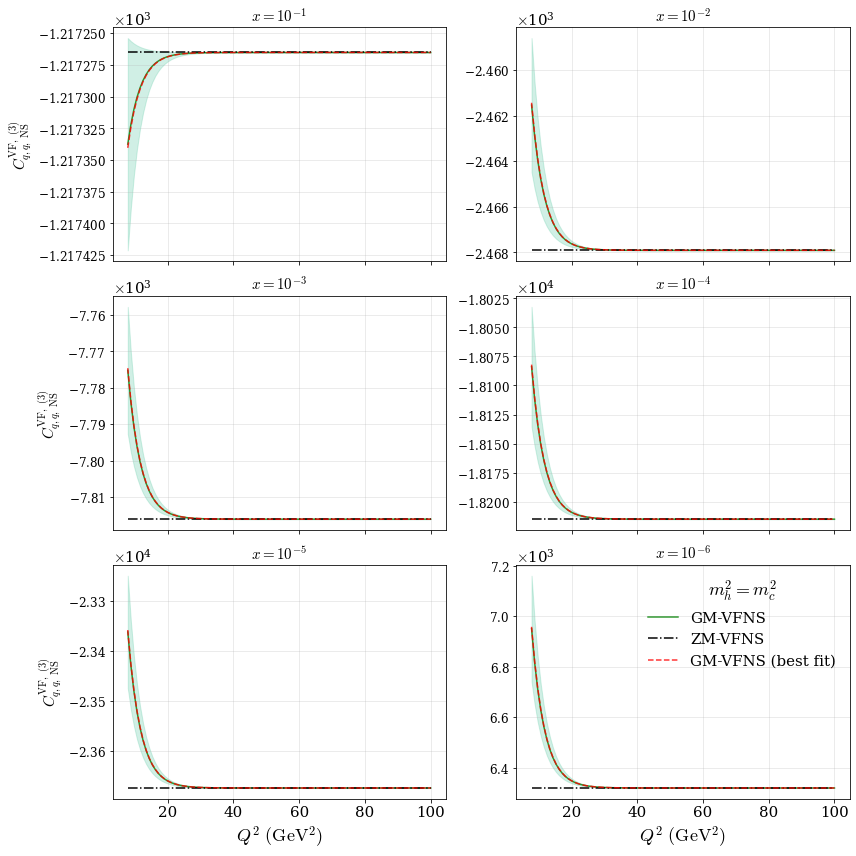}
\caption{\label{fig: Cqq_ns_N3LO}The N$^{3}$LO GM-VFNS function $C_{q, q,\ \mathrm{NS}}^{\mathrm{VF},\ (3)}$ compared with the N$^{3}$LO ZM-VFNS function $C_{q, q,\ \mathrm{NS}}^{\mathrm{ZM},\ (3)}$ across a variety of $x$ and $Q^{2}$ values. $C_{q, q,\ \mathrm{NS}}^{\mathrm{VF},\ (3)}$ is parameterised via Equations~\eqref{eq: NSlightquark - N3LO} and \eqref{eq: CqFF_param}. Mass threshold is set at the charm quark level ($m_{h}^{2} = m_{c}^{2} = 1.4\ \text{GeV}^{2}$).}
\end{figure}
Fig.~\ref{fig: Cqq_ns_N3LO} shows the result of this approximation for the full $C_{q, q,\ \mathrm{NS}}^{\mathrm{VF},\ (3)}$ function. We notice that the behaviour is similar to that of NNLO across all $(x, Q^{2})$ and appropriately larger in magnitude to account for the extra contributions obtained at N$^{3}$LO compared to NNLO. By definition, the parameterisation converges well to the ZM-VFNS scheme with the magnitude at high-$Q^{2}$ (ZM-VFNS regime) remaining similar to that at low-$Q^{2}$ for each specific value of $x$. This final point gives assurances that even if this low-$Q^{2}$ guess is not entirely representative of the actual N$^{3}$LO function, the effects of including this approximation are virtually negligible in a PDF fit. Also shown in Fig.~\ref{fig: Cqq_ns_N3LO} is the variation in the $C_{q, q,\ \mathrm{NS}}^{\mathrm{VF},\ (3)}$ function stemming solely from the $A_{qq,H}^{\mathrm{NS},\ (3)}$ function.

\subsection*{$C_{q,q}^{\mathrm{PS}}$}\label{subsec: Cqqps}

To complete the light-quark GM-VFNS coefficient function picture the pure-singlet contribution from Equation~\eqref{eq: FFexpanse} and Equation~\eqref{eq: expanse_q} is described by,
\begin{equation}
C_{q,q}^{\mathrm{FF},\ \mathrm{PS}} = C_{q,q}^{\mathrm{VF},\ \mathrm{PS}} \otimes A_{qq,H}^{\mathrm{PS}}\ +\ C_{q,g}^{\mathrm{VF}}\otimes A_{gq,H} +\ C_{q,H}^{\mathrm{VF},\ \mathrm{PS}}\otimes A_{Hq}
\end{equation}
\begin{subequations}
\begin{align}
\mathcal{O}(\alpha_{s}^{0}): & \hspace{1cm}
\begin{multlined}
C_{q,q}^{\mathrm{FF},\ \mathrm{PS},\ (0)}\ =\ 0
\end{multlined}
\\ \mathcal{O}(\alpha_{s}^{1}): & \hspace{1cm}
\begin{multlined}
C_{q,q}^{\mathrm{FF},\ \mathrm{PS},\ (1)}\ =\ 0
\end{multlined}
\\ \mathcal{O}(\alpha_{s}^{2}): & \hspace{1cm}
\begin{multlined}
C_{q,q}^{\mathrm{FF},\ \mathrm{PS},\ (2)}\ =\ C_{q,q,\ \mathrm{PS}}^{\mathrm{VF},\ (2)}
\end{multlined}
\\ \mathcal{O}(\alpha_{s}^{3}): & \hspace{1cm}
\begin{multlined}\label{eq: PSlightquark - N3LO}
C_{q,q}^{\mathrm{FF},\ \mathrm{PS},\ (3)}\ =\ C_{q,q,\ \mathrm{PS}}^{\mathrm{VF},\ (3)} + C_{q,g}^{\mathrm{VF},\ (1)}\otimes A_{gq,H}^{(2)}.
\end{multlined}
\end{align}
\label{eq: PSlightquark}
\end{subequations}
As with the non-singlet analysis the heavy flavour contributions to the pure-singlet appear at higher orders to allow for the possibility of heavy quark contributions. In the pure-singlet case, the heavy flavour contributions are pushed one order higher than the non-singlet
due to the requirement for an extra
intermediary gluon. 

\subsubsection*{N$^{3}$LO}

In the pure-singlet case, the FFNS function is non-existent up until N$^{3}$LO. Because of this, we choose to parameterise the pure-singlet with a weak constraint suppressing the FFNS function $C_{q,q}^{\mathrm{FF},\ \mathrm{PS},\ (3)}$ across all $x$ for very low-$Q^{2}$. The reason for this is that the coefficient functions acquire more contributions as they exist through higher orders. If $C_{q,q}^{\mathrm{FF},\ \mathrm{PS},\ (3)}$ is beginning at this order, then one could expect the low-$Q^{2}$ form to be relatively small compared to the known ZM-VFNS function~\cite{Vermaseren:2005qc}. This is somewhat justified by the low-$Q^{2}$ kinematic restrictions for the singlet distribution which broadly manifest into a suppression at low-$Q^{2}$. We reiterate here that the low-$Q^{2}$ form of this function is still essentially around the same magnitude across all $Q^{2}$. Therefore, as with $C_{q,q}^{\mathrm{FF},\ \mathrm{NS},\ (3)}$, it will be virtually negligible in the overall structure function.

After constructing the approximation for $C_{q,q}^{\mathrm{FF},\ \mathrm{PS},\ (3)}$ with Equation~\eqref{eq: CqFF_param}, Equation~\eqref{eq: PSlightquark - N3LO} is used to approximate the GM-VFNS function. The exact form of Equation~\eqref{eq: PSlightquark - N3LO} is chosen based on intuition and where the best fit quality can be achieved\footnote{The differences in fit quality for sensible choices are $<0.1\%$ compared to the overall $\chi^{2}$ for the light quark PS coefficient function.}.

\begin{figure}
\centering
\includegraphics[width=\textwidth]{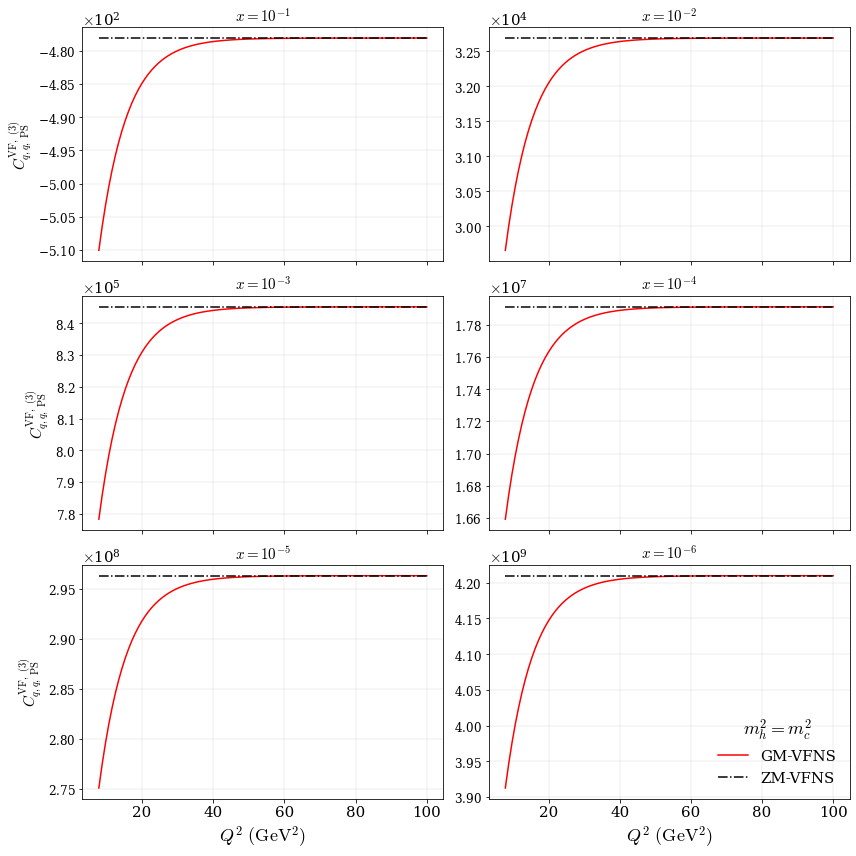}
\caption{\label{fig: Cqq_ps_N3LO}The N$^{3}$LO GM-VFNS function $C_{q, q,\ \mathrm{PS}}^{\mathrm{VF},\ (3)}$ compared with the N$^{3}$LO ZM-VFNS function $C_{q, q,\ \mathrm{PS}}^{\mathrm{ZM},\ (3)}$ across a variety of $x$ and $Q^{2}$ values. $C_{q, q,\ \mathrm{PS}}^{\mathrm{VF},\ (3)}$ is parameterised via Equations~\eqref{eq: PSlightquark - N3LO} and \eqref{eq: CqFF_param}. Mass threshold is set at the charm quark level ($m_{h}^{2} = m_{c}^{2} = 1.4\ \text{GeV}^{2}$).}
\end{figure}
It can be seen from Fig.~\ref{fig: Cqq_ps_N3LO} that the overall magnitude of $C_{q, q,\ \mathrm{PS}}^{\mathrm{VF},\ (3)}$ decreases substantially towards large-$x$ as one would expect from a pure-singlet function. Inspecting the predicted values of $C_{q, q,\ \mathrm{PS}}^{\mathrm{VF},\ (3)}$, we can confirm that the non-singlet function from Fig.~\ref{fig: Cqq_ns_N3LO} begins to dominate at large-$x$. Conversely towards small-$x$, $C_{q, q,\ \mathrm{PS}}^{\mathrm{VF},\ (3)}$ is much larger than $C_{q, q,\ \mathrm{NS}}^{\mathrm{VF},\ (3)}$, thereby preserving the familiar interplay between quark distributions. The suppression of the FFNS parameterisation towards low-$Q^{2}$ is also seen to give sensible results in terms of the expected percentage change in magnitude through the range of $Q^{2}$ values. Specifically we see $< 10\%$ difference in magnitude between low and high-$Q^{2}$. Since scale violating terms become more dominant at higher orders and we are essentially at leading order in terms of heavy flavour contributions, a high level of scaling with $Q^{2}$ is not expected at this order.

\subsection*{$C_{q,g}$}\label{subsec: Cqg}

Finally the gluon-light quark coefficient function is constructed from Equation~\eqref{eq: FFexpanse} and Equation~\eqref{eq: expanse_q} to be,
\begin{equation}
C_{q,g}^{\mathrm{FF}} = C_{q,q}^{\mathrm{VF}} \otimes A_{qg,H}\ +\ C_{q,g}^{\mathrm{VF}}\otimes A_{gg,H} +\ C_{q,H}^{\mathrm{VF},\ \mathrm{PS}}\otimes A_{Hg}
\end{equation}
\begin{subequations}
\begin{align}
\mathcal{O}(\alpha_{s}^{0}): & \hspace{1cm}
\begin{multlined}
C_{q,g}^{\mathrm{FF},\ (0)}\ =\ 0
\end{multlined}
\\ \mathcal{O}(\alpha_{s}^{1}): & \hspace{1cm}
\begin{multlined}
C_{q,g}^{\mathrm{FF},\ (1)}\ =\ C_{q,g}^{\mathrm{VF},\ (1)}
\end{multlined}
\\ \mathcal{O}(\alpha_{s}^{2}): & \hspace{1cm}
\begin{multlined}\label{eq: Glightquark - NNLO}
C_{q,g}^{\mathrm{FF},\ (2)}\ =\ C_{q,g}^{\mathrm{VF},\ (2)} + C_{q,g}^{\mathrm{VF},\ (1)}\otimes A_{gg, H}^{(1)} + A_{qg, H}^{(2)}
\end{multlined}
\\ \mathcal{O}(\alpha_{s}^{3}): & \hspace{1cm}
\begin{multlined}\label{eq: Glightquark - N3LO}
C_{q,g}^{\mathrm{FF},\ (3)}\ =\ C_{q,g}^{\mathrm{VF},\ (3)} + A_{qg,H}^{(3)} + C_{q,q}^{\mathrm{VF},\ (1)}\otimes A_{qg,H}^{(2)} \\+ C_{q,g}^{\mathrm{VF},\ (2)}\otimes A_{gg,H}^{(1)} + C_{q,g}^{\mathrm{VF},\ (1)}\otimes A_{gg,H}^{(2)} \\+ C_{q,H,\ \mathrm{PS}}^{\mathrm{VF},\ (2)}\otimes A_{Hg}^{(1)}.
\end{multlined}
\end{align}
\label{eq: Glightquark}
\end{subequations}
For $C_{q,g}$, the FFNS function is non-existent up to NNLO, similar to $C_{q,q,\ \mathrm{NS}}^{\mathrm{FF}, (3)}$. However, the $A_{qg,H}$ contribution at NNLO is sub-leading in $n_{f}$~\cite{Buza:OMENNLO} and is therefore not considered here.

\subsubsection*{N$^{3}$LO}

\begin{figure}
\centering
\includegraphics[width=\textwidth]{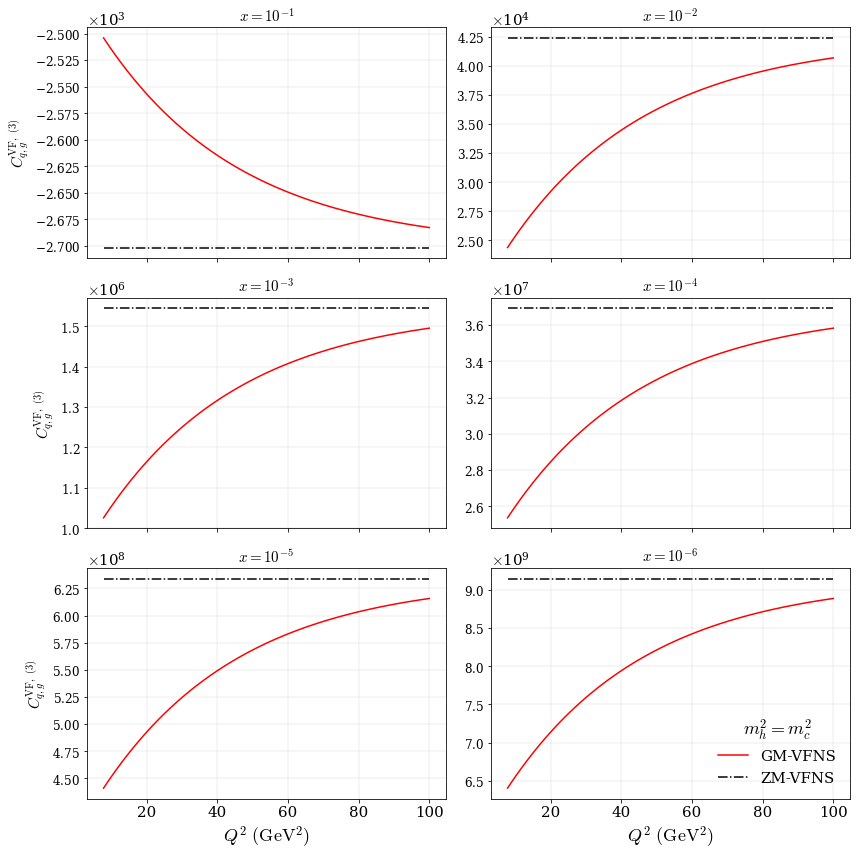}
\caption{\label{fig: Cqg_N3LO}The N$^{3}$LO GM-VFNS function $C_{q, g}^{\mathrm{VF},\ (3)}$ compared with the N$^{3}$LO ZM-VFNS function $C_{q, g}^{\mathrm{ZM},\ (3)}$ across a variety of $x$ and $Q^{2}$ values. $C_{q, g}^{\mathrm{VF},\ (3)}$ is parameterised via Equations~\eqref{eq: Glightquark - N3LO} and \eqref{eq: CqFF_param}. Mass threshold is set at the charm quark level ($m_{h}^{2} = m_{c}^{2} = 1.4\ \text{GeV}^{2}$).}
\end{figure}
At N$^{3}$LO in Equation~\eqref{eq: Glightquark - N3LO}, no information is available for the $C_{q, g}^{\mathrm{FF},\ (3)}$ at low-$Q^{2}$. Whereas at high-$Q^{2}$ the ZM-VFNS function is known~\cite{Vermaseren:2005qc}. To construct the parameterisation, we apply the same method described for $C_{q,q,\ \mathrm{PS}}^{\mathrm{FF},\ (3)}$. Specifically, by applying a suppression to the FFNS parameterisation in the low-$Q^{2}$ limit. After constructing the parameterisation for $C_{q, g}^{\mathrm{FF},\ (3)}$ with Equation~\eqref{eq: CqFF_param}, Equation~\eqref{eq: Glightquark - N3LO} is used to approximate the GM-VFNS function. Since there is no information in the low-$Q^{2}$ limit, the parameterisation in Equation~\eqref{eq: CqFF_param} is chosen roughly based on how the fit prefers the evolution in $Q^{2}$ to behave.

Fig.~\ref{fig: Cqg_N3LO} illustrates the GM-VFNS function in Equation~\eqref{eq: Glightquark - N3LO} with Equation~\eqref{eq: CqFF_param} as $C_{q, g}^{\mathrm{FF},\ (3)}$ across a range of $x$ and $Q^{2}$. $C_{q, g}^{\mathrm{VF},\ (3)}$ increases in magnitude when moving to smaller $x$ and by definition converges onto the ZM-VFNS function. The convergence in this case is chosen to be less steep than for the light quark convergences due to some minor tensions in the fit\footnote{The differences in fit quality for sensible choices of Equation~\eqref{eq: CqFF_param} are $<0.5\%$ compared to the overall $\chi^{2}$ for the light quark gluon coefficient function.}. The magnitude of $C_{q, g}^{\mathrm{VF},\ (3)}$ across the entire range of $Q^{2}$ is still relatively constant, although less flat than the behaviour predicted for $C_{q,q,\ \mathrm{PS/NS}}^{\mathrm{VF},\ (3)}$. However, considering Equation~\eqref{eq: Glightquark - N3LO}, some justification for this behaviour can be offered. When comparing the contributions to the FFNS functions in the NS, PS and gluon cases (Equations~\eqref{eq: NSlightquark - N3LO}, \eqref{eq: PSlightquark - N3LO} and \eqref{eq: Glightquark - N3LO} respectively), the $A_{Hg}$ and $A_{gg,H}$ contributions involved in $C_{q,g}^{\mathrm{FF}}$ are much larger than the contributions from $A_{gq,H}$, $A_{Hq}$ and $A^{\mathrm{NS}}_{qq,H}$. Therefore we can expect a larger difference across $Q^{2}$ for the $C_{q, g}^{\mathrm{VF},\ (3)}$ function. With this being said, the specific form at low-$Q^{2}$ is not very important in current PDF fits, only that the form is continuous and valid.

\section{\texorpdfstring{N$^{3}$LO}{N3LO} \texorpdfstring{$K$}{K}-factors}\label{sec: n3lo_K}

Thus far the primary concern has been the N$^{3}$LO additions to the theoretical form of the DIS cross section. However, to complement these changes it is necessary to extend other cross section data to the same order. With these ingredients it is possible to maintain a consistent approximate N$^{3}$LO treatment across all datasets. At the time of writing, $K$-factors which provide exact transformations for each dataset up to NNLO are available\footnote{An exception to this is the CMS $7\ \text{TeV}\ W + c$ \cite{CMS7Wpc} dataset where $K$-factors are available only up to NLO.}. Although there has been progress in N$^{3}$LO calculations for various processes including Drell-Yan (DY), top production and Higgs processes~\cite{DY_N3LO_Kfac, Gehrmann:DYN3LO, duhr:DY2021, Kidonakis:tt,Ball:2013bra,Bonvini:2014jma,Bonvini:2016frm,Ahmed:2016otz,Bonvini:2018ixe,Bonvini:2018iwt,Bonvini:2013kba,anastasiou2014higgs,anastasiou2016high,Mistlberger:2018}, there is still missing information on how these $K$-factors behave above NNLO. In this section we investigate the effects of the $K$-factors for each dataset when extended to N$^{3}$LO. Five process categories are considered separately: Drell-Yan, Jets, $p_{T}$ jets, $t\bar{t}$ production and Dimuon data. Inside each of these process categories we assume a perfect positive correlation between the behaviour of datasets i.e. all Drell-Yan $K$-factor shifts from NNLO are positively correlated. Clearly this treatment is a simplification, based on the expectation of a high degree of correlation between datasets concerned with the same processes. In practice, the uncertainty introduced from including these $K$-factors is already relatively small compared to the other sources of MHOUs already discussed, therefore any correction to this is guaranteed to be small (this will be shown more clearly in Section~\ref{sec: results}).

\subsection{Extension to \texorpdfstring{aN$^{3}$LO}{aN3LO}}\label{subsec: kN3LO}

The extension to aN$^{3}$LO is parameterised with a mixture of the NLO and NNLO $K$-factors. This allows control of the magnitude and shape of the transformation from NNLO to aN$^{3}$LO, using the known shifts from lower orders.

The basic idea is presented as,
\begin{equation}\label{eq: k_fac_param}
	K^{\mathrm{N}^{3}\mathrm{LO}/\mathrm{LO}} = a_{\mathrm{NNLO}}\ K^{\mathrm{NNLO}/\mathrm{LO}} + a_{\mathrm{NLO}}\ K^{\mathrm{NLO}/\mathrm{LO}},
\end{equation}
where $K^{\mathrm{N}^{3}\mathrm{LO}/\mathrm{LO}}, K^{\mathrm{NNLO}/\mathrm{LO}}\ \mathrm{and}\ K^{\mathrm{NLO}/\mathrm{LO}}$ are the relevant $K$-factors with respect to the LO cross section, and $a_{\mathrm{N(N)LO}}$ are variational parameters controlling the mixture of NNLO and NLO $K$-factors included in the N$^{3}$LO $K$-factor approximation. Hence we have 2 parameters for each of the five processes included in the fit, and now 20 theory nuisance parameters in total -- 10 controlling aN$^{3}$LO $K$-factors, 5 controlling aN$^{3}$LO splitting functions and 5 controlling heavy flavour aN$^{3}$LO contributions.

To describe this formalism in terms of physical observables we consider the cross section,
\begin{equation}
\sigma = \sigma_{0} + \sigma_{1} + \sigma_{2} + \dots \equiv \sigma_{\mathrm{NNLO}} + \dots,	
\end{equation}
where there is an implicit order of $\alpha_{s}^{p+i}$ absorbed into the definition of $\sigma_{i}$ beginning at the relevant LO for each process, i.e. $p = 0$ for DY.

$K^{\mathrm{NLO/LO}}$ is then the relative shift from $\sigma_{\mathrm{LO}}$ to $\sigma_{\mathrm{NLO}}$,
\begin{equation}\label{eq: KNLO}
K^{\mathrm{NLO/LO}} = \frac{\sigma_{0} + \sigma_{1}}{\sigma_{0}} = 1 + \frac{\sigma_{1}}{\sigma_{0}}.
\end{equation}
Similarly for NNLO we have,
\begin{equation}\label{eq: KNNLO}
K^{\mathrm{NNLO/LO}} = \frac{\sigma_{0} + \sigma_{1} + \sigma_{2}}{\sigma_{0}} = 1 + \frac{\sigma_{1}}{\sigma_{0}} + \frac{\sigma_{2}}{\sigma_{0}}.
\end{equation}

Moving to N$^{3}$LO, we write
\begin{equation}
	\sigma = \sigma_{0} + \sigma_{1} + \sigma_{2} + \sigma_{3} + \dots \equiv \sigma_{\mathrm{N^{3}LO}} + \dots,
\end{equation}
where $\sigma_{3} = a_{1} \sigma_{1} + a_{2} \sigma_{2}$ is approximated as some superposition of the two lower orders, with $(a_{1}, a_{2}) = (0, 0)$ reproducing the NNLO case.

Pushing forward with this approximation and using the definitions for $\sigma_{1,2}$ in terms of $K$-factors (Equations~\eqref{eq: KNLO} and \eqref{eq: KNNLO}) we have,
\begin{align}
\sigma_{\mathrm{N^3LO}} &= \sigma_{\mathrm{NNLO}} + a_{1}\sigma_{1} + a_{2}\sigma_{2}	\\
&= \sigma_{\mathrm{NNLO}} + a_{1}\sigma_{0} (K^{\mathrm{NLO/LO}} - 1) + a_{2} \sigma_{0} (K^{\mathrm{NNLO/LO}} - K^{\mathrm{NLO/LO}})
\end{align}
since,
\begin{align}
	\sigma_{1} &= \sigma_{0}\left(K^{\mathrm{NLO/LO}} - 1\right) \\
	\sigma_{2} &= \sigma_{0}\left(K^{\mathrm{NNLO/LO}} - \sigma_{1} - \sigma_{0}\right)\nonumber\\
	&= \sigma_{0}\left(K^{\mathrm{NNLO/LO}} - K^{\mathrm{NLO/LO}}\right).
\end{align}
From here one can obtain,
\begin{equation}\label{eq: Knn-kn}
	K^{\mathrm{NNLO/LO}} - K^{\mathrm{NLO/LO}} = \frac{\sigma_{2}}{\sigma_{0}} = \frac{\sigma_{2} + \sigma_{0}}{\sigma_{0}} - 1 \approx \frac{\sigma_{2} + \sigma_{1} + \sigma_{0}}{\sigma_{1} + \sigma_{0}} - 1 = K^{\mathrm{NNLO/NLO}} - 1,
\end{equation}
assuming $\sigma_{1} \ll \sigma_{0}$, which is in general true for a valid perturbative expansion. Using \eqref{eq: Knn-kn} $\sigma_{\mathrm{N^{3}LO}}$ can be expressed by,
\begin{equation}
\sigma_{\mathrm{N^3LO}} \simeq \sigma_{\mathrm{NNLO}}\left(1 + a_{1}(K^{\mathrm{NLO/LO}} - 1) + a_{2} (K^{\mathrm{NNLO/NLO}} - 1)\right),
\end{equation}
where $\sigma_{2} \ll \sigma_{1} \ll \sigma_{0}$.

This defines the proposed approximated N$^{3}$LO cross section. It is given in terms of extra contributions from lower order shifts, which are controlled by variational parameters $a_{1}$ and $a_{2}$. It is also true that the contributions to N$^{3}$LO are expected to be suppressed by $\alpha_{s} / \pi$ in the NNLO case and $(\alpha_{s}/\pi)^{2}$ in the NLO case to account for the strengths of each contribution. Currently this is taken into account within the variational parameters $a_{1}, a_{2}$. However for the purpose of this description, it is more appropriate to explicitly redefine $a_{1}, a_{2} = a_{s}^{2}\hat{a}_{1}, a_{s}\hat{a}_{2}$ where $a_{s} = \mathcal{N}\alpha_{s}$ and $\mathcal{N}$ is some normalisation factor. This then results in,
\begin{equation}\label{eq: meth2}
K^{\mathrm{N^3LO/LO}} = K^{\mathrm{NNLO/LO}}\left(1 + \hat{a}_{1}\mathcal{N}^{2}\alpha_{s}^{2}(K^{\mathrm{NLO/LO}} - 1) + \hat{a}_{2} \mathcal{N}\alpha_{s} (K^{\mathrm{NNLO/NLO}} - 1)\right).
\end{equation}
where the LO cross section $\sigma_{0}$ is cancelled and Equation~\eqref{eq: meth2} is written in terms of the $K$-factor shifts only. \eqref{eq: meth2} also implicitly includes the correct order $\mathcal{O}(\alpha_{s}^{3})$ in the parameterisation through \eqref{eq: KNLO} and \eqref{eq: KNNLO}. We can then choose $\mathcal{N}$ in order to set the approximate magnitude of our variational parameters $\hat{a}_{1}, \hat{a}_{2}$. Given $\alpha_{s} \sim 0.1$ for the processes considered, if we neglect $\mathcal{N}$ (i.e. choose $\mathcal{N} \sim 1$), then our order by order reduction in the magnitude of the $K$-factors would be $\sim 10\%$ for $\mathcal{O}(1)$ for variational parameters, however from previous orders we see that typically $K$-factors tend to be $30-40\%$ of the previous order, therefore we instead choose $\mathcal{N} = 3$. This then ensures the natural scale of variation allowed is also of this order with $\mathcal{O}(1)$ variational parameters describing the admixture of NLO and NNLO $K$-factors, with conservative penalties applied accordingly.

Reflecting on this, it is worth noting that these fitted $K$-factors will be sensitive to all orders, not just N$^{3}$LO. Considering these $K$-factors as approximating asymptotic behaviour to all orders in perturbation theory when assessing the stability of predictions, we can be less concerned with any somewhat large shifts from NNLO to aN$^{3}$LO, as we will specifically see in the case of Fig.'s~\ref{fig:Kfac_pTjet} and \ref{fig:Kfac_top}. Finally, we remind the reader that at higher orders, new terms with more divergent leading logarithms appear which are missed by the current theoretical description. 
Due to this, the all-orders asymptotic description will still remain approximate up to the inclusion of more divergent leading logarithms in $(x,Q^{2})$ limits at even higher orders. 

\subsection{Numerical Results}\label{subsec: kfac_res}

Using this formalism for the aN$^{3}$LO $K$-factors, we present the global fit results for each of the five process categories considered.

\subsubsection*{Drell-Yan Processes}

\begin{figure}
    \centering
    \includegraphics[width=\textwidth]{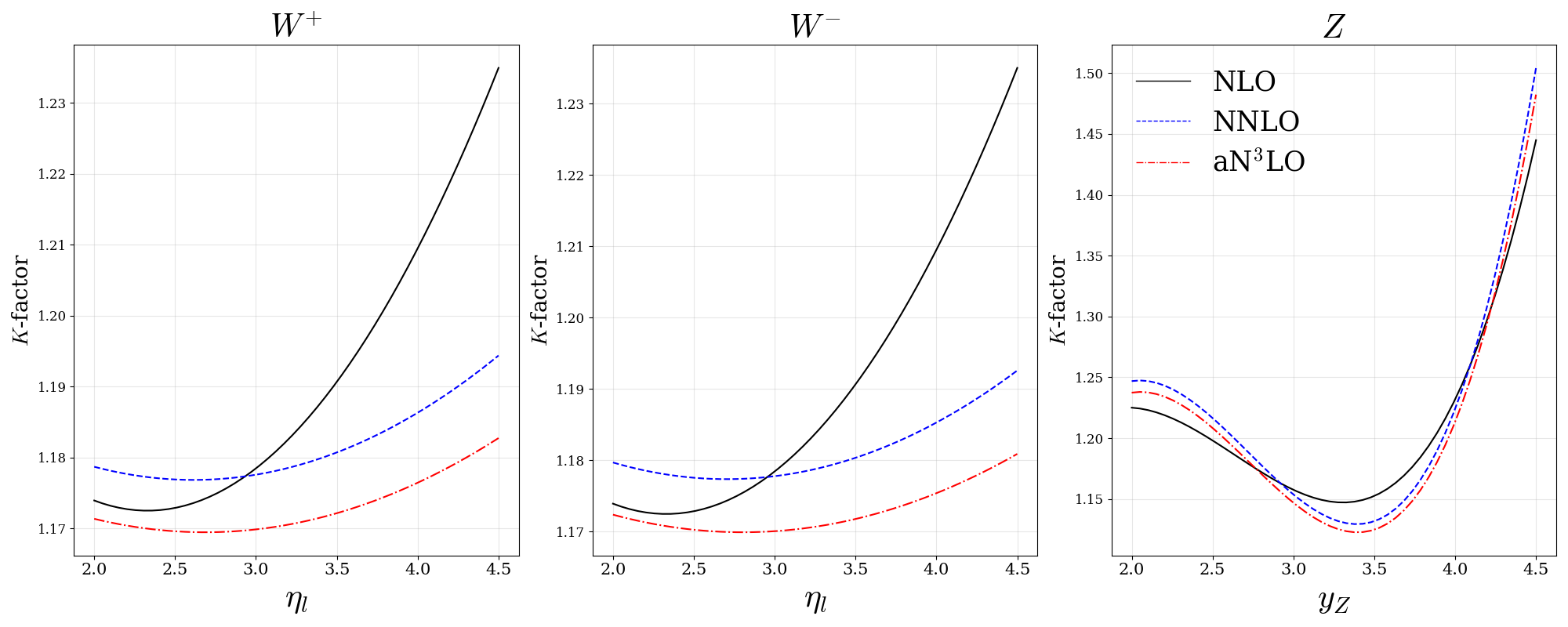}
    \caption{$K$-factor expansion up to aN$^{3}$LO shown for the LHCb 2015 $W,\ Z$ dataset \cite{LHCbZ7,LHCbWZ8}. The $K$-factors shown here are absolute i.e. all with respect to LO ($K^{\mathrm{N^{m}LO/LO}}\ \forall\ m \in \{1, 2, 3\}$).}
    \label{fig:Kfac_DY}
\end{figure}
For the Drell-Yan processes (all calculated at $\mu_{r,f}=m_{ll}/2$), a reduction of $\sim 1-2\%$ in the $K$-factor shift is predicted across most of the corresponding datasets at aN$^{3}$LO. This is in agreement with recent work~\cite{Gehrmann:DYN3LO}. An example of this reduction is shown in Fig.~\ref{fig:Kfac_DY}.

\begin{figure}
    \centering
    \includegraphics[width=\textwidth]{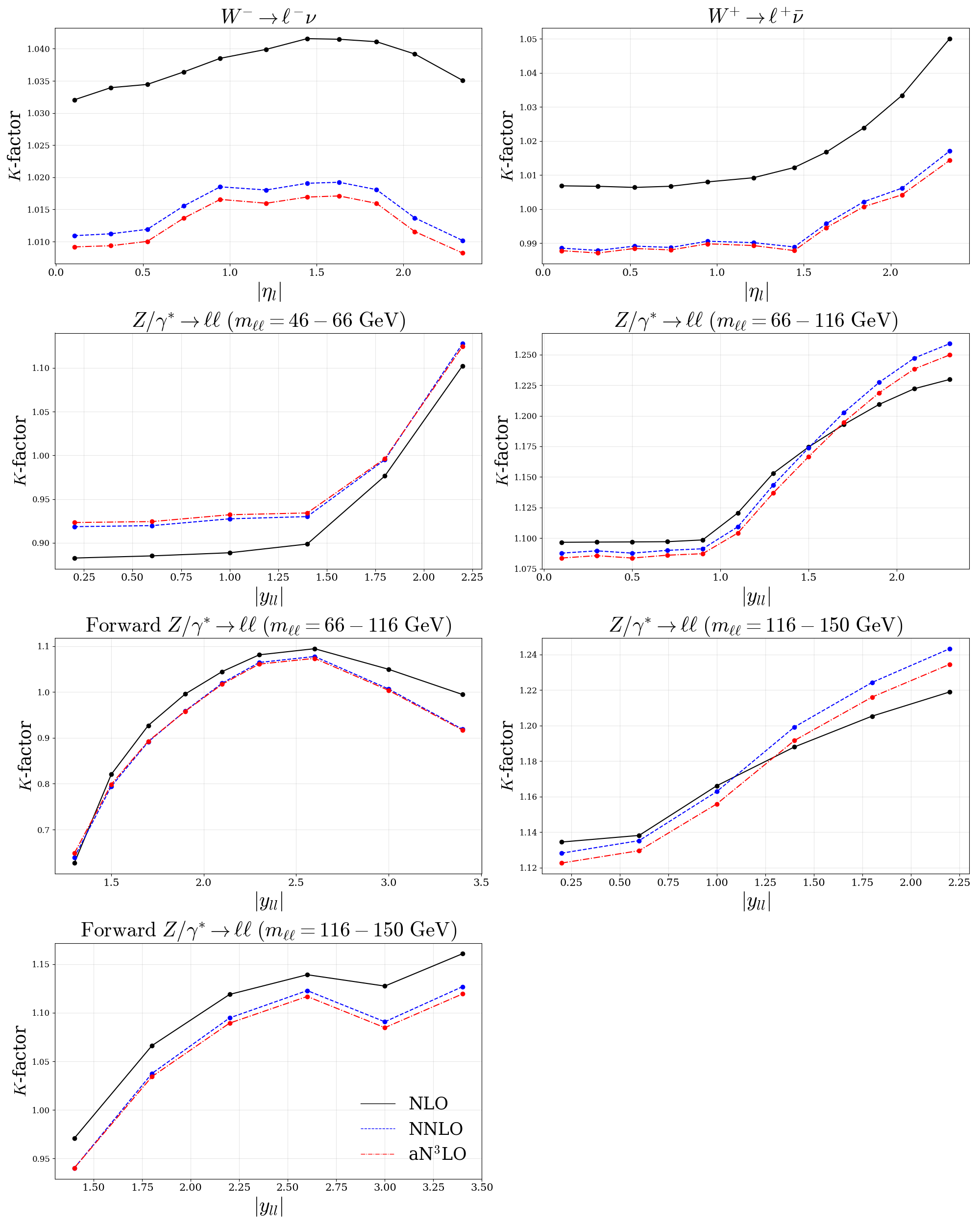}
    \caption{$K$-factor expansion up to aN$^{3}$LO shown for the ATLAS 7 TeV high precision $W,\ Z$ dataset \cite{ATLASWZ7f}. The $K$-factors shown here are absolute i.e. all with respect to LO ($K^{\mathrm{N^{m}LO/LO}}\ \forall\ m \in \{1, 2, 3\}$).}
    \label{fig:Kfac_DY_highprec}
\end{figure}
Conversely, Fig.~\ref{fig:Kfac_DY_highprec} displays an example where the $K$-factor shift has much less of a contribution at N$^{3}$LO. This is a feature of the ATLAS datasets included in the fit due to the impact of chosen $p_{T}$ cuts which reduce the sensitivity to higher orders. 

\begin{table}
\centerline{
\begin{tabular}{|c|c|c|c|}
\hline
DY Dataset & $\chi^{2}$ & $\Delta \chi^{2}$ & $\Delta \chi^{2}$ from NNLO\\
 & & from NNLO & (NNLO $K$-factors)\\
\hline
E866 / NuSea $pp$ DY \cite{E866DY} &   223.3 / 184  & $-1.8$  & $+2.7$ \\
E866 / NuSea $pd/pp$ DY \cite{E866DYrat} &   8.4 / 15  & $-2.0$  & $-1.1$ \\
D{\O} II $Z$ rap. \cite{D0Zrap} &   17.3 / 28  & $+0.9$  & $+0.6$ \\
CDF II $Z$ rap. \cite{CDFZrap} &   40.5 / 28  & $+3.3$  & $+1.3$ \\
D{\O} II $W \rightarrow \nu \mu$ asym. \cite{D0Wnumu} &   16.5 / 10  & $-0.8$  & $-1.1$ \\
CDF II $W$ asym. \cite{CDF-Wasym} &   18.2 / 13  & $-0.8$  & $-0.8$ \\
D{\O} II $W \rightarrow \nu e$ asym. \cite{D0Wnue} &   30.7 / 12  & $-3.2$  & $-3.2$ \\
ATLAS $W^{+},\ W^{-},\ Z$ \cite{ATLASWZ} &   30.0 / 30  & $+0.1$  & $+0.3$ \\
CMS W asym. $p_{T} > 35\ \text{GeV}$ \cite{CMS-easym} &   6.7 / 11  & $-1.1$  & $-1.1$ \\
CMS W asym. $p_{T} > 25, 30\ \text{GeV}$ \cite{CMS-Wasymm} &   7.7 / 24  & $+0.3$  & $+0.3$ \\
LHCb $Z \rightarrow e^{+}e^{-}$ \cite{LHCb-Zee} &   24.1 / 9  & $+1.4$  & $+0.7$ \\
LHCb W asym. $p_{T} > 20\ \text{GeV}$ \cite{LHCb-WZ} &   12.6 / 10  & $+0.1$  & $+0.3$ \\
CMS $Z \rightarrow e^{+}e^{-}$ \cite{CMS-Zee} &   17.5 / 35  & $-0.5$  & $-0.3$ \\
ATLAS High-mass Drell-Yan \cite{ATLAShighmass} &   18.1 / 13  & $-0.8$  & $-0.9$ \\
CMS double diff. Drell-Yan \cite{CMS-ddDY} &   129.5 / 132  & $-15.1$  & $+9.7$ \\
LHCb 2015 $W, Z$ \cite{LHCbZ7,LHCbWZ8} &   103.9 / 67  & $+4.5$  & $-0.1$ \\
LHCb $8 \text{TeV}$ $Z \rightarrow ee$ \cite{LHCbZ8} &   28.8 / 17  & $+2.6$  & $+1.7$ \\
CMS $8\ \text{TeV}\ W$ \cite{CMSW8} &   11.8 / 22  & $-0.9$  & $+0.1$ \\
ATLAS $7\ \text{TeV}$ high prec. $W, Z$ \cite{ATLASWZ7f} &   94.5 / 61  & $-22.1$  & $-8.3$ \\
D{\O} $W$ asym. \cite{D0Wasym} &   12.2 / 14  & $+0.1$  & $+1.3$ \\
ATLAS $8\ \text{TeV}$ High-mass DY \cite{ATLASHMDY8} &   63.0 / 48  & $+5.9$  & $+1.5$ \\
ATLAS $8\ \text{TeV}\ W$ \cite{ATLASW8} &   58.0 / 22  & $+0.4$  & $-0.1$ \\
ATLAS $8\ \text{TeV}$ double diff. $Z$ \cite{ATLAS8Z3D} &   91.6 / 59  & $+15.7$  & $+6.7$ \\
\hline
Total & 1065.4 / 864 & $-12.8$ & $+10.4$\\
\hline
\end{tabular}
}
\caption{\label{tab: DY_kfac} Table showing the relevant DY datasets and how the individual $\chi^{2}$ changes from NNLO by including the N$^{3}$LO treatment of $K$-factors, and theoretical N$^{3}$LO additions discussed earlier. The result with purely NNLO $K$-factors included for all data in the fit is also given.}
\end{table}
Table~\ref{tab: DY_kfac} demonstrates that in most cases, the new fitted DY aN$^{3}$LO $K$-factors are producing a slightly better fit with a moderate cumulative effect. We remind the reader that we have included a total of 20 extra parameters into the fit. These extra 20 parameters are fit across all datasets and multiple processes, whereas the decrease here is for a subset of datasets corresponding to the DY processes included in a global fit.

Across these datasets, the $K$-factors act to extend the description of these processes to approximate N$^{3}$LO. 
The result of including this procedure is a better fit in the DY regime while also relaxing tensions with other processes included in the fit. Comparing the $\Delta \chi^{2}$ results with and without aN$^{3}$LO $K$-factors, we can see the extent to which the $K$-factors and all other N$^{3}$LO additions are reducing the overall $\chi^{2}$.

In some individual cases, the dataset $\chi^{2}$ becomes somewhat worse relative to NNLO most notably the ATLAS $8\ \text{TeV}$ double differential $Z$ \cite{ATLAS8Z3D}, whilst in a few others the $\chi^{2}$ improvement upon addition of the aN$^{3}$LO splitting functions, transition matrix elements and coefficient function pieces is seen to deteriorate upon addition of the aN$^{3}$LO $K$-factors, e.g. the LHCb 2015 $W, Z$ \cite{LHCbZ7,LHCbWZ8}, which exhibits a mild preference for the N$^{3}$LO theory with NNLO $K$-factors. The addition of the aN$^{3}$LO $K$-factors do nonetheless result in a net reduction in $\chi^{2}$ and for a large number of cases the aN$^{3}$LO $K$-factors allow for a slight reduction in the individual $\chi^{2}$. Of particular note is the $7\ \text{TeV}$ high precision $W, Z$ \cite{ATLASWZ7f} data, which improves by over 20 points. This may indicate that the challenge in achieving a statistically good fit to this high precision data that is observed across all NNLO PDF fits is in part related to the lack of higher order corrections in the theory. On the other hand, for arguably the other most precise (and multi-differential) data, namely the ATLAS $8\ \text{TeV}$ double differential $Z$ \cite{ATLAS8Z3D}, the fit quality deteriorates. This may be due to the differing mass binning and/or cuts, but in general it is difficult to draw firm conclusions here, at least with the current $K$-factor treatment. The CMS double diff. Drell-Yan \cite{CMS-ddDY} also shows a particularly large reduction when these are added on top of the aN$^{3}$LO theory, this is a dataset which shows some tension with the DIS N$^{3}$LO additions which is then eased by the addition of the aN$^{3}$LO $K$-factors.

\subsubsection*{Jet Production Processes}

\begin{figure}
    \centering
    \includegraphics[width=\textwidth]{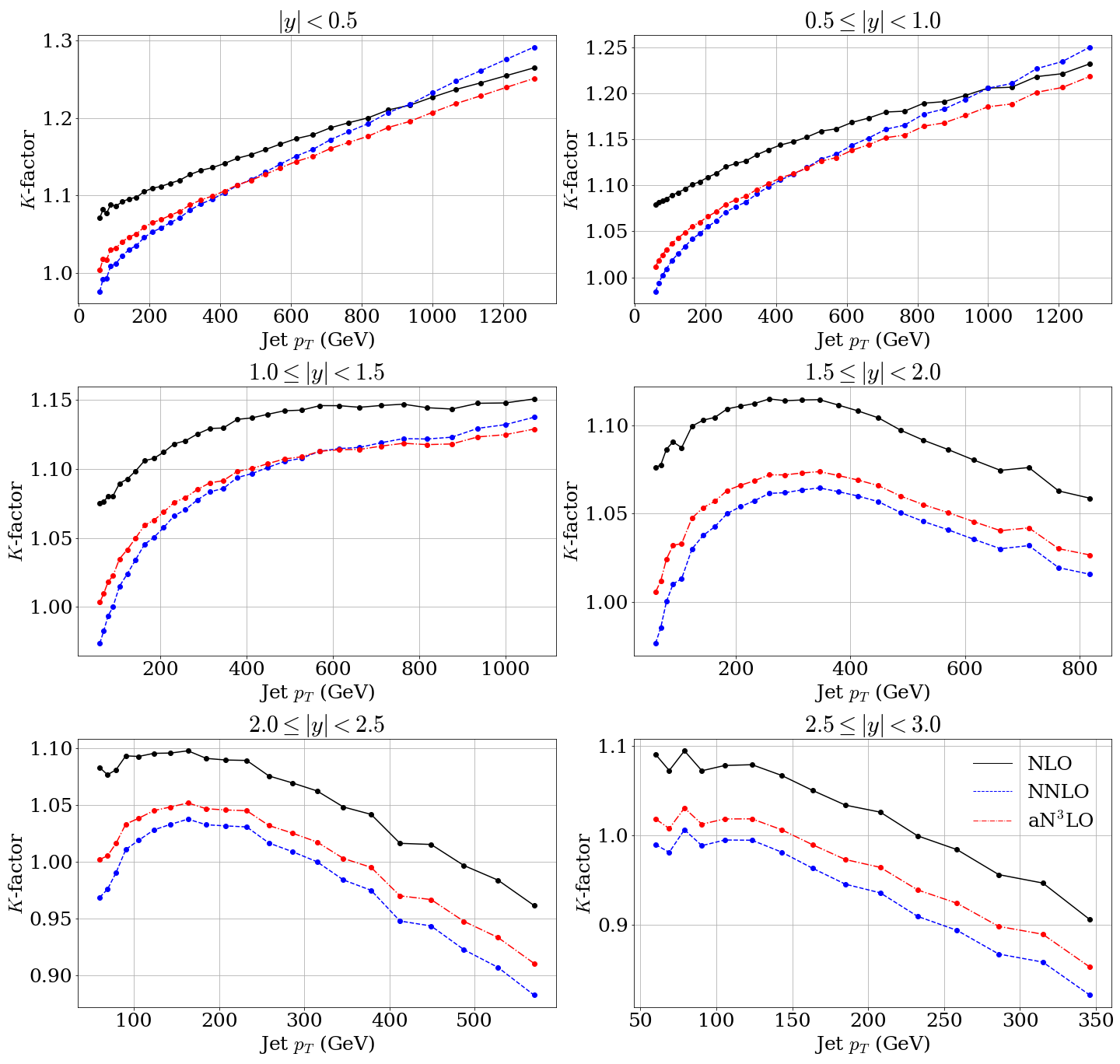}
    \caption{$K$-factor expansion up to aN$^{3}$LO shown for the CMS 7 TeV jets dataset ($R=0.7$) \cite{CMS7jetsfinal}. The $K$-factors shown here are absolute i.e. all with respect to LO ($K^{\mathrm{N^{m}LO/LO}}\ \forall\ m \in \{1, 2, 3\}$).}
    \label{fig:Kfac_jets}
\end{figure}
The jets processes (all calculated for $\mu_{r,f}=p_{T}^{jet}$) show a general increase in the $K$-factor shifts from NNLO as seen in Fig.~\ref{fig:Kfac_jets}, which displays the $K$-factor expansion up to aN$^{3}$LO for the CMS 7 TeV jets dataset~\cite{CMS7jetsfinal}. It is apparent that there is a mild shift to N$^{3}$LO from the NNLO $K$-factor. This behaviour follows what one might expect for a perturbative expansion considering the forms of the NLO and NNLO functions. 

\begin{table}
\centerline{
\begin{tabular}{|c|c|c|c|}
\hline
Jets Dataset & $\chi^{2}$ & $\Delta \chi^{2}$ & $\Delta \chi^{2}$ from NNLO\\
 & & from NNLO & (NNLO $K$-factors)\\
\hline
CDF II $p\bar{p}$ incl. jets \cite{CDFjet} &   66.5 / 76  & $+6.1$  & $+1.7$ \\
D{\O} II $p\bar{p}$ incl. jets \cite{D0jet} &   113.3 / 110  & $-6.9$  & $-3.0$ \\
ATLAS $7\ \text{TeV}$ jets \cite{ATLAS7jets} &   215.9 / 140  & $-5.6$  & $+3.7$ \\
CMS $7\ \text{TeV}$ jets \cite{CMS7jetsfinal} &   186.8 / 158  & $+11.0$  & $+9.3$ \\
CMS $8\ \text{TeV}$ jets \cite{CMS8jets} &   271.3 / 174  & $+10.0$  & $+21.5$ \\
CMS $2.76\ \text{TeV}$ jet \cite{CMS276jets} &   109.8 / 81  & $+6.9$  & $+9.0$ \\
\hline
Total & 963.6 / 739 & $+21.5$ & $+42.2$\\
\hline
\end{tabular}
}
\caption{\label{tab: Jets_kfac}Table showing the relevant jet datasets and how the individual $\chi^{2}$ changes from NNLO by including the N$^{3}$LO treatment of $K$-factors. The result with purely NNLO $K$-factors included for all data in the fit is also given.}
\end{table}
A $\chi^{2}$ summary of the Jets datasets is provided in Table~\ref{tab: Jets_kfac}. By combining the N$^{3}$LO structure function and DGLAP additions (Section's \ref{sec: n3lo_split} to \ref{sec: n3lo_coeff}) with NNLO $K$-factors, the fit exhibits a substantial increase in the $\chi^{2}$ from Jets data. Including aN$^{3}$LO $K$-factors acts to reduce some of this tension with around half the initial overall $\chi^{2}$ increase still remaining. 
We note that in the case of the ATLAS $7\ \text{TeV}$ jets~\cite{ATLAS7jets}, it is well known that there are issues in achieving a good fit quality across all rapidity bins (see~\cite{MMHTjets} for a detailed study as well as~\cite{Bogdan:decorr} where the 8 TeV data are presented and the same issues observed). In~\cite{MMHTjets,Bogdan:decorr} the possibility of decorrelating some of the systematic error sources where the degree of correlation is less well established, was considered and indeed in our study we follow such a procedure, as described in~\cite{Thorne:MSHT20}. Alternatively, however, it might be that the issues in fit quality could at least part be due to deficiencies in theoretical predictions, such as MHOs. To assess this, we revert to the default ATLAS correlation scenario and repeat the global fit. We find that the $\chi^{2}$ deteriorates by $+40.7$ points to 256.6, which is very close to the result found in a pure NNLO fit~\cite{Thorne:MSHT20}. In other words, in our framework the impact of MHOUs does not resolve this issue.

The $\chi^{2}$ results for datasets in Table~\ref{tab: Jets_kfac} show evidence for some tensions with the N$^{3}$LO form of the high-$x$ gluon. It is also apparent that the CMS data is in more tension than ATLAS datasets with N$^{3}$LO structure function and DGLAP theory. Therefore it will be interesting to see how this behaviour changes when considering this data as dijets in the global fit~\cite{AbdulKhalek:2020jut}. We do not consider the dijet data here, though this will be addressed in a future publication.

\subsubsection*{$Z\ p_{T}$ \& Vector Boson $+$ Jets Processes}

\begin{figure}
    \centering
    \includegraphics[width=\textwidth]{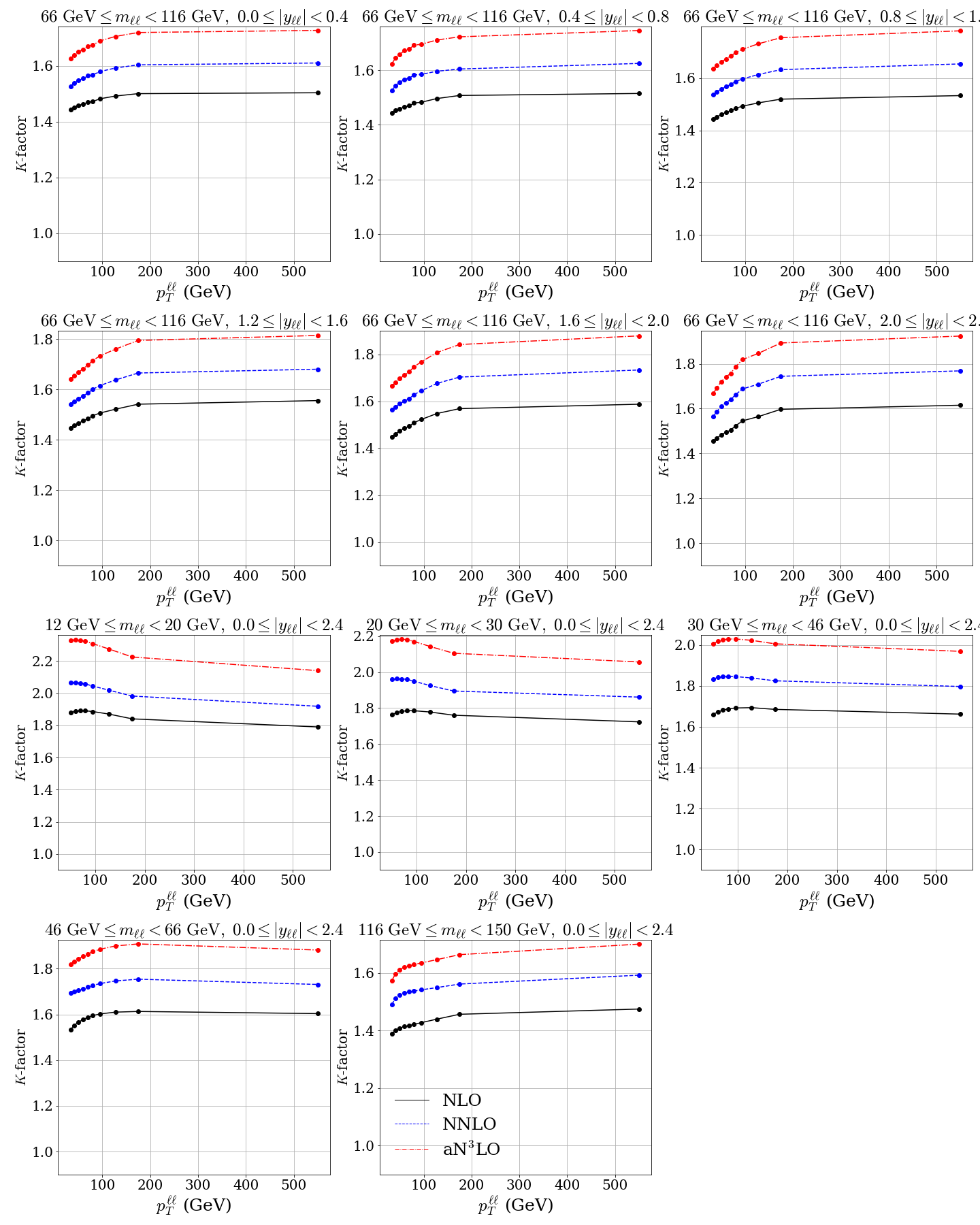}
    \caption{$K$-factor expansion up to aN$^{3}$LO shown for the ATLAS 8 TeV $Z\ p_{T}$ dataset \cite{ATLASZpT}. The $K$-factors shown here are absolute i.e. all with respect to LO ($K^{\mathrm{N^{m}LO/LO}}\ \forall\ m \in \{1, 2, 3\}$).}
    \label{fig:Kfac_pTjet}
\end{figure}
In the case of $Z\ p_{T}$ \& vector boson $+$ jet processes (all calculated at $\mu_{r,f}=\sqrt{p_{T,ll}^2 + m_{ll}^2}$), the $K$-factor shift is almost completely dominated by the ATLAS 8 TeV $Z\ p_{T}$ dataset~\cite{ATLASZpT} (due to the larger number of data points included in this dataset) shown in Fig.~\ref{fig:Kfac_pTjet}. 
The gluon is less directly constrained than the quarks in a global fit. Therefore it can be expected that the significant modifications at small-$x$ will indirectly affect the high-$x$ gluon, where these processes are most sensitive. Considering the jet production processes in Table~\ref{tab: Jets_kfac}, when performing separate PDF fits not including ATLAS 8 TeV $Z\ p_{T}$ data~\cite{ATLASZpT}, we find a reduction of $\Delta\chi^{2} = - 7.0$ in CMS $8\ \text{TeV}$ jets data~\cite{CMS8jets} eliminating most of the tension for this dataset (similar to MSHT20 NNLO results in Table 16 of \cite{Thorne:MSHT20}). Further to this, when not including HERA and ATLAS 8 TeV $Z\ p_{T}$ data we find a reduction of $\Delta\chi^{2} = - 26.4$ in CMS $8\ \text{TeV}$ jet data~\cite{CMS8jets} and $\Delta\chi^{2} = - 12.7$ in CMS $2.76\ \text{TeV}$ jet data~\cite{CMS276jets}.

Although the overall magnitude of the $K$-factor in Fig.~\ref{fig:Kfac_pTjet} may seem large, this new shift is contained within a $15\%$ increase from NNLO (due to the NLO and NNLO $K$-factors also being significant). Moreover, not only does the size of this shift have some dependence on the central scale, but this shift may be more correctly interpreted as the preferred all-orders cross section rather than simply the pure ${\rm N}^3{\rm LO}$ result.

\begin{table}
\centerline{
\begin{tabular}{|c|c|c|c|}
\hline
$p_{T}$ Jets Dataset & $\chi^{2}$ & $\Delta \chi^{2}$ & $\Delta \chi^{2}$ from NNLO\\
 & & from NNLO & (NNLO $K$-factors)\\
\hline
CMS $7\ \text{TeV}\ W + c$ \cite{CMS7Wpc} &   10.8 / 10  & $+2.2$  & $+0.2$ \\
ATLAS $8\ \text{TeV}\ Z\ p_{T}$ \cite{ATLASZpT} &   108.4 / 104  & $-80.0$  & $-54.5$ \\
ATLAS $8\ \text{TeV}\ W + \text{jets}$ \cite{ATLASWjet} &   18.8 / 30  & $+0.7$  & $-0.3$ \\
\hline
Total & 138.0 / 144 & $-77.2$ & $-54.7$\\
\hline
\end{tabular}
}
\caption{\label{tab: pt_jet_kfac}Table showing the relevant $Z\ p_{T}$ \& Vector Boson jet datasets and how the individual $\chi^{2}$ changes from NNLO by including the N$^{3}$LO treatment of $K$-factors, and theoretical N$^{3}$LO additions discussed earlier. The result with purely NNLO $K$-factors included for all data in the fit is also given.}
\end{table}
The extent of the $\chi^{2}$ reduction in the $Z\ p_{T}$ datasets is shown in Table~\ref{tab: pt_jet_kfac}. Note that around $\sim 68\%$ of the improvement to the ATLAS $8\ \text{TeV}\ Z\ p_{T}$~\cite{ATLASZpT} $\chi^{2}$ is due to the extra N$^{3}$LO theory included in the DGLAP and DIS descriptions. 
It is also known the ATLAS $8\ \text{TeV}\ Z\ p_{T}$ data~\cite{ATLASZpT} previously exhibited a significant level of tension with many datasets (including HERA data) at NNLO~\cite{Thorne:MSHT20}. This was investigated by performing a global PDF fit with and without HERA data and comparing the individual $\chi^{2}$'s from each dataset. At NNLO it was found that the ATLAS $8\ \text{TeV}\ Z\ p_{T}$ dataset~\cite{ATLASZpT} reduced by $\Delta\chi^{2} = -39.2$ when fitting to all non-HERA data (see Table~\ref{tab: no_HERA_fullNNLO}). At aN$^{3}$LO we observe that the ATLAS $8\ \text{TeV}\ Z\ p_{T}$ dataset~\cite{ATLASZpT} actually increased by $\Delta\chi^{2} = +12.8$ when fitting to all non-HERA data (see Table~\ref{tab: no_HERA_fullN3LO}). The aN$^{3}$LO additions therefore eliminate this tension previously observed at NNLO,  suggesting that this issue at NNLO in fitting the ATLAS $8 \text{TeV} Z\ p_{T}$ dataset~\cite{ATLASZpT} was a sign of MHOs. This is in contrast with the result observed for ATLAS $7\ \text{TeV}$ jets~\cite{ATLAS7jets} where the issues with fit quality were not alleviated by the inclusion of known higher order N3LO information and approximations for the remaining missing pieces.

Finally we remind the reader that the CMS $7\ \text{TeV}\ W + c$ dataset~\cite{CMS7Wpc} does not include a $K$-factor at NNLO. To overcome this, we tie the overall N$^{3}$LO $K$-factor shift to the NLO value ($K^{\mathrm{NNLO/NLO}} = 1$ in Equation~\eqref{eq: meth2}), therefore contributing as an overall normalisation effect.

\subsubsection*{Top Quark Processes}

\begin{figure}[t]
    \centering
    \includegraphics[width=0.7\textwidth]{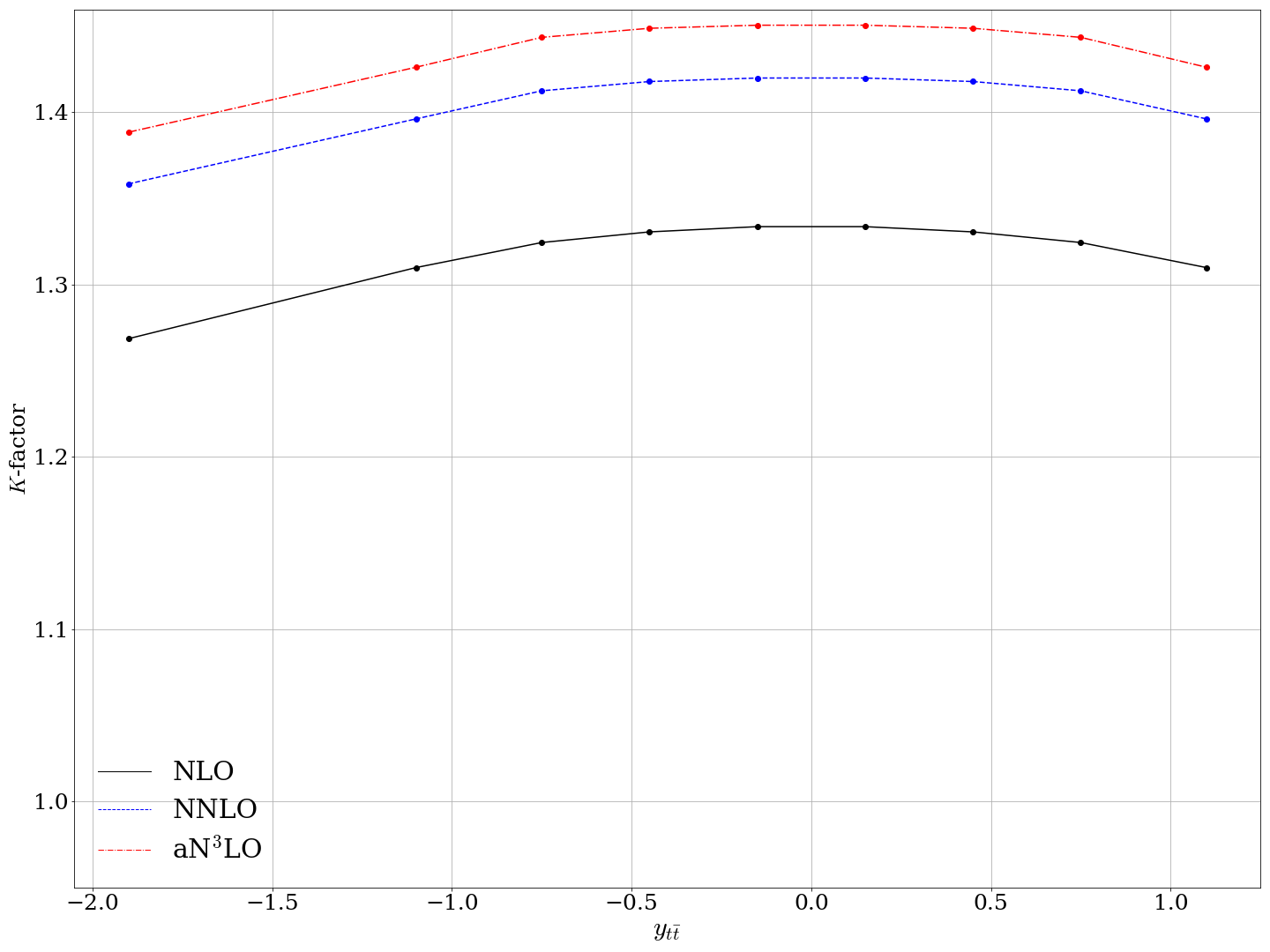}
    \caption{$K$-factor expansion up to aN$^{3}$LO shown for the CMS 8 TeV single diff. $t\bar{t}$ dataset \cite{CMSttbar08_ytt}. The $K$-factors shown here are absolute i.e. all with respect to LO ($K^{\mathrm{N^{m}LO/LO}}\ \forall\ m \in \{1, 2, 3\}$).}
    \label{fig:Kfac_top}
\end{figure}
Moving to top quark processes, for the single differential datasets the scale choice for $\mu_{r,f}$ is $H_T/4$ with the exception of data differential in the average transverse momentum of the top or anti-top, $p_T^t,p_T^{\bar{t}}$, for which $m_T/2$ is used. For the double diff. dataset the scale choice is $H_T/4$ and for the inclusive top $\sigma_{t\bar{t}}$ a scale of $m_{t}$ is chosen. Fig.~\ref{fig:Kfac_top} displays the $K$-factor shifts up to N$^{3}$LO for the CMS 8 TeV single diff. $t\bar{t}$ dataset~\cite{CMSttbar08_ytt}, which shows the greatest reduction in its $\chi^2$. A familiar perturbative pattern can be seen for this process's $K$-factors, with the shift at aN$^{3}$LO increasing by around 3--4\% from NNLO. This is in agreement with a recent $\sim3.5\%$ predicted increase in the N$^3$LO $t\bar{t}$ production $K$-factor at $8\ \text{TeV}$ in~\cite{Kidonakis:tt}, whereby an approximate N$^{3}$LO cross section for $t\bar{t}$ production in proton-proton collisions has been calculated employing a resummation formalism~\cite{Kidonakis:1997gm,Kidonakis:2009ev,Kidonakis:2010dk,Kidonakis:2021yrh}.

\begin{table}
\centerline{
\begin{tabular}{|c|c|c|c|}
\hline
Top Dataset & $\chi^{2}$ & $\Delta \chi^{2}$ & $\Delta \chi^{2}$ from NNLO\\
 & & from NNLO & (NNLO $K$-factors)\\
\hline
Tevatron, ATLAS, CMS $\sigma_{t\bar{t}}$ \cite{Tevatron-top,ATLAS-top7-1,ATLAS-top7-2,ATLAS-top7-3,ATLAS-top7-4,ATLAS-top7-5,ATLAS-top7-6,CMS-top7-1,CMS-top7-2,CMS-top7-3,CMS-top7-4,CMS-top7-5,CMS-top8} &   14.3 / 17  & $-0.2$  & $-1.0$ \\
ATLAS $8\ \text{TeV}$ single diff. $t\bar{t}$ \cite{ATLASsdtop} &   24.2 / 25  & $-1.4$  & $-0.8$ \\
ATLAS $8\ \text{TeV}$ single diff. $t\bar{t}$ dilep. \cite{ATLASttbarDilep08_ytt} &   2.7 / 5  & $-0.7$  & $-0.7$ \\
CMS $8\ \text{TeV}$ double diff. $t\bar{t}$ \cite{CMS8ttDD} &   23.6 / 15  & $+1.0$  & $+4.0$ \\
CMS $8\ \text{TeV}$ single diff. $t\bar{t}$ \cite{CMSttbar08_ytt} &   10.3 / 9  & $-2.9$  & $-4.0$ \\
\hline
Total & 75.1 / 71 & $-4.2$ & $-2.5$\\
\hline
\end{tabular}
}
\caption{\label{tab: top_kfac}Table showing the relevant Top Quark datasets and how the individual $\chi^{2}$ changes from NNLO by including the N$^{3}$LO treatment of $K$-factors, and theoretical N$^{3}$LO additions discussed earlier. The result with purely NNLO $K$-factors included for all data in the fit is also given.}
\end{table}
The $\chi^{2}$ results in Table~\ref{tab: top_kfac} display a mildly better fit for top processes, with most datasets not feeling a large overall effect from the N$^{3}$LO additions. Comparing with and without aN$^{3}$LO $K$-factors, we see a slightly better fit overall, with most of the reduction in overall $\chi^{2}$ stemming from CMS $8\ \text{TeV}$ double diff. $t\bar{t}$ data~\cite{CMS8ttDD}. 

\subsubsection*{Semi-Inclusive DIS Dimuon Processes}

\begin{figure}
    \centering
    \includegraphics[width=\textwidth]{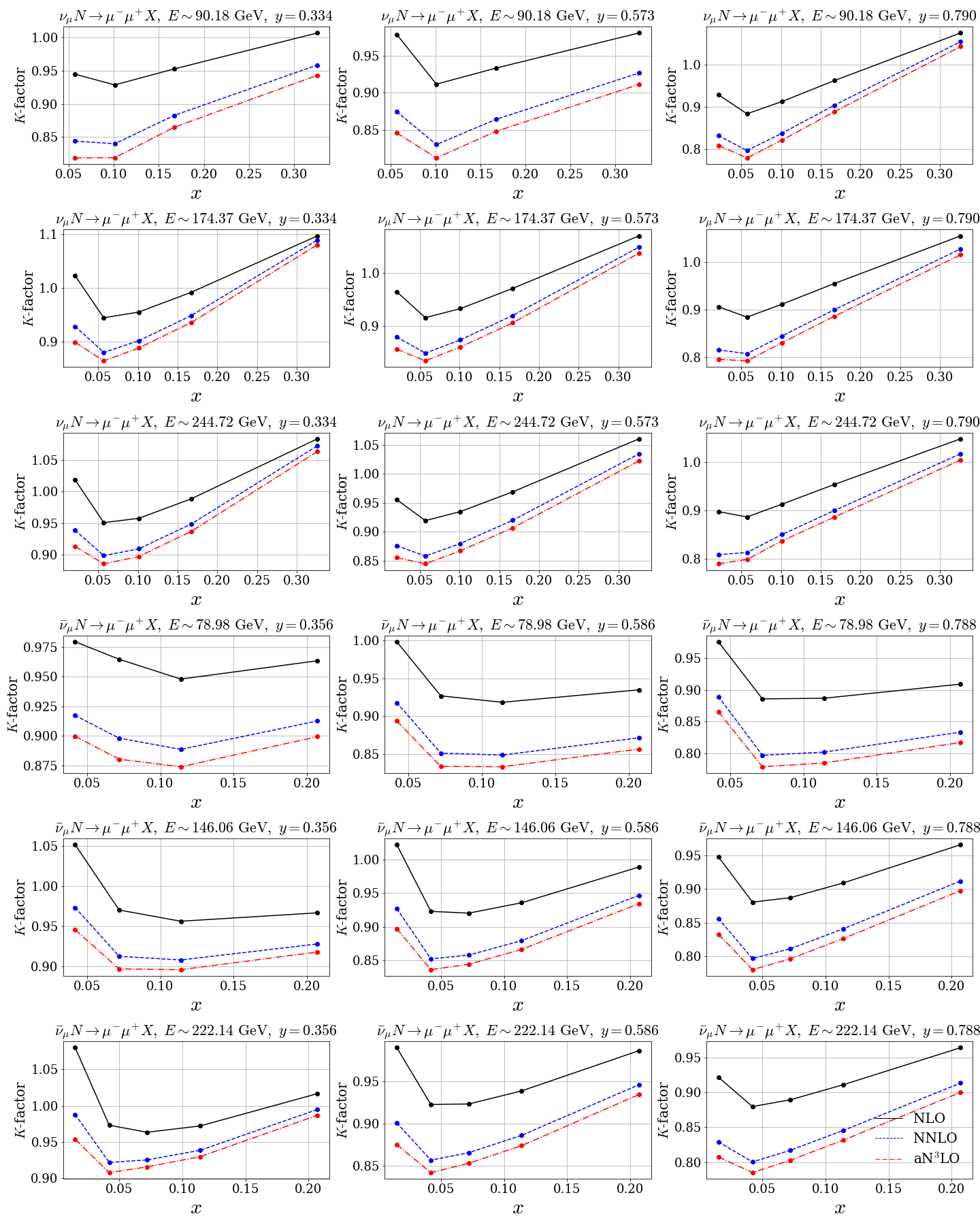}
    \caption{$K$-factor expansion up to aN$^{3}$LO shown for the NuTeV $\nu N \rightarrow \mu \mu X$ dataset \cite{Dimuon}. The $K$-factors shown here are absolute i.e. all with respect to LO ($K^{\mathrm{N^{m}LO/LO}}\ \forall\ m \in \{1, 2, 3\}$).}
    \label{fig:Kfac_dimuon}
\end{figure}
The final set of results to consider in this Section are the aN$^{3}$LO $K$-factors associated with semi-inclusive DIS dimuon cross sections (with $\mu_{r,f}^2=Q^{2}$). Although the dimuon cross section is associated with the DIS process described from our approximate N$^{3}$LO structure function picture, it is a semi-inclusive DIS process. Therefore it is sensible to treat this process as entirely separate from DIS. The NNLO cross-sections used in this case are a general-mass variable flavour number scheme extension of the results in \cite{Berger:2016inr}, as described in more detail in \cite{Thorne:MSHT20}. 
The $K$-factors shown in Fig.~\ref{fig:Kfac_dimuon} (for the NuTeV $\nu N \rightarrow \mu \mu X$ data~\cite{Dimuon}) are somewhat similar to NNLO. The reason for this is mostly due to these datasets also including a branching ratio (${\rm BR}(D\to \mu)$) which absorbs any overall normalisation shifts. 
This behaviour is not a concern since in practice these two work in tandem and when combined together it makes no difference where the normalisation factors are absorbed into.
\begin{table}
\centerline{
\begin{tabular}{|c|c|c|c|c|}
\hline
\multirow{2}{*}{} & \multirow{2}{*}{NLO} & \multirow{2}{*}{NNLO} & \multirow{2}{*}{aN$^{3}$LO} & aN$^{3}$LO\\
 & &  &  & (NNLO $K$-factors)\\
\hline
 & \multirow{2}{*}{0.095} & \multirow{2}{*}{0.088}  & \multirow{2}{*}{0.082}  & \multirow{2}{*}{0.081} \\
${\rm BR}(D\to \mu)$ & & & & \\
\hline
\end{tabular}
}
\caption{\label{tab: dimuon_br}Table displaying dimuon branching ratios (BRs) at NLO, NNLO, aN$^{3}$LO and aN$^{3}$LO with NNLO $K$-factors.}
\end{table}
Investigating the change in the BR's with the addition of N$^{3}$LO contributions in Table~\ref{tab: dimuon_br}, the BR at N$^{3}$LO decreases substantially from NNLO, with little difference from the addition of aN$^{3}$LO $K$-factors. The predicted dimuon BR at aN$^{3}$LO is inside the allowed $\pm 1 \sigma$ range of $0.092 \pm 0.010$. When performing a fit with the BR fixed at its central value (BR = $0.092$), one is able to observe the effect of manually forcing the normalisation into the $K$-factor variation alone. The result of this is a worse global fit quality $\Delta \chi^{2} = + 11.2$, where $+3.9$ units arise from an increased penalty for the Dimuon $K$-factor description and $+2.3$ units from a slightly worse fit to the Dimuon datasets listed in Table~\ref{tab: dimuon_kfac}. The rest of the observed increase in $\chi^{2}_{\mathrm{global}}$ is dominated by a $+4.1$ increase in the ATLAS $7\ \text{TeV}$ high prec.~$W, Z$ \cite{ATLASWZ7f} data due to a smaller strange quark PDF (compensating the higher BR in dimuon datasets). Returning to consider the case of the $K$-factors and BR together, the predicted effect on dimuon datasets is very similar. However due to the errors accounting for a larger allowed shift in the BR relative to the $K$-factors, the fit favours moving the BR by a larger amount to reduce the penalty $\chi^{2}$ contribution from $K$-factors which explains the results shown in Table~\ref{tab: dimuon_br}.

\begin{table}
\centerline{
\begin{tabular}{|c|c|c|c|}
\hline
Dimuon Dataset & $\chi^{2}$ & $\Delta \chi^{2}$ & $\Delta \chi^{2}$ from NNLO\\
 & & from NNLO & (NNLO $K$-factors)\\
\hline
CCFR $\nu N \rightarrow \mu\mu X$ \cite{Dimuon} &   68.3 / 86  & $+0.6$  & $+1.1$ \\
NuTeV $\nu N \rightarrow \mu\mu X$ \cite{Dimuon} &   56.7 / 84  & $-1.8$  & $-0.6$ \\
\hline
Total & 125.0 / 170 & $-1.2$ & $+0.5$\\
\hline
\end{tabular}
}
\caption{\label{tab: dimuon_kfac}Table showing the relevant Dimuon datasets and how the individual $\chi^{2}$ changes from NNLO by including the N$^{3}$LO treatment of $K$-factors, and theoretical N$^{3}$LO additions discussed earlier. The result with purely NNLO $K$-factors included for all data in the fit is also given.}
\end{table}
Table~\ref{tab: dimuon_kfac} further confirms the expectation that the Dimuon datasets are not too sensitive to N$^{3}$LO additions. The results with and without a full treatment of aN$^{3}$LO $K$-factors are also similar in magnitude. 
It is therefore clear that the dimuon BR's are compensating for any indirect normalisation effects from the form of the PDFs in the full aN$^{3}$LO fit, as opposed to the aN$^{3}$LO $K$-factors.

\section{MSHT20 Approximate \texorpdfstring{N$^{3}$LO}{N3LO} Global Analysis}\label{sec: results}

With the inclusion of all N$^{3}$LO approximations discussed in earlier sections resulting in 20 extra free parameters from the NNLO MSHT20 fit, we now present the results for the first approximate N$^{3}$LO global PDF fit with theoretical uncertainties from missing N$^{3}$LO contributions and implicitly some MHOs beyond this. This includes the results for the best fit for the nuisance parameters describing the theoretical uncertainty. We remind the reader that these are parameterised specifically to represent the missing uncertainty at N$^3$LO, which is currently the dominant source of uncertainty due to missing higher orders. However, the fit will also be influenced, to a limited extent, by effects at even higher orders. Later in the Section we discuss this in more detail.

\subsection{\texorpdfstring{$\chi^{2}$}{Chi-squared} Breakdown}\label{subsec: chi2}

\begin{table}
\centerline{
\begin{tabular}{|P{6.5cm}|P{1cm}|P{2cm}|P{2cm}|}
\hline
Dataset & $N_{\mathrm{pts}}$ & $\chi^{2}$ & $\Delta \chi^{2}$ from NNLO \\
\hline
BCDMS $\mu p$ $F_{2}$ \cite{BCDMS} &   163  &  174.4  &  $-5.8$ \\
BCDMS $\mu d$ $F_{2}$ \cite{BCDMS} &   151  &  144.3  &  $-1.7$ \\
NMC $\mu p$ $F_{2}$ \cite{NMC} &   123  &  121.5  &  $-2.6$ \\
NMC $\mu d$ $F_{2}$ \cite{NMC} &   123  &  104.2  &  $-8.4$ \\
SLAC $ep$ $F_{2}$ \cite{SLAC,SLAC1990} &   37  &  31.6  &  $-0.4$ \\
SLAC $ed$ $F_{2}$ \cite{SLAC,SLAC1990} &   38  &  22.8  &  $-0.2$ \\
E665 $\mu d$ $F_{2}$ \cite{E665} &   53  &  63.9  &  $+4.2$ \\
E665 $\mu p$ $F_{2}$ \cite{E665} &   53  &  67.5  &  $+2.9$ \\
NuTeV $\nu N$ $F_{2}$ \cite{NuTev} &   53  &  35.7  &  $-2.6$ \\
NuTeV $\nu N$ $xF_{3}$ \cite{NuTev} &   42  &  34.8  &  $+4.1$ \\
NMC $\mu n / \mu p$ \cite{NMCn/p} &   148  &  131.6  &  $+0.8$ \\
E866 / NuSea $pp$ DY \cite{E866DY} &   184  &  223.3  &  $-1.8$ \\
E866 / NuSea $pd/pp$ DY \cite{E866DYrat} &   15  &  8.4  &  $-2.0$ \\
HERA $ep$ $F_{2}^{\text{charm}}$ \cite{H1+ZEUScharm} &   79  &  143.7  &  $+11.4$ \\
NMC/BCDMS/SLAC/HERA $F_{L}$ \cite{BCDMS,NMC,SLAC1990,H1FL,H1-FL,ZEUS-FL} &   57  &  45.6  &  $-22.9$ \\
CCFR $\nu N \rightarrow \mu\mu X$ \cite{Dimuon} &   86  &  68.3  &  $+0.6$ \\
NuTeV $\nu N \rightarrow \mu\mu X$ \cite{Dimuon} &   84  &  56.7  &  $-1.8$ \\
CHORUS $\nu N$ $F_{2}$ \cite{CHORUS} &   42  &  29.2  &  $-1.0$ \\
CHORUS $\nu N$ $xF_{3}$ \cite{CHORUS} &   28  &  18.1  &  $-0.3$ \\
HERA $e^{+}p$ CC \cite{H1+ZEUS} &   39  &  49.7  &  $-2.3$ \\
HERA $e^{-}p$ CC \cite{H1+ZEUS} &   42  &  64.9  &  $-5.3$ \\
HERA $e^{+}p$ NC $820\ \text{GeV}$ \cite{H1+ZEUS} &   75  &  84.3  &  $-5.6$ \\
HERA $e^{-}p$ NC $460\ \text{GeV}$ \cite{H1+ZEUS} &   209  &  247.7  &  $-0.6$ \\
HERA $e^{+}p$ NC $920\ \text{GeV}$ \cite{H1+ZEUS} &   402  &  474.0  &  $-38.7$ \\
HERA $e^{-}p$ NC $575\ \text{GeV}$ \cite{H1+ZEUS} &   259  &  248.5  &  $-14.5$ \\
HERA $e^{-}p$ NC $920\ \text{GeV}$ \cite{H1+ZEUS} &   159  &  243.0  &  $-1.4$ \\
CDF II $p\bar{p}$ incl. jets \cite{CDFjet} &   76  &  66.5  &  $+6.1$ \\
D{\O} II $Z$ rap. \cite{D0Zrap} &   28  &  17.3  &  $+0.9$ \\
CDF II $Z$ rap. \cite{CDFZrap} &   28  &  40.5  &  $+3.3$ \\
D{\O} II $W \rightarrow \nu \mu$ asym. \cite{D0Wnumu} &   10  &  16.5  &  $-0.8$ \\
CDF II $W$ asym. \cite{CDF-Wasym} &   13  &  18.2  &  $-0.8$ \\
\hline
\end{tabular}
}
\caption{\label{tab: K-fac_firstresults}Full breakdown of $\chi^{2}$ results for the aN$^{3}$LO PDF fit. The global fit includes the N$^{3}$LO treatment for transition matrix elements, coefficient functions, splitting functions and $K$-factor additions with their variational parameters determined by the fit.}
\end{table}
\begin{table}\ContinuedFloat
\centerline{
\begin{tabular}{|P{6.5cm}|P{1cm}|P{2cm}|P{2cm}|}
\hline
Dataset & $N_{\mathrm{pts}}$ & $\chi^{2}$ & $\Delta \chi^{2}$ from NNLO \\
\hline
D{\O} II $W \rightarrow \nu e$ asym. \cite{D0Wnue} &   12  &  30.7  &   $-3.2$ \\
D{\O} II $p\bar{p}$ incl. jets \cite{D0jet} &   110  &  113.3  &   $-6.9$ \\
ATLAS $W^{+},\ W^{-},\ Z$ \cite{ATLASWZ} &   30  &  30.0  &   $+0.1$ \\
CMS W asym. $p_{T} > 35\ \text{GeV}$ \cite{CMS-easym} &   11  &  6.7  &   $-1.1$ \\
CMS W asym. $p_{T} > 25, 30\ \text{GeV}$ \cite{CMS-Wasymm} &   24  &  7.7  &   $+0.3$ \\
LHCb $Z \rightarrow e^{+}e^{-}$ \cite{LHCb-Zee} &   9  &  24.1  &   $+1.4$ \\
LHCb W asym. $p_{T} > 20\ \text{GeV}$ \cite{LHCb-WZ} &   10  &  12.6  &   $+0.1$ \\
CMS $Z \rightarrow e^{+}e^{-}$ \cite{CMS-Zee} &   35  &  17.5  &   $-0.5$ \\
ATLAS High-mass Drell-Yan \cite{ATLAShighmass} &   13  &  18.1  &   $-0.8$ \\
Tevatron, ATLAS, CMS $\sigma_{t\bar{t}}$ \cite{Tevatron-top,ATLAS-top7-1,ATLAS-top7-2,ATLAS-top7-3,ATLAS-top7-4,ATLAS-top7-5,ATLAS-top7-6,CMS-top7-1,CMS-top7-2,CMS-top7-3,CMS-top7-4,CMS-top7-5,CMS-top8} &   17  &  14.3  &   $-0.2$ \\
CMS double diff. Drell-Yan \cite{CMS-ddDY} &   132  &  129.5  &   $-15.1$ \\
LHCb 2015 $W, Z$ \cite{LHCbZ7,LHCbWZ8} &   67  &  103.9  &   $+4.5$ \\
LHCb $8 \text{TeV}$ $Z \rightarrow ee$ \cite{LHCbZ8} &   17  &  28.8  &   $+2.6$ \\
CMS $8\ \text{TeV}\ W$ \cite{CMSW8} &   22  &  11.8  &   $-0.9$ \\
ATLAS $7\ \text{TeV}$ jets \cite{ATLAS7jets} &   140  &  215.9  &   $-5.6$ \\
CMS $7\ \text{TeV}\ W + c$ \cite{CMS7Wpc} &   10  &  10.8  &   $+2.2$ \\
ATLAS $7\ \text{TeV}$ high prec. $W, Z$ \cite{ATLASWZ7f} &   61  &  94.5  &   $-22.1$ \\
CMS $7\ \text{TeV}$ jets \cite{CMS7jetsfinal} &   158  &  186.8  &   $+11.0$ \\
D{\O} $W$ asym. \cite{D0Wasym} &   14  &  12.2  &   $+0.1$ \\
ATLAS $8\ \text{TeV}\ Z\ p_{T}$ \cite{ATLASZpT} &   104  &  108.4  &   $-80.0$ \\
CMS $8\ \text{TeV}$ jets \cite{CMS8jets} &   174  &  271.3  &   $+10.0$ \\
ATLAS $8\ \text{TeV}$ sing. diff. $t\bar{t}$ \cite{ATLASsdtop} &   25  &  24.2  &   $-1.4$ \\
ATLAS $8\ \text{TeV}$ sing. diff. $t\bar{t}$ dilep. \cite{ATLASttbarDilep08_ytt} &   5  &  2.7  &   $-0.7$ \\
ATLAS $8\ \text{TeV}$ High-mass DY \cite{ATLASHMDY8} &   48  &  63.0  &   $+5.9$ \\
ATLAS $8\ \text{TeV}\ W + \text{jets}$ \cite{ATLASWjet} &   30  &  18.8  &   $+0.7$ \\
CMS $8\ \text{TeV}$ double diff. $t\bar{t}$ \cite{CMS8ttDD} &   15  &  23.6  &   $+1.0$ \\
ATLAS $8\ \text{TeV}\ W$ \cite{ATLASW8} &   22  &  58.0  &   $+0.4$ \\
CMS $2.76\ \text{TeV}$ jet \cite{CMS276jets} &   81  &  109.8  &   $+6.9$ \\
CMS $8\ \text{TeV}$ sing. diff. $t\bar{t}$ \cite{CMSttbar08_ytt} &   9  &  10.3  &   $-2.9$ \\
ATLAS $8\ \text{TeV}$ double diff. $Z$ \cite{ATLAS8Z3D} &   59  &  91.6  &   $+15.7$ \\
\hline
\end{tabular}
}
\caption{\textit{(Continued)} Full breakdown of $\chi^{2}$ results for the aN$^{3}$LO PDF fit. The global fit includes the N$^{3}$LO treatment for transition matrix elements, coefficient functions, splitting functions and $K$-factor additions with their variational parameters determined by the fit.}
\end{table}
\begin{table}[t]\ContinuedFloat
\centerline{
\begin{tabular}{|c|c|c|c|}
\hline
Low-$Q^{2}$ Coefficient & & & \\
\hline
$c_{q}^{\mathrm{NLL}}$ $ = -3.868$  &   0.004  &  $c_{g}^{\mathrm{NLL}}$ $ = -5.837$  &   0.844 \\
\hline
Transition Matrix Elements & & & \\
\hline
$a_{Hg}$ $ = 12214.000$  &   0.601  &  $a_{qq,H}^{\mathrm{NS}}$ $ = -64.411$  &   0.001 \\
$a_{gg,H}$ $ = -1951.600$  &   0.857  & & \\
\hline
Splitting Functions & & & \\
\hline
$\rho_{qq}^{NS}$ $ = 0.007$  &   0.000  &  $\rho_{gq}$ $ = -1.784$  &   0.802 \\
$\rho_{qq}^{PS}$ $ = -0.501$  &   0.186  &  $\rho_{gg}$ $ = 19.245$  &   3.419 \\
$\rho_{qg}$ $ = -1.754$  &   0.015  & & \\
\hline
K-factors & & & \\
\hline
$\mathrm{DY}_{\mathrm{NLO}}$ $ = -0.282$  &   0.080  &  $\mathrm{DY}_{\mathrm{NNLO}}$ $ = 0.079$  &   0.006 \\
$\mathrm{Top}_{\mathrm{NLO}}$ $ = 0.041$  &   0.002  &  $\mathrm{Top}_{\mathrm{NNLO}}$ $ = 0.651$  &   0.424 \\
$\mathrm{Jet}_{\mathrm{NLO}}$ $ = -0.300$  &   0.090  &  $\mathrm{Jet}_{\mathrm{NNLO}}$ $ = -0.691$  &   0.478 \\
$p_{T}\mathrm{Jets}_{\mathrm{NLO}}$ $ = 0.583$  &   0.339  &  $p_{T}\mathrm{Jets}_{\mathrm{NNLO}}$ $ = -0.080$  &   0.006 \\
$\mathrm{Dimuon}_{\mathrm{NLO}}$ $ = -0.444$  &   0.197  &  $\mathrm{Dimuon}_{\mathrm{NNLO}}$ $ = 0.922$  &   0.850 \\
\hline
N$^{3}$LO Penalty Total & 9.201 / 20 & Average Penalty & 0.460 \\
\hline
\multicolumn{2}{c|}{} & Total & 4957.2 / 4363\\
\multicolumn{2}{c|}{} & $\Delta \chi^{2}$ from NNLO & $-154.4$ \\
\cline{3-4}
\end{tabular}
}
\caption{\textit{(Continued)} Full breakdown of $\chi^{2}$ results for the aN$^{3}$LO PDF fit. The global fit includes the N$^{3}$LO treatment for transition matrix elements, coefficient functions, splitting functions and $K$-factor additions with their variational parameters determined by the fit.}
\end{table}
Table~\ref{tab: K-fac_firstresults} shows the global $\chi^{2}$ results for an aN$^{3}$LO best fit, inclusive of penalties associated with the new theory variational parameters (from Equation~\eqref{eq: penalty_example}). The theory parameters are labelled as: $A_{Hg}(a_{Hg})$, $A_{gg,H}(a_{gg,H})$, $A_{qq,H}^{\mathrm{NS}}(a_{qq,H}^{\mathrm{NS}})$ for the transition matrix elements; $P_{qq}^{\mathrm{NS}}(\rho_{qq}^{\mathrm{NS}})$, $P_{qq}^{\mathrm{PS}}(\rho_{qq}^{\mathrm{PS}})$, $P_{qg}(\rho_{qg})$, $P_{gq}(\rho_{gq})$ and $P_{gg}(\rho_{gg})$ for the splitting functions; and $c_{q}^{\mathrm{NLL}}$ and $c_{g}^{\mathrm{NLL}}$ correspond to the NLL parameters discussed in Section~\ref{subsec: NLL_coeff}. These are supplemented by the 10 additional nuisance parameters for the NLO and NNLO $K$-factors for the five process categories. These 20 additional parameters and their associated penalties are also shown in Table~\ref{tab: K-fac_firstresults}.

The extra N$^{3}$LO theory and level of freedom introduced has allowed the fit to achieve a total $\Delta \chi^{2} = - 154.4$ compared to MSHT20 NNLO total $\chi^{2}$ (Table 7 from \cite{Thorne:MSHT20}). 
Comparing with lower order PDF fits, we find a smooth convergence in the fit quality which follows what one may expect from an increase in the accuracy of a perturbative expansion ($\chi^{2} / N_{\mathrm{pts}}$ = LO: 2.57, NLO: 1.33, NNLO: 1.17, N$^{3}$LO: 1.14). In part, this is due to the extra freedom in the $K$-factors, which will almost always act to reduce this $\chi^{2}$ due to the minimisation procedure. However, even with this freedom, in most cases the N$^{3}$LO theory (non $K$-factor) contributions include large divergences from NNLO. With this in mind, we must conclude that the fit is preferring a description different from the current NNLO standard. 

At NNLO (Table~\ref{tab: no_HERA_fullNNLO}), the tension between HERA and non-HERA datasets accounted for $\Delta \chi^{2} = -61.6$ reduction in the overall fit quality when the former was removed, with the majority of this tension between HERA and ATLAS $8\ \text{TeV}\ Z\ p_{T}$~\cite{ATLASZpT} data. Whereas comparing fit results with and without HERA data at N$^{3}$LO, we find $\Delta \chi^{2} = -49.0$.  Although the overall difference is not too substantial we do report a substantial shift in the leading tensions, where most of the tension with HERA data is now residing with NMC $F_{2}$~\cite{NMC} and CMS $8\ \text{TeV}$ jets~\cite{CMS8jets} data. Tensions with NMC $F_{2}$~\cite{NMC} data are also seen to some extent at NNLO where we show a $\Delta \chi^{2} = - 20.6$ in a fit omitting HERA data (combining the NMC $F_{2}$ datasets shown in Table~\ref{tab: no_HERA_fullNNLO}). However at N$^{3}$LO, Table~\ref{tab: no_HERA_fullN3LO} shows a $\Delta \chi^{2} = - 24.4$ reduction from NMC $F_{2}$ data in a fit omitting HERA data. Therefore whilst the N$^{3}$LO additions remove tensions with $Z\ p_{T}$ data, it remains that the HERA data is preferring the high-$x$ quarks to be lower than favoured by NMC data. This is suggestive of higher twist effects for NMC data at low-$Q^{2}$ (as we observe a worse fit to low-$Q^{2}$ data). We also emphasise that when conducting a fit at NNLO with $Z\ p_{T}$ data removed, an improvement of $\Delta \chi^{2} = -41.3$ is observed in the rest of the data, whereas at N$^{3}$LO an improvement of $\Delta \chi^{2} = -65.2$ is observed in all other datasets without removing $Z\ p_{T}$, therefore these results are not purely an effect of removing any $Z\ p_{T}$ tension. Considering tensions with CMS $8\ \text{TeV}$ jets~\cite{CMS8jets} data, as discussed in Section~\ref{sec: n3lo_K}, in general the jets datasets show tensions with the N$^{3}$LO description (especially for CMS $8\ \text{TeV}$ jets~\cite{CMS8jets}), therefore it will be interesting to observe how this picture evolves when considering this data in the form of dijets. 

Since a naturally richer description of the small-$x$ regime is being included at N$^{3}$LO, which has a direct effect on the HERA datasets, the reduction of important tensions from NNLO is even further justification for the inclusion of the N$^{3}$LO theory. The extra N$^{3}$LO additions are allowing the large-$x$ behaviour of the PDFs to be less dominated by data at small-$x$, while also producing a better fit quality at small-$x$ (i.e. for HERA data). Some of the above observations are also made in \cite{NNPDFsx,xFitterDevelopersTeam:2018hym} where studies of including small-$x$ resummation results into a PDF fit have been reported.

Reflecting on the chosen prior distributions for each of the sources of missing N$^{3}$LO uncertainty, Table~\ref{tab: K-fac_firstresults} confirms that no especially large penalties are being incurred in this new description. 
These results therefore demonstrate that the fit is succeeding in leveraging contributions (such as $P_{qq}^{(3)}$ and $P_{qg}^{(3)}$ in the quark evolution part of Equation~\eqref{eq: DGLAP}) to produce a better overall fit.

\subsubsection*{DIS Processes}

\begin{table}[t]
\centerline{
\begin{tabular}{|P{6.5cm}|P{3cm}|P{3cm}|P{3.4cm}|}
\hline
DIS Dataset & $\chi^{2}$ & $\Delta \chi^{2}$ & $\Delta \chi^{2}$ from NNLO\\
 & & from NNLO & (NNLO $K$-factors)\\
\hline
BCDMS $\mu p$ $F_{2}$ \cite{BCDMS} &   174.4 / 163  & $-5.8$  & $-5.6$ \\
BCDMS $\mu d$ $F_{2}$ \cite{BCDMS} &   144.3 / 151  & $-1.7$  & $+0.6$ \\
NMC $\mu p$ $F_{2}$ \cite{NMC} &   121.5 / 123  & $-2.6$  & $-3.8$ \\
NMC $\mu d$ $F_{2}$ \cite{NMC} &   104.2 / 123  & $-8.4$  & $-11.5$ \\
SLAC $ep$ $F_{2}$ \cite{SLAC,SLAC1990} &   31.6 / 37  & $-0.4$  & $-0.0$ \\
SLAC $ed$ $F_{2}$ \cite{SLAC,SLAC1990} &   22.8 / 38  & $-0.2$  & $-0.8$ \\
E665 $\mu p$ $F_{2}$ \cite{E665} &   63.9 / 53  & $+4.2$  & $+4.5$ \\
E665 $\mu d$ $F_{2}$ \cite{E665} &   67.5 / 53  & $+2.9$  & $+2.5$ \\
NuTeV $\nu N$ $F_{2}$ \cite{NuTev} &   35.7 / 53  & $-2.6$  & $-1.3$ \\
NuTeV $\nu N$ $xF_{3}$ \cite{NuTev} &   34.8 / 42  & $+4.1$  & $+2.1$ \\
NMC $\mu n / \mu p$ \cite{NMCn/p} &   131.6 / 148  & $+0.8$  & $+2.2$ \\
HERA $ep$ $F_{2}^{\text{charm}}$ \cite{H1+ZEUScharm} &   143.7 / 79  & $+11.4$  & $+13.8$ \\
NMC/BCDMS/SLAC/HERA $F_{L}$ \cite{BCDMS,NMC,SLAC1990,H1FL,H1-FL,ZEUS-FL} &   45.6 / 57  & $-22.9$  & $-23.2$ \\
CHORUS $\nu N$ $F_{2}$ \cite{CHORUS} &   29.2 / 42  & $-1.0$  & $-0.8$ \\
CHORUS $\nu N$ $xF_{3}$ \cite{CHORUS} &   18.1 / 28  & $-0.3$  & $-0.5$ \\
HERA $e^{+}p$ CC \cite{H1+ZEUS} &   49.7 / 39  & $-2.3$  & $-1.0$ \\
HERA $e^{-}p$ CC \cite{H1+ZEUS} &   64.9 / 42  & $-5.3$  & $-4.9$ \\
HERA $e^{+}p$ NC $820\ \text{GeV}$ \cite{H1+ZEUS} &   84.3 / 75  & $-5.6$  & $-5.1$ \\
HERA $e^{-}p$ NC $460\ \text{GeV}$ \cite{H1+ZEUS} &   247.7 / 209  & $-0.6$  & $-0.7$ \\
HERA $e^{+}p$ NC $920\ \text{GeV}$ \cite{H1+ZEUS} &   474.0 / 402  & $-38.7$  & $-36.4$ \\
HERA $e^{-}p$ NC $575\ \text{GeV}$ \cite{H1+ZEUS} &   248.5 / 259  & $-14.5$  & $-14.1$ \\
HERA $e^{-}p$ NC $920\ \text{GeV}$ \cite{H1+ZEUS} &   243.0 / 159  & $-1.4$  & $-2.1$ \\
\hline
Total & 2580.9 / 2375 & $-90.8$ & $-86.2$\\
\hline
\end{tabular}
}
\caption{\label{tab: DIS_results}Table showing the relevant DIS datasets and how the individual $\chi^{2}$ changes from NNLO by including the N$^{3}$LO contributions to the structure function $F_{2}(x, Q^{2})$. The result within purely NNLO $K$-factors included for all data in the fit is also given.}
\end{table}
To complement the discussions in Section~\ref{sec: n3lo_K}, we isolate the $\chi^{2}$ results from DIS data in Table~\ref{tab: DIS_results}. This data is directly affected by the N$^{3}$LO structure functions constructed approximately in Section's~\ref{sec: structure} to \ref{sec: n3lo_coeff}. A substantial decrease in the total $\chi^{2}$ from NNLO is observed across DIS datasets. Considering the results in Table~\ref{tab: DIS_results} in the context of Table's~\ref{tab: DY_kfac} to \ref{tab: dimuon_kfac}, a better fit quality is observed for all DIS and non-DIS datasets than at NNLO with the inclusion of N$^{3}$LO contributions. As the DIS data makes up over half of the total data included in a global fit, it is the dominant force in deciding the overall form of the PDFs, especially at small-$x$ (discussed further in Section~\ref{subsec: pdf_results}). Table~\ref{tab: DIS_results} further reinforces the point that the N$^{3}$LO description is flexible enough to fit to HERA and non-HERA data, without being largely constrained by tensions between the small-$x$ (HERA dominated) and large-$x$ (non-HERA dominated) regions.

\subsection{Correlation Results} \label{subsec: correlations}

\begin{figure}
\begin{center}
\includegraphics[width=0.9\textwidth]{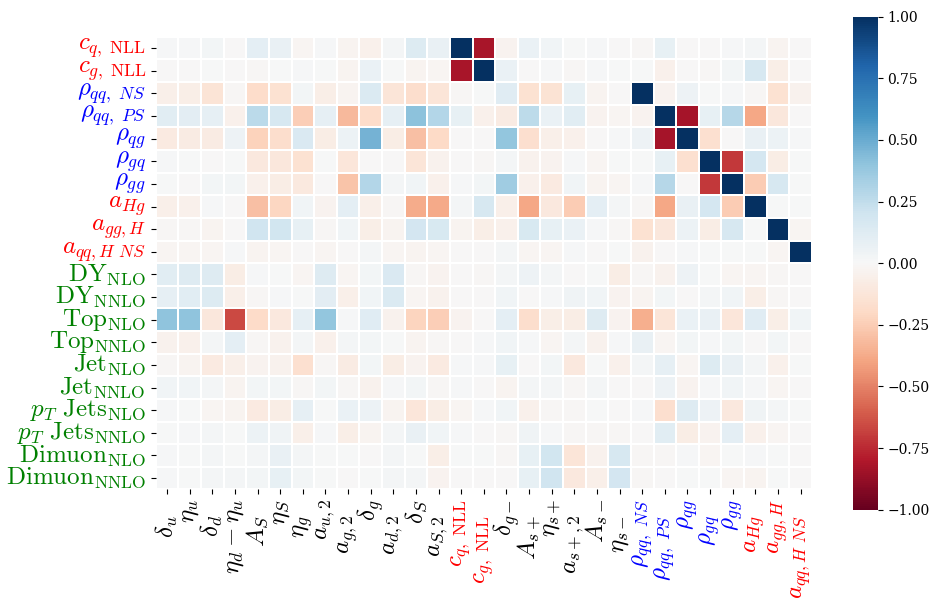}
\includegraphics[width=0.9\textwidth]{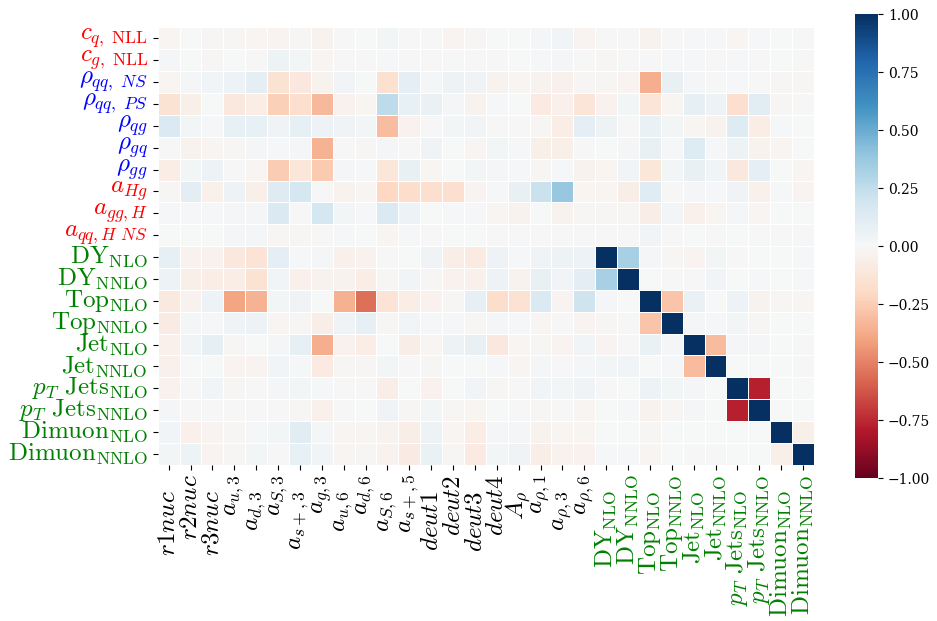}
\end{center}
\caption{\label{fig: corr_theory_full}Correlation matrix for all N$^{3}$LO theory parameters included in the fit against the subset of the MSHT20 parameters (shown in black) used in constructing the Hessian eigenvectors. This is shown for the case where the $K$-factors correlations with the first 42 parameters are included. N$^{3}$LO theory parameters associated with the splitting functions are coloured blue, the parameters affecting the transition matrix elements and coefficient functions are in red and the $K$-factor parameters are in green.}
\end{figure}
The correlation matrix shown in Fig.~\ref{fig: corr_theory_full} illustrates the correlations between extra N$^{3}$LO theory parameters and the subset of the MSHT20 parameters which are included in the construction of Hessian eigenvectors (see Section~\ref{subsec: eigenvector_results} and \cite{Thorne:MSHT20} for details). 
It is apparent that the correlations between $K$-factor parameters for each process (shown in green) and other PDF and theory parameters are usually small, with some exceptions e.g. for the Top$_{\text{NLO}}$ parameter. Due to this there is an argument that each process' $K$-factor parameters could be treated separately from all other parameters in the Hessian prescription (see Section~\ref{subsec: genframe}) which allows for a more flexible PDF set that can be decorrelated from a process. By using the uncorrelated Hessian results for a process NNLO hard cross sections can be transformed to aN$^{3}$LO and therefore provide more reliable predictions (more details in Section~\ref{sec: availability}). This is a fairly intuitive result, since most correlations are showing a natural separation between the process dependent and process independent physics in the DIS picture~\footnote{The same pattern can be seen for $c_{\{q,g\}}^{\mathrm{NLL}}$ parameters which are involved in the DIS hard cross section.}. Mathematically, the $K$-factors are directly associated with the hard cross section, whereas other N$^{3}$LO theory parameters ($\rho_{ij}$ and $a_{ij}$) are having a direct effect on the PDFs. Fig.~\ref{fig: corr_theory_full} therefore begins to motivate the inclusion of the `pure' theory (splitting functions and transition matrix elements) parameters within the standard MSHT eigenvector analysis~\cite{Thorne:MSHT20}, with the decorrelation of the $K$-factor parameters, as discussed in Section~\ref{subsec: decorr_params}. We investigate and compare both treatments (complete correlation and $K$-factor decorrelation) throughout the rest of this section. We show in Section~\ref{subsec: pdf_results} that while the decorrelation of $K$-factors is not complete, both treatments result in similar uncertainty bands, therefore confirming that the effect of making the assumption of full decorrelation is minimal in practice. Note that although the $c_{i}^{\mathrm{NLL}}$ parameters also show minimal correlation with other parameters, we include these within the `pure' theory group of parameters (i.e. correlated with $\rho_{ij}$ and $a_{ij}$) as they are essential ingredients in the underlying DIS theory.

\subsection{Eigenvector Results}\label{subsec: eigenvector_results}

In the MSHT fitting procedure (described in \cite{Thorne:MSHT20}) the eigenvectors of a Hessian matrix are found, which encapsulate the sources of uncertainties and corresponding correlations. Combining these with the central PDFs, forms the entire PDF set with uncertainties. In this eigenvector analysis a dynamical rescaling of each eigenvector $e_{i}$ is performed via a tolerance factor $t$ to encapsulate the $68\%$ confidence limit (C.L.).
\begin{equation}
    a_{i} = a_{i}^{0} \pm t e_{i},
\end{equation}
where $a_{i}^{0}$ is the best fit parameter.
$t$ is then adjusted to give the desired tolerance $T$ for the required confidence interval defined as $T = \sqrt{\Delta \chi_{\mathrm{global}}^{2}}$ (for $68\%$ C.L.). In a quadratic approximation, for suitably well-behaved eigenvectors, $t = T$ is true. Although for eigenvectors with larger eigenvalues, it is possible to observe significant deviations from $t=T$. The standard MSHT fitting procedure involves allowing all relevant parameters from \cite{Thorne:MSHT20} to vary when finding the best fit, now including all N$^{3}$LO theory parameters ($\rho_{ij}, a_{ij}, c_{i}^{\mathrm{NLL}}, K_{\mathrm{NLO/NNLO}}$) discussed in this work. 
After accounting for high degrees of correlation between parameters (described in \cite{Thorne:MSTW09}), the result is a Hessian matrix which in general, depends on a subset of the parameters that were allowed to vary in a best fit and provides a set of suitably well-behaved eigenvectors. The standard MSHT NNLO PDF eigenvectors are based on a set of 32 parameters, reduced from the 52 parameters allowed to vary in the full fit. In the following analysis we are therefore concerned with a smaller number of parameters, specifically the 32 parameters from the standard MSHT fitting procedure plus an extra 20 N$^{3}$LO parameters (shown in Fig.~\ref{fig: corr_theory_full}).

A standard choice of tolerance $T$ is $T = \sqrt{\Delta \chi^{2}_{\mathrm{global}}} = 1$ for a 68\% C.L. limit. However, this assumes all datasets are consistent with Gaussian errors. In practice, due to incomplete theory, tensions between datasets and parameterisation inflexibility, this is known not to be the case in a global PDF fit. To overcome this, a 68\%  C.L. region for each dataset is defined. Then for each eigenvector, the value of $\sqrt{\chi^{2}_{\mathrm{global}}}$ for each chosen $t$ is recorded (ideally showing a quadratic behaviour). Finally, a value of $T$ is chosen to ensure that all datasets are described within their 68\% CL in each eigenvector direction. For a fuller mathematical description of the dynamical tolerance procedure used in MSHT PDF fits, the reader is referred to \cite{Thorne:MSTW09}. In this section we present a demonstration of how well the resultant eigenvectors follow the quadratic assumption based on $t=T$, including the specific choices of dynamical tolerances and which dataset/penalty constrains this tolerance in each eigenvector direction.

\subsubsection*{PDF + N$^{3}$LO DIS Theory + N$^{3}$LO $K$-factor (decorrelated) Parameters}

As discussed in Section~\ref{subsec: correlations}, when determining the eigenvectors and therefore PDF uncertainties, we can choose to either include the correlations between the 10 $K$-factor parameters added with the other 42 parameters (encompassing the standard 32 MSHT eigenvector parameters and the 10 new theory parameters from the splitting functions, transition matrix elements and coefficient functions) or to decorrelate the 10 $K$-factor parameters. 

In this section we address the scenario where we decorrelate the $K$-factors as
\begin{equation}\label{eq: decorr}
	H_{ij}^{-1} + \sum_{p = 1}^{N_{p}}K_{ij,p}^{-1}
\end{equation}
and consider each term individually.

\begin{figure}
\begin{center}
\includegraphics[width=\textwidth]{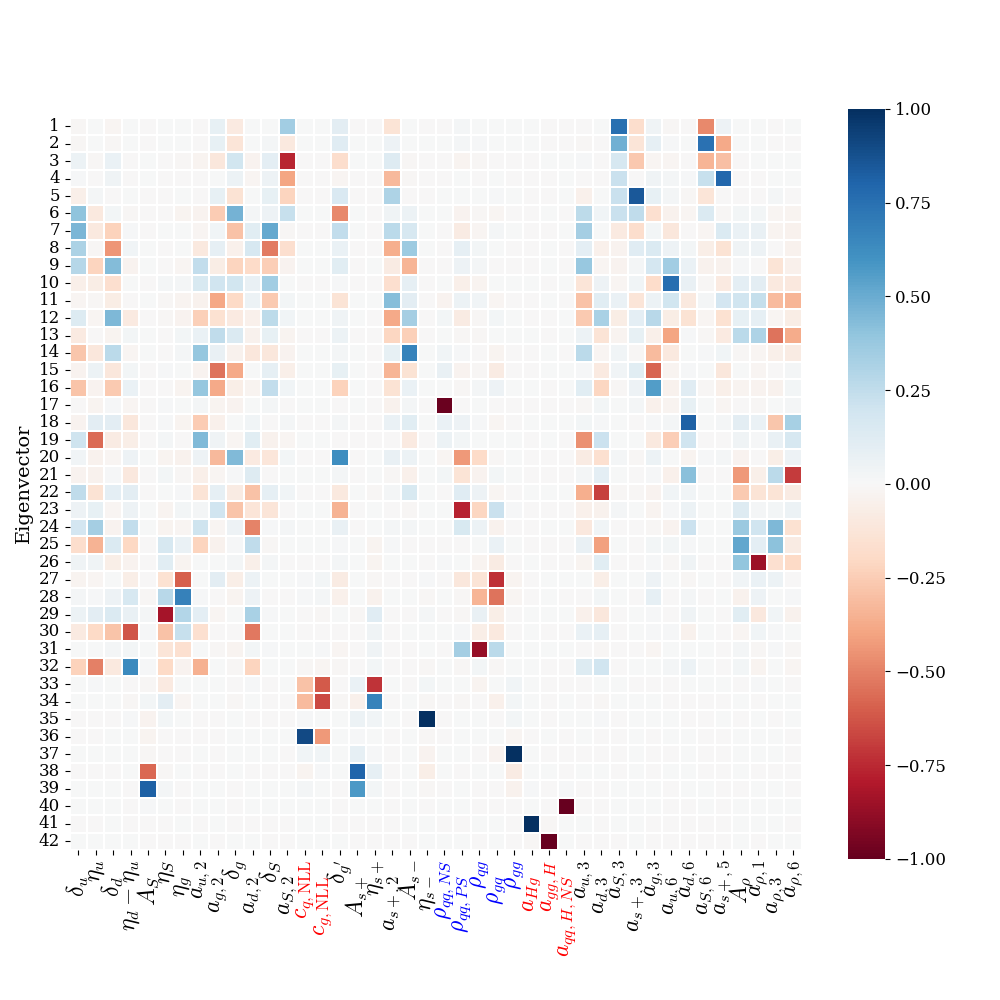}
\end{center}
\caption{\label{fig: eigenmap} Correlation matrix of the first 42 (total 52) eigenvectors found with the N$^{3}$LO parameters added into the analysis in the case where the $K$-factors are decorrelated from these first 42 parameters. Parameters associated with the splitting functions are coloured blue, those affecting the transition matrix elements and coefficient functions are in red.}
\end{figure}
\begin{figure}
\begin{center}
\includegraphics[width=0.7\textwidth]{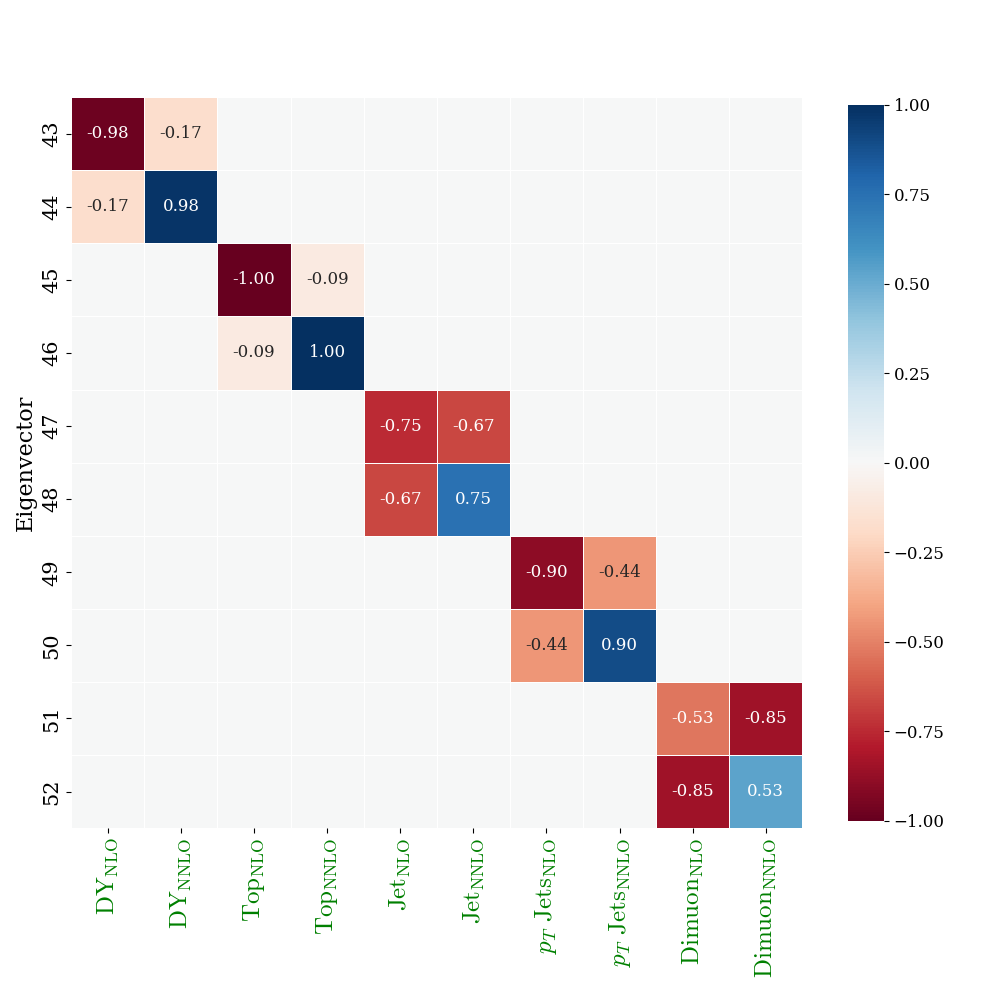}
\end{center}
\caption{\label{fig: eigenmap_K} Map of the 10 $K$-factor eigenvectors found with the N$^{3}$LO parameters added into the analysis in the case where the $K$-factors are decorrelated from these first 42 parameters. Combined with the 42 eigenvectors shown in Fig.~\ref{fig: eigenmap}, these form the total 52 eigenvectors in the decorrelated case. Parameters associated with the $K$-factor parameters are in green.}
\end{figure}
Fig.~\ref{fig: eigenmap} shows the map of eigenvectors produced from $H_{ij}$ in Equation~\eqref{eq: decorr}, where we have included the new N$^{3}$LO DIS theory parameters (splitting functions in blue and coefficient functions/transition matrix elements in red) correlated with the PDF parameters. Eigenvectors 35 and 36 are prime examples of where the eigenvectors have specifically encompassed the correlation/anti-correlation between the two NLL FFNS coefficient function parameters $c_{i}^{\mathrm{NLL}}$ ($i \in \{q,g\}$). Whereas the splitting functions naturally give rise to a much more complicated mixing with other PDF parameters as these  directly affect the evolution of the PDFs. Due to the direct impact of $\rho_{ij}$'s on the PDFs (via  DGLAP evolution), combined with the large contributions to the evolution shown at N$^{3}$LO, this result is as expected.

Another somewhat pleasing aspect is the recovery of a natural separation between eigenvectors associated with the N$^{3}$LO coefficient function/transition matrix elements and our original PDF parameters (incl. N$^{3}$LO splitting functions). This separation is reminiscent of our DIS picture, whereby the splitting functions are much more intertwined with the raw PDFs and the transition matrix elements have a symbiotic relationship with the coefficient functions (see GM-VFNS description in Section~\ref{sec: structure}). Due to this, the form of these eigenvectors has not only some level of physical interpretation inherited from our underlying theory, but also offers a useful way to access the different sources of N$^{3}$LO additions within the PDF set.

In Fig.~\ref{fig: eigenmap_K}  the eigenvectors resulting from the $\sum_{p = 1}^{N_{p}}K_{ij,p}^{-1}$ terms in Equation~\eqref{eq: decorr} are shown. These eigenvectors are constructed in pairs, describing the correlation and anti-correlation of the two $K$-factor parameters (controlling the NLO and NNLO contributions to N$^{3}$LO) for each process $p$ contained within the corresponding $K_{ij,p}$ correlation matrix.

Table~\ref{tab: Kij_limits} shows further information regarding the $K$-factor parameter limits from each eigenvector. In most cases the parameter limits are well within the allowed variation ($-1 < a < 1$), which is an indication that the data included in the fit is constraining these parameters rather than the individual penalties for each parameter\footnote{We remind the reader that the Dimuon datasets also include a branching ratio factor which is providing some compensation with these $K$-factor parameters (as discussed in Section \ref{sec: n3lo_K}).}.

To assess whether the eigenvectors are violating the quadratic treatment, four examples displaying this behaviour are shown in Fig.~\ref{fig: tol_decorr}, with a full analysis provided in Appendix~\ref{app: tolerance_decorr}. Additionally, Table~\ref{tab: eigenanalysis} provides a summary of all tolerances found within the eigenvector scans.
\begin{table}[t]
\centering
\begin{tabular}{c|c|c|c|c|c|c|c|c}
\hline
\multirow{2}{*}{Matrix} & \multicolumn{2}{c|}{Central Values} & \multirow{2}{*}{Eigenvector} & \multicolumn{2}{c|}{+ Limit} & \multicolumn{2}{c|}{- Limit} & \multirow{2}{*}{Scale} \\
\cline{2-3} \cline{5-8}
& $a_{\mathrm{NLO}}$ &  $a_{\mathrm{NNLO}}$ & & $a_{\mathrm{NLO}}$ &  $a_{\mathrm{NNLO}}$ & $a_{\mathrm{NLO}}$ &  $a_{\mathrm{NNLO}}$ & \\
\hline
\multirow{2}{*}{$K_{ij}^{\mathrm{DY}}$} & \multirow{2}{*}{-0.282} & \multirow{2}{*}{0.079} & 43 & -0.378 & 0.062 & -0.145 & 0.103 & \multirow{2}{*}{$m/2$}\\
& & & 44 & -0.334 & 0.374 & -0.256 & -0.071 & \\
\hline
\multirow{2}{*}{$K_{ij}^{\mathrm{Top}}$} & \multirow{2}{*}{0.041} & \multirow{2}{*}{0.651} & 45 & -0.564 & 0.455 & 0.692 & 0.862 & \multirow{2}{*}{Section~\ref{subsec: kfac_res}} \\
& & & 46 & 0.026 & 1.210 & 0.070 & -0.456 & \\
\hline
\multirow{2}{*}{$K_{ij}^{\mathrm{Jets}}$} & \multirow{2}{*}{-0.300} & \multirow{2}{*}{-0.691} & 47 & -0.515 & -0.957 & 0.105 & -0.189 &  \multirow{2}{*}{$p_{T}^{\mathrm{jet}}$}\\
& & & 48 & -0.725 & -0.033 & 0.036 & -1.212 & \\
\hline
\multirow{2}{*}{$K_{ij}^{p_{T}\ \mathrm{Jets}}$} & \multirow{2}{*}{0.583} & \multirow{2}{*}{-0.080} & 49 & 0.388 & -0.406 & 0.812 & 0.301 &  \multirow{2}{*}{$p_{T}$}\\
& & & 50 & 0.480 & 0.624 & 0.680 & -0.742 & \\
\hline
\multirow{2}{*}{$K_{ij}^{\mathrm{Dimuon}}$} & \multirow{2}{*}{-0.444} & \multirow{2}{*}{0.922} & 51 & -1.109 & -0.208 & -0.103 & 1.502 & \multirow{2}{*}{$Q^{2}$}\\
& & & 52 & -1.091 & 1.359 & 0.981 & -0.039 & \\
\hline
\end{tabular}
\caption{Limiting values for specific $K$-factor parameters for each of the processes considered in the decorrelated case. Parameter values are shown in the positive and negative limits for each eigenvector. The scale choices for top quark processes are described in Section~\ref{subsec: kfac_res} to be $H_T/4$ for the single differential datasets with the exception of data differential in the average transverse momentum of the top or antitop, $p_T^t,p_T^{\bar{t}}$, for which $m_T/2$ is used. For the double diff. dataset the scale choice is $H_T/4$ and for the inclusive top $\sigma_{t\bar{t}}$ a scale of $m_{t}$ is chosen.}
\label{tab: Kij_limits}
\end{table}
\begin{figure}
    \begin{center}
    \includegraphics[width=0.49\textwidth]{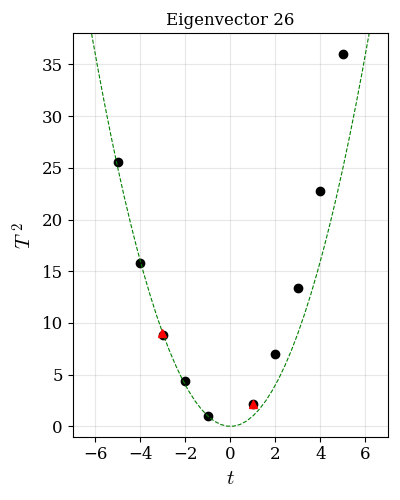}
    \includegraphics[width=0.49\textwidth]{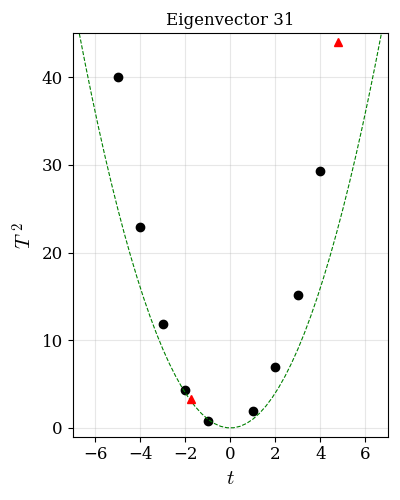}
    \\
    \includegraphics[width=0.49\textwidth]{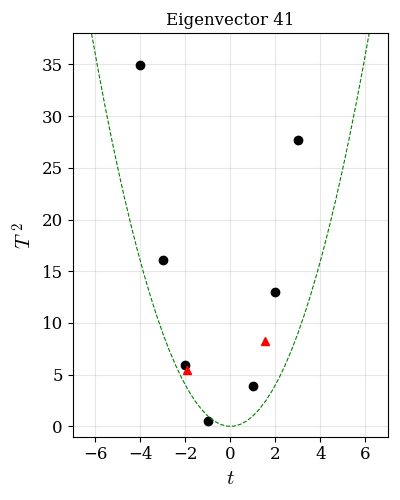}
    \includegraphics[width=0.49\textwidth]{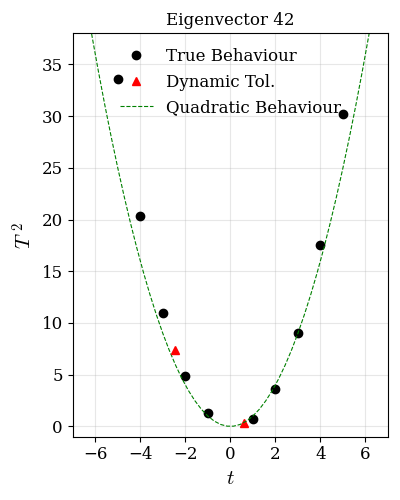}
    \caption{\label{fig: tol_decorr}Dynamic tolerance behaviour for 4 selected eigenvectors in the case of decorrelated $K$-factor parameters. The black dots show the fixed tolerance relations found for integer values of $t$, whereas the red triangles show the final chosen dynamical tolerances for each eigenvector direction. For an exhaustive analysis of all eigenvectors see Fig.~\ref{fig: full_tol_decorr}.}
    \end{center}
\end{figure}
\begin{table}
\centerline{
\begin{tabular}{|p{0.5cm}|p{0.7cm}|p{0.7cm}|p{4.5cm}|p{0.7cm}|p{0.7cm}|p{4.5cm}|p{2.4cm}|}
\hline
\#  &  $t+$  &  $T+$  &  Limiting Factor ($+$) &  $t-$  &  $T-$  &  Limiting Factor ($-$) & Primary \\
& & & & & & & Parameter \\ \hline
\hline
1 & 3.88 & 3.86 & ATLAS $7\ \text{TeV}$ high prec. $W, Z$ \cite{ATLASWZ7f} & 2.94 & 3.13 & ATLAS $8\ \text{TeV}$ double diff. $Z$ \cite{ATLAS8Z3D} & $a_{S,3}$ \\
2 & 5.35 & 5.56 & HERA $e^{+}p$ NC $920\ \text{GeV}$ \cite{H1+ZEUS} & 3.10 & 2.98 & NMC $\mu d$ $F_{2}$ \cite{NMC} & $a_{S,6}$ \\
3 & 4.20 & 4.47 & NuTeV $\nu N \rightarrow \mu\mu X$ \cite{Dimuon} & 2.30 & 2.17 & NMC $\mu d$ $F_{2}$ \cite{NMC} & $a_{S,2}$ \\
4 & 4.87 & 4.85 & ATLAS $8\ \text{TeV}$ double diff. $Z$ \cite{ATLAS8Z3D} & 1.81 & 1.80 & NuTeV $\nu N \rightarrow \mu\mu X$ \cite{Dimuon} & $a_{s+,5}$ \\
5 & 3.30 & 3.63 & ATLAS $7\ \text{TeV}$ high prec. $W, Z$ \cite{ATLASWZ7f} & 2.95 & 2.77 & NuTeV $\nu N \rightarrow \mu\mu X$ \cite{Dimuon} & $a_{s+,3}$ \\
6 & 4.89 & 5.38 & ATLAS $7\ \text{TeV}$ high prec. $W, Z$ \cite{ATLASWZ7f} & 5.47 & 5.27 & NMC $\mu d$ $F_{2}$ \cite{NMC} & $\delta_{g^{\prime}}$ \\
7 & 5.30 & 5.47 & D{\O} II $W \rightarrow \nu e$ asym. \cite{D0Wnue} & 3.44 & 3.41 & D{\O} $W$ asym. \cite{D0Wasym} & $\delta_{S}$ \\
8 & 4.12 & 4.05 & D{\O} II $W \rightarrow \nu e$ asym. \cite{D0Wnue} & 2.19 & 2.64 & D{\O} $W$ asym. \cite{D0Wasym} & $\delta_{S}$ \\
9 & 3.88 & 3.93 & BCDMS $\mu p$ $F_{2}$ \cite{BCDMS} & 6.78 & 7.11 & CMS W asym. $p_{T} > 25, 30\ \text{GeV}$ \cite{CMS-Wasymm} & $\delta_{d}$ \\
10 & 3.24 & 3.44 & BCDMS $\mu p$ $F_{2}$ \cite{BCDMS} & 5.64 & 5.83 & D{\O} II $W \rightarrow \nu \mu$ asym. \cite{D0Wnumu} & $a_{u,6}$ \\
11 & 2.22 & 4.41 & E866 / NuSea $pp$ DY \cite{E866DY} & 5.31 & 4.79 & ATLAS $7\ \text{TeV}$ high prec. $W, Z$ \cite{ATLASWZ7f} & $a_{s+,2}$ \\
12 & 2.62 & 2.81 & D{\O} $W$ asym. \cite{D0Wasym} & 5.49 & 5.71 & ATLAS $8\ \text{TeV}\ W$ \cite{ATLASW8} & $\delta_{d}$ \\
13 & 1.12 & 3.37 & E866 / NuSea $pp$ DY \cite{E866DY} & 3.87 & 3.19 & D{\O} $W$ asym. \cite{D0Wasym} & $a_{\rho,3}$ \\
14 & 2.01 & 2.34 & NuTeV $\nu N \rightarrow \mu\mu X$ \cite{Dimuon} & 3.54 & 3.60 & NuTeV $\nu N$ $xF_{3}$ \cite{NuTev} & $A_{s-}$ \\
15 & 3.69 & 3.75 & ATLAS $8\ \text{TeV}\ Z\ p_{T}$ \cite{ATLASZpT} & 3.65 & 4.05 & NMC $\mu d$ $F_{2}$ \cite{NMC} & $a_{g,3}$ \\
16 & 4.42 & 5.08 & ATLAS $8\ \text{TeV}\ W$ \cite{ATLASW8} & 4.96 & 4.84 & ATLAS $8\ \text{TeV}\ Z\ p_{T}$ \cite{ATLASZpT} & $a_{g,3}$ \\
17 & 1.23 & 1.02 & $\rho_{qq}^{NS}$ & 1.24 & 1.05 & $\rho_{qq}^{NS}$ & $\rho_{qq}^{NS}$ \\
18 & 3.09 & 3.50 & D{\O} $W$ asym. \cite{D0Wasym} & 2.93 & 3.06 & D{\O} $W$ asym. \cite{D0Wasym} & $a_{d,6}$ \\
19 & 4.69 & 4.57 & BCDMS $\mu p$ $F_{2}$ \cite{BCDMS} & 3.85 & 5.06 & CMS $7\ \text{TeV}$ jets \cite{CMS7jetsfinal} & $\eta_{u}$ \\
20 & 5.40 & 5.37 & NuTeV $\nu N \rightarrow \mu\mu X$ \cite{Dimuon} & 5.04 & 5.60 & HERA $e^{+}p$ NC $920\ \text{GeV}$ \cite{H1+ZEUS} & $\delta_{g^{\prime}}$ \\
21 & 2.10 & 2.36 & E866 / NuSea $pd/pp$ DY \cite{E866DYrat} & 1.26 & 1.69 & E866 / NuSea $pd/pp$ DY \cite{E866DYrat} & $a_{\rho,6}$ \\
22 & 3.04 & 3.27 & NuTeV $\nu N \rightarrow \mu\mu X$ \cite{Dimuon} & 1.90 & 2.27 & D{\O} $W$ asym. \cite{D0Wasym} & $a_{d,3}$ \\
\hline
\end{tabular}
}
\caption{\label{tab: eigenanalysis}Tolerances resulting from eigenvector scans with decorrelated $K$-factors for each process. The average tolerance for this set of eigenvectors is $T=3.34$.}
\end{table}
\begin{table}\ContinuedFloat
\centerline{
\begin{tabular}{|p{0.5cm}|p{0.7cm}|p{0.7cm}|p{4.5cm}|p{0.7cm}|p{0.7cm}|p{4.5cm}|p{2.4cm}|}
\hline
\#  &  $t+$  &  $T+$  &  Limiting Factor ($+$) &  $t-$  &  $T-$  &  Limiting Factor ($-$) & Primary \\
& & & & & & & Parameter \\ \hline
\hline
23 & 5.61 & 6.20 & HERA $e^{+}p$ NC $920\ \text{GeV}$ \cite{H1+ZEUS} & 5.43 & 6.13 & HERA $e^{+}p$ NC $920\ \text{GeV}$ \cite{H1+ZEUS} & $\rho_{qq}^{PS}$ \\
24 & 3.53 & 3.70 & E866 / NuSea $pd/pp$ DY \cite{E866DYrat} & 1.47 & 1.82 & D{\O} $W$ asym. \cite{D0Wasym} & $a_{d,2}$ \\
25 & 1.60 & 2.03 & E866 / NuSea $pd/pp$ DY \cite{E866DYrat} & 4.37 & 4.88 & CMS W asym. $p_{T} > 35\ \text{GeV}$ \cite{CMS-easym} & $A_{\rho}$ \\
26 & 1.00 & 1.46 & E866 / NuSea $pd/pp$ DY \cite{E866DYrat} & 3.02 & 3.00 & D{\O} $W$ asym. \cite{D0Wasym} & $a_{\rho,1}$ \\
27 & 1.60 & 2.09 & $\rho_{gq}$ & 4.16 & 5.53 & ATLAS $7\ \text{TeV}$ high prec. $W, Z$ \cite{ATLASWZ7f} & $\rho_{gq}$ \\
28 & 1.51 & 2.16 & $\rho_{gq}$ & 3.20 & 4.07 & ATLAS $8\ \text{TeV}$ sing. diff. $t\bar{t}$ dilep. \cite{ATLASttbarDilep08_ytt} & $\eta_{g}$ \\
29 & 2.99 & 3.21 & CMS $8\ \text{TeV}\ W$ \cite{CMSW8} & 2.04 & 2.52 & NuTeV $\nu N$ $xF_{3}$ \cite{NuTev} & $\eta_{S}$ \\
30 & 0.97 & 1.30 & D{\O} $W$ asym. \cite{D0Wasym} & 3.56 & 4.90 & ATLAS $8\ \text{TeV}\ W$ \cite{ATLASW8} & $\eta_{d} - \eta_{u}$ \\
31 & 4.78 & 6.64 & HERA $e^{+}p$ NC $920\ \text{GeV}$ \cite{H1+ZEUS} & 1.77 & 1.81 & $\rho_{gq}$ & $\rho_{qg}$ \\
32 & 2.51 & 7.32 & BCDMS $\mu p$ $F_{2}$ \cite{BCDMS} & 1.80 & 3.96 & D{\O} $W$ asym. \cite{D0Wasym} & $\eta_{d} - \eta_{u}$ \\
33 & 2.71 & 3.35 & CMS $7\ \text{TeV}\ W + c$ \cite{CMS7Wpc} & 2.92 & 3.37 & NuTeV $\nu N \rightarrow \mu\mu X$ \cite{Dimuon} & $\eta_{s+}$ \\
34 & 3.53 & 3.91 & HERA $ep$ $F_{2}^{\text{charm}}$ \cite{H1+ZEUScharm} & 3.87 & 4.82 & CMS $7\ \text{TeV}\ W + c$ \cite{CMS7Wpc} & $\eta_{s+}$ \\
35 & 3.67 & 3.50 & NuTeV $\nu N \rightarrow \mu\mu X$ \cite{Dimuon} & 2.72 & 3.62 & NuTeV $\nu N \rightarrow \mu\mu X$ \cite{Dimuon} & $\eta_{s-}$ \\
36 & 1.25 & 1.26 & $c_{q}^{\mathrm{NLL}}$ & 1.41 & 1.57 & $c_{q}^{\mathrm{NLL}}$ & $c_{q}^{\mathrm{NLL}}$ \\
37 & 0.71 & 0.77 & $\rho_{gg}$ & 4.25 & 5.33 & NuTeV $\nu N \rightarrow \mu\mu X$ \cite{Dimuon} & $\rho_{gg}$ \\
38 & 2.72 & 2.71 & CMS $8\ \text{TeV}\ W$ \cite{CMSW8} & 3.60 & 4.21 & $\rho_{gq}$ & $A_{s+}$ \\
39 & 1.56 & 5.03 & ATLAS $7\ \text{TeV}$ high prec. $W, Z$ \cite{ATLASWZ7f} & 1.60 & 5.82 & ATLAS $7\ \text{TeV}$ high prec. $W, Z$ \cite{ATLASWZ7f} & $A_{S}$ \\
40 & 0.97 & 1.00 & $a_{qq,H}^{\mathrm{NS}}$ & 1.04 & 1.01 & $a_{qq,H}^{\mathrm{NS}}$ & $a_{qq,H}^{\mathrm{NS}}$ \\
41 & 1.56 & 2.87 & HERA $ep$ $F_{2}^{\text{charm}}$ \cite{H1+ZEUScharm} & 1.93 & 2.34 & $\rho_{gg}$ & $a_{Hg}$ \\
42 & 0.63 & 0.53 & $a_{gg,H}$ & 2.46 & 2.72 & $a_{gg,H}$ & $a_{gg,H}$ \\
43 & 1.78 & 2.30 & ATLAS $8\ \text{TeV}$ double diff. $Z$ \cite{ATLAS8Z3D} & 2.56 & 1.94 & CMS double diff. Drell-Yan \cite{CMS-ddDY} & $\mathrm{DY}_{\mathrm{NLO}}$ \\
44 & 4.63 & 4.88 & ATLAS $8\ \text{TeV}\ W$ \cite{ATLASW8} & 2.35 & 2.09 & ATLAS $8\ \text{TeV}$ double diff. $Z$ \cite{ATLAS8Z3D}&$\mathrm{DY}_{\mathrm{NNLO}}$ \\
45 & 5.05 & 1.46 & ATLAS $8\ \text{TeV}$ sing. diff. $t\bar{t}$ dilep. \cite{ATLASttbarDilep08_ytt} & 5.44 & 1.46 & Tevatron, ATLAS, CMS $\sigma_{t\bar{t}}$ \cite{Tevatron-top,ATLAS-top7-1,ATLAS-top7-2,ATLAS-top7-3,ATLAS-top7-4,ATLAS-top7-5,ATLAS-top7-6,CMS-top7-1,CMS-top7-2,CMS-top7-3,CMS-top7-4,CMS-top7-5,CMS-top8} & $\mathrm{Top}_{\mathrm{NLO}}$ \\
\hline
\end{tabular}
}
\caption{\textit{(Continued)} Tolerances resulting from eigenvector scans with decorrelated $K$-factors for each process. The average tolerance for this set of eigenvectors is $T=3.34$.}
\end{table}
\begin{table}\ContinuedFloat
\centerline{
\begin{tabular}{|p{0.5cm}|p{0.7cm}|p{0.7cm}|p{4.5cm}|p{0.7cm}|p{0.7cm}|p{4.5cm}|p{2.4cm}|}
\hline
\#  &  $t+$  &  $T+$  &  Limiting Factor ($+$) &  $t-$  &  $T-$  &  Limiting Factor ($-$) & Primary \\
& & & & & & & Parameter \\ \hline
\hline
46 & 1.34 & 1.31 & $\mathrm{Top}_{\mathrm{NLO}}$ & 2.65 & 2.68 & ATLAS $8\ \text{TeV}$ sing. diff. $t\bar{t}$ \cite{ATLASsdtop} & $\mathrm{Top}_{\mathrm{NNLO}}$ \\
47 & 1.47 & 1.57 & CDF II $p\bar{p}$ incl. jets \cite{CDFjet} & 2.78 & 2.96 & $a_{gg,H}$ & $\mathrm{Jet}_{\mathrm{NLO}}$ \\
48 & 3.26 & 3.38 & CMS $2.76\ \text{TeV}$ jet \cite{CMS276jets} & 2.58 & 3.01 & CMS $7\ \text{TeV}$ jets \cite{CMS7jetsfinal} & $\mathrm{Jet}_{\mathrm{NNLO}}$ \\
49 & 2.20 & 2.39 & ATLAS $8\ \text{TeV}\ Z\ p_{T}$ \cite{ATLASZpT} & 2.58 & 2.56 & ATLAS $8\ \text{TeV}\ W + \text{jets}$ \cite{ATLASWjet} & $p_{T}\ \mathrm{Jets}_{\mathrm{NLO}}$ \\
50 & 2.36 & 2.43 & ATLAS $8\ \text{TeV}\ Z\ p_{T}$ \cite{ATLASZpT} & 2.22 & 2.23 & ATLAS $8\ \text{TeV}\ W + \text{jets}$ \cite{ATLASWjet} & $p_{T}\ \mathrm{Jets}_{\mathrm{NNLO}}$ \\
51 & 1.38 & 1.55 & $\mathrm{Dimuon}_{\mathrm{NLO}}$ & 0.71 & 0.81 & $\mathrm{Dimuon}_{\mathrm{NNLO}}$ & $\mathrm{Dimuon}_{\mathrm{NNLO}}$ \\
52 & 0.85 & 0.85 & $\mathrm{Dimuon}_{\mathrm{NLO}}$ & 1.86 & 1.87 & $\mathrm{Dimuon}_{\mathrm{NLO}}$ & $\mathrm{Dimuon}_{\mathrm{NLO}}$ \\
\hline
\end{tabular}
}
\caption{\textit{(Continued)} Tolerances resulting from eigenvector scans with decorrelated $K$-factors for each process. The average tolerance for this set of eigenvectors is $T=3.34$.}
\end{table}
There is relatively consistent agreement between $t$ and $T$ across all eigenvectors with later eigenvectors (i.e. higher \#) generally becoming less quadratic (a feature which is built into the fit). Eigenvectors 31, 41 and 42 displayed in Fig.~\ref{fig: tol_decorr} are shown in Table~\ref{tab: eigenanalysis} to be either dominated or limited by at least one new N$^{3}$LO parameter. Conversely, eigenvector 26 is much more dominated by the original PDF parameters from MSHT20 NNLO. Comparing these cases, the eigenvectors associated more strongly with the N$^{3}$LO parameters exhibit a similar level of agreement (and occasionally better) with the desired quadratic behaviour as eigenvectors more closely associated with the original PDF parameters.

The last 5 sets of eigenvectors (i.e. the last 10 where a set contains 2 eigenvectors for a particular process) we see in Table~\ref{tab: eigenanalysis} are the decorrelated $K$-factor eigenvectors, where there are correlated/anti-correlated eigenvectors for each process. For all $K$-factor cases, Table~\ref{tab: eigenanalysis} provides sensible results with either the dominant datasets or parameter penalties constraining each eigenvector direction. One interesting feature one can observe here is a sign of tension between the ATLAS $8\ \text{TeV}\ Z\ p_{T}$~\cite{ATLASZpT} and ATLAS $8\ \text{TeV}\ W + \text{jets}$~\cite{ATLASWjet} datasets where the limiting factors in Table~\ref{tab: eigenanalysis} for eigenvectors 49 and 50 show that these datasets are preferring a slightly different $K$-factor.

To provide some extra level of comparison between the eigenvectors shown here and the eigenvectors found in the NNLO case, the average tolerance $T$ for aN$^{3}$LO (decorrelated $K$-factors) set is 3.34, compared to the NNLO average $T$ of 3.37.

\pagebreak
\subsubsection*{PDF + N$^{3}$LO DIS Theory + N$^{3}$LO $K$-factor (correlated) Parameters}

In this section we address the scenario,
\begin{equation}
	H_{ij}^{\prime} = \left(H_{ij}^{-1} + \sum_{p = 1}^{N_{p}}K_{ij,p}^{-1} \right)^{-1}.
\end{equation}
Moving to an analysis including aN$^{3}$LO $K$-factors as correlated parameters with PDF and other N$^{3}$LO theory parameters. This provides a comparison to the case of decorrelated $K$-factors and justification for treating the cross section behaviour separately to the PDF theory behaviour.

\begin{figure}
\begin{center}
\includegraphics[width=\textwidth]{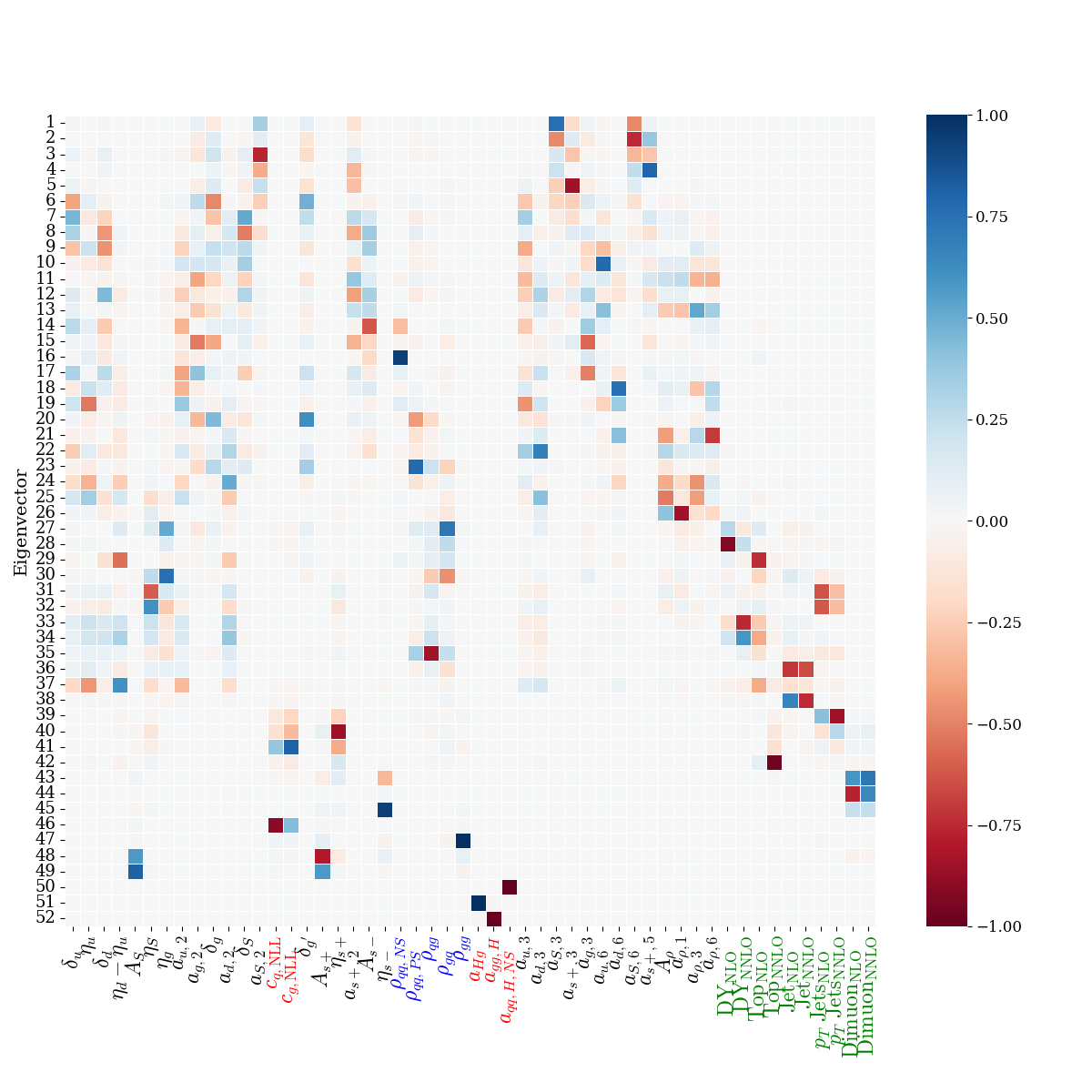}
\end{center}
\caption{\label{fig: eigenmap52}Map of eigenvectors found with the N$^{3}$LO theory and $K$-factor parameters added into the analysis. Parameters associated with the splitting functions are coloured blue, those affecting the transition matrix elements and coefficient functions are in red and the $K$-factor parameters are in green.}
\end{figure}

Fig.~\ref{fig: eigenmap52} shows a map of eigenvectors with the extra 10 N$^{3}$LO $K$-factor parameters (shown in green) included into the correlations considered. As expected, the result of including the correlations between PDF parameters and aN$^{3}$LO $K$-factors results in a slightly more intertwined set of eigenvectors (although a high level of decorrelation remains). Specifically, due to the much higher number of DY datasets included in the global fit, these N$^{3}$LO $K$-factor parameters tend to be included across more of a spread of eigenvectors. On the other hand, the Dimuon $K$-factors are almost entirely isolated within two eigenvectors, similar to the decorrelated case.

\begin{figure}
    \begin{center}
    \includegraphics[width=0.49\textwidth]{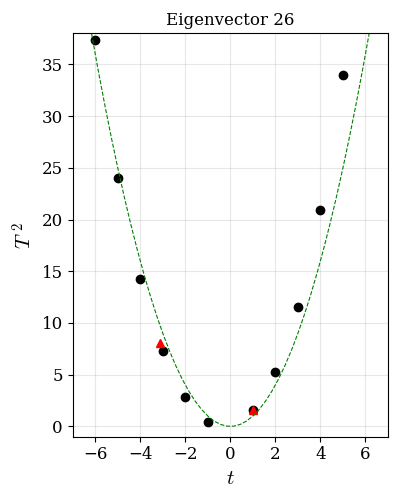}
    \includegraphics[width=0.49\textwidth]{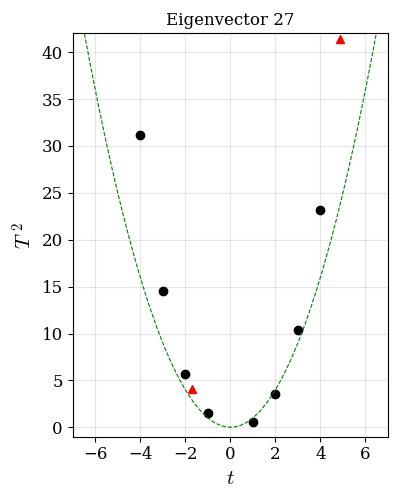}
    \includegraphics[width=0.49\textwidth]{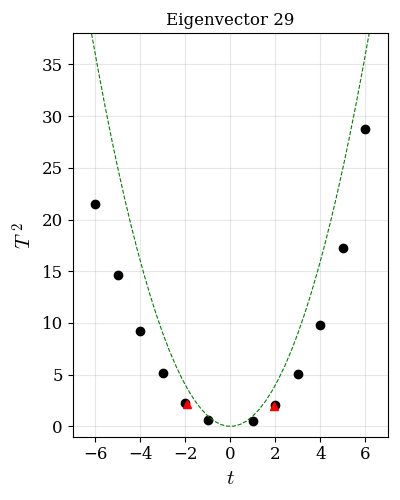}
    \includegraphics[width=0.49\textwidth]{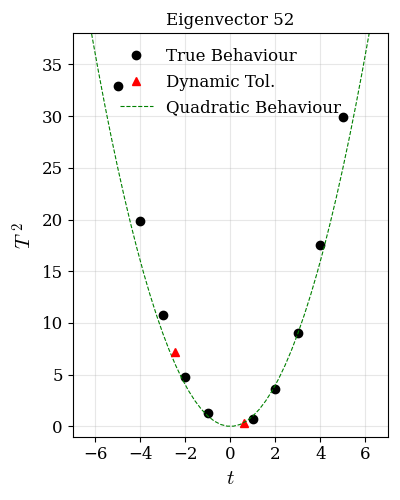}
    \caption{\label{fig: tol_corr}Dynamic tolerance behaviour for 4 selected eigenvectors in the case of correlated $K$-factor parameters. The black dots show the fixed tolerance relations found for integer values of $t$, whereas the red triangles show the final chosen dynamical tolerances for each eigenvector direction. For an exhaustive analysis of all eigenvectors see Fig.~\ref{fig: full_tol_corr}.}
    \end{center}
\end{figure}
\begin{table}
\centerline{
\begin{tabular}{|p{0.5cm}|p{0.7cm}|p{0.7cm}|p{4.5cm}|p{0.7cm}|p{0.7cm}|p{4.5cm}|p{2.4cm}|}
\hline
\#  &  $t+$  &  $T+$  &  Limiting Factor ($+$) &  $t-$  &  $T-$  &  Limiting Factor ($-$) & Primary \\
& & & & & & & Parameter \\ \hline
\hline
1 & 3.76 & 3.67 & ATLAS $7\ \text{TeV}$ high prec. $W, Z$ \cite{ATLASWZ7f} & 2.95 & 3.06 & ATLAS $8\ \text{TeV}$ double diff. $Z$ \cite{ATLAS8Z3D} & $a_{S,3}$ \\
2 & 3.67 & 3.52 & NMC $\mu d$ $F_{2}$ \cite{NMC} & 5.37 & 5.54 & HERA $e^{+}p$ NC $920\ \text{GeV}$ \cite{H1+ZEUS} & $a_{S,6}$ \\
3 & 4.15 & 4.33 & NuTeV $\nu N \rightarrow \mu\mu X$ \cite{Dimuon} & 2.58 & 2.39 & NMC $\mu d$ $F_{2}$ \cite{NMC} & $a_{S,2}$ \\
4 & 4.90 & 4.82 & ATLAS $8\ \text{TeV}$ double diff. $Z$ \cite{ATLAS8Z3D} & 2.21 & 2.16 & NuTeV $\nu N \rightarrow \mu\mu X$ \cite{Dimuon} & $a_{s+,5}$ \\
5 & 2.89 & 2.61 & NuTeV $\nu N \rightarrow \mu\mu X$ \cite{Dimuon} & 3.34 & 3.55 & CCFR $\nu N \rightarrow \mu\mu X$ \cite{Dimuon} & $a_{s+,3}$ \\
6 & 5.58 & 5.23 & NMC $\mu d$ $F_{2}$ \cite{NMC} & 5.49 & 5.84 & ATLAS $7\ \text{TeV}$ high prec. $W, Z$ \cite{ATLASWZ7f} & $\delta_{g^{\prime}}$ \\
7 & 5.34 & 5.44 & D{\O} II $W \rightarrow \nu e$ asym. \cite{D0Wnue} & 3.46 & 3.37 & D{\O} $W$ asym. \cite{D0Wasym} & $\delta_{S}$ \\
8 & 4.11 & 3.82 & D{\O} II $W \rightarrow \nu e$ asym. \cite{D0Wnue} & 2.18 & 2.48 & D{\O} $W$ asym. \cite{D0Wasym} & $\delta_{S}$ \\
9 & 6.39 & 6.59 & CMS W asym. $p_{T} > 25, 30\ \text{GeV}$ \cite{CMS-Wasymm} & 4.20 & 4.07 & BCDMS $\mu p$ $F_{2}$ \cite{BCDMS} & $\delta_{d}$ \\
10 & 3.14 & 3.11 & BCDMS $\mu p$ $F_{2}$ \cite{BCDMS} & 5.66 & 5.66 & D{\O} II $W \rightarrow \nu \mu$ asym. \cite{D0Wnumu} & $a_{u,6}$ \\
11 & 2.07 & 3.99 & E866 / NuSea $pp$ DY \cite{E866DY} & 5.88 & 5.00 & ATLAS $7\ \text{TeV}$ high prec. $W, Z$ \cite{ATLASWZ7f} & $a_{g,2}$ \\
12 & 2.76 & 2.76 & D{\O} $W$ asym. \cite{D0Wasym} & 5.60 & 5.67 & CMS $8\ \text{TeV}\ W$ \cite{CMSW8} & $\delta_{d}$ \\
13 & 3.73 & 3.01 & D{\O} $W$ asym. \cite{D0Wasym} & 1.26 & 3.96 & E866 / NuSea $pp$ DY \cite{E866DY} & $a_{\rho,3}$ \\
14 & 3.85 & 3.55 & NuTeV $\nu N$ $xF_{3}$ \cite{NuTev} & 2.24 & 2.30 & NuTeV $\nu N \rightarrow \mu\mu X$ \cite{Dimuon} & $A_{s-}$ \\
15 & 3.71 & 3.51 & ATLAS $8\ \text{TeV}\ Z\ p_{T}$ \cite{ATLASZpT} & 3.44 & 3.72 & NuTeV $\nu N \rightarrow \mu\mu X$ \cite{Dimuon} & $a_{g,3}$ \\
16 & 1.78 & 1.18 & $\rho_{qq}^{NS}$ & 1.78 & 1.24 & $\rho_{qq}^{NS}$ & $\rho_{qq}^{NS}$ \\
17 & 4.76 & 4.49 & D{\O} $W$ asym. \cite{D0Wasym} & 4.37 & 4.81 & ATLAS $8\ \text{TeV}\ W$ \cite{ATLASW8} & $a_{g,3}$ \\
18 & 3.44 & 3.81 & D{\O} $W$ asym. \cite{D0Wasym} & 3.18 & 2.94 & D{\O} $W$ asym. \cite{D0Wasym} & $a_{d,6}$ \\
19 & 5.33 & 5.02 & D{\O} $W$ asym. \cite{D0Wasym} & 3.77 & 4.42 & CMS $7\ \text{TeV}$ jets \cite{CMS7jetsfinal} & $\eta_{u}$ \\
20 & 6.04 & 5.82 & HERA $e^{+}p$ NC $920\ \text{GeV}$ \cite{H1+ZEUS} & 5.08 & 5.63 & HERA $e^{+}p$ NC $920\ \text{GeV}$ \cite{H1+ZEUS} & $\delta_{g^{\prime}}$ \\
21 & 2.14 & 2.02 & E866 / NuSea $pd/pp$ DY \cite{E866DYrat} & 1.29 & 1.39 & E866 / NuSea $pd/pp$ DY \cite{E866DYrat} & $a_{\rho,6}$ \\
22 & 1.95 & 2.03 & D{\O} $W$ asym. \cite{D0Wasym} & 2.88 & 2.83 & NuTeV $\nu N \rightarrow \mu\mu X$ \cite{Dimuon} & $a_{d,3}$ \\
\hline
\end{tabular}
}
\caption{\label{tab: eigenanalysis_2}Tolerances resulting from eigenvector scans with correlated $K$-factors for each process. The average tolerance for this set of eigenvectors is $T=3.57$.}
\end{table}
\begin{table}\ContinuedFloat
\centerline{
\begin{tabular}{|p{0.5cm}|p{0.7cm}|p{0.7cm}|p{4.5cm}|p{0.7cm}|p{0.7cm}|p{4.5cm}|p{2.4cm}|}
\hline
\#  &  $t+$  &  $T+$  &  Limiting Factor ($+$) &  $t-$  &  $T-$  &  Limiting Factor ($-$) & Primary \\
& & & & & & & Parameter \\ \hline
\hline
23 & 5.33 & 5.87 & HERA $e^{+}p$ NC $920\ \text{GeV}$ \cite{H1+ZEUS} & 5.66 & 6.18 & HERA $e^{+}p$ NC $920\ \text{GeV}$ \cite{H1+ZEUS} & $\rho_{qq}^{PS}$ \\
24 & 1.47 & 1.52 & D{\O} $W$ asym. \cite{D0Wasym} & 3.51 & 3.48 & E866 / NuSea $pd/pp$ DY \cite{E866DYrat} & $a_{d,2}$ \\
25 & 4.34 & 4.71 & CMS W asym. $p_{T} > 35\ \text{GeV}$ \cite{CMS-easym} & 1.64 & 1.59 & E866 / NuSea $pd/pp$ DY \cite{E866DYrat} & $A_{\rho}$ \\
26 & 1.00 & 1.24 & E866 / NuSea $pd/pp$ DY \cite{E866DYrat} & 3.12 & 2.84 & D{\O} $W$ asym. \cite{D0Wasym} & $a_{\rho,1}$ \\
27 & 4.86 & 6.43 & CMS double diff. Drell-Yan \cite{CMS-ddDY} & 1.68 & 2.01 & $\rho_{gq}$ & $\rho_{gq}$ \\
28 & 2.36 & 2.66 & ATLAS $7\ \text{TeV}$ high prec. $W, Z$ \cite{ATLASWZ7f} & 1.64 & 1.36 & CMS double diff. Drell-Yan \cite{CMS-ddDY} & $\mathrm{DY}_{\mathrm{NLO}}$ \\
29 & 1.95 & 1.41 & $\rho_{qq}^{NS}$ & 1.94 & 1.47 & $\rho_{qq}^{NS}$ & $\mathrm{Top}_{\mathrm{NLO}}$ \\
30 & 1.84 & 2.22 & $\rho_{gq}$ & 2.70 & 2.56 & ATLAS $8\ \text{TeV}$ sing. diff. $t\bar{t}$ dilep. \cite{ATLASttbarDilep08_ytt} & $\eta_{g}$ \\
31 & 3.53 & 3.52 & ATLAS $8\ \text{TeV}\ Z\ p_{T}$ \cite{ATLASZpT} & 2.60 & 2.73 & NuTeV $\nu N$ $xF_{3}$ \cite{NuTev} & $p_{T}\ \mathrm{Jets}_{\mathrm{NLO}}$ \\
32 & 2.85 & 3.15 & NuTeV $\nu N$ $xF_{3}$ \cite{NuTev} & 3.84 & 3.57 & ATLAS $8\ \text{TeV}\ W + \text{jets}$ \cite{ATLASWjet} & $p_{T}\ \mathrm{Jets}_{\mathrm{NLO}}$ \\
33 & 5.40 & 6.16 & ATLAS $7\ \text{TeV}$ high prec. $W, Z$ \cite{ATLASWZ7f} & 1.63 & 1.68 & D{\O} $W$ asym. \cite{D0Wasym} & $\mathrm{DY}_{\mathrm{NNLO}}$ \\
34 & 3.94 & 4.19 & ATLAS $8\ \text{TeV}\ W$ \cite{ATLASW8} & 1.56 & 1.86 & D{\O} $W$ asym. \cite{D0Wasym} & $\mathrm{DY}_{\mathrm{NNLO}}$ \\
35 & 5.06 & 6.73 & HERA $e^{+}p$ NC $920\ \text{GeV}$ \cite{H1+ZEUS} & 2.00 & 1.84 & $\rho_{gq}$ & $\rho_{qg}$ \\
36 & 2.31 & 2.48 & CDF II $p\bar{p}$ incl. jets \cite{CDFjet} & 4.73 & 4.88 & ATLAS $7\ \text{TeV}$ jets \cite{ATLAS7jets} & $\mathrm{Jet}_{\mathrm{NLO}}$ \\
37 & 2.72 & 7.05 & BCDMS $\mu p$ $F_{2}$ \cite{BCDMS} & 1.67 & 3.04 & D{\O} $W$ asym. \cite{D0Wasym} & $\eta_{d} - \eta_{u}$ \\
38 & 2.63 & 2.90 & $\mathrm{Jet}_{\mathrm{NLO}}$ & 3.24 & 3.35 & CMS $2.76\ \text{TeV}$ jet \cite{CMS276jets} & $\mathrm{Jet}_{\mathrm{NNLO}}$ \\
39 & 2.41 & 2.34 & ATLAS $8\ \text{TeV}\ W + \text{jets}$ \cite{ATLASWjet} & 2.88 & 2.99 & ATLAS $8\ \text{TeV}\ Z\ p_{T}$ \cite{ATLASZpT} & $p_{T}\ \mathrm{Jets}_{\mathrm{NNLO}}$ \\
40 & 2.98 & 3.44 & CMS $7\ \text{TeV}\ W + c$ \cite{CMS7Wpc} & 2.43 & 2.36 & NuTeV $\nu N \rightarrow \mu\mu X$ \cite{Dimuon} & $\eta_{s+}$ \\
41 & 4.59 & 5.42 & HERA $ep$ $F_{2}^{\text{charm}}$ \cite{H1+ZEUScharm} & 2.93 & 3.20 & HERA $ep$ $F_{2}^{\text{charm}}$ \cite{H1+ZEUScharm} & $c_{g}^{\mathrm{NLL}}$ \\
42 & 2.74 & 2.83 & CMS $8\ \text{TeV}$ double diff. $t\bar{t}$ \cite{CMS8ttDD} & 1.36 & 1.28 & $\mathrm{Top}_{\mathrm{NLO}}$ & $\mathrm{Top}_{\mathrm{NNLO}}$ \\
43 & 0.83 & 0.71 & $\mathrm{Dimuon}_{\mathrm{NLO}}$ & 1.36 & 1.46 & $p_{T}\ \mathrm{Jets}_{\mathrm{NNLO}}$ & $\mathrm{Dimuon}_{\mathrm{NNLO}}$ \\
44 & 0.81 & 0.80 & $\mathrm{Dimuon}_{\mathrm{NLO}}$ & 2.07 & 2.05 & $p_{T}\ \mathrm{Jets}_{\mathrm{NNLO}}$ & $\mathrm{Dimuon}_{\mathrm{NLO}}$ \\
45 & 1.45 & 1.46 & $\mathrm{Dimuon}_{\mathrm{NLO}}$ & 2.09 & 2.38 & $p_{T}\ \mathrm{Jets}_{\mathrm{NNLO}}$ & $\eta_{s-}$ \\
\hline
\end{tabular}
}
\caption{\textit{(Continued)} Tolerances resulting from eigenvector scans with correlated $K$-factors for each process. The average tolerance for this set of eigenvectors is $T=3.57$.}
\end{table}
\begin{table}\ContinuedFloat
\centerline{
\begin{tabular}{|p{0.5cm}|p{0.7cm}|p{0.7cm}|p{4.5cm}|p{0.7cm}|p{0.7cm}|p{4.5cm}|p{2.4cm}|}
\hline
\#  &  $t+$  &  $T+$  &  Limiting Factor ($+$) &  $t-$  &  $T-$  &  Limiting Factor ($-$) & Primary \\
& & & & & & & Parameter \\ \hline
\hline
46 & 1.41 & 1.57 & $c_{q}^{\mathrm{NLL}}$ & 1.25 & 1.26 & $c_{q}^{\mathrm{NLL}}$ & $c_{q}^{\mathrm{NLL}}$ \\
47 & 0.71 & 0.77 & $\rho_{gg}$ & 4.27 & 5.21 & NuTeV $\nu N \rightarrow \mu\mu X$ \cite{Dimuon} & $\rho_{gg}$ \\
48 & 3.62 & 4.07 & $\rho_{gq}$ & 2.72 & 2.63 & CMS $8\ \text{TeV}\ W$ \cite{CMSW8} & $A_{s+}$ \\
49 & 1.53 & 4.81 & ATLAS $7\ \text{TeV}$ high prec. $W, Z$ \cite{ATLASWZ7f} & 1.56 & 5.54 & ATLAS $7\ \text{TeV}$ high prec. $W, Z$ \cite{ATLASWZ7f} & $A_{S}$ \\
50 & 0.97 & 1.00 & $a_{qq,H}^{\mathrm{NS}}$ & 1.04 & 1.00 & $a_{qq,H}^{\mathrm{NS}}$ & $a_{qq,H}^{\mathrm{NS}}$ \\
51 & 1.56 & 2.69 & HERA $ep$ $F_{2}^{\text{charm}}$ \cite{H1+ZEUScharm} & 1.93 & 1.98 & $\rho_{gg}$ & $a_{Hg}$ \\
52 & 0.63 & 0.54 & $a_{gg,H}$ & 2.46 & 2.68 & $a_{gg,H}$ & $a_{gg,H}$ \\
\hline
\end{tabular}
}
\caption{\textit{(Continued)} Tolerances resulting from eigenvector scans with correlated $K$-factors for each process. The average tolerance for this set of eigenvectors is $T=3.57$.}
\end{table}
Once again, to investigate deviations from the quadratic behaviour, Fig.~\ref{fig: tol_corr} illustrates examples of the tolerance behaviours of selected eigenvectors, with a full analysis provided in Appendix~\ref{app: tolerance_corr}. Further to this, Table~\ref{tab: eigenanalysis_2} displays the tolerances and limiting datasets/parameters for the 52 correlated eigenvectors. It is difficult to compare and contrast these results with the decorrelated case, since the eigenvectors are inherently different. However in both cases, the eigenvectors are similarly well behaved, exhibit relatively good consistency between $t$ and $T$ and are therefore providing valid descriptions for a PDF fit. 

For most of the 12 eigenvectors with N$^{3}$LO $K$-factors as primary parameters, there is expected behaviour, with the eigenvectors constrained either by their own penalties or by dominant datasets for the associated process. However, due to the extra correlations considered, there are a small number of eigenvector directions which are not as trivial to explain (e.g. eigenvector 31). We therefore recover the lack of correlation between $K$-factor parameters seen within Fig.~\ref{fig: corr_theory_full} in the set of correlated PDF eigenvectors presented here. Further to this, comparing the $t$ and $T$ values found for eigenvectors associated with N$^{3}$LO $K$-factors in Table's~\ref{tab: eigenanalysis} and \ref{tab: eigenanalysis_2}, one can observe clear similarities between eigenvectors. This suggests that even when correlating the $K$-factor parameters, the fit succeeds in decorrelating the individual processes, thereby motivating our original assumption that the correlations with $K$-factors can be ignored. Another similarity one can observe between Table~\ref{tab: eigenanalysis} and Table~\ref{tab: eigenanalysis_2} is the suggestion of some tension between ATLAS $8\ \text{TeV}\ Z\ p_{T}$~\cite{ATLASZpT} and ATLAS $8\ \text{TeV}\ W + \text{jets}$ \cite{ATLASWjet} datasets seen in the limiting factors of eigenvector 39 in the correlated case.

Eigenvectors 27, 29 and 52 displayed in Fig.~\ref{fig: tol_corr} can be seen from Table~\ref{tab: eigenanalysis_2} to be associated with the new N$^{3}$LO theory parameters. Whereas eigenvector 37 is primarily focused on an original PDF parameter. One can observe a similar level of quadratic behaviour across all four of these eigenvector tolerances. Comparing all eigenvectors in the decorrelated/correlated cases, the behaviours are similarly well behaved. The average tolerance $T$ for the aN$^{3}$LO (with correlated $K$-factors) case is 3.57, slightly higher than the NNLO average of 3.37 and the aN$^{3}$LO (with decorrelated $K$-factors) average of 3.34.

\subsection{PDF Results}\label{subsec: pdf_results}

\begin{figure}
\begin{center}
\includegraphics[width=\textwidth]{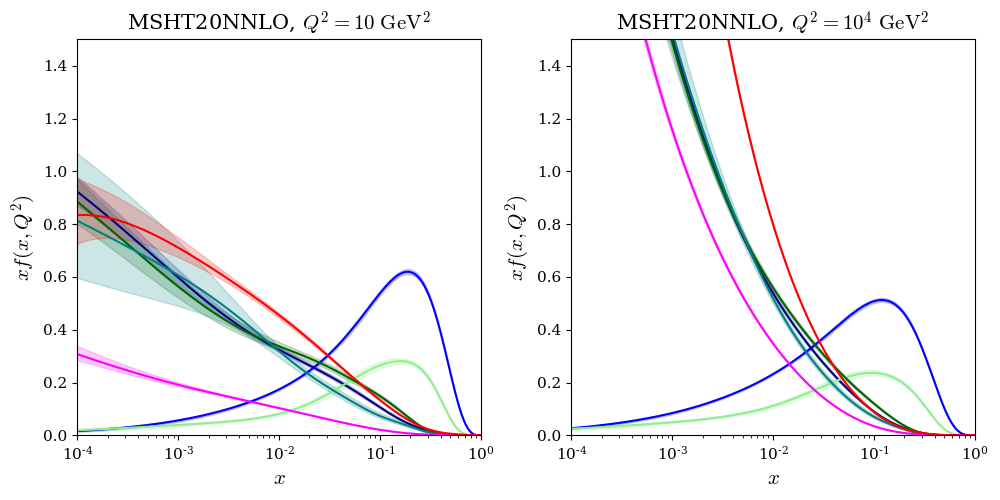}
\includegraphics[width=\textwidth]{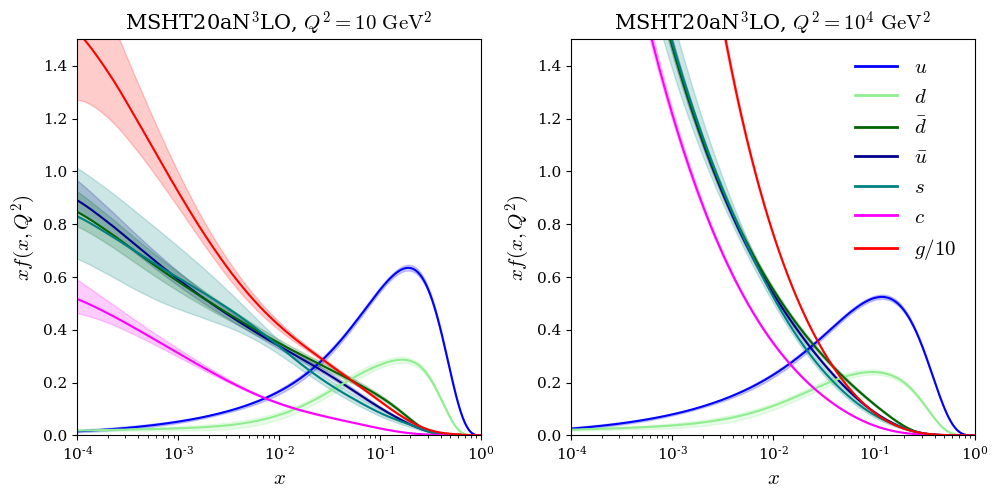}
\end{center}
\caption{\label{fig: pdfs}General forms of NNLO (top) and aN$^{3}$LO (bottom) PDFs at low (left) and high (right) $Q^{2}$. Several main features can be compared and contrasted such as the marked increase in the gluon and charm at small-$x$ (note the difference in y-axis scale between NNLO (top) and aN$^{3}$LO (bottom)).}
\end{figure}
Fig.~\ref{fig: pdfs} displays the overall shape of the PDFs including the N$^{3}$LO additions compared to the standard NNLO set. We provide this comparison to accompany the results described in earlier sections. At small-$x$ and low-$Q^{2}$ the gluon exhibits a marked enhancement due to the large small-$x$ logarithms inserted at N$^{3}$LO. The changes induced from specific N$^{3}$LO contributions are investigated in Section~\ref{subsec: n3lo_contrib}.

\begin{figure}
\begin{center}
\includegraphics[width=0.49\textwidth]{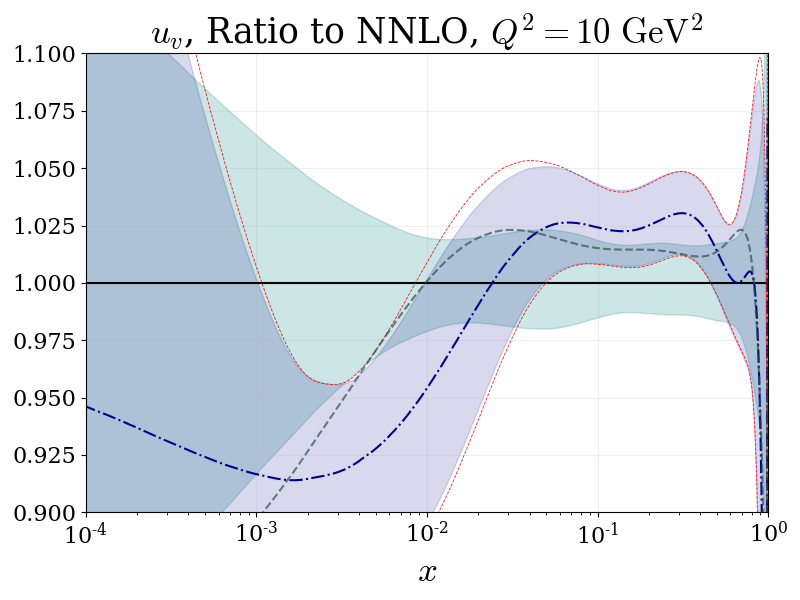}
\includegraphics[width=0.49\textwidth]{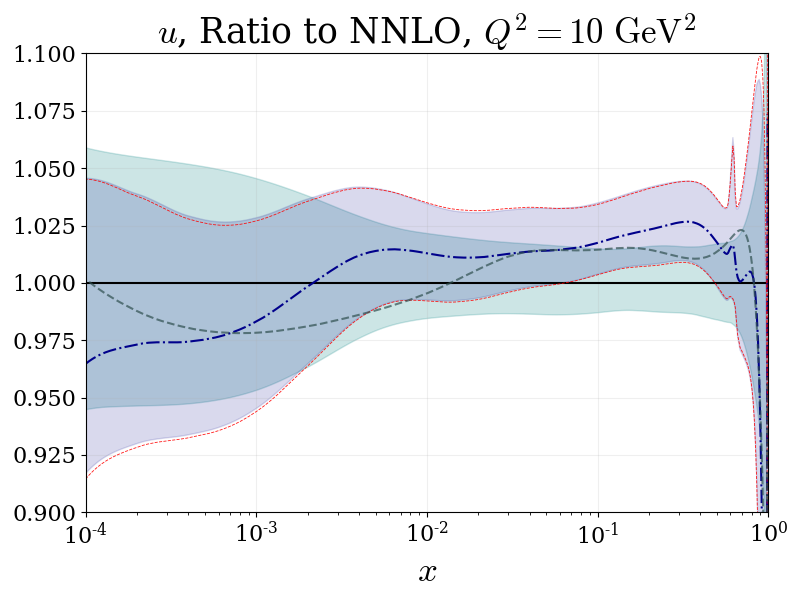}
\includegraphics[width=0.49\textwidth]{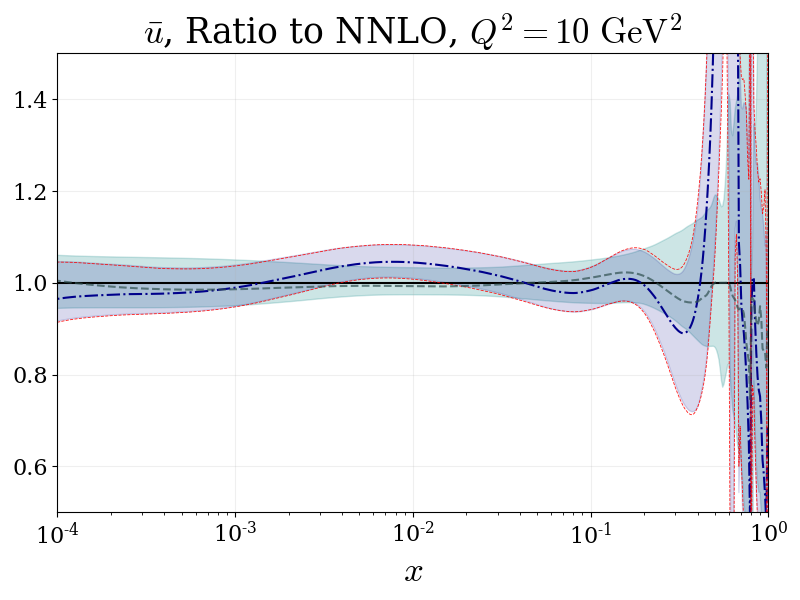}
\includegraphics[width=0.49\textwidth]{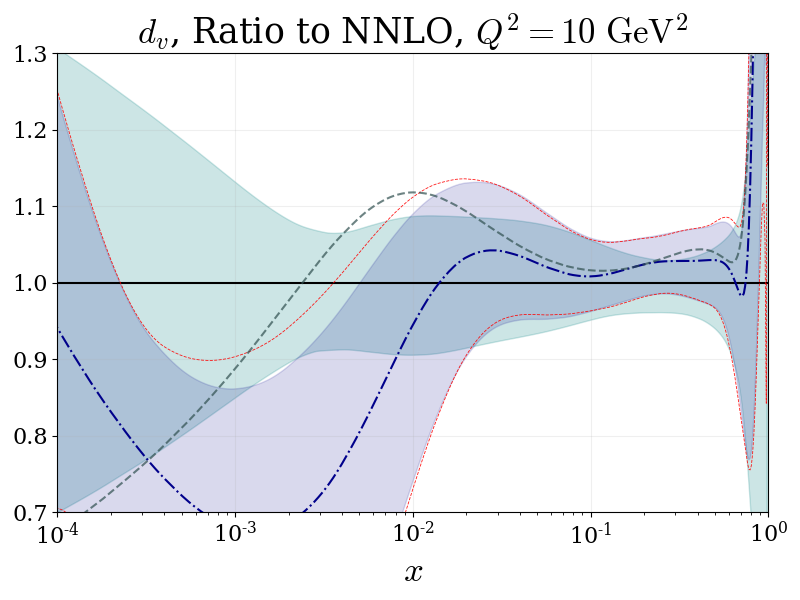}
\includegraphics[width=0.49\textwidth]{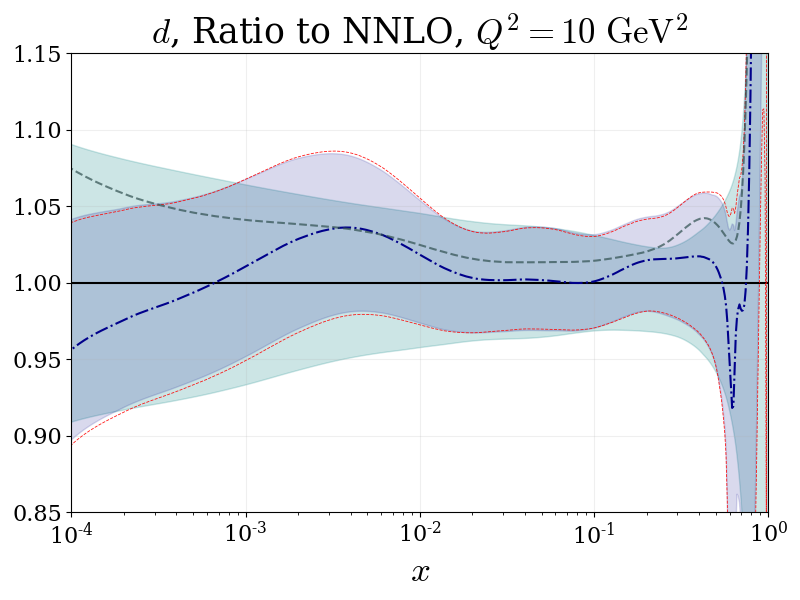}
\includegraphics[width=0.49\textwidth]{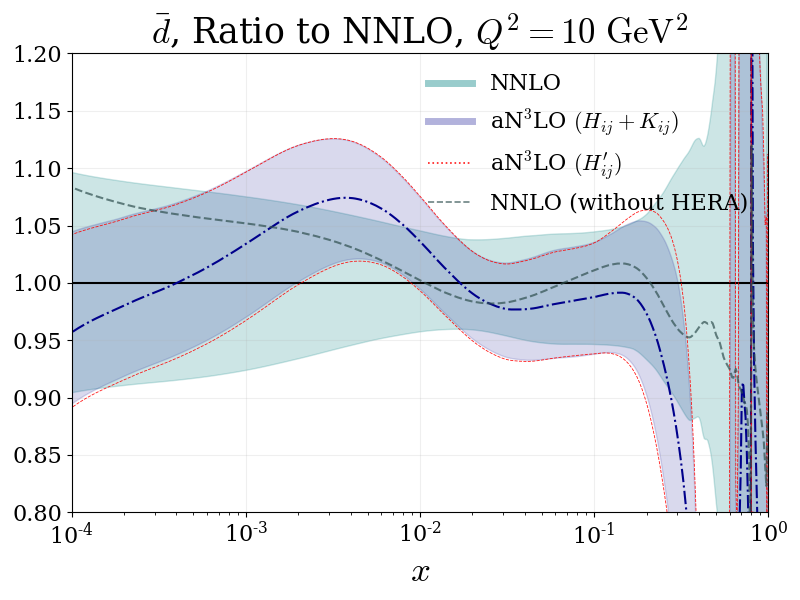}
\end{center}
\caption{\label{fig: pdf_ratios_qlow}Low-$Q^{2}$ ratio plots showing the aN$^{3}$LO 68\% confidence intervals with decorrelated ($H_{ij} + K_{ij}$) and correlated ($H_{ij}^{\prime}$) $K$-factor parameters, compared to NNLO 68\% confidence intervals. Also shown are the central values at NNLO when fit to all non-HERA datasets which show similarities with N$^{3}$LO in the large-$x$ region of selected PDF flavours. All plots are shown for $Q^{2} = 10\ \mathrm{GeV}^{2}$ with the exception of the bottom quark shown for $Q^{2} = 25\ \mathrm{GeV}^{2}$.}
\end{figure}
\begin{figure}[t]\ContinuedFloat
\begin{center}
\includegraphics[width=0.49\textwidth]{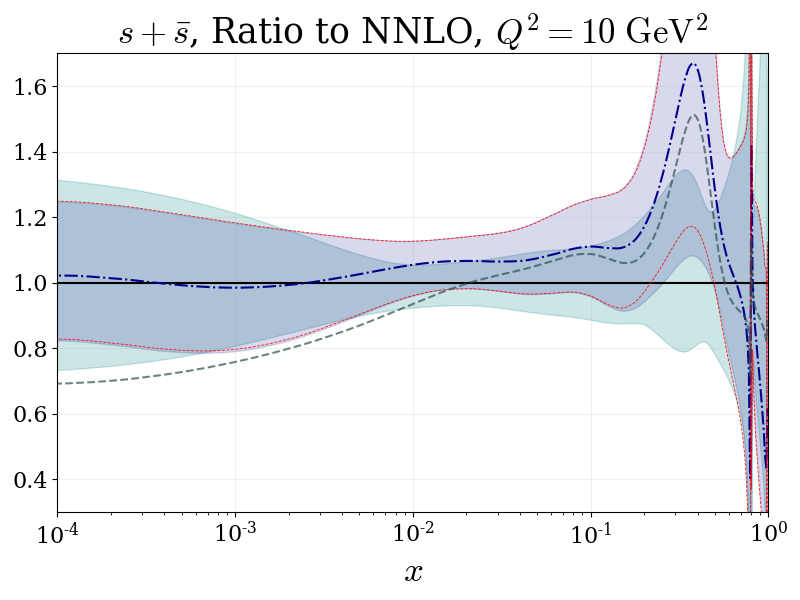}
\includegraphics[width=0.49\textwidth]{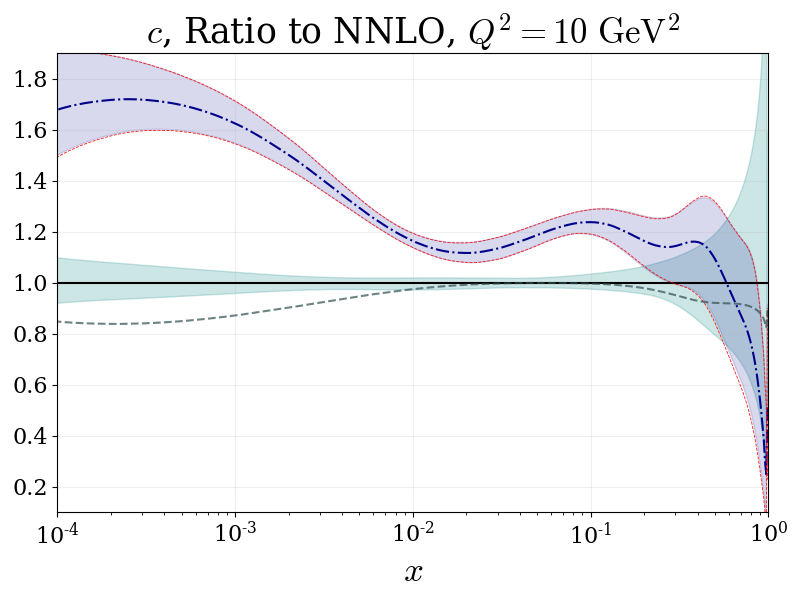}
\includegraphics[width=0.49\textwidth]{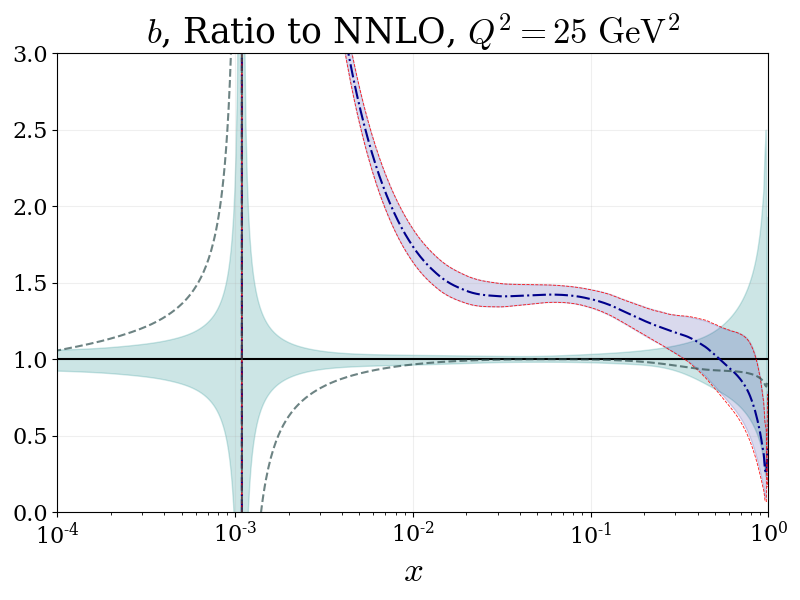}
\includegraphics[width=0.49\textwidth]{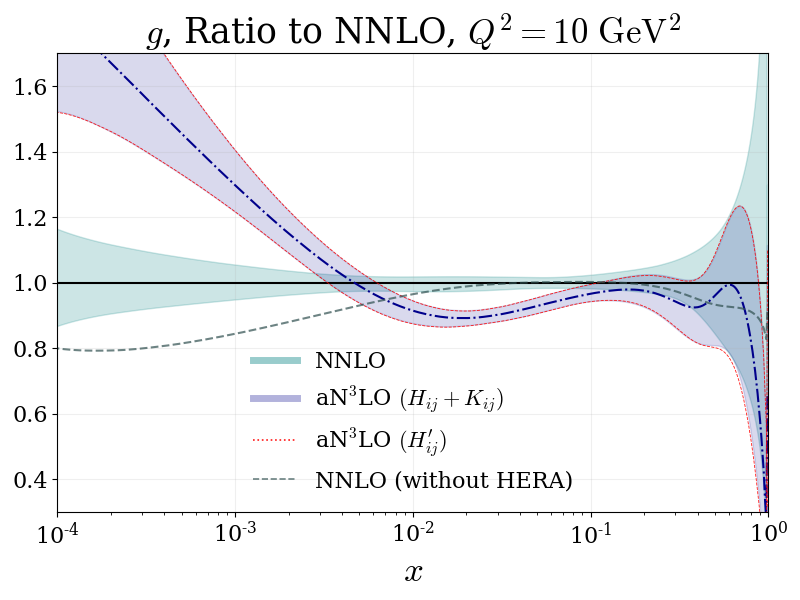}
\end{center}
\caption{\textit{(Continued)} Low-$Q^{2}$ ratio plots showing the aN$^{3}$LO 68\% confidence intervals with decorrelated ($H_{ij} + K_{ij}$) and correlated ($H_{ij}^{\prime}$) $K$-factor parameters, compared to NNLO 68\% confidence intervals. Also shown are the central values at NNLO when fit to all non-HERA datasets which show similarities with N$^{3}$LO in the large-$x$ region of selected PDF flavours. All plots are shown for $Q^{2} = 10\ \mathrm{GeV}^{2}$ with the exception of the bottom quark shown for $Q^{2} = 25\ \mathrm{GeV}^{2}$.}
\end{figure}
\begin{figure}
\begin{center}
\includegraphics[width=0.49\textwidth]{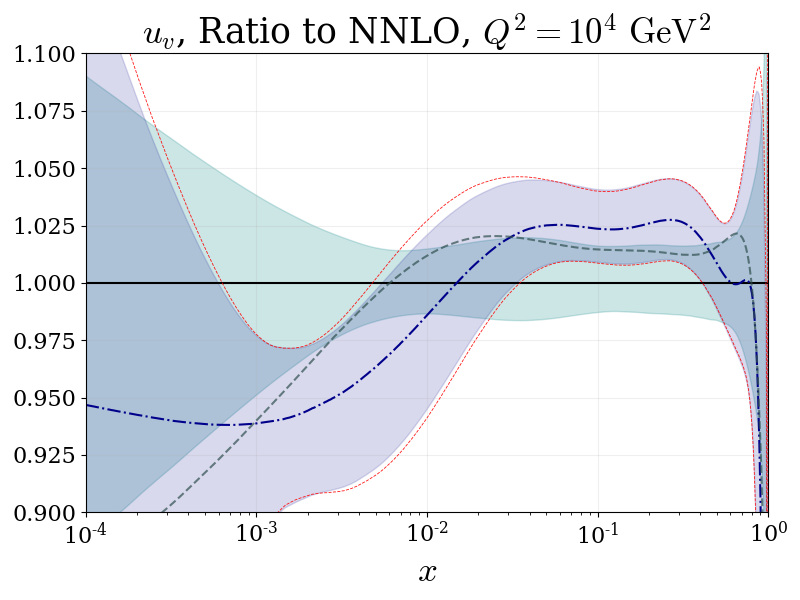}
\includegraphics[width=0.49\textwidth]{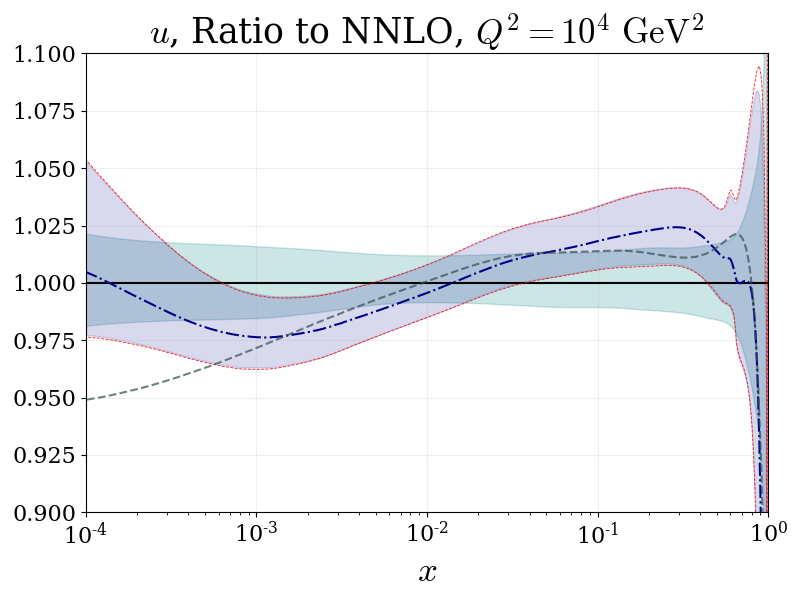}
\includegraphics[width=0.49\textwidth]{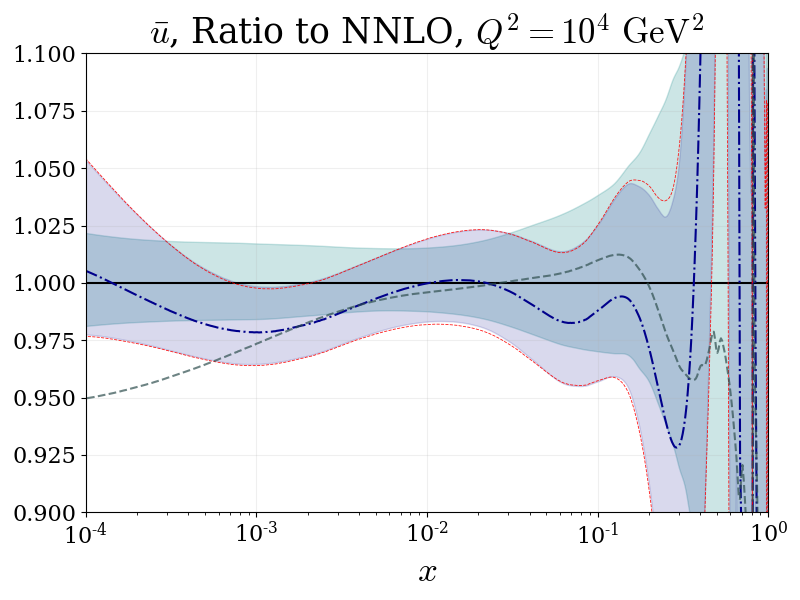}
\includegraphics[width=0.49\textwidth]{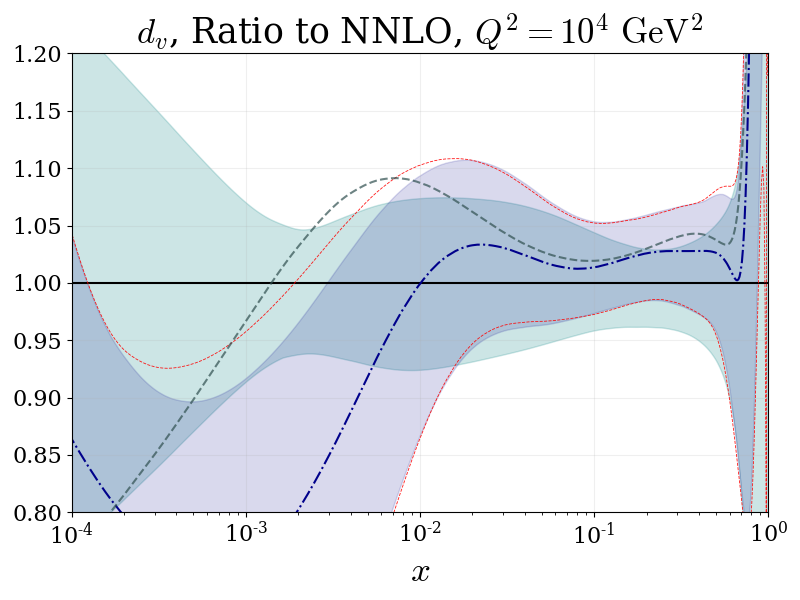}
\includegraphics[width=0.49\textwidth]{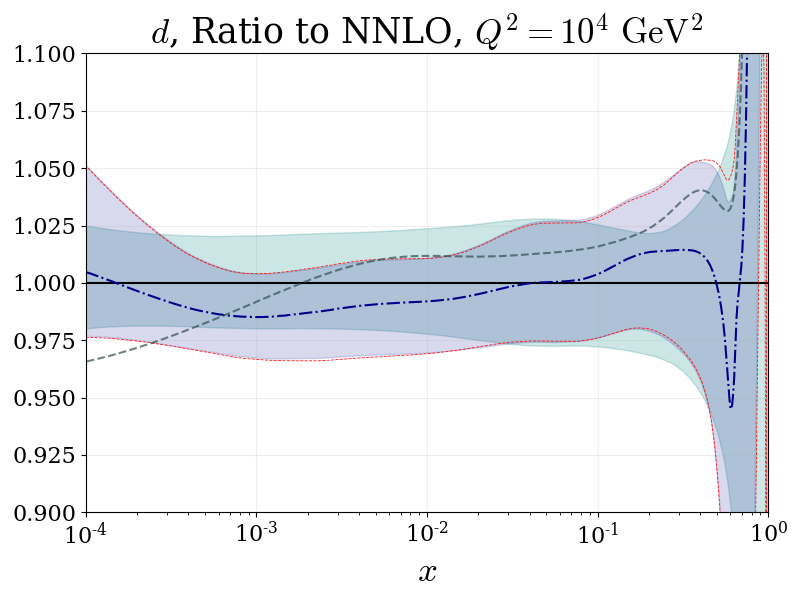}
\includegraphics[width=0.49\textwidth]{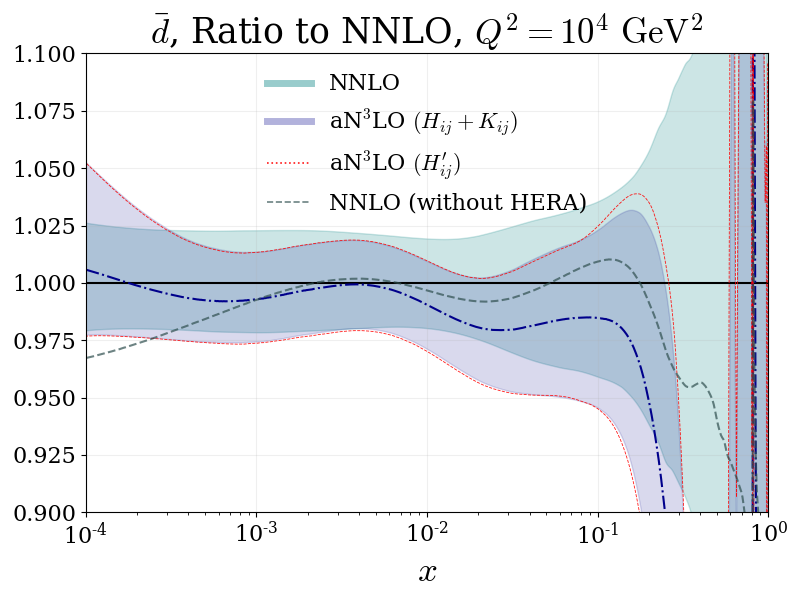}
\end{center}
\caption{\label{fig: pdf_ratios_qhigh}High-$Q^{2}$ ratio plots showing the aN$^{3}$LO 68\% confidence intervals with decorrelated ($H_{ij} + K_{ij}$) and correlated ($H_{ij}^{\prime}$) $K$-factor parameters, compared to NNLO 68\% confidence intervals. Also shown are the central values at NNLO when fit to all non-HERA datasets which show similarities with N$^{3}$LO in the large-$x$ region of selected PDF flavours. All plots are shown for $Q^{2} = 10^{4}\ \mathrm{GeV}^{2}$.}
\end{figure}
\begin{figure}[t]\ContinuedFloat
\begin{center}
\includegraphics[width=0.49\textwidth]{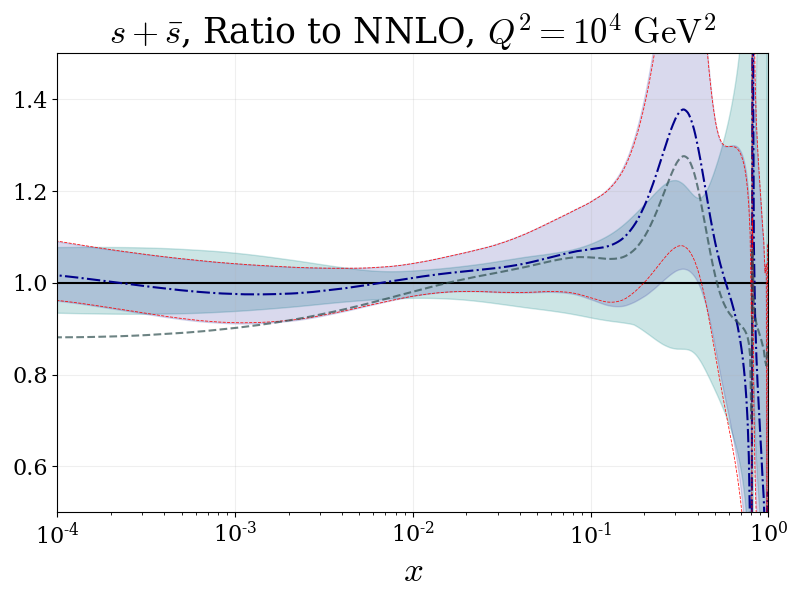}
\includegraphics[width=0.49\textwidth]{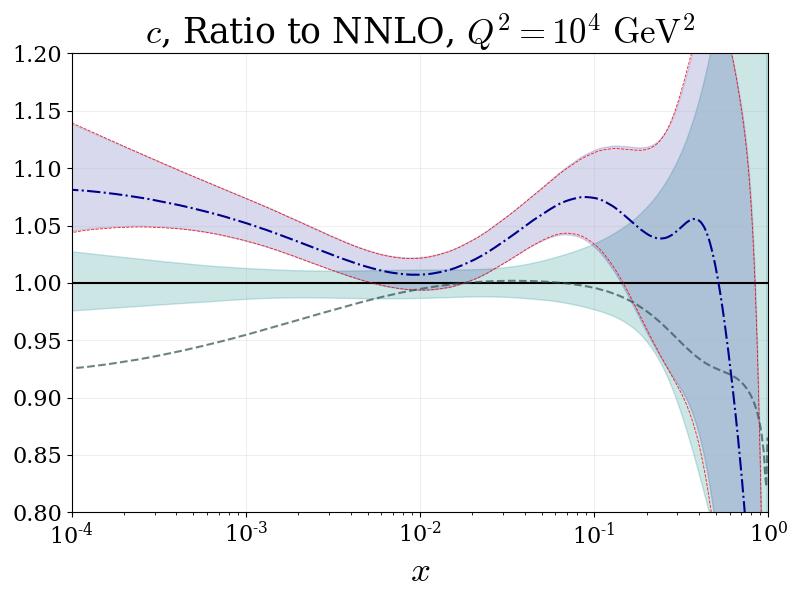}
\includegraphics[width=0.49\textwidth]{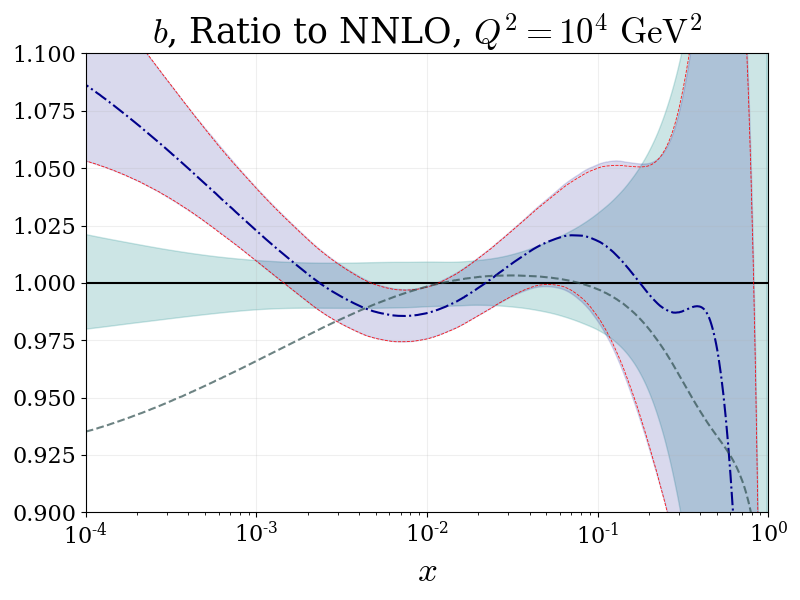}
\includegraphics[width=0.49\textwidth]{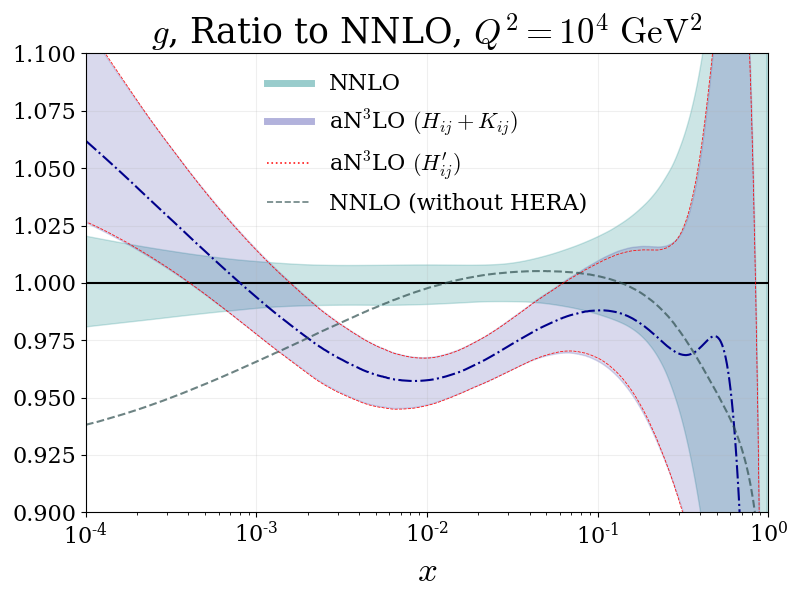}
\end{center}
\caption{\textit{(Continued)} High-$Q^{2}$ ratio plots showing the aN$^{3}$LO 68\% confidence intervals with decorrelated ($H_{ij} + K_{ij}$) and correlated ($H_{ij}^{\prime}$) $K$-factor parameters, compared to NNLO 68\% confidence intervals. Also shown are the central values at NNLO when fit to all non-HERA datasets which show similarities with N$^{3}$LO in the large-$x$ region of selected PDF flavours. All plots are shown for $Q^{2} = 10^{4}\ \mathrm{GeV}^{2}$.}
\end{figure}
Shown in Fig.'s~\ref{fig: pdf_ratios_qlow} and \ref{fig: pdf_ratios_qhigh} are the ratios for each flavour of aN$^{3}$LO PDF compared to the NNLO set with their 68\% confidence intervals at low and high-$Q^{2}$ respectively. The shaded aN$^{3}$LO regions indicate the PDF uncertainty produced with the decorrelated ($(H^{-1}_{ij} + \sum_{p=1}^{N_{p}}K^{-1}_{ij,\ p})^{-1}$) aN$^{3}$LO $K$-factors for each process. As a comparison to these shaded regions, the bounds of uncertainty for the fully correlated ($H_{ij}^{\prime}$) N$^{3}$LO $K$-factor parameters are also provided (red dashed line).

Considering Fig.~\ref{fig: pdf_ratios_qlow} we present the aN$^{3}$LO PDF set at $Q^{2} = 10\ \mathrm{GeV}^{2}$ with the bottom quark PDF at $Q^{2} = 25\ \mathrm{GeV}^{2}$. These PDF ratios better display the substantial increase in the gluon at small-$x$, reminiscent of the gluon PDF presented in \cite{NNPDFsx,xFitterDevelopersTeam:2018hym}. The predicted harder small-$x$ gluon is then accommodated for by reductions in the PDFs at large and small-$x$ (particularly the gluon near $x=10^{-2}$) from NNLO. Another prominent feature is the enhanced charm and bottom quark at N$^{3}$LO. Since the heavy flavour quarks are perturbatively calculated in the MSHT framework, this amplification is a feature of the transition matrix element $A_{Hg}^{(3)}$ at high-$x$, combined with the increase in the gluon PDF at small-$x$ (as these two ingredients are convoluted together). Comparing with Fig.~95 in \cite{Thorne:MSHT20}, we observe that the approximate N$^{3}$LO charm quark now follows a much closer trend to the CT18 PDF and is therefore even more significantly different from the NNPDF NNLO fitted charm at large-$x$ than MSHT20 at NNLO. In the high-$Q^{2}$ setting shown in Fig.~\ref{fig: pdf_ratios_qhigh} we observe similar albeit less drastic effects to those described above.

Also contained in Fig.'s~\ref{fig: pdf_ratios_qlow} and \ref{fig: pdf_ratios_qhigh} are the relative forms of NNLO PDFs when fit to all non-HERA data (full $\chi^{2}$ results are provided in Appendix~\ref{app: noHERA}). Comparing the \textit{non}-HERA NNLO PDFs with aN$^{3}$LO PDFs, there are some similarities in the shapes and magnitudes of a handful of PDFs in the intermediate to large-$x$ regime, most noticeably the light quarks. At small-$x$ the HERA data heavily constrains the PDF fit and therefore these similarities rapidly break down. However, this analysis displays further evidence that including N$^{3}$LO contributions, even though approximate, reduces tensions between the HERA and non-HERA data (when considering the reduction in tension seen in Table~\ref{tab: no_HERA_fullNNLO}). The aN$^{3}$LO PDFs are seemingly able to fit to HERA and non-HERA datasets with superior flexibility than at NNLO. 

While in principle the negativity of quarks is possible in the $\overline{MS}$ scheme, it is unlikely to be correct at very high scales and the behaviour can lead to issues concerning negative cross section predictions~\cite{Candido:2020yat,Collins:2021vke}. In the case of the $\overline{d}$, the form of this PDF has a negative central value above $x \sim 0.5$ with a minimum of $\sim -0.001$ at $x \sim 0.6$. It is also noted that although the $\overline{d}$ central value becomes negative in this region, it is still positive within PDF uncertainties. These features are not uncommon in PDF analyses and are discussed in detail in \cite{PDF4LHC22}. The proposed smoothing of parameterisations employed in \cite{PDF4LHC22} ensures the definite positive nature of PDFs in the high-$x$ region. Comparing the negativity of the approximate N$^{3}$LO $\overline{d}$ PDF with that in \cite{PDF4LHC22}, the $\overline{d}$ PDF presented here is much less negative and positive within PDF uncertainties. Due to this and the fact that this effect is only apparent in the $\overline{d}$, we present these PDFs as they are. We also note that in the current MSHT20 fit, recent results surrounding the $\overline{d}/\overline{u}$ from the SeaQuest collaboration~\cite{SeaQuest:2021zxb} are not included at the time of writing. It is therefore only the E866 / NuSea $pd/pp$ DY dataset~\cite{E866DYrat} that is constraining this ratio, which is not as precise as the more recent results. However, SeaQuest results suggest a preference for a higher $\overline{d}$ at large-$x$, therefore including this data may in fact help constrain the high-$x$ $\overline{d}$ behaviour seen here.

\begin{figure}
\begin{center}
\includegraphics[width=0.49\textwidth]{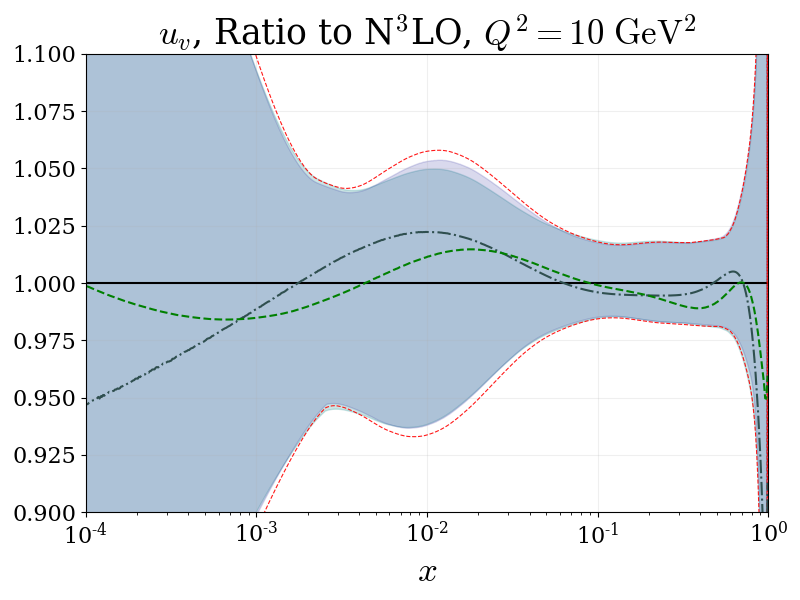}
\includegraphics[width=0.49\textwidth]{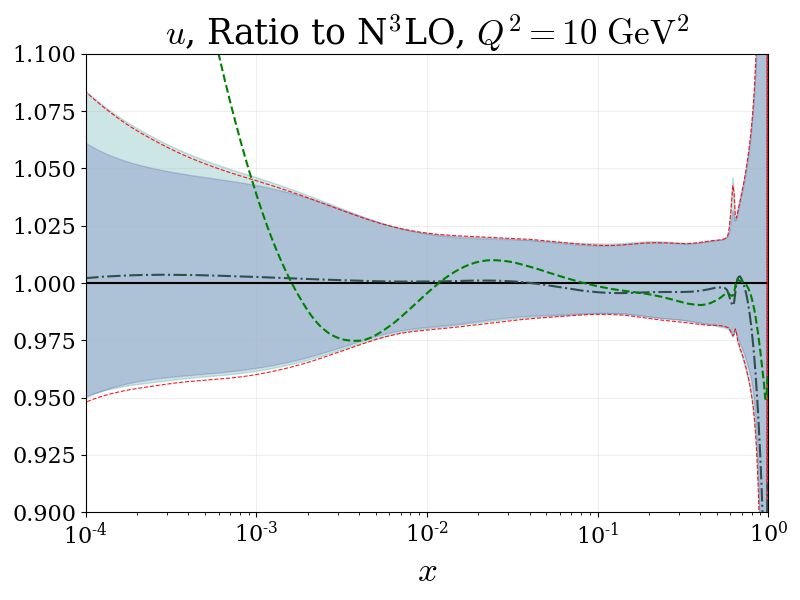}
\includegraphics[width=0.49\textwidth]{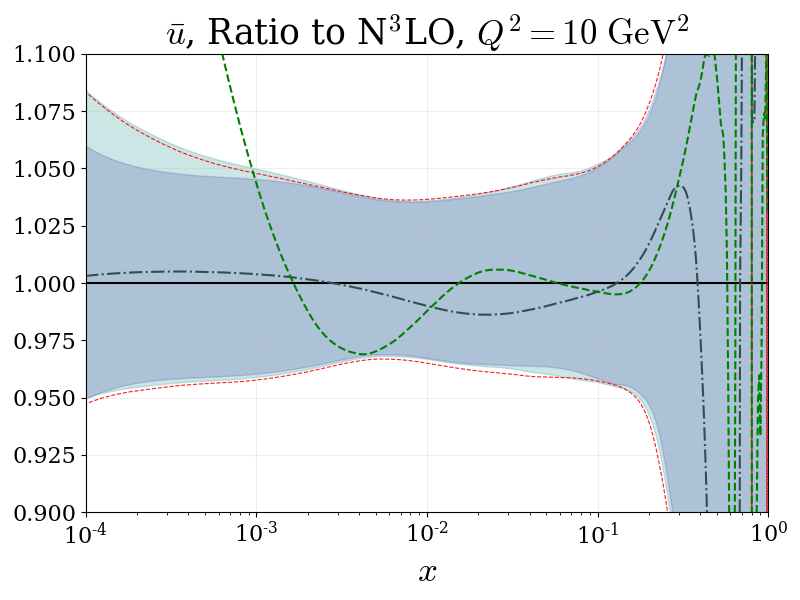}
\includegraphics[width=0.49\textwidth]{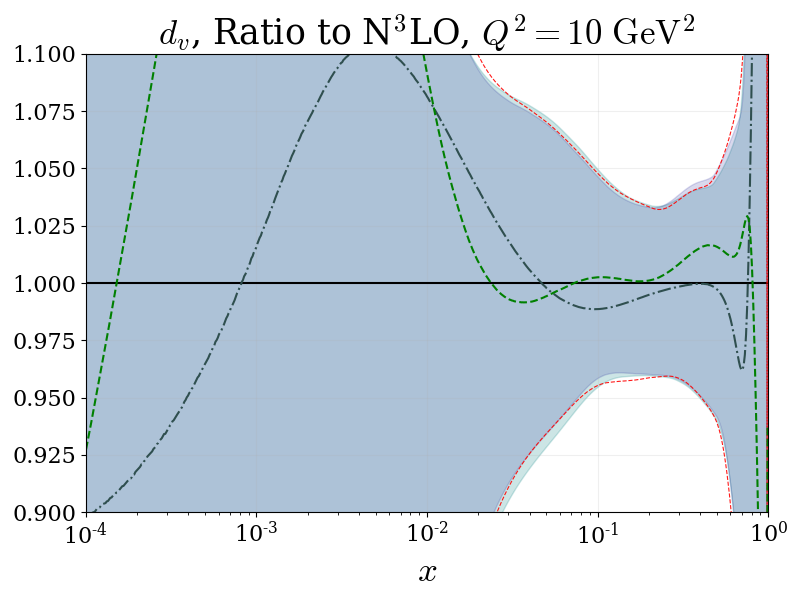}
\includegraphics[width=0.49\textwidth]{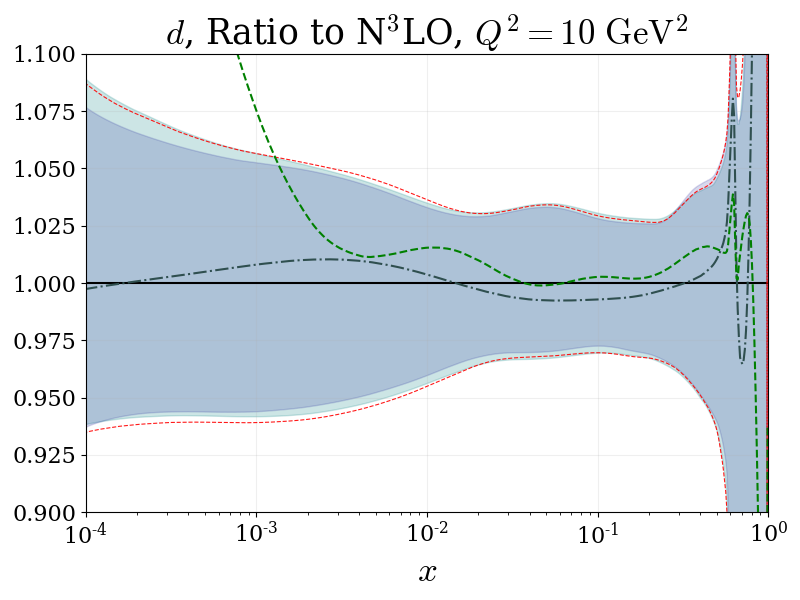}
\includegraphics[width=0.49\textwidth]{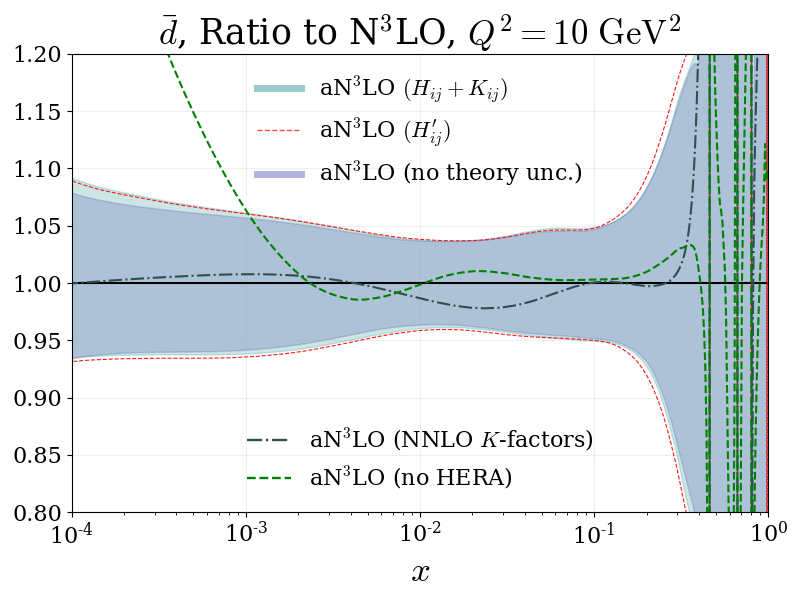}
\end{center}
\caption{\label{fig: pdf_ratios_N3LO_qlow}Low-$Q^{2}$ ratio plots showing the aN$^{3}$LO 68\% confidence intervals with decorrelated and correlated $K$-factor parameters, compared to the aN$^{3}$LO central value. Also shown are the central values at aN$^{3}$LO when fit to all non-HERA datasets and the central values with all $K$-factors set at NNLO. All plots are shown for $Q^{2} = 10\ \mathrm{GeV}^{2}$ with the exception of the bottom quark shown for $Q^{2} = 25\ \mathrm{GeV}^{2}$.}
\end{figure}
\begin{figure}[t]\ContinuedFloat
\begin{center}
\includegraphics[width=0.49\textwidth]{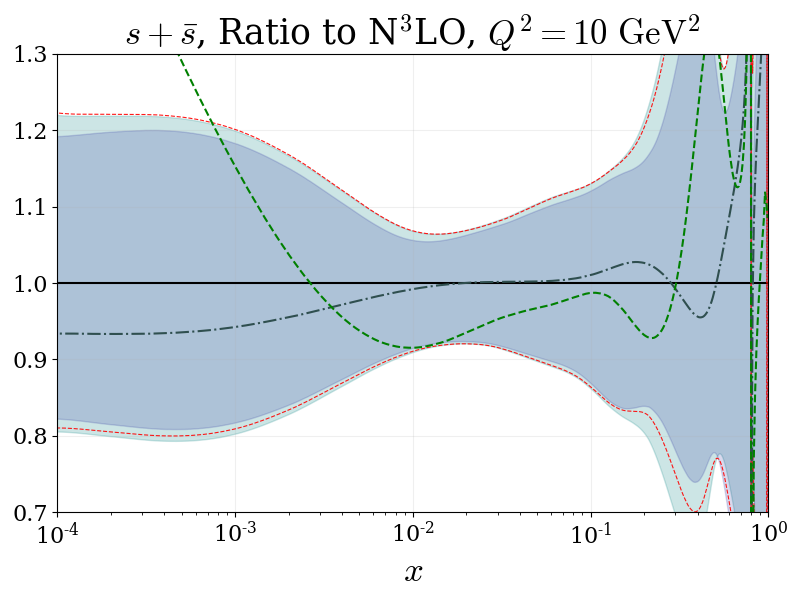}
\includegraphics[width=0.49\textwidth]{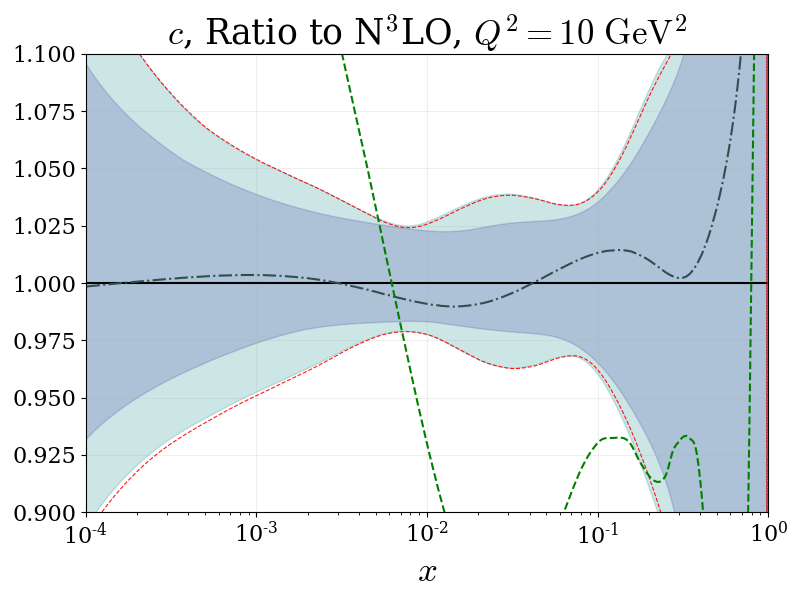}
\includegraphics[width=0.49\textwidth]{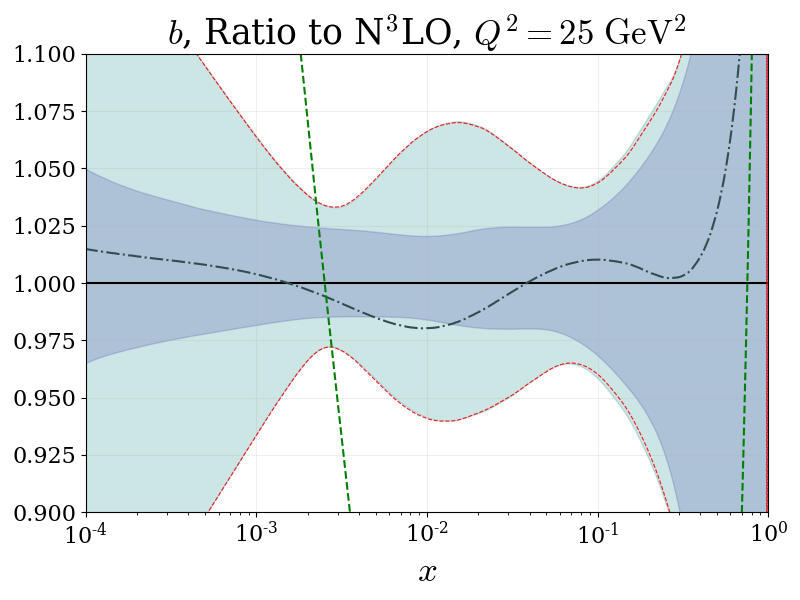}
\includegraphics[width=0.49\textwidth]{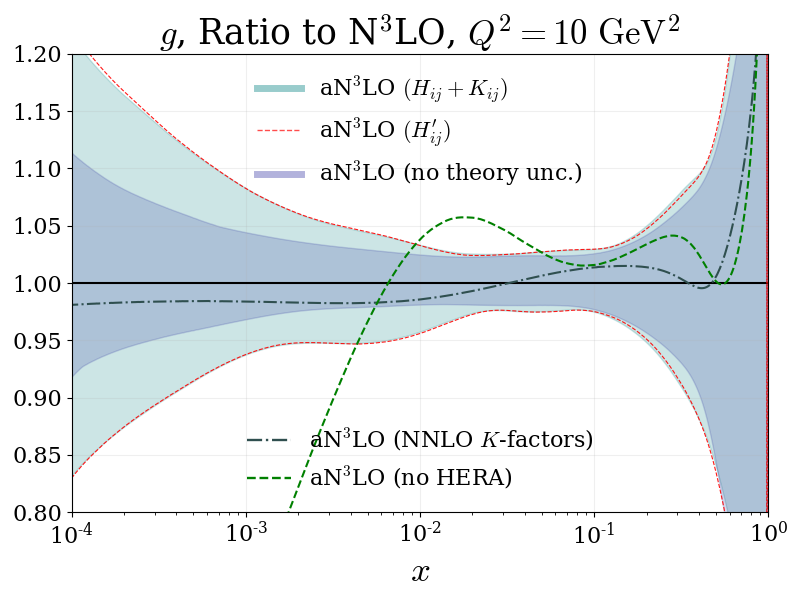}
\end{center}
\caption{\textit{(Continued)} Low-$Q^{2}$ ratio plots showing the aN$^{3}$LO 68\% confidence intervals with decorrelated and correlated $K$-factor parameters, compared to the aN$^{3}$LO central value. Also shown are the central values at aN$^{3}$LO when fit to all non-HERA datasets and the central values with all $K$-factors set at NNLO. All plots are shown for $Q^{2} = 10\ \mathrm{GeV}^{2}$ with the exception of the bottom quark shown for $Q^{2} = 25\ \mathrm{GeV}^{2}$.}
\end{figure}
\begin{figure}
\begin{center}
\includegraphics[width=0.49\textwidth]{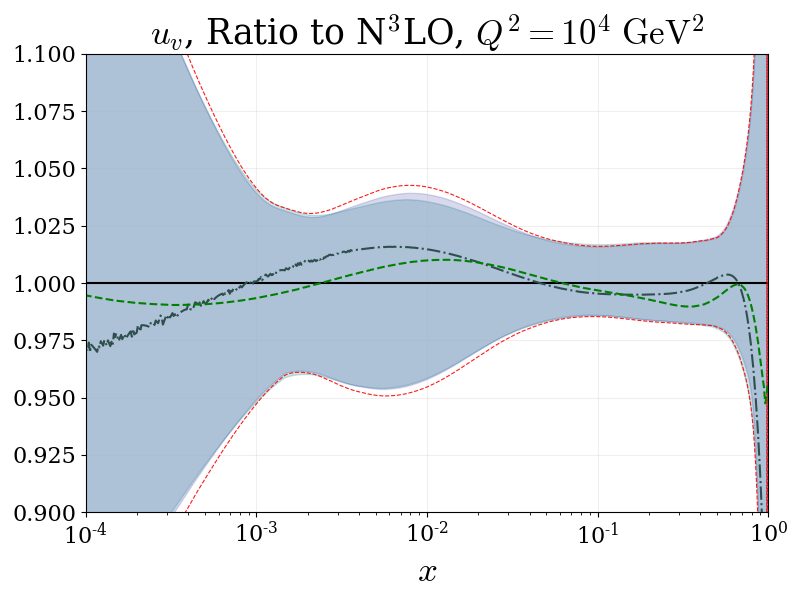}
\includegraphics[width=0.49\textwidth]{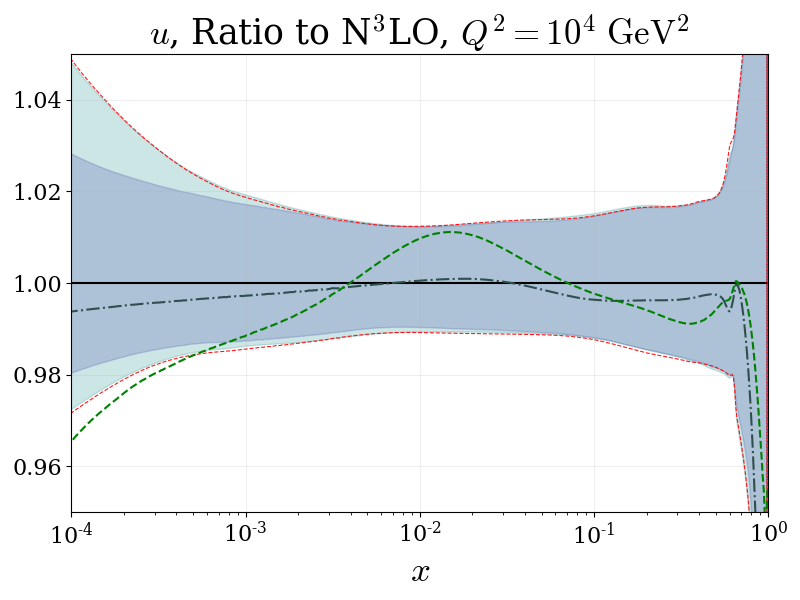}
\includegraphics[width=0.49\textwidth]{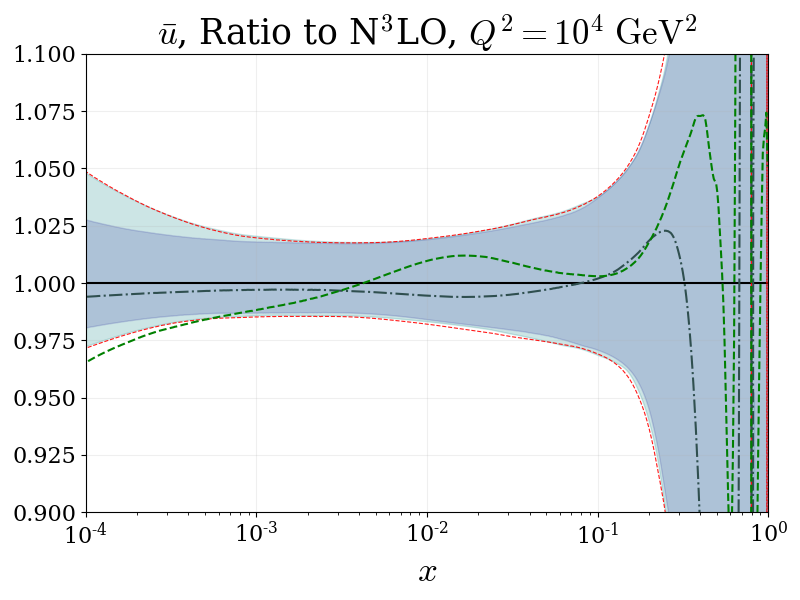}
\includegraphics[width=0.49\textwidth]{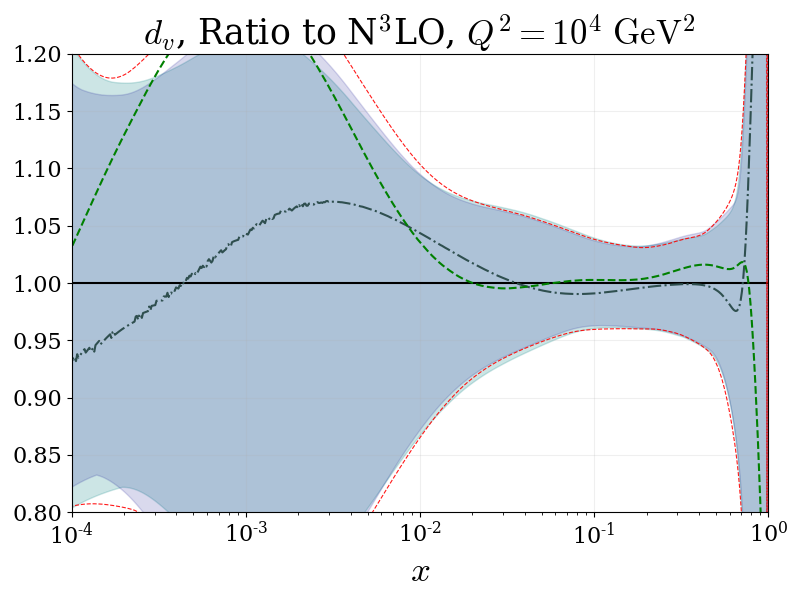}
\includegraphics[width=0.49\textwidth]{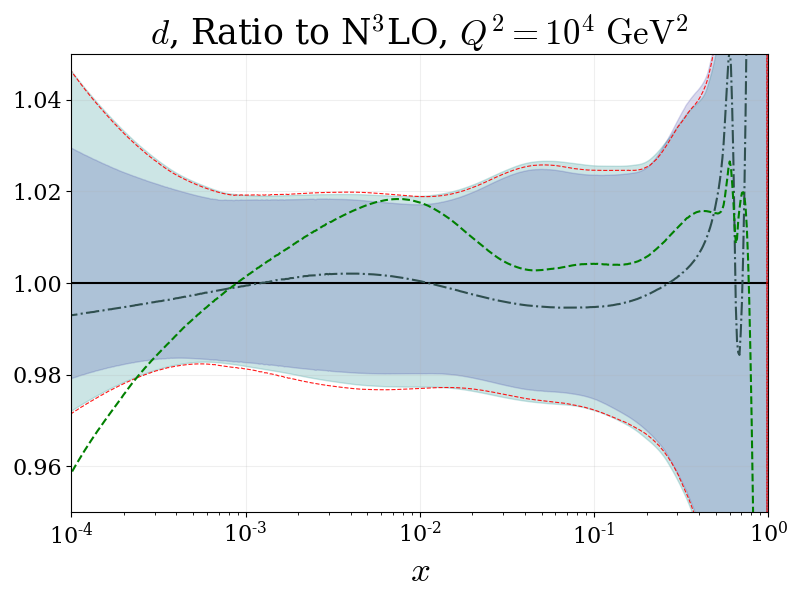}
\includegraphics[width=0.49\textwidth]{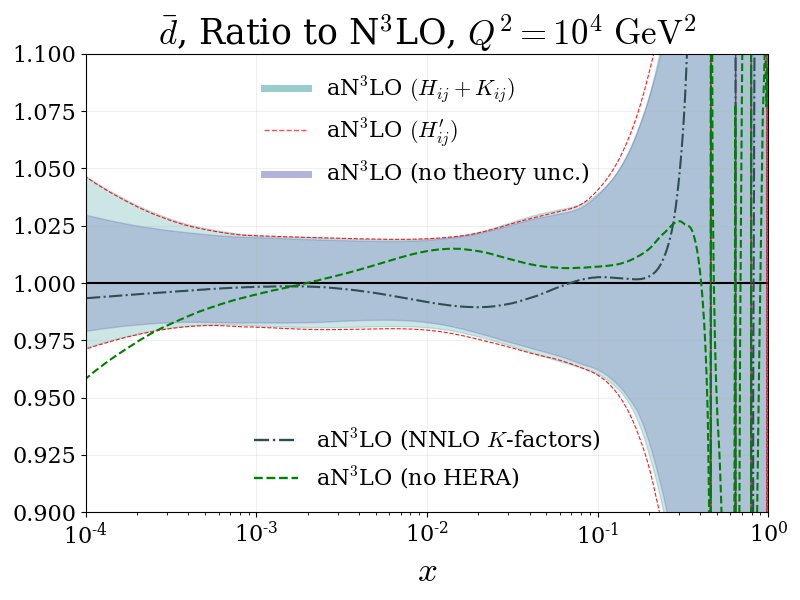}
\end{center}
\caption{\label{fig: pdf_ratios_N3LO_qhigh} High-$Q^{2}$ ratio plots showing the aN$^{3}$LO 68\% confidence intervals with decorrelated and correlated $K$-factor parameters, compared to the aN$^{3}$LO central value. Also shown are the central values at aN$^{3}$LO when fit to all non-HERA datasets and the central values with all $K$-factors set at NNLO. All plots are shown for $Q^{2} = 10^{4}\ \mathrm{GeV}^{2}$.}
\end{figure}
\begin{figure}[t]\ContinuedFloat
\begin{center}
\includegraphics[width=0.49\textwidth]{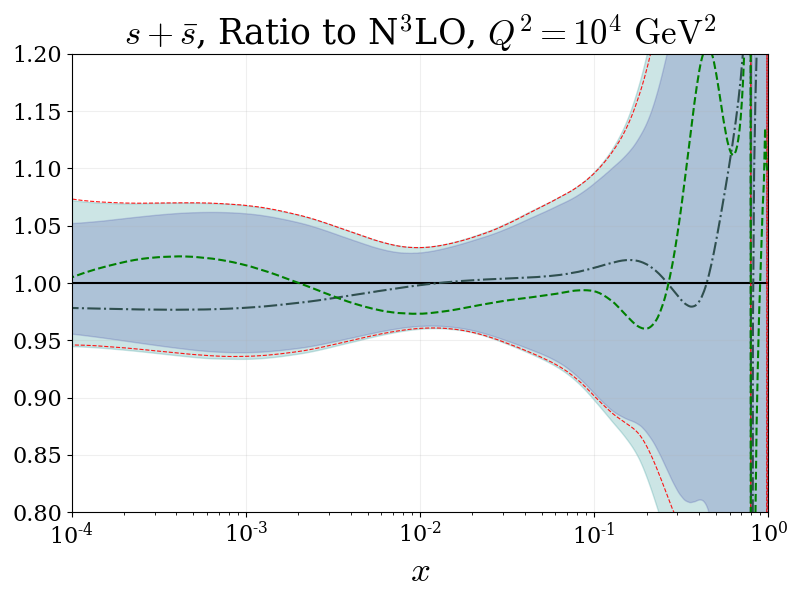}
\includegraphics[width=0.49\textwidth]{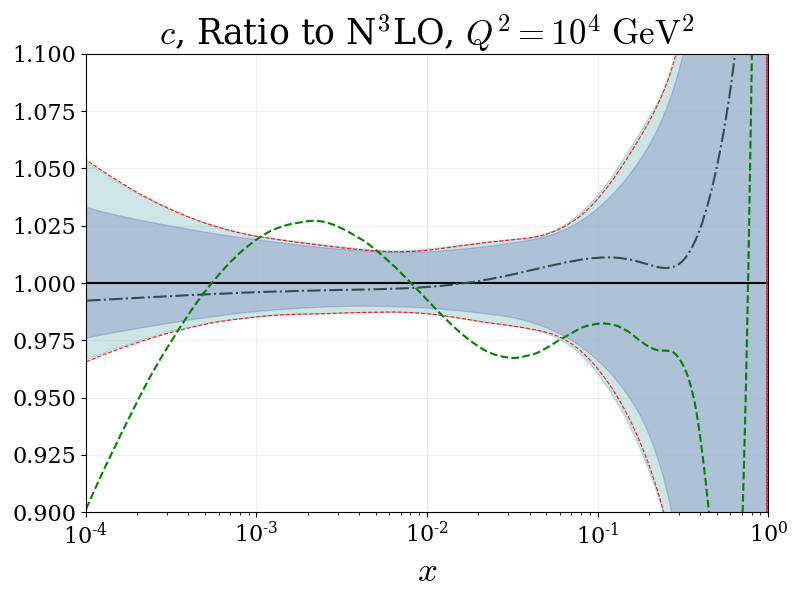}
\includegraphics[width=0.49\textwidth]{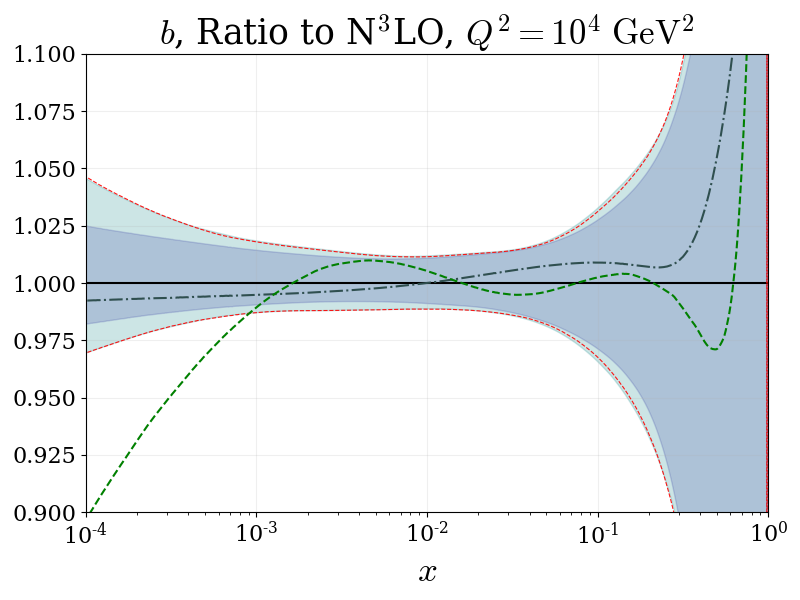}
\includegraphics[width=0.49\textwidth]{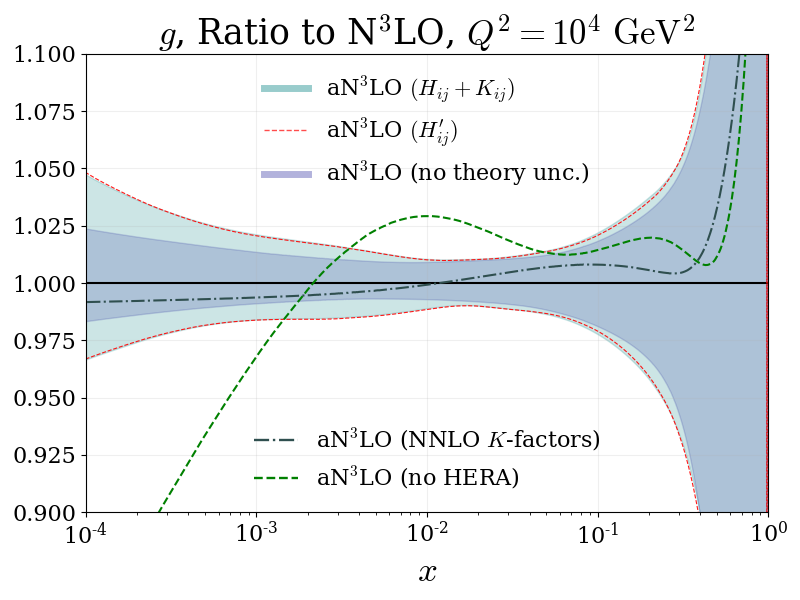}
\end{center}
\caption{\textit{(Continued)} High-$Q^{2}$ ratio plots showing the aN$^{3}$LO 68\% confidence intervals with decorrelated and correlated $K$-factor parameters, compared to the aN$^{3}$LO central value. Also shown are the central values at aN$^{3}$LO when fit to all non-HERA datasets and the central values with all $K$-factors set at NNLO. All plots are shown for $Q^{2} = 10^{4}\ \mathrm{GeV}^{2}$.}
\end{figure}
Fig.'s~\ref{fig: pdf_ratios_N3LO_qlow} and \ref{fig: pdf_ratios_N3LO_qhigh} express the aN$^{3}$LO PDFs with decorrelated (green shaded region) and correlated (red dashed lines) aN$^{3}$LO $K$-factors at low and high-$Q^{2}$ respectively (again with the bottom quark provided at $Q^{2} = 25\ \mathrm{GeV}^{2}$ at low-$Q^{2}$) as a ratio to the N$^{3}$LO central value. For comparison we also include the level of uncertainty predicted with all N$^{3}$LO theory fixed (blue shaded region) i.e. only considering the variation \textit{without} N$^{3}$LO theoretical uncertainty. 

Comparing the two different aN$^{3}$LO sets in Fig.'s~\ref{fig: pdf_ratios_N3LO_qlow} and \ref{fig: pdf_ratios_N3LO_qhigh}, in general there is good agreement between the total uncertainties considering the cases with correlated (red dash) and decorrelated (green shaded) aN$^{3}$LO $K$-factors. The differences that are apparent between between the two aN$^{3}$LO cases, are relatively small across all PDFs, with slightly larger effects only where the PDF itself tends towards zero i.e. valence quarks at small-$x$.

A larger distinction is observed when comparing the sets \textit{with} and \textit{without} theoretical uncertainty (where N$^{3}$LO theory is fixed at the best fit value). In general there is an expected substantial increase in the PDF uncertainties when taking into account the missing N$^{3}$LO uncertainty for the gluon (and therefore the heavy quarks). In particular, the form of the N$^{3}$LO bottom quark uncertainty is reminiscent of the $(H+\overline{H})$ prediction from Fig.~\ref{fig: step_results}. One can therefore directly observe the effect of the $A_{Hg}$ theoretical uncertainty on the bottom quark directly above its mass threshold. In other areas, the \textit{without} theoretical uncertainty PDF set exhibits a comparable uncertainty to aN$^{3}$LO and is even shown to increase the overall 68\% confidence intervals in certain regions of $(x,Q^{2})$ due to N$^{3}$LO parameters being fixed (i.e. $u_{v}$ and $d_{v}$ PDFs in Fig.~\ref{fig: pdf_ratios_N3LO_qlow} and Fig.~\ref{fig: pdf_ratios_N3LO_qhigh}). As the fit now resides in a different $\chi^{2}$ landscape where a best fit has been achieved through fitting the N$^{3}$LO theory, fixing the aN$^{3}$LO theory parameters is likely to have a substantial effect across all PDFs.

An important point made by Fig.'s~\ref{fig: pdf_ratios_N3LO_qlow} and \ref{fig: pdf_ratios_N3LO_qhigh} is that that the difference between the decorrelated and correlated cases is much smaller than the difference of not including theoretical uncertainties at all (blue shaded region). This analysis therefore provides evidence to support the original assumption of being able to decorrelate the cross section (aN$^{3}$LO $K$-factors) and PDF theory (including other N$^{3}$LO theory). 

Along with the separate cases of uncertainty illustrated in Fig.'s~\ref{fig: pdf_ratios_N3LO_qlow} and \ref{fig: pdf_ratios_N3LO_qhigh}, we also display the central values of an aN$^{3}$LO fit to all non-HERA data and an aN$^{3}$LO fit with NNLO $K$-factors. Examining the form of the no HERA aN$^{3}$LO PDFs for $x > 10^{-2}$, we show some agreement with the standard N$^{3}$LO central value across most PDFs (more so at high-$Q^{2}$ than low-$Q^{2}$). Whereas the form at small-$x$ gives some insight into the importance of HERA data in constraining PDFs in this region. In slightly better agreement across all $x$ are the aN$^{3}$LO PDFs with NNLO $K$-factors, which compliment the $\chi^{2}$ results in Section~\ref{sec: n3lo_K} and Section~\ref{subsec: chi2} arguing that the form (and fit results) of aN$^{3}$LO PDFs is mostly determined from the extra PDF $+$ DIS coefficient function N$^{3}$LO additions i.e. not aN$^{3}$LO $K$-factors which prefer a softer high-$x$ gluon (similar to the N$^{3}$LO no HERA case -- also shown in Fig.'s~\ref{fig: pdf_ratios_N3LO_qlow} and \ref{fig: pdf_ratios_N3LO_qhigh}).

\subsection{MSHT20\texorpdfstring{aN$^{3}$LO}{aN3LO} PDFs at \texorpdfstring{$Q^{2} = 2\ \mathrm{GeV}^{2}$}{Q2 = 2 GeV2}}

\begin{figure}
\begin{center}
\includegraphics[width=0.49\textwidth]{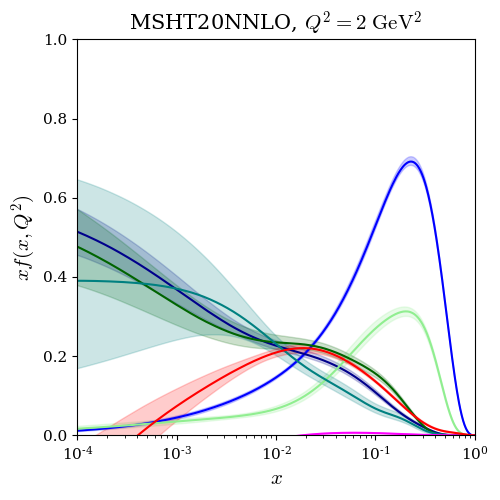}
\includegraphics[width=0.49\textwidth]{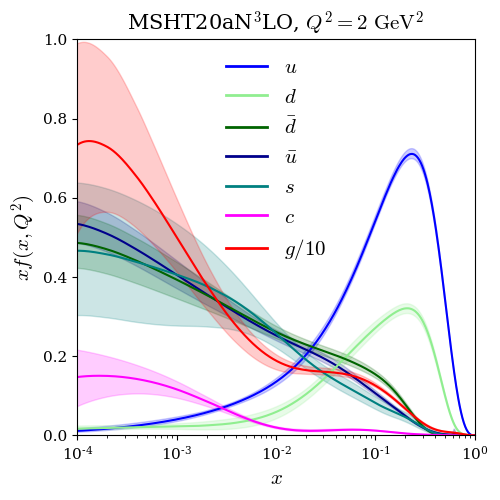}
\end{center}
\caption{\label{fig: pdfs_low} General forms of NNLO (left) and aN$^{3}$LO (right) PDFs at $Q^{2} = 2\ \mathrm{GeV}^{2}$. Axis are set to the same scale to highlight the main differences between NNLO and aN$^{3}$LO. Specifically in the gluon and heavy flavour sectors.}
\end{figure}
Fig.~\ref{fig: pdfs_low} compares the MSHT NNLO and aN$^{3}$LO PDF sets at $Q^{2} = 2\ \mathrm{GeV}^{2}$. In this very low-$Q^{2}$ regime, some major differences are evident between NNLO and aN$^{3}$LO sets at $Q^{2} = 2\ \mathrm{GeV}^{2}$, especially towards small-$x$. For example, the gluon PDF is predicted to be much harder across this region, such that it is now positive across all $x$ values considered here. The  effect of this can be immediately seen in the sea and heavy quarks. 

Since the charm quark is directly coupled to the gluon PDF (through a convolution with $A_{Hg}$), the charm PDF receives a notable enhancement at small-$x$ and also remains positive across all $x$ values considered\footnote{Since this is a convolution, it is the higher small-$x$ gluon, combined with the high-$x$ enhancement of $A_{Hg}$ at N$^{3}$LO which gives rise to this increase in the charm PDF.}. Another interesting feature is the reduction in uncertainty of the strange quark at small-$x$. It may seem counter intuitive to have an uncertainty reduction by adding sources of theoretical uncertainty, however we should recall that the underlying theory has also been altered. Although one can expect an uncertainty increase in PDFs across $(x,Q^{2})$, there are exceptions to this e.g. where tensions are relieved by introducing the N$^{3}$LO theory. The shift in the $\chi^{2}$ landscape then has the potential to result in more precise regions of $(x, Q^{2})$ (in this case manifesting in an uncertainty reduction for the strange quark towards small-$x$). 

\begin{figure}
\begin{center}
\includegraphics[width=0.49\textwidth]{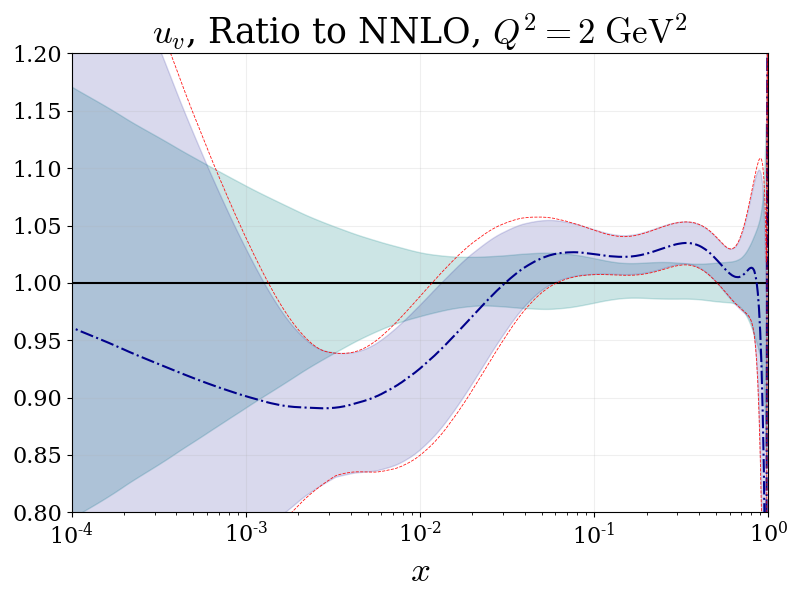}
\includegraphics[width=0.49\textwidth]{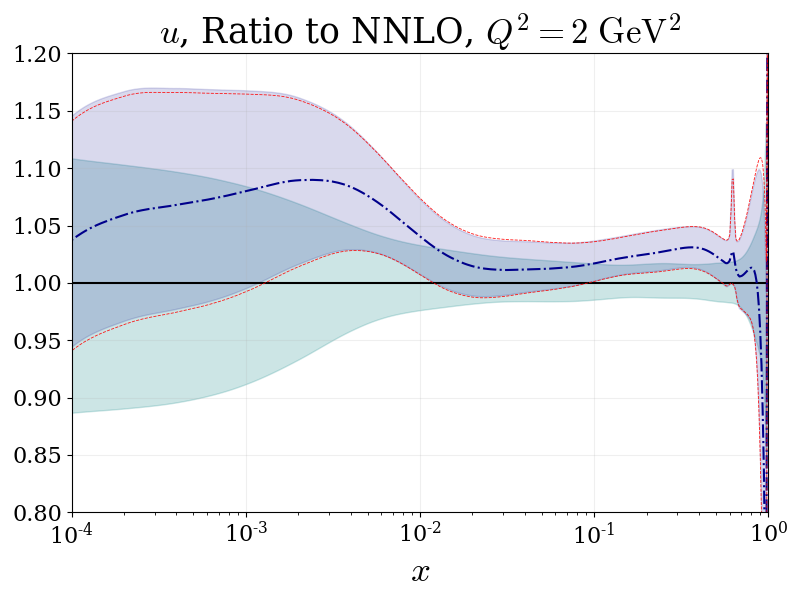}
\includegraphics[width=0.49\textwidth]{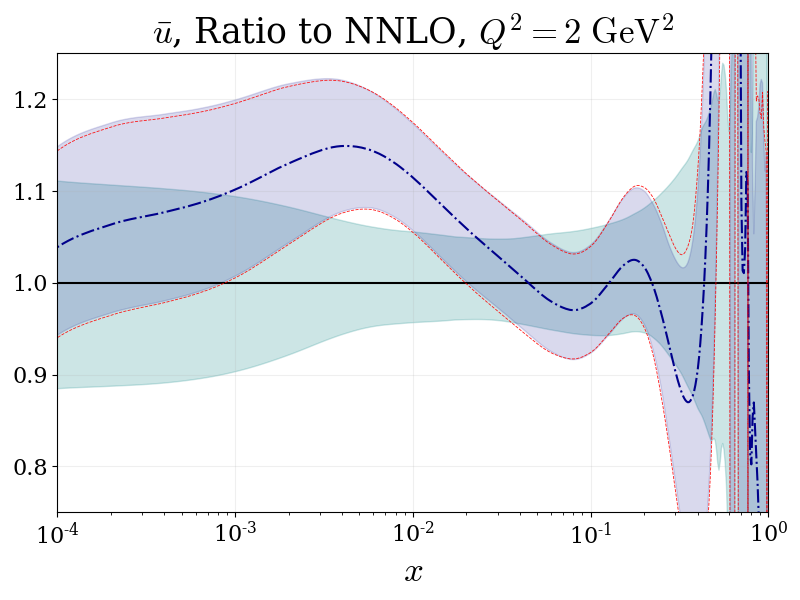}
\includegraphics[width=0.49\textwidth]{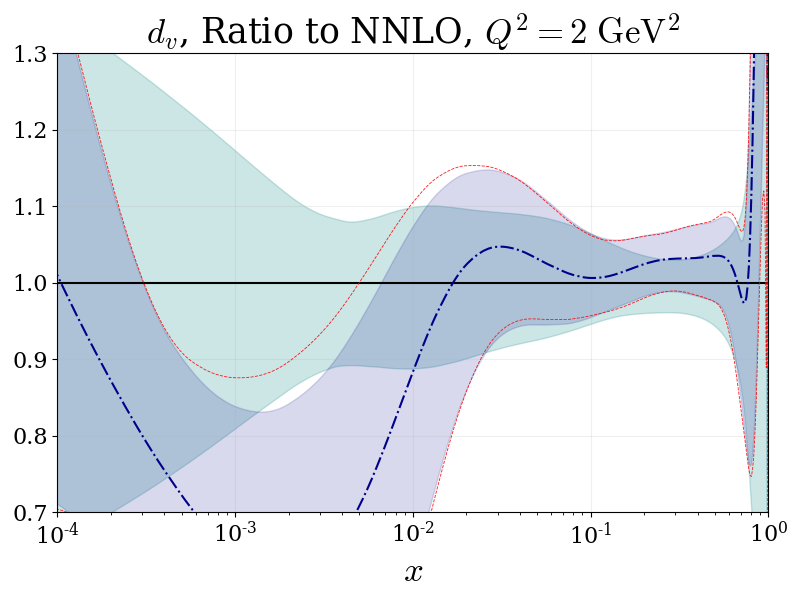}
\includegraphics[width=0.49\textwidth]{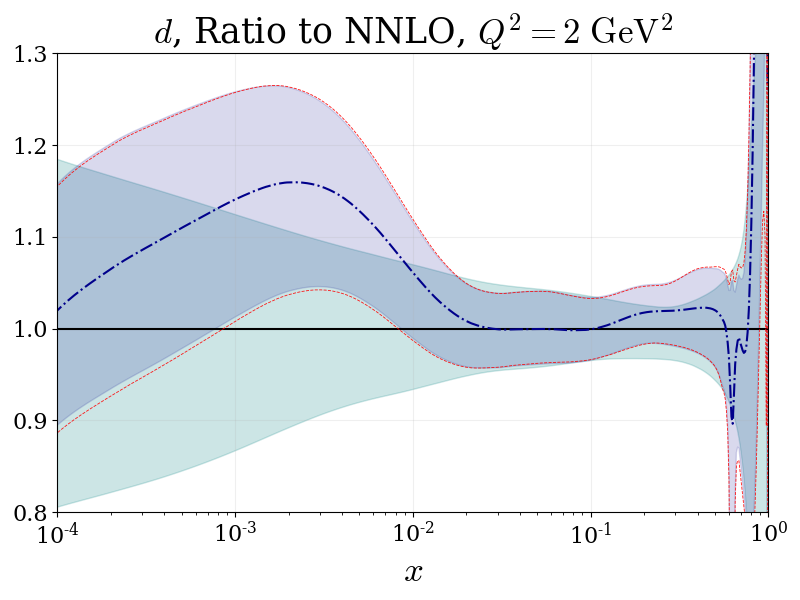}
\includegraphics[width=0.49\textwidth]{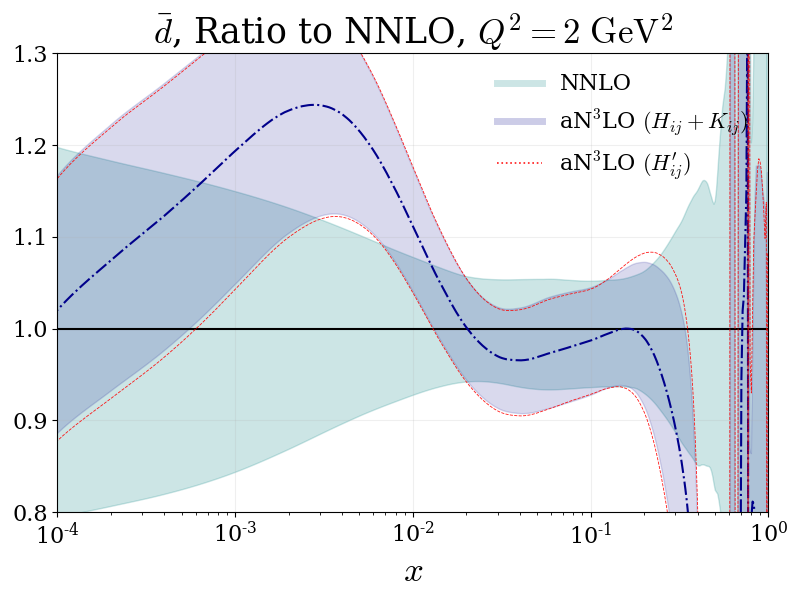}
\end{center}
\caption{\label{fig: pdf_ratios_qvlow}Very low-$Q^{2}$ ratio plots showing the aN$^{3}$LO 68\% confidence intervals with decorrelated and correlated $K$-factor parameters, compared to NNLO 68\% confidence intervals. All plots are shown for $Q^{2} = 2\ \mathrm{GeV}^{2}$.}
\end{figure}
\begin{figure}[t]\ContinuedFloat
\begin{center}
\includegraphics[width=0.49\textwidth]{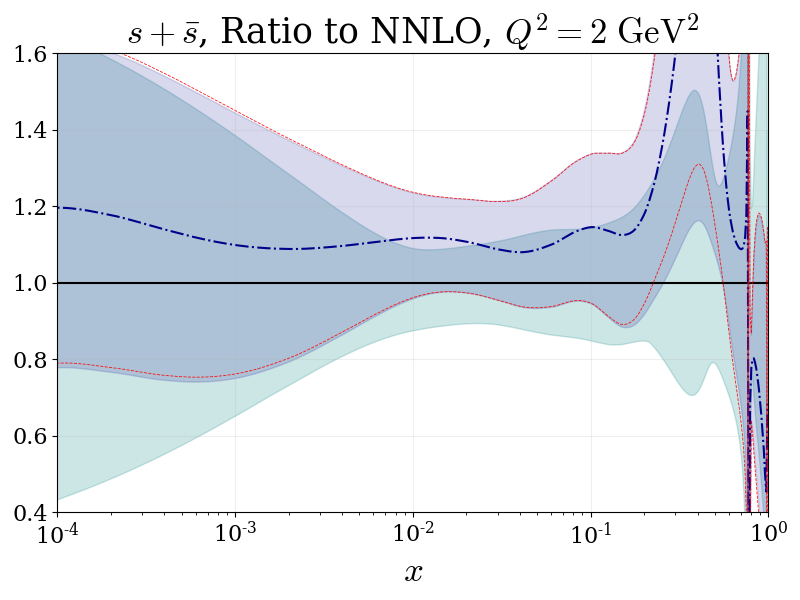}
\includegraphics[width=0.49\textwidth]{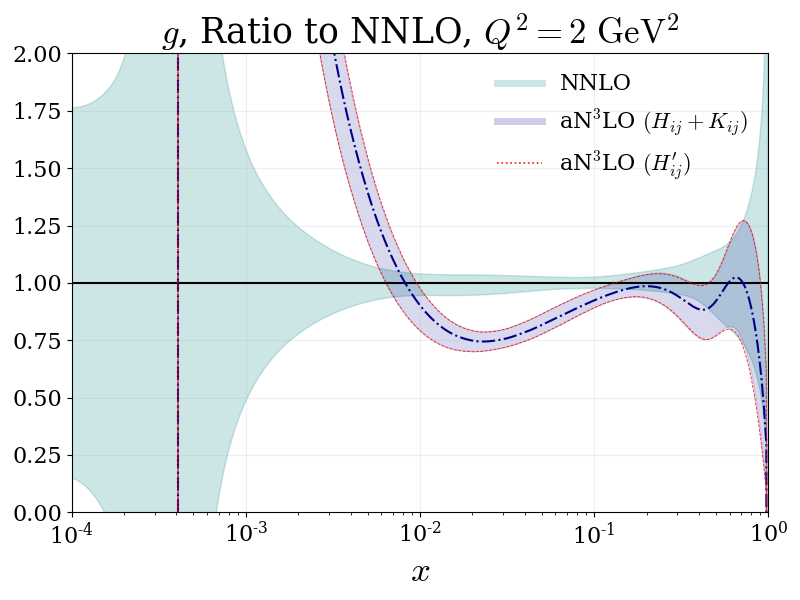}
\end{center}
\caption{\textit{(Continued)} Very low-$Q^{2}$ ratio plots showing the aN$^{3}$LO 68\% confidence intervals with decorrelated and correlated $K$-factor parameters, compared to NNLO 68\% confidence intervals. All plots are shown for $Q^{2} = 2\ \mathrm{GeV}^{2}$.}
\end{figure}
Fig.~\ref{fig: pdf_ratios_qvlow} displays the ratios of the aN$^{3}$LO MSHT PDFs to their NNLO counterparts at $Q^{2} = 2\ \mathrm{GeV}^{2}$. Here the specific shifts of each PDF are displayed more clearly. We note that there are many similar features shown here to those discussed for Fig.'s~\ref{fig: pdf_ratios_qlow} and \ref{fig: pdf_ratios_qhigh}. Even in this very low-$Q^{2}$ regime, the uncertainty difference between correlated and decorrelated aN$^{3}$LO $K$-factor PDF sets is minimal in all relevant regions of $x$.

\subsection{Effect of a \texorpdfstring{$x < 10^{-3}$}{x < 0.001} cut at \texorpdfstring{aN$^{3}$LO}{aN3LO}}\label{subsec:low-x}

In this section we include results from a global PDF fit with small-$x$ ($x < 10^{-3}$) data omitted. This analysis is provided to shed some light on the tensions between regions of $x$ at aN$^{3}$LO while also providing some context with regards to the form of the PDFs in different regions of $x$.

\begin{figure}
\begin{center}
\includegraphics[width=0.49\textwidth]{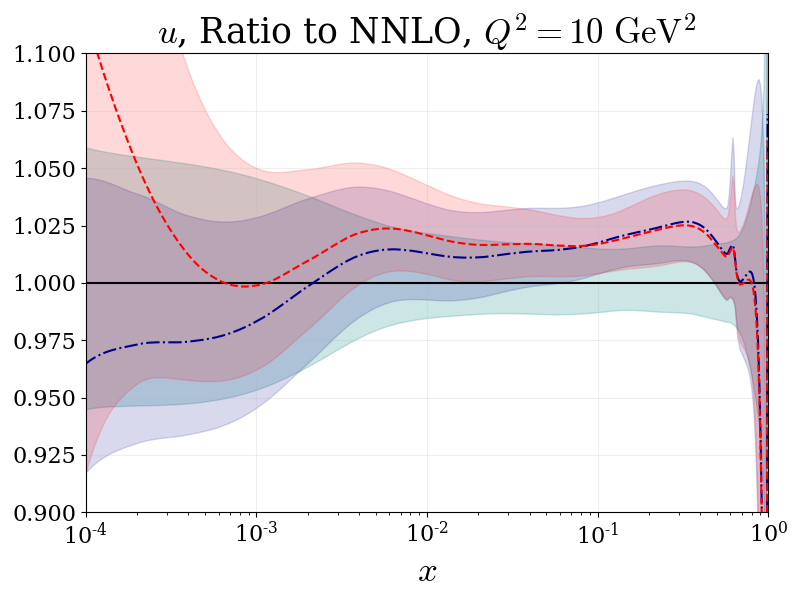}
\includegraphics[width=0.49\textwidth]{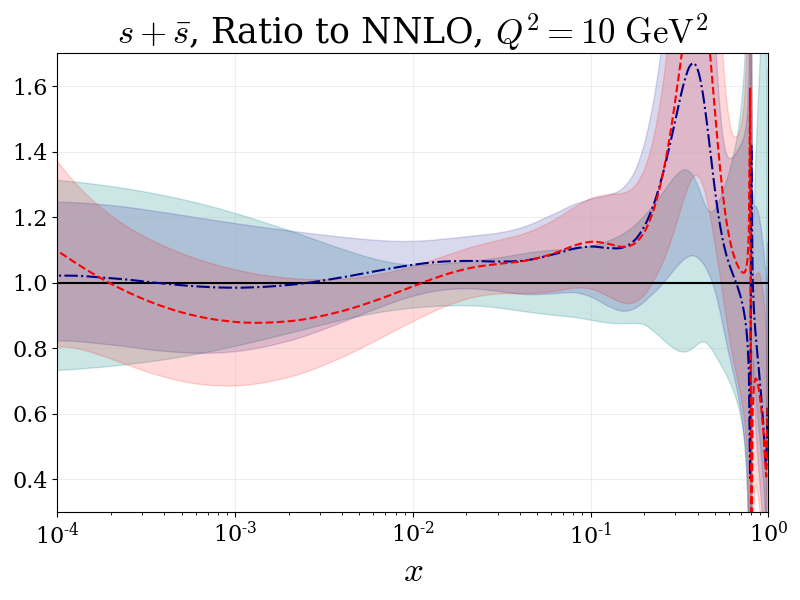}
\includegraphics[width=0.49\textwidth]{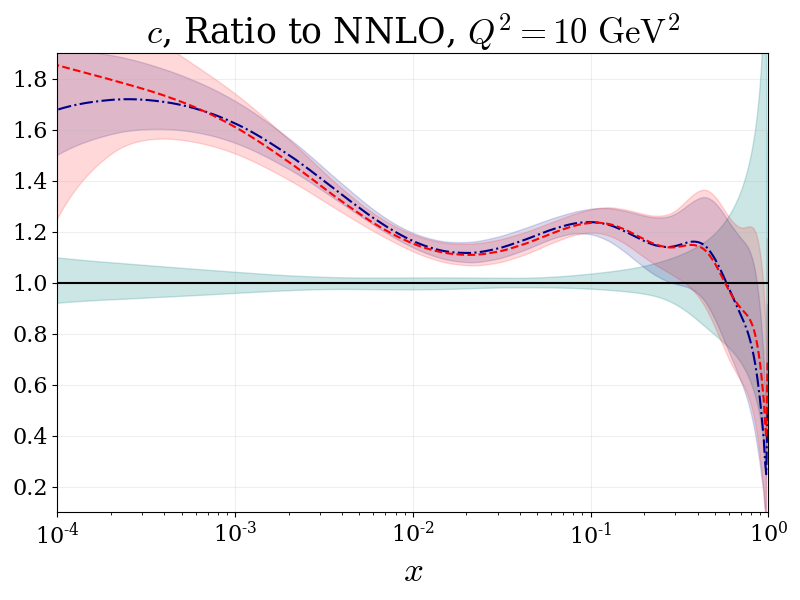}
\includegraphics[width=0.49\textwidth]{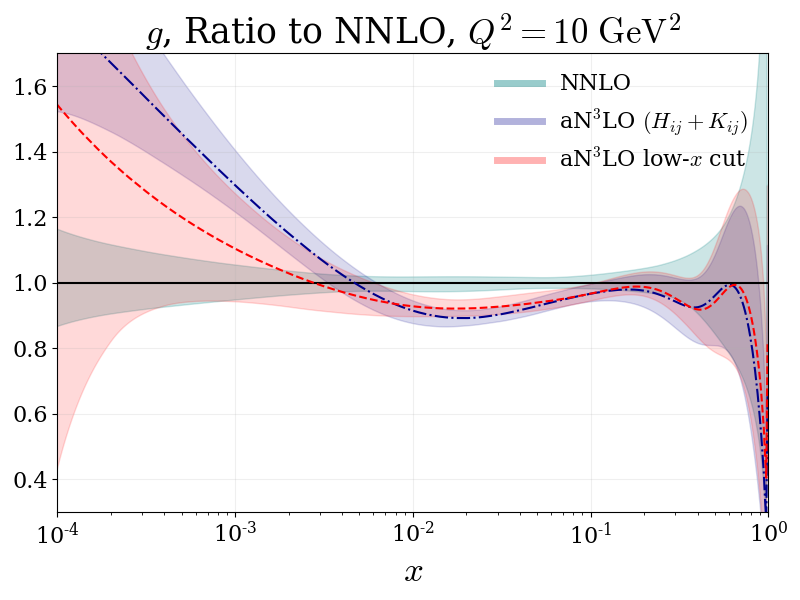}
\end{center}
\caption{\label{fig:low-x}Low-$Q^{2}$ PDF ratios showing aN$^{3}$LO PDFs fitted with and without small-$x$ ($<10^{-3}$) data included in a global fit. All plots are shown for $Q^{2} = 10\ \mathrm{GeV}^{2}$ with the exception of the bottom quark shown for $Q^{2} = 25\ \mathrm{GeV}^{2}$.}
\end{figure}
Immediately, in Fig.~\ref{fig:low-x}, one can observe that omitting all small-$x$ data results in a set of less constrained PDFs for $x < 10^{-3}$ (most notably in the gluon sector). However, also in Fig.~\ref{fig:low-x} it can be observed that overall, the large-$x$ behaviour of these PDFs is very similar across both fits, indicating that the full fit is able to sufficiently fit both large and small $x$ regions simultaneously. We provide this analysis as a cross-check to further support the reliability of our procedure, showing that the small-$x$ behaviour is not overwhelmingly attempting to fit to any all order, or specifically small-$x$ resummation, result at the expense of the large-$x$ description. We also note that while there is some definite change in the central values of the PDFs at small $x$, in most cases this is very well within uncertainties, and at most at the level of about one standard deviation, particularly for the gluon for $x$ just below $10^{-3}$. This distinct, but limited shift in the best fit PDFs suggests that effects beyond N$^3$LO are certainly not insignificant at very low $x$, they are also not dominating the pull on the fit. 

\subsection{Posterior \texorpdfstring{N$^{3}$LO}{N3LO} Theory Parameters}\label{subsec: an3lo_approx}

Following from the previous section, it is also interesting to examine where the aN$^{3}$LO theory contributions and their uncertainties reside after a global PDF fit.

\begin{table}[]
    \footnotesize
    \centering
    \begin{tabular}{|c|c|c|c|c|c|c|}
    \hline
        \multirow{2}{*}{Parameter} &  \multicolumn{3}{c|}{Default Fit} & \multicolumn{3}{c|}{Small-$x$ cut Fit} \\
        \cline{2-7}
         & Central & $+$ Limit & $-$ Limit & Central & $+$ Limit & $-$ Limit \\
         \hline
        Low-$Q^{2}$ Coefficient & & & & & & \\
        \hline
        $c_{q}^{\mathrm{NLL}}$ & $-3.868$  &   $-1.891$  & $-6.132$ & $-5.822$ & $-4.333$ & $-10.373$ \\
        $c_{g}^{\mathrm{NLL}}$ & $-5.837$  &   $-4.444$  & $-7.429$ & $-6.995$ & $-3.701$ & $-7.991$ \\
        \hline
        Transition Matrix Elements & & & & & & \\
        \hline
        $a_{qq,H}^{\mathrm{NS}}$ & $-64.411$  &  $-38.778$ & $-91.850$ & $-65.103$ & $-40.073$ & $-93.225$ \\
        $a_{Hg}$ & $12214.000$  &   $12966.856$  & $11279.376$ & $12524.000$ & $13831.976$ & $11286.843$ \\  
        $a_{gg,H}$ & $-1951.600$  &   $-1577.155$  & $-3418.568$ & $-1392.600$ & $-512.817$ & $-2190.354$ \\
        \hline
        Splitting Functions & & & & & &\\
        \hline
        $\rho_{qq}^{NS}$ & $0.007$  &   $0.015$  &  $-0.002$  & $0.006$  & $0.020$ & $-0.005$ \\
        $\rho_{qq}^{PS}$ & $-0.501$  &   $-0.254$  & $-0.644$ &  $-0.505$  & $-0.285$ & $-0.692$ \\
        $\rho_{qg}$ & $-1.754$  &   $-1.157$  & $-1.897$ & $-1.309$ & $-0.620$ & $-1.881$ \\
        $\rho_{gq}$ & $-1.784$  &   $-1.548$  & $-2.212$  & $-1.622$ & $-1.367$ & $-1.877$ \\
        $\rho_{gg}$ & $19.245$  &   $21.505$  & $9.025$  & $12.997$ & $16.142$ & $6.611$ \\
        \hline
    \end{tabular}
    \caption{Posterior predicted $\pm 1\sigma$ limits on aN$^{3}$LO theoretical nuisance parameters for splitting functions, transition matrix elements and coefficient functions.}
    \label{tab:post_limits}
\end{table}
Displayed in Table~\ref{tab:post_limits} are the predicted posterior limits on each (non $K$-factor) aN$^{3}$LO theory parameter. Here one can directly compare these variations with the prior variations decided in earlier Sections. Also provided is a comparison of these posterior limits across a fit with and without small-$x$ ($x < 10^{-3}$) data included. This comparison compliments the previous section by showing a similar trend in the central values predicted in both cases (i.e. with an overlap of uncertainties). As with the PDFs, there is some significant evidence of tensions, but these are not severe, and the central values of many parameters are extremely stable. Furthermore this is evidence that the small-$x$ behaviour is influencing, but not likely to be dominating the behaviour of aN$^{3}$LO parameters in a manner which is significantly adversarial to the preference of data at $x > 10^{-3}$.

\begin{figure}
\begin{center}
\includegraphics[width=0.49\textwidth]{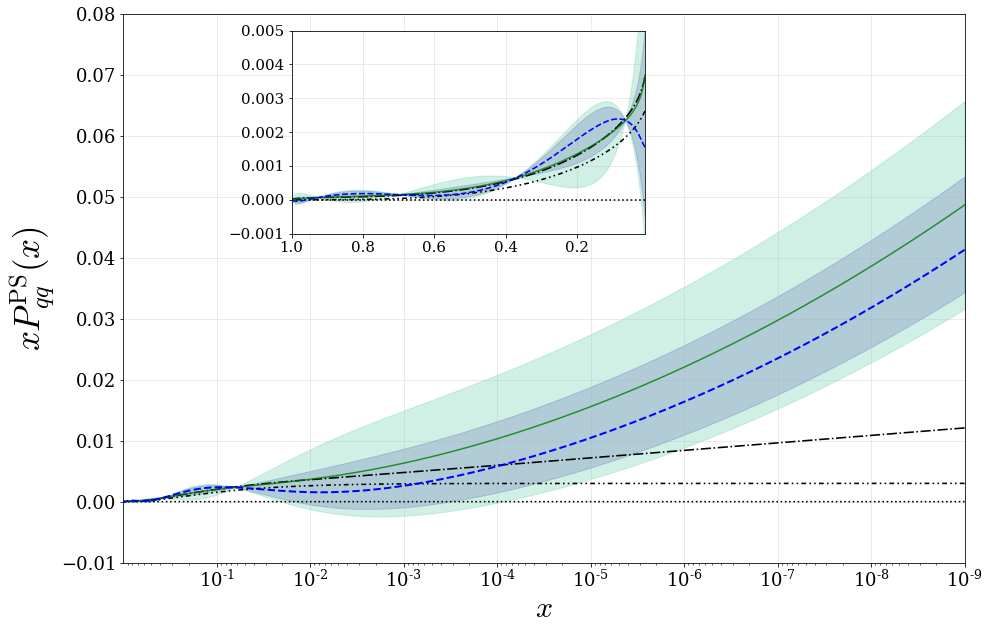}
\includegraphics[width=0.49\textwidth]{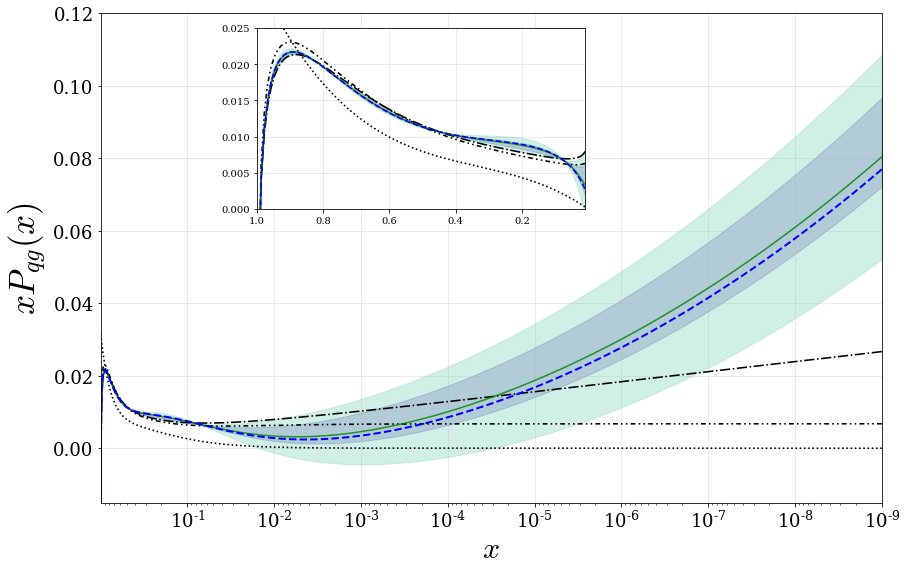}
\includegraphics[width=0.49\textwidth]{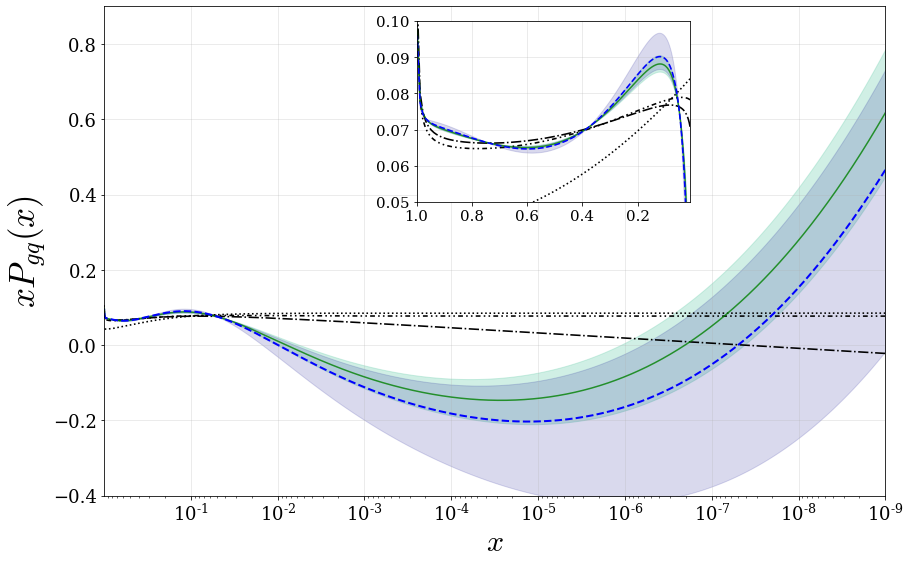}
\includegraphics[width=0.49\textwidth]{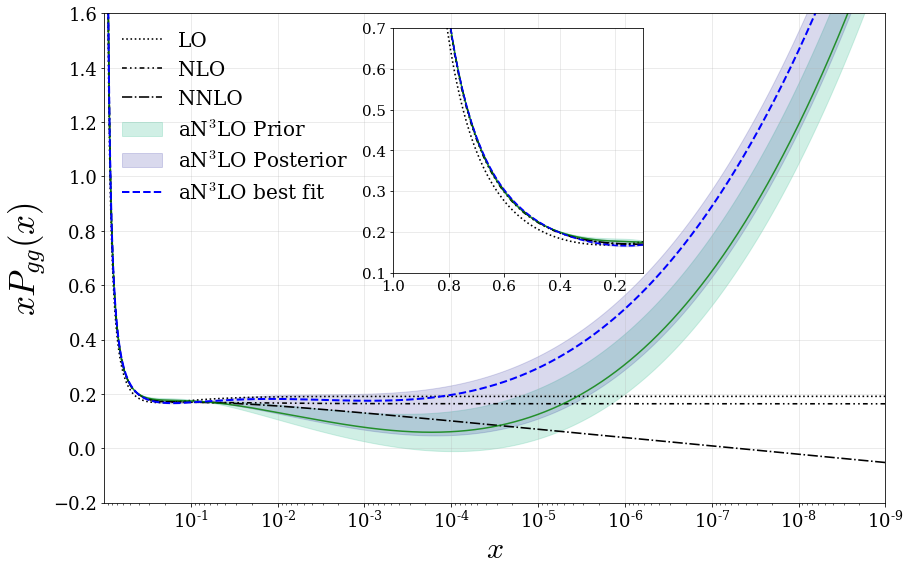}
\includegraphics[width=0.49\textwidth]{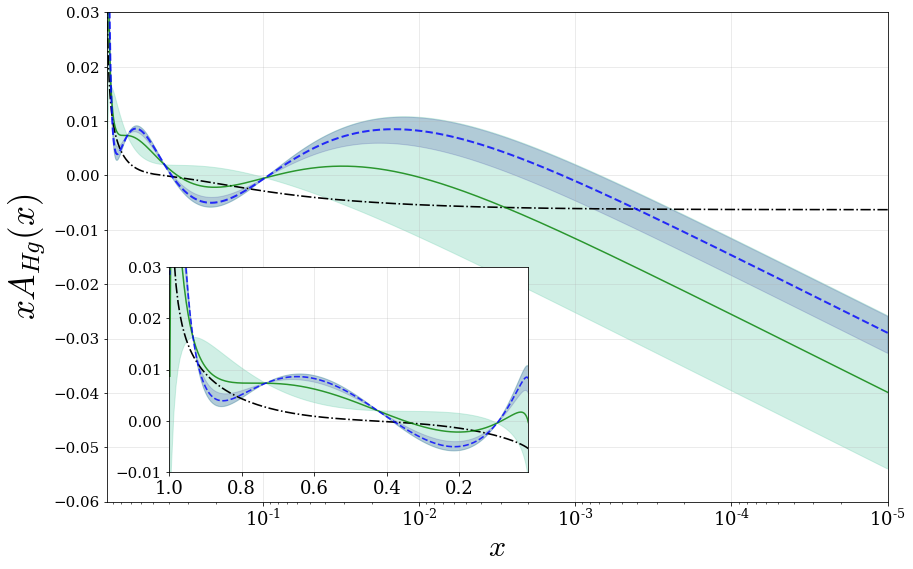}
\includegraphics[width=0.49\textwidth]{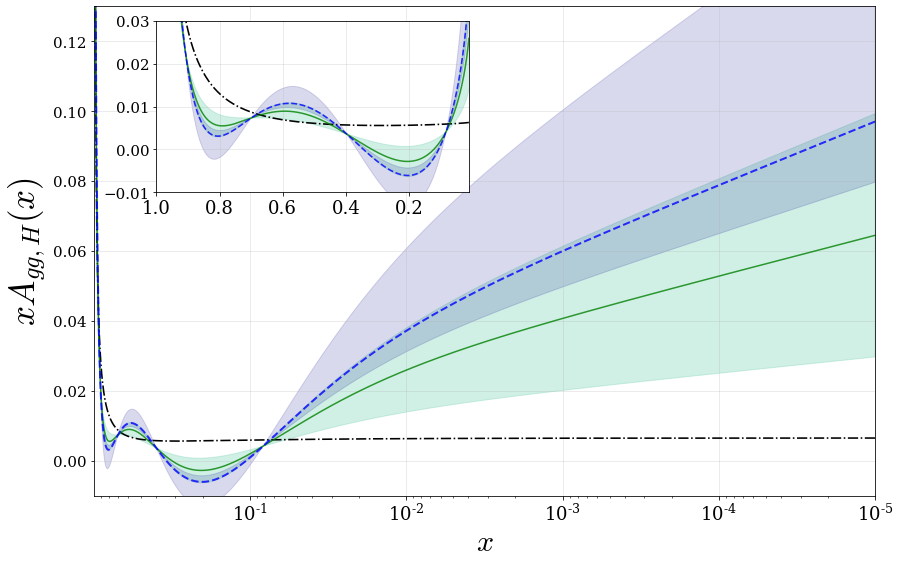}
\end{center}
\caption{\label{fig:an3loapp}Posterior variations of the aN$^{3}$LO splitting functions and transition matrix elements predicted from a full global fit (blue shaded band) compared to the prior variations in each case (green shaded band).}
\end{figure}
Fig.~\ref{fig:an3loapp} displays a comparison of the prior and posterior variations predicted for the perturbative expansions of the relevant splitting functions and transition matrix elements discussed in Section's~\ref{sec: n3lo_split} and \ref{sec: n3lo_OME}. We exclude the non-singlet quantities from this comparison as the variations predicted for these quantities are very similar to their priors (as can be seen in Table~\ref{tab:post_limits}) and have a small overall effect on the PDFs. It is true that once a fit is performed, the variation of the aN$^{3}$LO theoretical nuisance parameters becomes less sensitive to the prior variation, suggesting that the the initial uncertainty estimate was conservative. Nevertheless in Fig.~\ref{fig:an3loapp}, one can observe that all posterior variations overlap with their corresponding priors, in most cases quite considerably. We also note that the most drastic differences between prior and posterior variations are as expected relating to the gluon PDF.

\begin{figure}
\begin{center}
\includegraphics[width=0.49\textwidth]{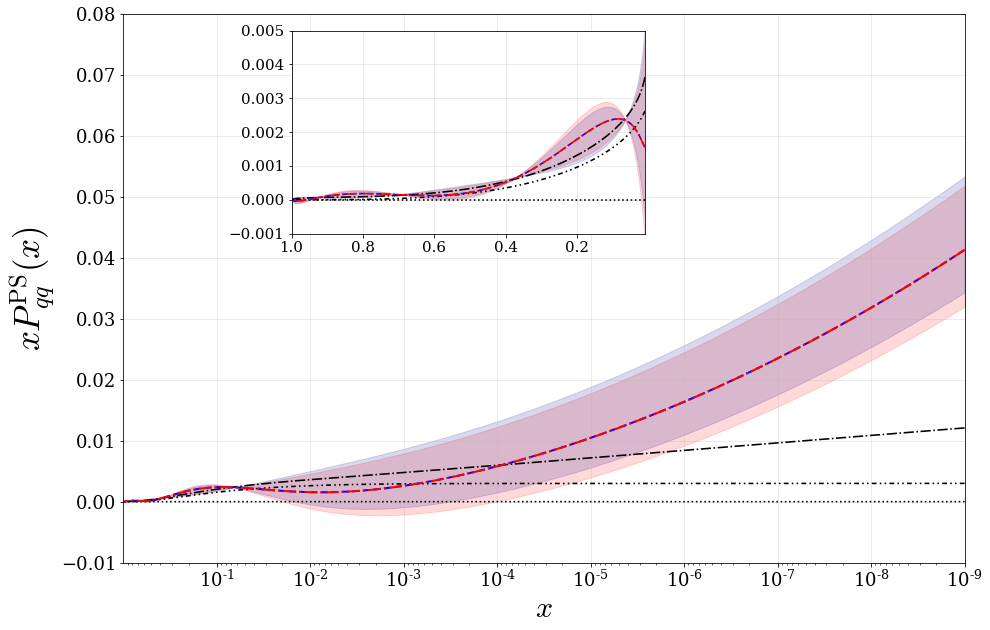}
\includegraphics[width=0.49\textwidth]{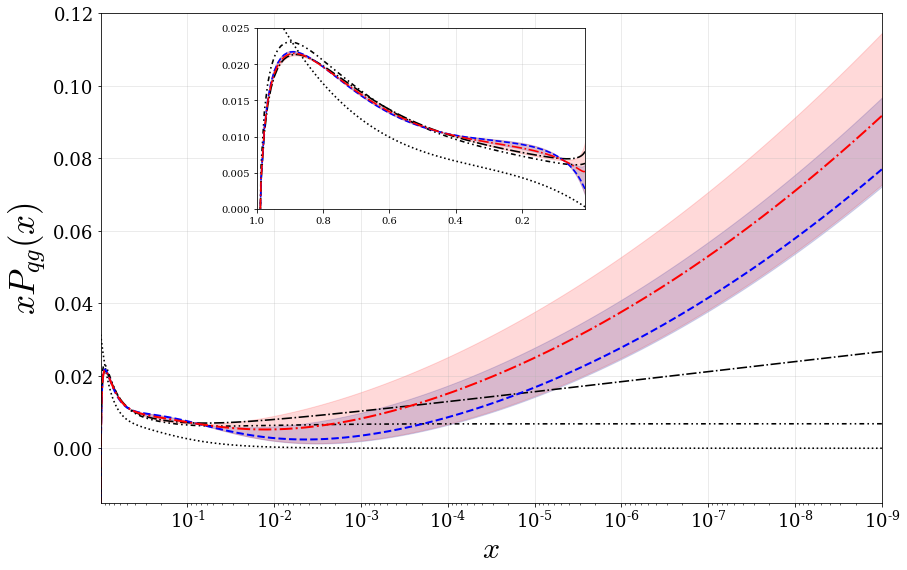}
\includegraphics[width=0.49\textwidth]{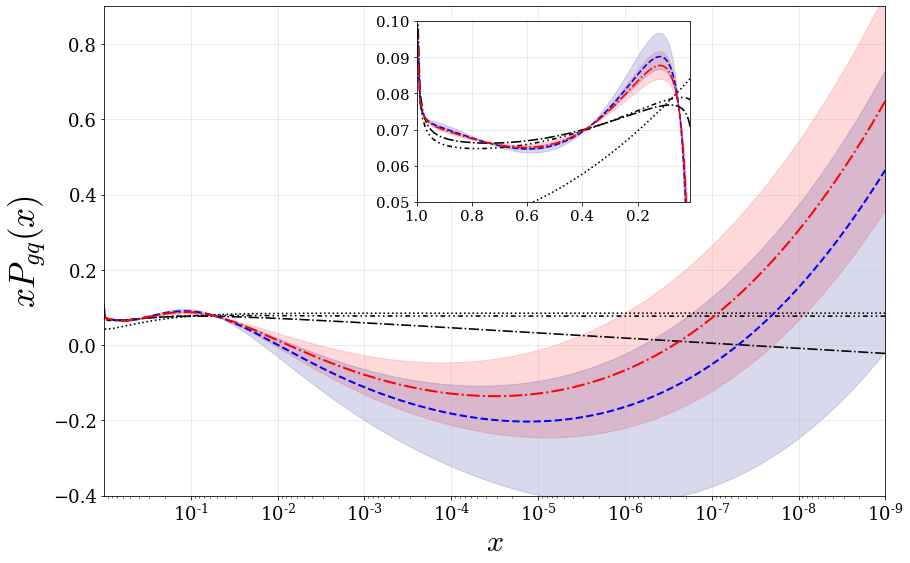}
\includegraphics[width=0.49\textwidth]{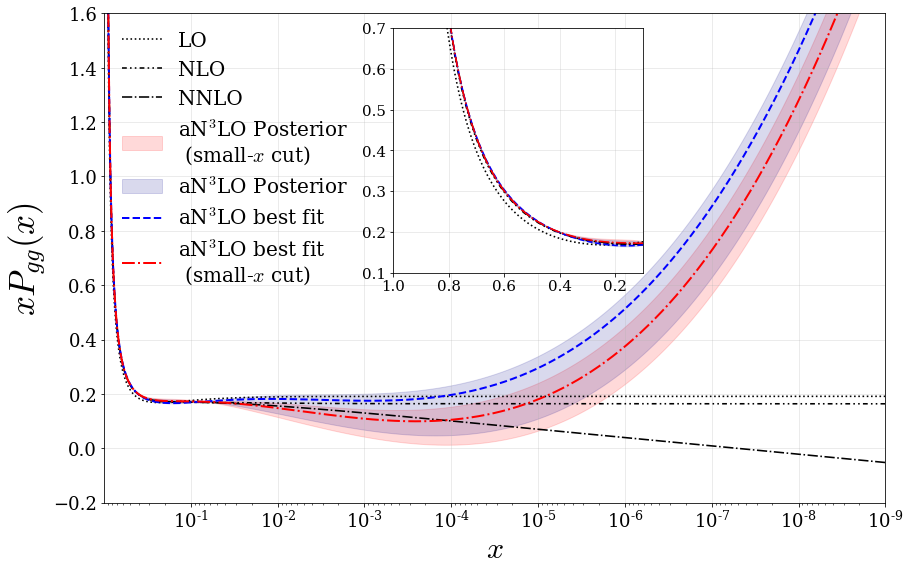}
\includegraphics[width=0.49\textwidth]{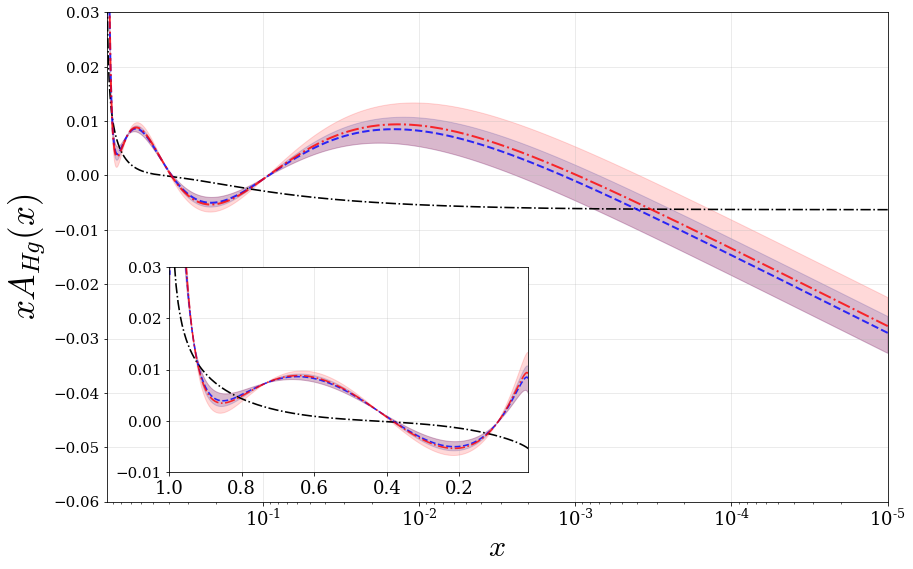}
\includegraphics[width=0.49\textwidth]{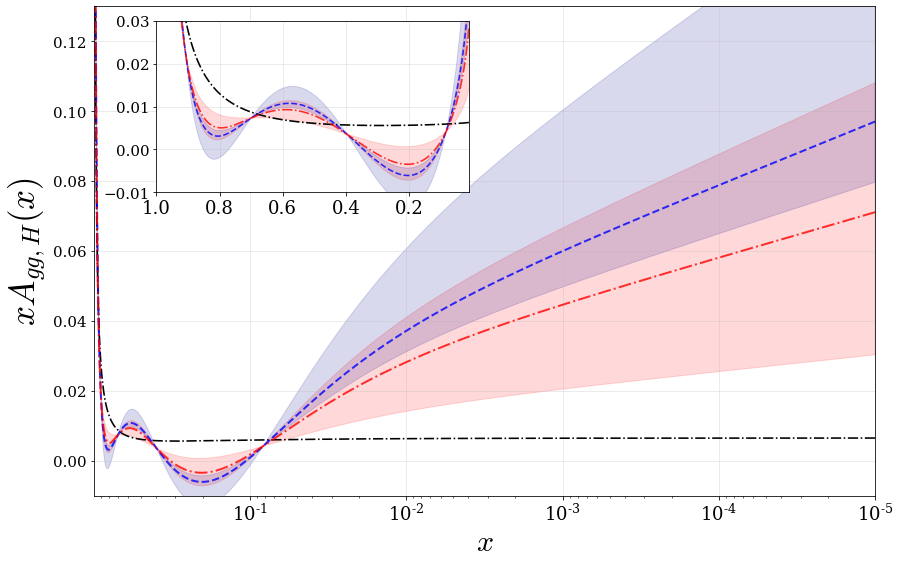}
\end{center}
\caption{\label{fig:an3loapp_post}Posterior variations of the aN$^{3}$LO splitting functions and transition matrix elements predicted from a full global fit (blue shaded band) compared to a fit with small-$x$ ($x < 10^{-3}$) data removed from a fit (red shaded band).}
\end{figure}
Fig.~\ref{fig:an3loapp_post} contains a comparison between the aN$^{3}$LO functions posterior variations with and without small-$x$ ($x < 10^{-3}$) data included in a global fit. These results accompany those presented in Table~\ref{tab:post_limits} and further display the reasonable agreement between the two fits, but there is some degree of tension occurring mainly in the cases of $P_{gg}$, $A_{Hg}$ and $A_{gg,H}$. In all cases the predicted variations overlap, with most central values being stable (i.e. contained well within the uncertainty predictions). However, for $P_{gg}$, $A_{Hg}$ and $A_{gg,H}$ the fit with the small $x$ cut does result in posterior functions which are more consistent with the prior functions, again suggesting that for these functions the posterior values are influenced, to a significant, but not overwhelming extent by terms beyond N$^3$LO, most likely those associated with small-$x$ resummation. Hence, as with the PDFs this provides evidence that the aN$^{3}$LO predictions are reasonably consistent across all values of $x$ but are influenced to a limited extent by the small-$x$ region. This supports our view that while we are explicitly determining the missing N$^3$LO corrections, which are indeed overall the dominant part of the missing higher order corrections, the fit is also probing some even higher order corrections, particularly at small $x$.

\subsection{\texorpdfstring{N$^{3}$LO}{N3LO} Contributions}\label{subsec: n3lo_contrib}

In this section all but one N$^{3}$LO contribution will be switched off, in particular only splitting functions, or only heavy or light flavour coefficient functions with their relevant transition matrix elements. In all cases the aN$^{3}$LO $K$-factors are left free to allow the fit some freedom in manipulating the cross sections of other datasets. In practice however, fixing these $K$-factors at the NNLO values has a minimal effect on the shape of the PDFs in all cases (as demonstrated in Fig.~\ref{fig: pdf_ratios_N3LO_qlow} and \ref{fig: pdf_ratios_N3LO_qhigh}).

\begin{figure}
\begin{center}
\includegraphics[width=0.49\textwidth]{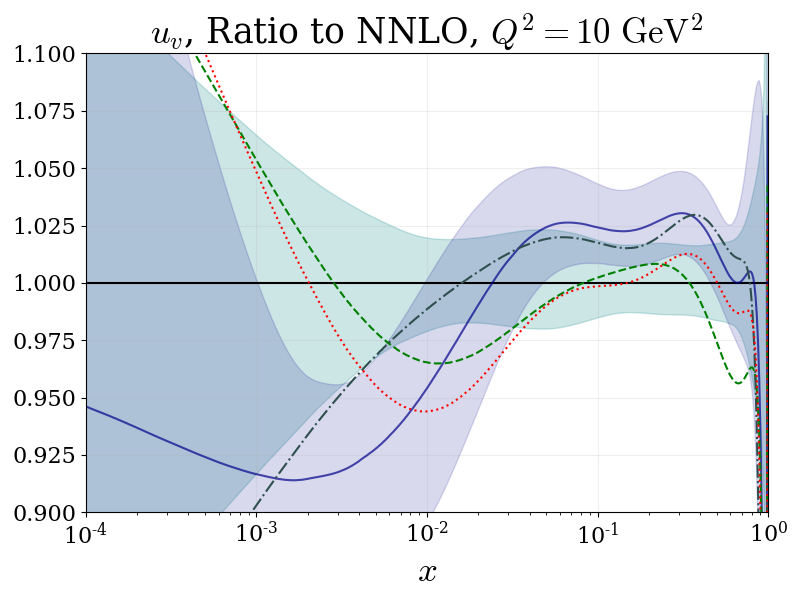}
\includegraphics[width=0.49\textwidth]{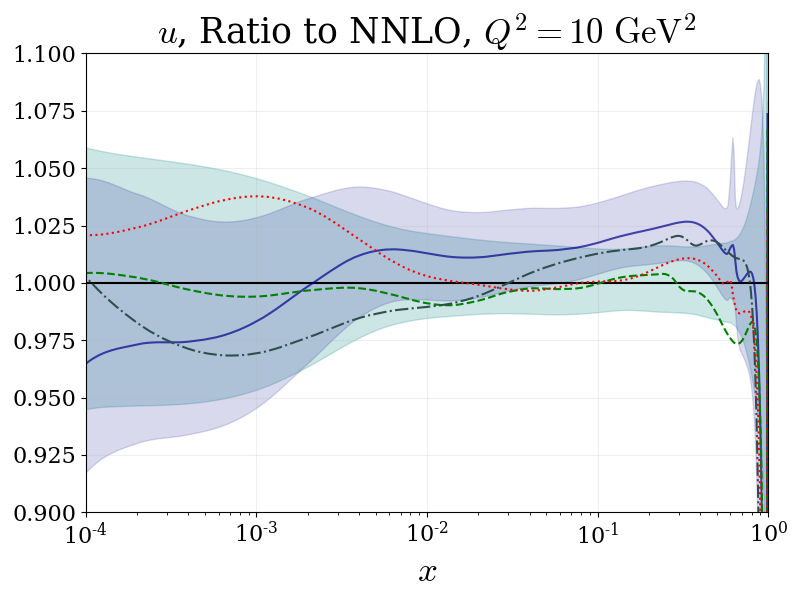}
\includegraphics[width=0.49\textwidth]{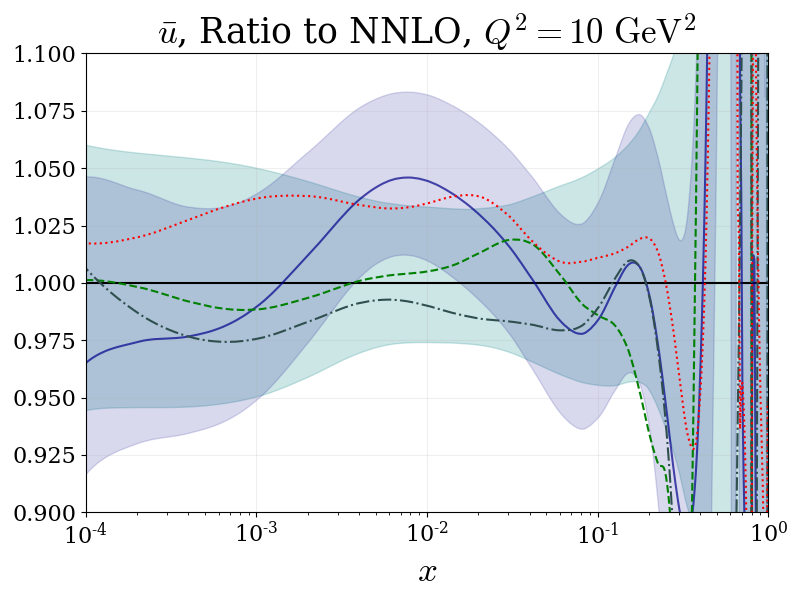}
\includegraphics[width=0.49\textwidth]{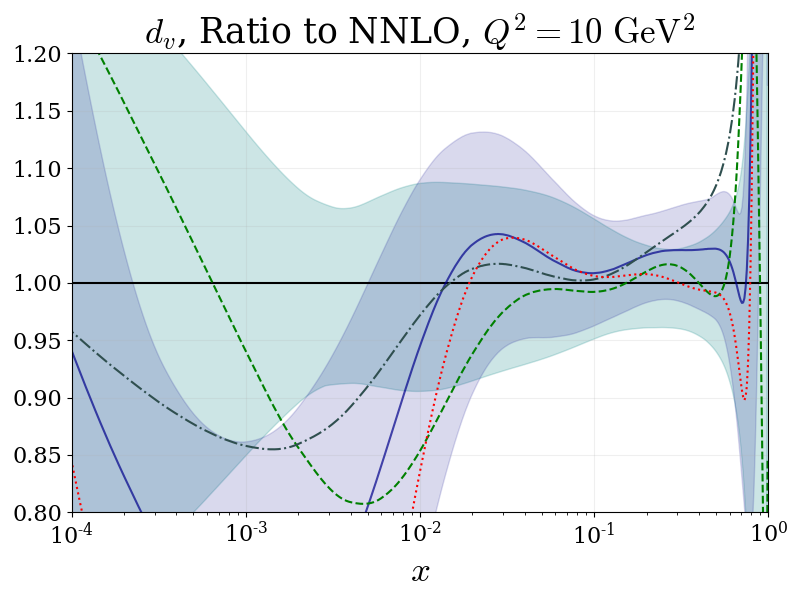}
\includegraphics[width=0.49\textwidth]{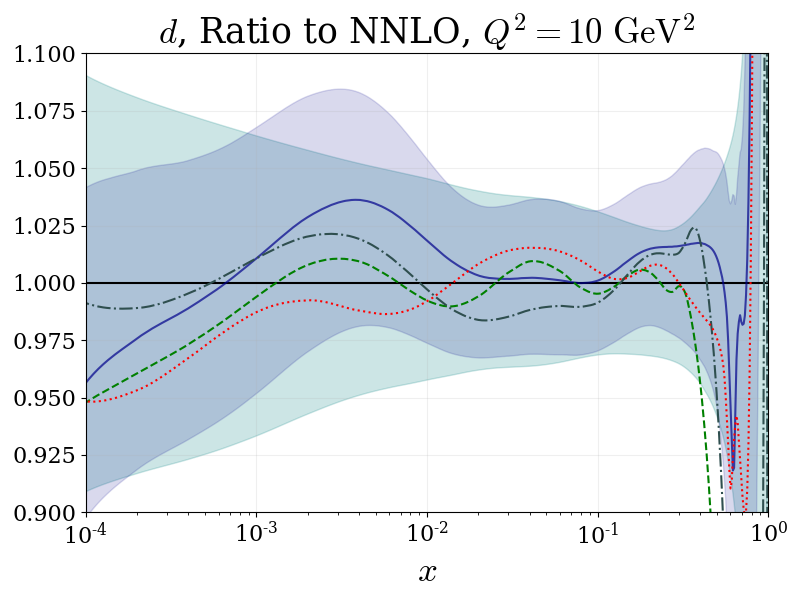}
\includegraphics[width=0.49\textwidth]{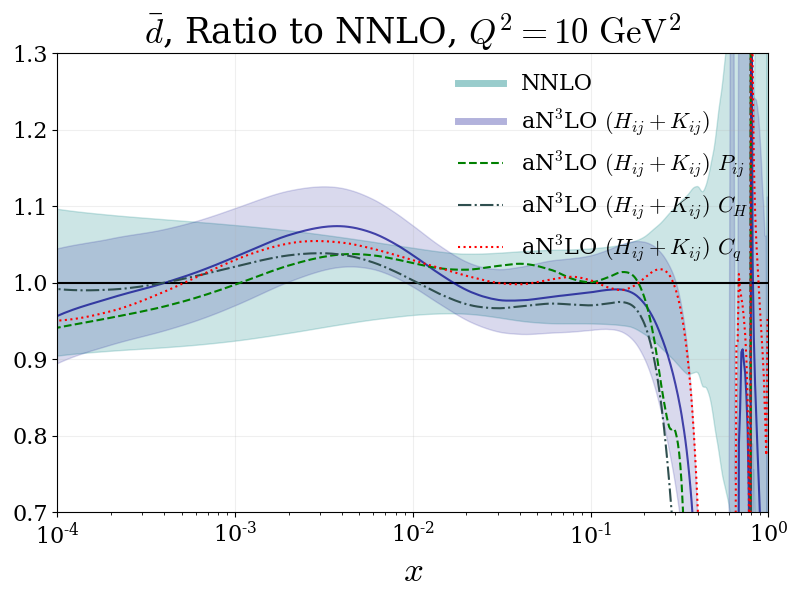}
\end{center}
\caption{\label{fig: pdf_contributions_ratios_qlow}Low-$Q^{2}$ PDF ratios showing aN$^{3}$LO (with decorrelated $K$-factors) 68\% confidence intervals compared to NNLO 68\% confidence intervals with varying theory contributions. All plots are shown for $Q^{2} = 10\ \mathrm{GeV}^{2}$ with the exception of the bottom quark shown for $Q^{2} = 25\ \mathrm{GeV}^{2}$. The PDFs included are: NNLO (green shaded), All N$^{3}$LO contributions (blue shaded), only splitting functions (green dashed), only heavy flavour coefficient functions and transition matrix elements (dark grey dash-dot) and only light flavour coefficient functions and transition matrix elements (red dotted).}
\end{figure}
\begin{figure}\ContinuedFloat
\begin{center}
\includegraphics[width=0.49\textwidth]{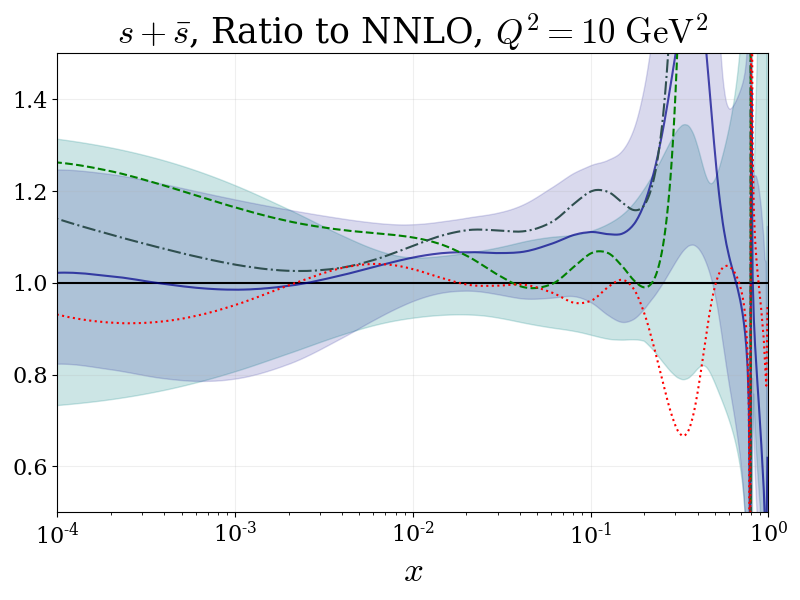}
\includegraphics[width=0.49\textwidth]{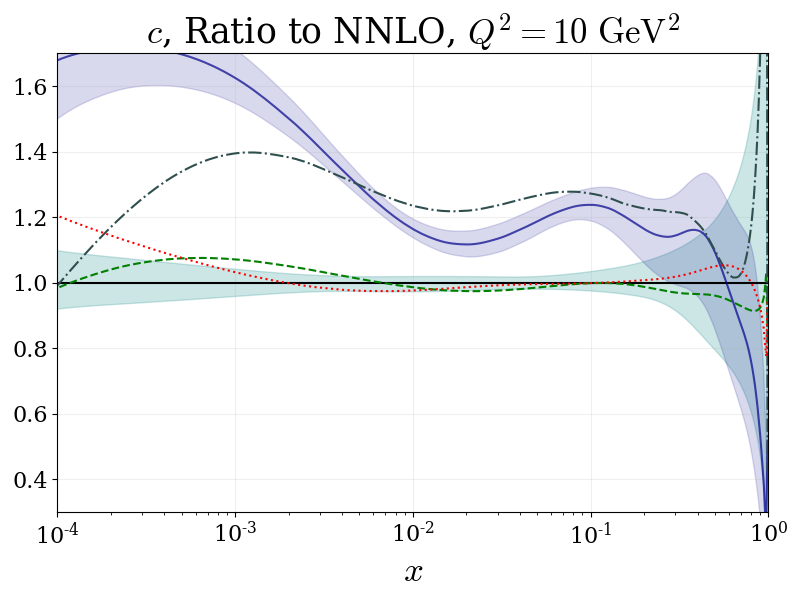}
\includegraphics[width=0.49\textwidth]{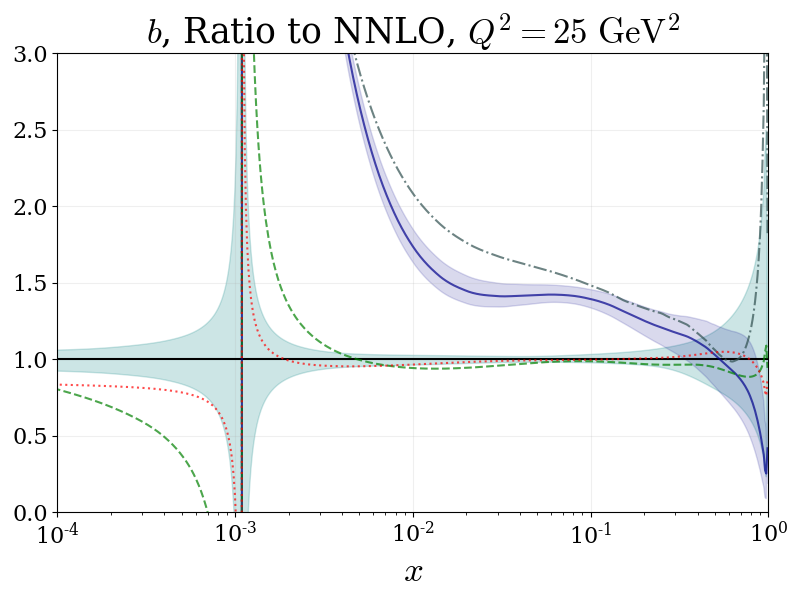}
\includegraphics[width=0.49\textwidth]{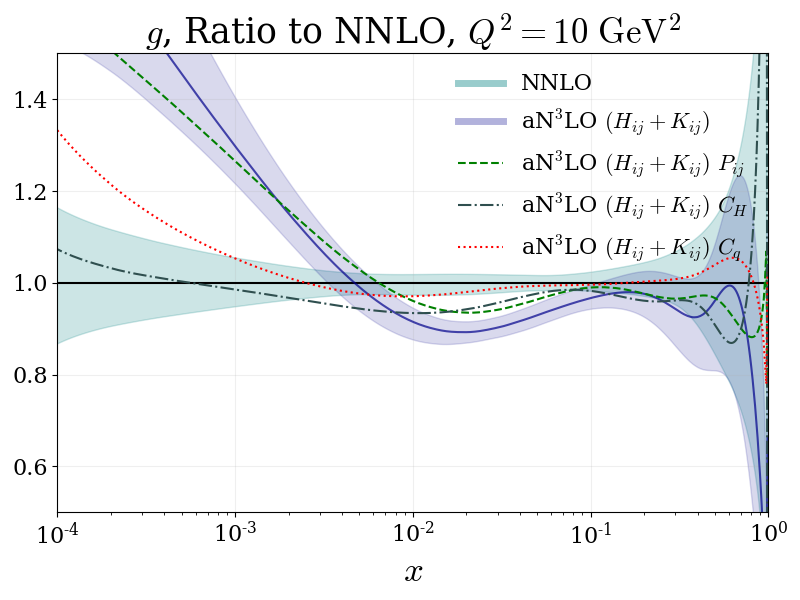}
\end{center}
\caption{\textit{(Continued)} Low-$Q^{2}$ PDF ratios showing aN$^{3}$LO (with decorrelated $K$-factors) 68\% confidence intervals compared to NNLO 68\% confidence intervals with varying theory contributions. All plots are shown for $Q^{2} = 10\ \mathrm{GeV}^{2}$ with the exception of the bottom quark shown for $Q^{2} = 25\ \mathrm{GeV}^{2}$. The PDFs included are: NNLO (green shaded), All N$^{3}$LO contributions (blue shaded), only splitting functions (green dashed), only heavy flavour coefficient functions and transition matrix elements (dark grey dash-dot) and only light flavour coefficient functions and transition matrix elements (red dotted).}
\end{figure}
The deconstructed aN$^{3}$LO PDFs as a ratio to the NNLO MSHT PDFs for various flavours at $Q^{2} = 10\ \mathrm{GeV}^{2}$ (with the bottom quark given at $Q^{2} = 25\ \mathrm{GeV}^{2}$) are shown in Fig.~\ref{fig: pdf_contributions_ratios_qlow}. Across the more tightly constrained light quark PDFs, all contributions lie very close to the aN$^{3}$LO $\pm 1\sigma$ uncertainty bands (blue shaded region and solid line). The additive and compensating nature of these contributions is also clear in a handful of the ratios from Fig.~\ref{fig: pdf_contributions_ratios_qlow}. In other areas the full description is biased towards a single contribution, for example the charm and bottom quarks follow the contribution from heavy flavours as one may expect. 
Conversely, to some extent the gluon follows the splitting functions much more closely as these contributions indirectly couple the gluon to the more constraining data\footnote{An exception to this can be seen around $x\sim 10^{-2}$ where the contributions act cumulatively. We make this point as this region of $x$ is of interest for Higgs calculations such as those discussed in Section~\ref{sec: predictions}}.

\subsection{\texorpdfstring{$\alpha_{s}$}{Strong Coupling Constant} Variation}

\begin{figure}[t]
\begin{center}
\includegraphics[width=0.7\textwidth]{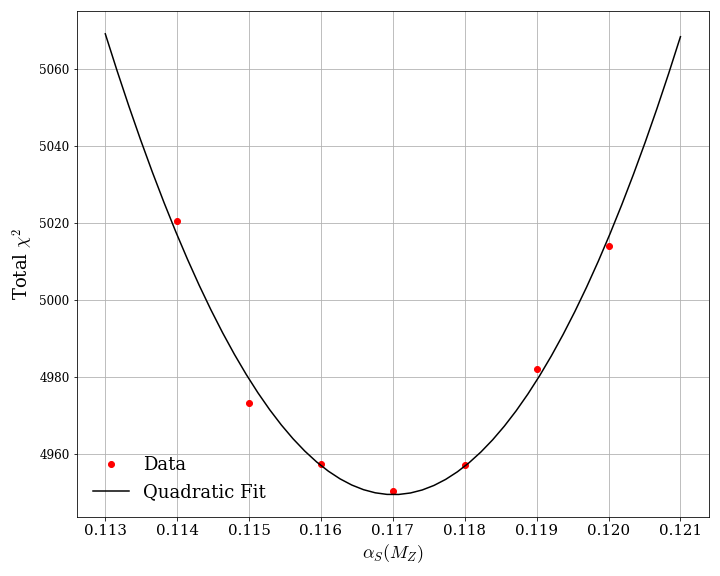}
\end{center}
\caption{\label{fig: alphaS_fit}Quadratic fit to the total $\chi^{2}$ results from various $\alpha_{s}(m_{Z})$ starting scales. The minimum of the quadratic fit provides a rough estimate of $\alpha_{s}(m_{Z}) = 0.1170$ at aN$^{3}$LO.}
\end{figure}
As in the standard MSHT20 NNLO PDF fit, we present the best fit aN$^{3}$LO PDFs with $\alpha_{s}(m_{Z}) = 0.118$, the common value chosen in the PDF4LHC combination~\cite{PDF4LHC22}. However, investigating the true minima in $\alpha_{s}(m_{Z})$, the $\chi^{2}$ profiles in Fig.~\ref{fig: alphaS_fit} prefer a value of around $\alpha_{s}(m_{Z}) = 0.1170$. This result follows the trend from lower orders whereby the best fit values are $\alpha_{s}(m_{Z}) = 0.1174 \pm 0.0013$ at NNLO and $\alpha_{s}(m_{Z}) = 0.1203 \pm 0.0015$ at NLO~\cite{Thorne:MSHT20_alphaS}. Following from NNLO, the aN$^{3}$LO $\alpha_{s}(m_{Z})$ prediction is also slightly lower than the NNLO world average central value at around $\alpha_{s}(m_{Z}) = 0.1179 \pm 0.0010$~\cite{PDG2019}. In any case, the preferred aN$^{3}$LO $\alpha_{s}(m_{Z})$ value stated here is in agreement with the MSHT20 NNLO result and the world average within uncertainties. A full analysis is left for a future publication.

\subsection{Charm Mass Dependence}

\begin{figure}[t]
\begin{center}
\includegraphics[width=0.7\textwidth]{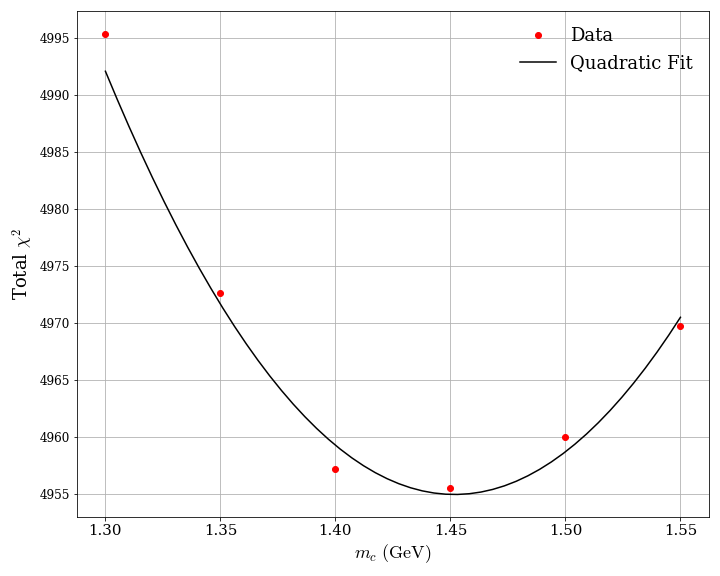}
\end{center}
\caption{\label{fig: chm_mass_fit}Quadratic fit to the total $\chi^{2}$ results from various charm masses ($m_{c}$). The minimum of the quadratic fit provides a rough estimate of $m_{c} = 1.45\ \mathrm{GeV}$ at aN$^{3}$LO.}
\end{figure}
\begin{figure}
\begin{center}
\includegraphics[width=0.49\textwidth]{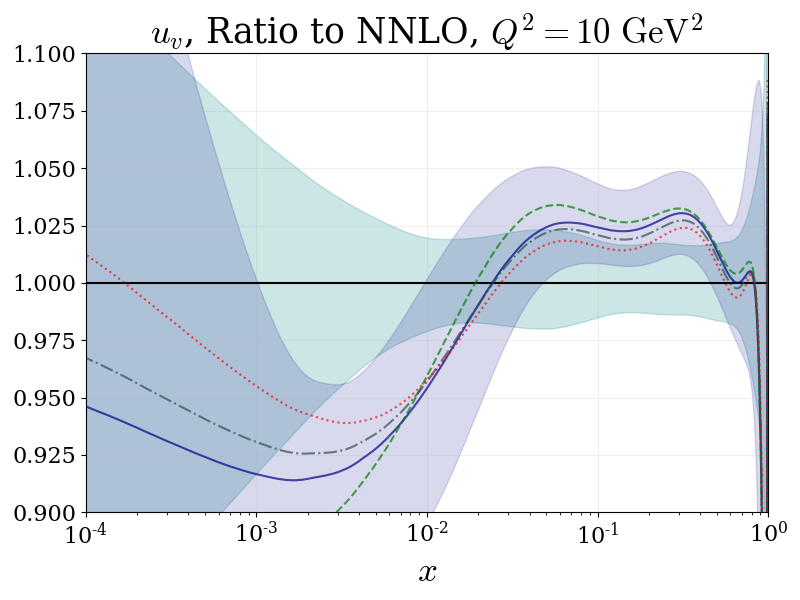}
\includegraphics[width=0.49\textwidth]{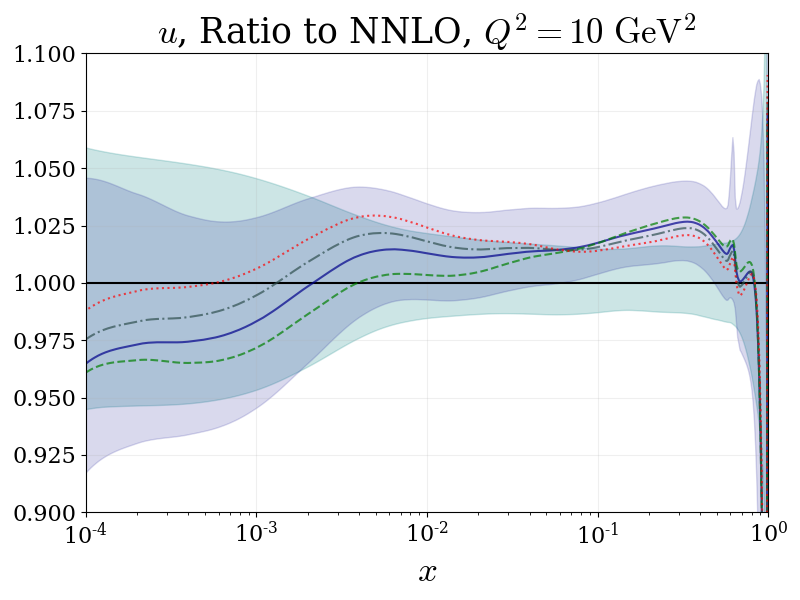}
\includegraphics[width=0.49\textwidth]{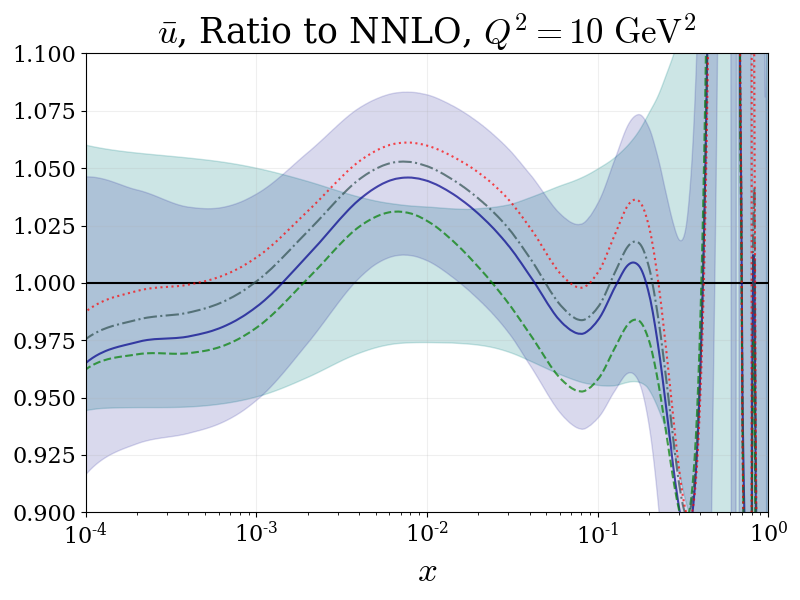}
\includegraphics[width=0.49\textwidth]{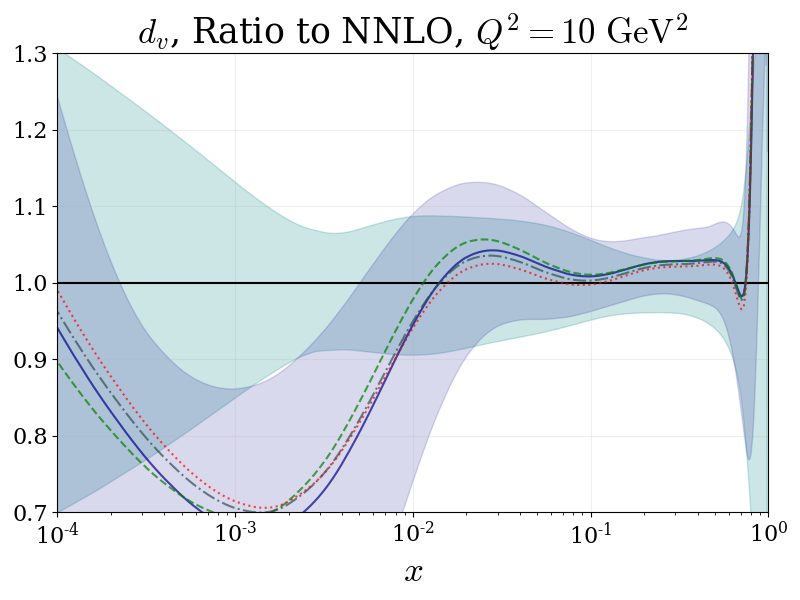}
\includegraphics[width=0.49\textwidth]{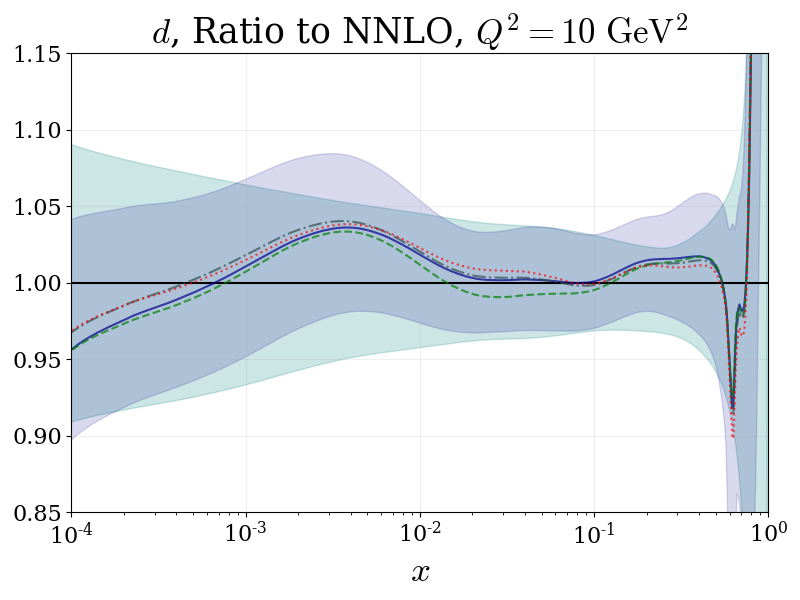}
\includegraphics[width=0.49\textwidth]{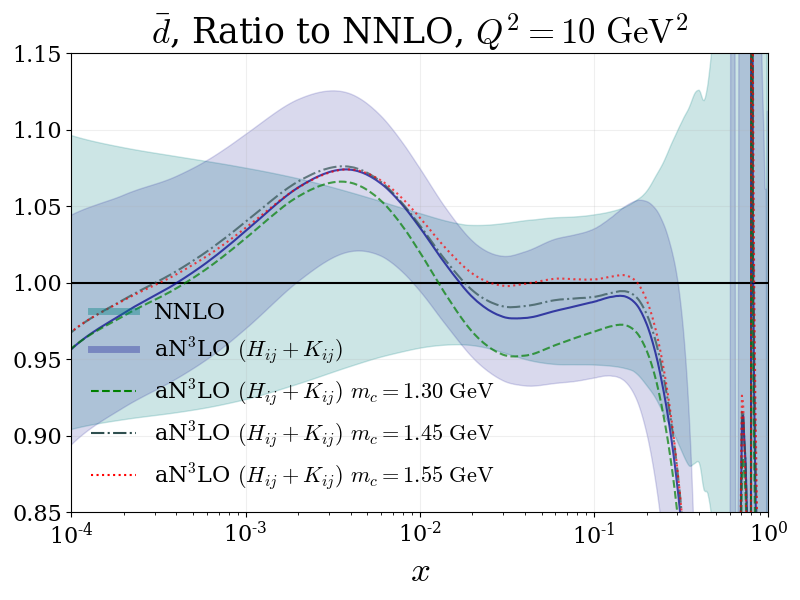}
\end{center}
\caption{\label{fig: pdf_charm_ratios_qlow}Low-$Q^{2}$ PDF ratios showing aN$^{3}$LO (with decorrelated $K$-factors) 68\% confidence intervals compared to NNLO 68\% confidence intervals with varying fixed values for the charm mass. All plots are shown for $Q^{2} = 10\ \mathrm{GeV}^{2}$ with the exception of the bottom quark shown for $Q^{2} = 25\ \mathrm{GeV}^{2}$. The PDFs included are: $m_{c} = 1.40\ \mathrm{GeV}$ (standard MSHT20 choice) (blue solid),  $m_{c} = 1.30\ \mathrm{GeV}$ (green dashed),  $m_{c} = 1.45\ \mathrm{GeV}$ (grey dotted dashed)  $m_{c} = 1.50\ \mathrm{GeV}$ (red dotted).}
\end{figure}
\begin{figure}[t]\ContinuedFloat
\begin{center}
\includegraphics[width=0.49\textwidth]{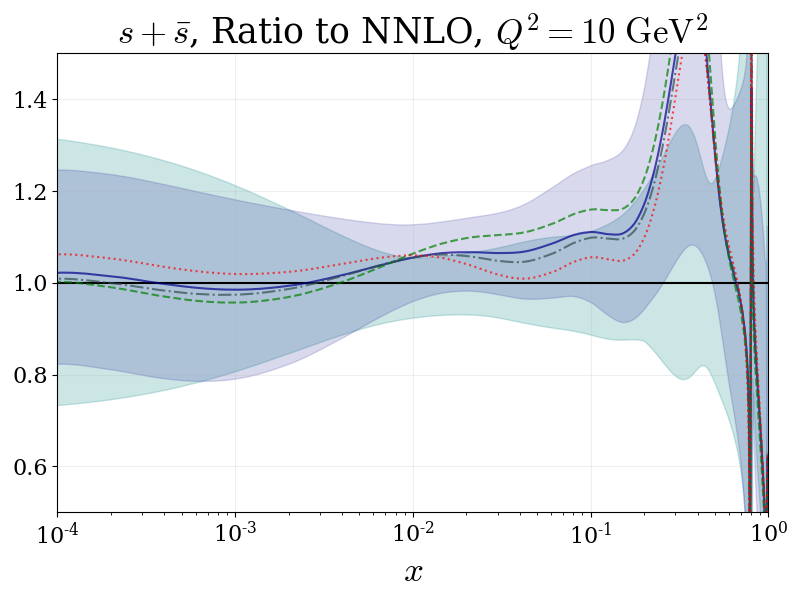}
\includegraphics[width=0.49\textwidth]{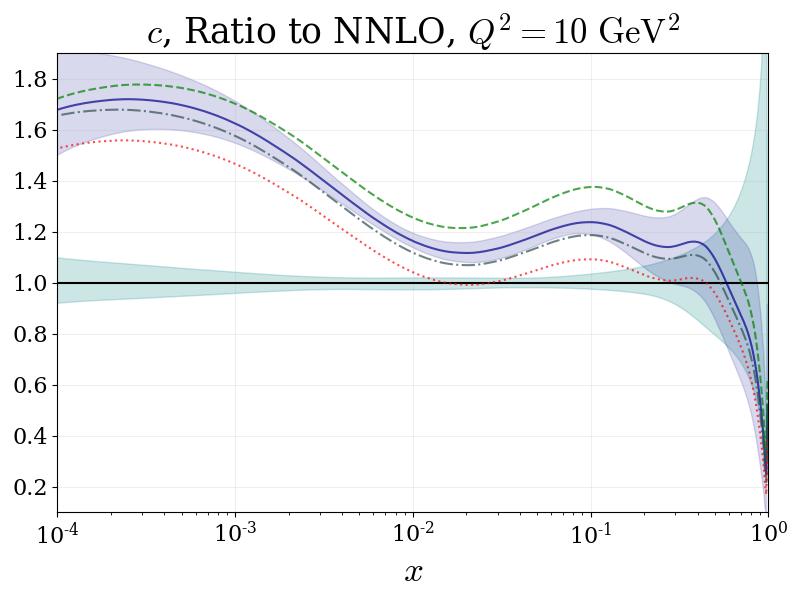}
\includegraphics[width=0.49\textwidth]{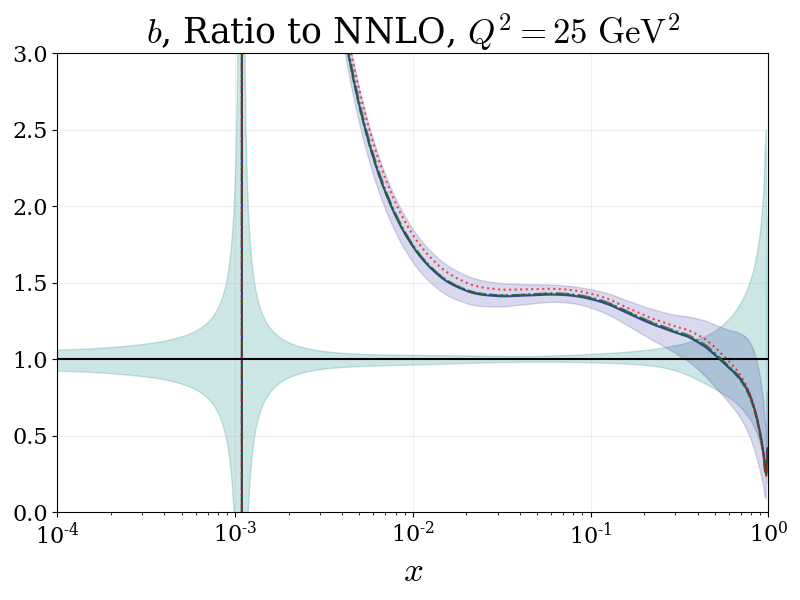}
\includegraphics[width=0.49\textwidth]{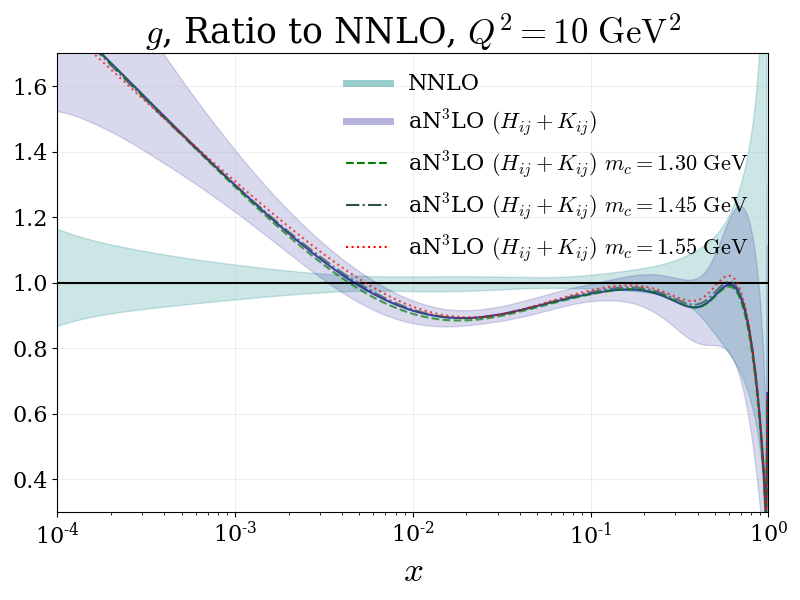}
\end{center}
\caption{\textit{(Continued)} Low-$Q^{2}$ PDF ratios showing aN$^{3}$LO (with decorrelated $K$-factors) 68\% confidence intervals compared to NNLO 68\% confidence intervals with varying fixed values for the charm mass. All plots are shown for $Q^{2} = 10\ \mathrm{GeV}^{2}$ with the exception of the bottom quark shown for $Q^{2} = 25\ \mathrm{GeV}^{2}$. The PDFs included are: $m_{c} = 1.40\ \mathrm{GeV}$ (standard MSHT20 choice) (blue solid),  $m_{c} = 1.30\ \mathrm{GeV}$ (green dashed),  $m_{c} = 1.45\ \mathrm{GeV}$ (grey dotted dashed)  $m_{c} = 1.50\ \mathrm{GeV}$ (red dotted).}
\end{figure}
In a standard MSHT fit~\cite{Thorne:MSHT20}, aN$^{3}$LO PDFs are produced with the charm pole mass $m_{c} = 1.40\ \mathrm{GeV}$. Fig.~\ref{fig: chm_mass_fit} displays the $\chi^{2}$ results when varying this charm mass. The predicted minimum at NNLO (for MSHT20 PDFs) is in the range $m_{c} = 1.35-1.40\ \mathrm{GeV}$~\cite{Thorne:MSHT20_alphaS}, whereas at aN$^{3}$LO we show a minimum in the region of $m_{c} = 1.42-1.47\ \mathrm{GeV}$. This aN$^{3}$LO result therefore shows a slightly better agreement with the world average~\cite{PDG2019}\footnote{There is some ambiguity in this value since the transformation from $\overline{MS}$ to the pole mass definition is not well-defined (see \cite{Thorne:MSHT20_alphaS} for more details).} of $m_{c} = 1.5 \pm 0.2\ \mathrm{GeV}$.

Considering Fig.~\ref{fig: pdf_charm_ratios_qlow}, one is then able to analyse the effect of this slightly higher charm mass on the form of the PDFs. As one can expect, the charm PDF is subject to the largest difference and is suppressed by a higher $m_{c}$. The extra suppression from a higher charm mass allows the fit to suppress the $c + \bar{c}$ sea contribution. This is then compensated by an increase in the $\bar{u}$ and $\bar{d}$ distributions which stabilises the overall sea contribution.
\section{\texorpdfstring{N$^{3}$LO}{N3LO} Predictions}\label{sec: predictions}

With the increasing number of hard cross section calculations at N$^{3}$LO, there is a growing demand for N$^{3}$LO accuracy in PDFs. In this section we investigate the effect of the MSHT approximate N$^{3}$LO PDFs on Higgs production via gluon fusion and vector boson fusion (VBF). The hard cross sections for these processes have been calculated to N$^{3}$LO accuracy~\cite{Ball:2013bra,Bonvini:2014jma,Bonvini:2016frm,Ahmed:2016otz,Bonvini:2018ixe,Bonvini:2018iwt,Bonvini:2013kba,anastasiou2014higgs,anastasiou2016high,Mistlberger:2018,Cacciari:2015,Dreyer:2016}. We present a full N$^{3}$LO computation for each prediction with our approximate N$^{3}$LO PDFs, including theoretical uncertainties. In future work, the intention will be to expand this analysis to include results for N$^{3}$LO DY~\cite{DY_N3LO_Kfac} and approximate N$^{3}$LO top production~\cite{Kidonakis:tt} cross sections.

Note that in this Section we follow the notation used previously and denote the aN$^{3}$LO results with decorrelated $K$-factors as $(H_{ij} + K_{ij})^{-1}$ and those with correlated $K$-factors with $H_{ij}^{\prime\ -1}$. In all cases, scale variations are found via the 9-point prescription~\cite{NNPDFscales} for results with NNLO PDFs. Whereas for aN$^{3}$LO PDFs, although the extra information introduced is at N$^{3}$LO, the data (and therefore all relevant theory nuisance parameters) which are included in the global fit are sensitive to all orders. 
In particular, we include theoretical uncertainties into our aN$^3$LO fit which incorporate MHO effects on the PDFs. Therefore we argue (and in these cases demonstrate) that the factorisation scale variation is contained within the PDF uncertainties. Due to this, it is only
the renormalisation scale which requires variation in predictions involving aN$^{3}$LO PDFs\footnote{This is to quantify the theoretical MHOU in the hard cross section, whereas the aN$^{3}$LO PDFs now come with an estimated MHOU.}.

\subsection{Higgs Production -- Gluon Fusion: \texorpdfstring{$gg \rightarrow H$}{gg to Higgs}}\label{subsec: ggH}

\begin{table}
\centerline{
\begin{tabular}{c|ccc}
\hline
$\sigma$ order & PDF order & $\sigma + \Delta \sigma_{+} - \Delta \sigma_{-}$ (pb) & $\sigma$ (pb) $+\ \Delta \sigma_{+} - \Delta \sigma_{-}$ (\%) \\
\hline
\multicolumn{4}{c}{PDF uncertainties} \\
\hline
\multirow{4}{*}{N$^{3}$LO} 
 & aN$^{3}$LO (no theory unc.) & $45.296 + 0.723 - 0.545$ & $45.296 + 1.60\% - 1.22\%$ \\
 & aN$^{3}$LO ($H_{ij} + K_{ij}$) & $45.296 + 0.832 - 0.755$ & $45.296 + 1.84\% - 1.67\%$ \\
 & aN$^{3}$LO ($H_{ij}^{\prime}$) & $45.296 + 0.821 - 0.761$ & $45.296 + 1.81\% - 1.68\%$ \\
 & NNLO & $47.817 + 0.558 - 0.581$  & $47.817 + 1.17\% - 1.22\%$ \\
\hline
NNLO & NNLO & $46.206 + 0.541 - 0.564$ & $46.206 + 1.17\% - 1.22\%$ \\
\hline
\multicolumn{4}{c}{PDF + Scale uncertainties} \\
\hline
\multirow{4}{*}{N$^{3}$LO}  & aN$^{3}$LO (no theory unc.) & $45.296 + 0.723 - 1.851$ & $45.296 + 1.60\% - 4.09\%$ \\
 & aN$^{3}$LO ($H_{ij} + K_{ij}$) & $45.296 + 0.832 - 1.923$ & $45.296 + 1.84\% - 4.25\%$ \\
 & aN$^{3}$LO ($H_{ij}^{\prime}$) & $45.296 + 0.821 - 1.926$ & $45.296 + 1.81\% - 4.25\%$ \\
 & NNLO & $47.817 + 0.577 - 2.210$  & $47.817 + 1.21\% - 4.62\%$ \\
\hline
NNLO & NNLO & $46.206 + 4.284 - 5.414$  & $46.206 + 9.27\% - 11.72\%$ \\
\end{tabular}
}
\caption{\label{tab: ggH_results_mh2}Higgs production cross section results via gluon fusion (with $\sqrt{s} = 13\ \text{TeV}$) using N$^{3}$LO and NNLO hard cross sections combined with NNLO and aN$^{3}$LO PDFs. All PDFs are at the standard choice $\alpha_{s}(m_{Z}) = 0.118$. These results are found with $\mu = m_{H}/2$ unless stated otherwise, with the values for $\mu = m_{H}$ supplied in Table~\ref{tab: ggH_results_mh}.}
\end{table}
\begin{figure}
\begin{center}
\includegraphics[width=0.49\textwidth]{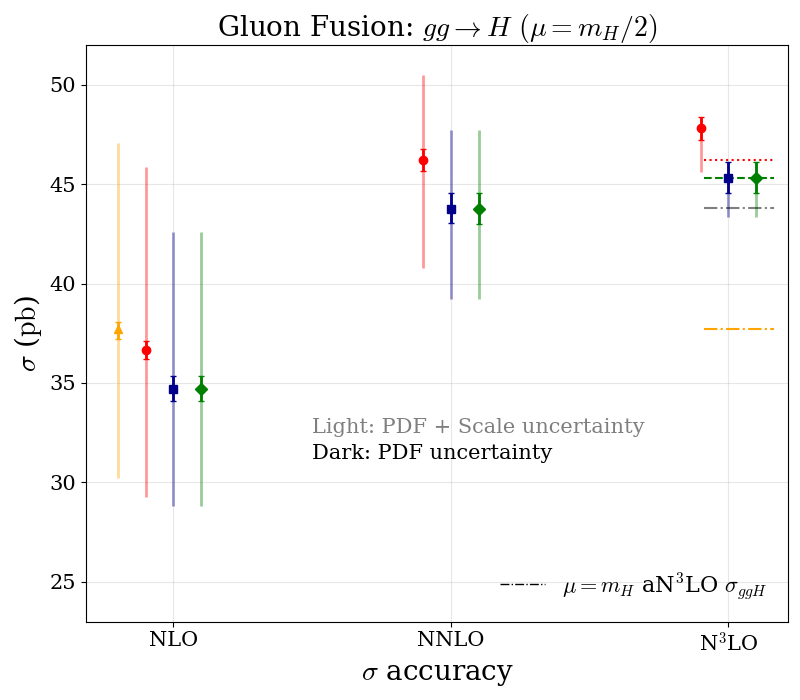}
\includegraphics[width=0.49\textwidth]{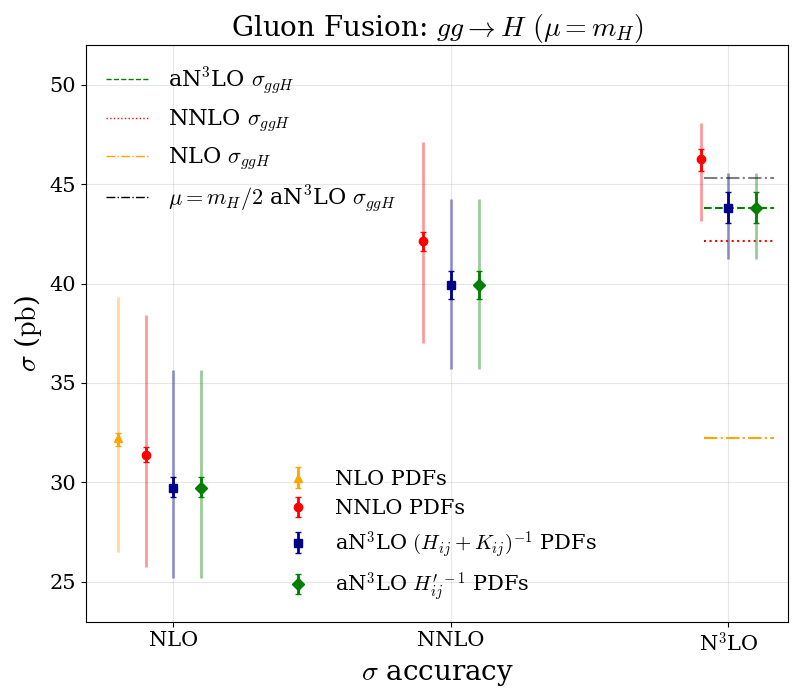}
\end{center}
\caption{\label{fig: ggH_predictions}Higgs production cross section results via gluon fusion (with $\sqrt{s} = 13\ \text{TeV}$) at two central scales: $\mu = m_{H}/2$ (left) and $\mu = m_{H}$ (right). Displayed are the results for aN$^{3}$LO PDFs with decorrelated $K$-factors ($(H_{ij} + K_{ij})^{-1}$), correlated $K$-factors ($H_{ij}^{\prime\ -1} = (H_{ij} + K_{ij})^{-1}$) each with a scale variation band from varying $\mu_{r}$ by a factor of 2. In the NNLO and NLO PDF cases, both scales $\mu_{f}$ and $\mu_{r}$ are varied by a factor of 2 following the 9-point convention~\cite{NNPDFscales}.}
\end{figure}
Table~\ref{tab: ggH_results_mh2} and Fig.~\ref{fig: ggH_predictions} (left) show predictions at a central scale of $\mu = \mu_{f}=\mu_{r}=m_{H}/2$ for the Higgs production cross section via gluon fusion\footnote{Results are obtained with the code \href{https://www.ge.infn.it/~bonvini/higgs/}{\texttt{ggHiggs}}\cite{Ball:2013bra,Bonvini:2014jma,Bonvini:2016frm,Ahmed:2016otz,Bonvini:2018ixe,Bonvini:2018iwt,Bonvini:2013kba,anastasiou2014higgs,anastasiou2016high,Mistlberger:2018,ggHiggs}.} at the LHC for $\sqrt{s} = 13\ \text{TeV}$, where $m_{H} = 125\ \text{GeV}$ is the Higgs mass and no fiducial cuts are applied. Fig.~\ref{fig: ggH_predictions} (right) displays the same analysis for the gluon fusion cross section with $\mu = \mu_{f}=\mu_{r}=m_{H}$ (numerical results provided in Table~\ref{tab: ggH_results_mh}).

Considering the $\mu = m_{H}/2$ and $\mu = m_{H}$ central value results displayed in Table~\ref{tab: ggH_results_mh2} and Fig.~\ref{fig: ggH_predictions}, it can be observed that aN$^{3}$LO PDFs predict a lower central value than NNLO PDFs across all hard cross section orders. One can also notice an overlap in all cases between predictions from NNLO and aN$^{3}$LO PDFs. However for $\mu = m_{H}/2$, whilst the error bands for predictions with N$^{3}$LO hard cross section and NNLO and N$^{3}$LO PDFs overlap, their central values are outside each other's respective error bands. Since estimating MHOUs via scale variations is a somewhat ambiguous procedure (and is therefore estimated conservatively to reflect this), these results highlight the benefit of being able to exploit a higher level of control over MHOUs i.e. via nuisance parameters. By predicting a different central value we include a more accurate estimation for higher order predictions which may not be contained within scale variations, especially at unmatched orders in perturbation theory.

Examining the predicted central values further, Fig.~\ref{fig: ggH_predictions} suggests that the increase in the cross section theory at N$^{3}$LO is compensated by the PDF theory at N$^{3}$LO, suggesting a cancellation between terms in the PDF and cross section theory at N$^{3}$LO. This point is important to consider when combining unmatched orders in physical calculations, since we must be open to the possibility that unmatched cancellations in physical calculations can lead to inaccurate predictions, as our results suggest here.

Further to this, the change in the gluon PDF is largely driven by the predicted form of $P_{qg}$ at aN$^{3}$LO and DIS data. Therefore the relevant changes in the gluon at aN$^{3}$LO are most likely due to indirect effects i.e. not directly related to gluon fusion predictions. Due to this, there is no reason to believe that the observed level of convergence should happen at aN$^{3}$LO for both choices of $\mu$.
However, owing to the inclusion of known information at higher orders, one can be confident that the prediction is more accurate than NNLO, whichever way it moves.

Comparing PDF uncertainty values calculated using NNLO and aN$^{3}$LO PDFs, another prominent feature one can notice in Table~\ref{tab: ggH_results_mh2} is an increase in PDF uncertainties. We find that the PDF uncertainty without N$^{3}$LO theory uncertainties included (i.e. using only the eigenvector description from the first 32 eigenvectors and with N$^{3}$LO parameters fixed at the best fit) also includes a marginal increase in the positive direction compared to NNLO.
Mathematically, the reason for this comes back to the fact that the best fit is inherently different from the NNLO theory, residing in a completely novel $\chi^{2}$ landscape. In turn, this means it is not guaranteed that the PDF uncertainty will remain consistent across the distinct PDF sets\footnote{As we can see from Section~\ref{subsec: pdf_results}, the theory uncertainty is also not guaranteed to add to the total uncertainty (and in fact acts to reduce the uncertainty in some areas of $(x, Q^{2})$).}. In the case of gluon fusion, the leading contribution to the positive uncertainty direction is an eigenvector primarily dominated by PDF parameters, while in the negative direction a N$^{3}$LO splitting function parameter dominates (eigenvector 9 and 31 in the $(H_{ij} + K_{ij})^{-1}$ N$^{3}$LO case -- see Table~\ref{tab: eigenanalysis}). As discussed in Section~\ref{subsec: n3lo_contrib}, the gluon predominantly follows the splitting function contributions, therefore it is not surprising that this eigenvector is having a noticeable effect.
Phenomenologically, the increase in predicted uncertainties from the inclusion of the theoretical uncertainties is a reflection of the estimated PDF MHOUs in this particular cross section, and acts to replace factorisation scale variation. As a consistency check, we find that when performing a 9-point scale variation procedure with aN$^{3}$LO PDFs, the values calculated (for both choices of $\mu$) are within the predicted PDF uncertainties. This is therefore a further verification of our MHOUs and that the $\mu_{f}$ variation is intrinsic in the PDF uncertainties.

Finally Fig.~\ref{fig: ggH_predictions} also demonstrates the increased stability of predictions when considering the two different central scales $\mu$ at N$^{3}$LO. As predicted from perturbation theory, the scale dependence is reduced and central values become more in agreement when increasing the order of either the PDFs or hard cross section. Furthermore, the aN$^{3}$LO $\sigma$ central predictions for both choices of $\mu$ are contained within the uncertainty bands of each other. This is true by definition for the NNLO PDFs since the factorisation scale $\mu_{f}$ variation includes both choices of $\mu$, whereas for aN$^{3}$LO PDFs this result is not guaranteed and is therefore intrinsic in the PDF (and renormalisation scale $\mu_{r}$ variation) uncertainty.

\subsection{Higgs Production -- Vector Boson Fusion: \texorpdfstring{$qq \rightarrow H$}{qq to Higgs}}

\begin{figure}
\begin{center}
\includegraphics[width=0.49\textwidth]{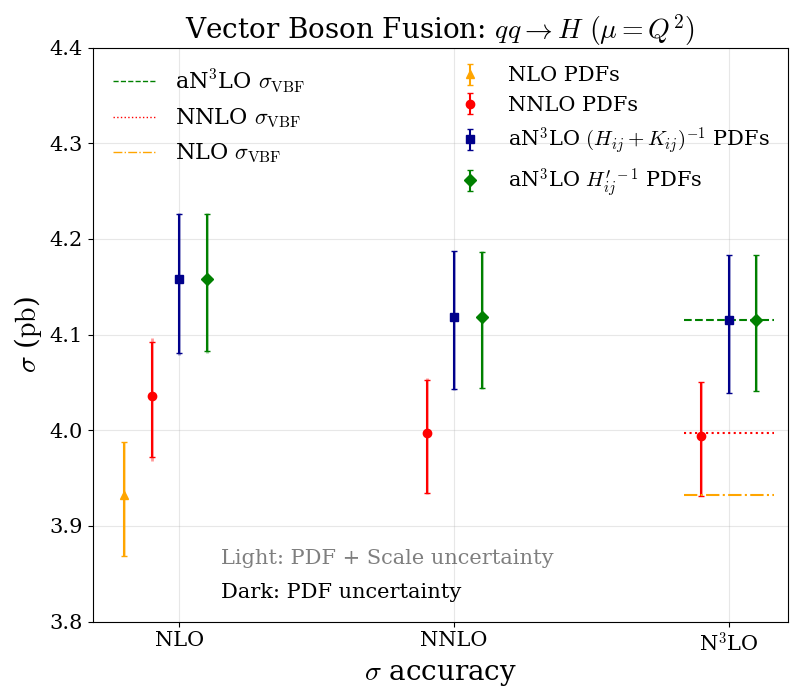}
\end{center}
\caption{\label{fig: vbf_predictions}Higgs production cross section results via vector boson fusion (with $\sqrt{s} = 13\ \text{TeV}$) at a central scale set to the vector boson momentum. Displayed are the results for aN$^{3}$LO PDFs with decorrelated $K$-factors ($(H_{ij} + K_{ij})^{-1}$), correlated $K$-factors ($H_{ij}^{\prime\ -1} = (H_{ij} + K_{ij})^{-1}$) each with a scale variation band from varying $\mu_{r}$ by a factor of 2. In the NNLO and NLO PDF cases, both scales $\mu_{f}$ and $\mu_{r}$ are varied by a factor of 2 following the 9-point convention~\cite{NNPDFscales}.}
\end{figure}
\begin{table}
\centerline{
\begin{tabular}{c|ccc}
\hline
$\sigma$ order & PDF order & $\sigma + \Delta \sigma_{+} - \Delta \sigma_{-}$ (pb) & $\sigma$ (pb) $+\ \Delta \sigma_{+} - \Delta \sigma_{-}$ (\%) \\
\hline
\multicolumn{4}{c}{PDF uncertainties} \\
\hline
\multirow{4}{*}{N$^{3}$LO} & aN$^{3}$LO (no theory unc.) & $4.1150 + 0.0638 - 0.0724$ & $4.1150 + 1.55\% - 1.76\%$ \\
& aN$^{3}$LO ($H_{ij} + K_{ij}$) & $4.1150 + 0.0682 - 0.0755$ & $4.1150 + 1.66\% - 1.83\%$ \\
& aN$^{3}$LO ($H_{ij}^{\prime}$) & $4.1150 + 0.0678 - 0.0742$ & $4.1150 + 1.65\% - 1.80\%$ \\
& NNLO & $3.9941 + 0.0558 - 0.0631$ & $3.9941 + 1.40\% - 1.58\%$ \\
\hline
NNLO & NNLO & $3.9974 + 0.0557 - 0.0633$ & $3.9974 + 1.39\% - 1.58\%$ \\
\hline
\multicolumn{4}{c}{PDF + Scale uncertainties} \\
\hline
\multirow{4}{*}{N$^{3}$LO} & aN$^{3}$LO (no theory unc.) & $4.1150 + 0.0638 - 0.0724$ & $4.1150 + 1.55\% - 1.76\%$ \\
& aN$^{3}$LO ($H_{ij} + K_{ij}$) & $4.1150 + 0.0683 - 0.0755$ & $4.1150 + 1.66\% - 1.83\%$ \\
& aN$^{3}$LO ($H_{ij}^{\prime}$) & $4.1150 + 0.0678 - 0.0742$ & $4.1150 + 1.65\% - 1.80\%$ \\
& NNLO & $3.9941 + 0.0560 - 0.0631$ & $3.9941 + 1.40\% - 1.58\%$ \\
\hline
NNLO & NNLO & $3.9974 + 0.0576 - 0.0642$ & $3.9974 + 1.44\% - 1.61\%$ \\
\end{tabular}
}
\caption{\label{tab: vbf_results}Higgs production cross section results via the vector boson fusion process (with $\sqrt{s} = 13\ \text{TeV}$) using N$^{3}$LO and NNLO hard cross sections combined with NNLO and aN$^{3}$LO PDFs. All PDFs are at the standard choice $\alpha_{s}(m_{Z}) = 0.118$. These results are found with $\mu = Q^{2}$ where $Q^{2}$ is the vector boson momentum.}
\end{table}
\begin{table}
\centerline{
\begin{tabular}{c|ccc}
\hline
$\sigma$ order & PDF order & $\sigma + \Delta \sigma_{+} - \Delta \sigma_{-}$ (pb) & $\sigma$ (pb) $+\ \Delta \sigma_{+} - \Delta \sigma_{-}$ (\%) \\
\hline
\multirow{3}{*}{N$^{3}$LO} & aN$^{3}$LO $n_{f}=5$ & 4.1150 + 0.0683 - 0.0755 & 4.1150 + 1.66\% - 1.83\% \\
 & aN$^{3}$LO $n_{f}=4$ & 4.0270 + 0.0685 - 0.0765 & 4.0270 + 1.70\% - 1.90\% \\
& aN$^{3}$LO $n_{f}=3$ & 2.7248 + 0.0653 - 0.0673 & 2.7248 + 2.40\% - 2.47\%  \\
\hline
\multirow{3}{*}{NNLO} & NNLO $n_{f}=5$ & 3.9974 + 0.0557 - 0.0633 & 3.9974 + 1.39\% - 1.58\% \\
 & NNLO $n_{f}=4$ & 3.9118 + 0.0561 - 0.0634 & 3.9118 + 1.44\% - 1.62\% \\
& NNLO $n_{f}=3$ & 2.6845 + 0.0539 - 0.0641 & 2.6845 + 2.01\% - 2.39\% \\
\end{tabular}
}
\caption{\label{tab: vbf_results_nf}Higgs production cross section results via the vector boson fusion process (with $\sqrt{s} = 13\ \text{TeV}$) using N$^{3}$LO and NNLO hard cross sections combined with NNLO and decorrelated aN$^{3}$LO PDFs whilst varying the number of active flavours $n_{f}$. All PDFs are at the standard choice $\alpha_{s}(m_{Z}) = 0.118$. These results are found with $\mu = Q^{2}$ where $Q^{2}$ is the vector boson momentum.}
\end{table}
Table~\ref{tab: vbf_results} and Fig.~\ref{fig: vbf_predictions} show the predictions at various orders in $\alpha_{s}$ for Higgs production cross sections via vector boson fusion\footnote{Results are obtained with the inclusive part of the code \href{https://provbfh.hepforge.org/}{\texttt{proVBFH}}~\cite{Cacciari:2015,Dreyer:2016,provbfh}.} at the LHC for $\sqrt{s} = 13\ \text{TeV}$ up to N$^{3}$LO~\cite{Cacciari:2015,Dreyer:2016}, again no fiducial cuts are applied in this comparison. The predictions shown are calculated with $\mu_{f}^2 = \mu_{r}^2 = Q^{2}$ as the central scale where $Q^{2}$ is the vector boson squared momentum. 

For this process one can follow the increase in the cross section as higher order PDFs are used. Contrasting with the case of gluon fusion, Fig.~\ref{fig: vbf_predictions} displays little cancellation between the terms added in the aN$^{3}$LO PDF description and the N$^{3}$LO cross section. However, the cross section for VBF produces around a $\sim 3-4 \%$ change order by order and is therefore fairly constant. Considering this relatively small difference between orders, this lack of cancellation is not a major concern. Further to this, the vector boson fusion process is much more reliant on the quark sector which, compared to the gluon, is relatively constant order by order (see Section~\ref{subsec: pdf_results}). The reason for this stems from the more direct data constraints on the shape of quark PDFs. 

Comparing the aN$^{3}$LO VBF cross section (with MHO theoretical uncertainties) with the NNLO cross section result (with NNLO PDFs) including MHOUs via scale variations, we see that the scale variation MHOUs are negligible against the PDF uncertainties at aN$^{3}$LO. This result is in part due to the fact that the scale variation for aN$^{3}$LO is only being included for the renormalisation scale. However at NNLO, the extra MHOU predicted was still only a small contribution. Therefore considering these results further, the effects of higher orders in both cases are expected to be small, which provides some agreement with the argument that there is little scope for cancellation between orders for VBF. As for the gluon fusion prediction in Section~\ref{subsec: ggH}, we confirm that any further factorisation scale variation (i.e. using the 9-point prescription) is contained within the predicted PDF uncertainties; hence further motivating our previous argument that factorisation scale variation is not necessary with aN$^{3}$LO PDFs.

Another feature of the VBF results is that the level of uncertainty at full aN$^{3}$LO is only increased slightly from the calculation involving NNLO PDFs. Comparing this to the gluon fusion results, where the uncertainty was more noticeably increased in both directions, it is evident that these approximate N$^{3}$LO additions are having a smaller effect on the VBF calculation. Once again, the origin of this is due to the nature of the process. VBF involves mostly the quark sector and is therefore much less affected by the extra N$^{3}$LO theory we have introduced (due to direct constraints from data). As we have presented in previous sections, most of the uncertainty in the N$^{3}$LO theory resides in the small-$x$ regime which is more directly probed by the gluon sector than in the quark sector. 

Lastly we briefly discuss the $n_{f}$ dependence of the VBF cross section. In VBF the scaling of contributions follows as $n_{f}^{2}$ due to the presence of two input quark flavours in the process. In Table~\ref{tab: vbf_results_nf} we observe that the VBF cross section receives a large contribution when including the charm quark ($n_{f} = 3 \rightarrow 4$) due to this scaling. We also show that at aN$^{3}$LO, this is where most of the difference in the central value and uncertainty from NNLO is accounted for. This is a consequence of the predicted enhancement of the charm PDF at aN$^{3}$LO, discussed in Section~\ref{subsec: pdf_results}. Beyond $n_{f} = 4$ the bottom contribution to VBF in the $W^{\pm}$ channel (the dominant channel) is heavily suppressed, since due to the CKM elements $b$ must transition to $t$ most of the time. Therefore the VBF cross section only receives a small contribution moving from $n_{f}=4$ to $n_{f}=5$.

\section{Availability and Recommended Usage of MSHT20 \texorpdfstring{aN$^{3}$LO}{aN3LO} PDFs}\label{sec: availability}

We provide the MSHT20 aN$^{3}$LO PDFs in \texttt{LHAPDF} format~\cite{Buckley:2014ana}:
\\
\\
\href{http://lhapdf.hepforge.org/}{\texttt{http://lhapdf.hepforge.org/}}
\\
\\
\noindent as well as on the repository:
\\
\\ 
\href{http://www.hep.ucl.ac.uk/msht/}{\texttt{http://www.hep.ucl.ac.uk/msht/}}
\\
\\
\noindent The approximate N$^{3}$LO functions (for $P_{ij}(x)$ and $A_{ij}(x)$) are provided as lightweight FORTRAN functions or as part of a Python framework in the repository:
\\
\\
\href{https://github.com/MSHTPDF/N3LO_additions}{\texttt{https://github.com/MSHTPDF/N3LO\_additions}}
\\
\\
\noindent We present the aN$^{3}$LO eigenvector sets with and without correlated $K$-factors as discussed in Section~\ref{sec: results}, with the default set being provided with decorrelated $K$-factors\footnote{These grids are updated from a previous version of the MSHT20 aN$^{3}$LO PDFs and should be used in favour of any sets downloaded before the latest upload date of this submission.}.
\\
\\
\href{http://www.hep.ucl.ac.uk/msht/Grids/MSHT20an3lo_as118.tar.gz}{\texttt{MSHT20an3lo\_as118}}\\
\href{http://www.hep.ucl.ac.uk/msht/Grids/MSHT20an3lo_as118_Kcorr.tar.gz}{\texttt{MSHT20an3lo\_as118\_Kcorr}}
\\
\\
Both these PDF sets contain a central PDF accompanied by 104 eigenvector directions (describing 52 eigenvectors) and can be used in exactly the same way as previous MSHT PDF sets i.e. the MSHT20 NNLO PDFs with 64 eigenvector directions.

As presented in this work, the aN$^{3}$LO PDFs include an estimation for missing N$^{3}$LO contributions (the leading theoretical uncertainty) and implicitly some MHOU beyond this within their PDF uncertainties. Due to this, we argue and motivate in Section~\ref{sec: predictions} that factorisation scale variations are no longer necessary in calculations involving aN$^{3}$LO PDFs. However the renormalisation scale should continue to be varied to provide estimates of MHOUs in the hard cross section piece of physical calculations.

In the case that the hard cross section for a process is available up to N$^{3}$LO the recommendation is to use the aN$^{3}$LO PDFs, since unmatched ingredients in cross section calculations can ignore important cancellations (between the PDFs and hard cross section).

If a process is included within the global fit and the hard cross section is known only up to NNLO (i.e. those discussed in Section~\ref{sec: n3lo_K}), we recommend the use of the decorrelated version of the aN$^{3}$LO PDF set. Using these PDFs and the details provided in Table~\ref{tab: Kij_limits}, the hard cross section can be transformed from NNLO to approximate N$^{3}$LO. From here the two approximate N$^{3}$LO ingredients can be used together to give a full approximate N$^{3}$LO result.

If a process is not included in the global PDF fit and the hard cross section is known only up to NNLO, the standard NNLO PDF set remains the default choice. However, we recommend the use of these aN$^{3}$LO PDFs as an estimate of potential MHOUs. In this case the aN$^{3}$LO PDF set + NNLO hard cross section prediction should be reflected in any MHOU estimates for the full NNLO prediction. For example, when the hard cross section is known only up to NNLO Equation~(3.13) from \cite{anastasiou2016high} can be adapted to be,
\begin{equation}
    \delta(\mathrm{PDF} - \mathrm{TH}) = \frac{1}{2}\abs*{\frac{\sigma_{\mathrm{aN}^{3}\mathrm{LO}}^{(2)} - \sigma_{\mathrm{NNLO}}^{(2)}}{\sigma_{\mathrm{aN}^{3}\mathrm{LO}}^{(2)}}}
\end{equation}
where $\delta(\mathrm{PDF} - \mathrm{TH})$ is the predicted PDF theory uncertainty on the $\sigma$ prediction, $\sigma_{\mathrm{aN}^{3}\mathrm{LO}}^{(2)}$ is the NNLO hard cross section with aN$^{3}$LO PDFs and $\sigma_{\mathrm{NNLO}}^{(2)}$ is the full NNLO result. A caveat to this treatment is that the theory uncertainty is sensitive to unmatched cancellations and should therefore be used with care (and caution), therefore the NNLO set remains the default in evaluating PDF uncertainties.

\section{Conclusions}\label{sec: conclusion}

In this paper we have presented the first approximate N$^{3}$LO global PDF fit. This follows the MSHT20 framework~\cite{Thorne:MSHT20}, where the aN$^{3}$LO PDF set also incorporates estimates for theoretical uncertainties from missing N$^{3}$LO contributions and implicitly some MHOU beyond this. In addition, the framework presented for obtaining these PDFs provides a means of utilising higher order information as and when it is available. In contrast, previously, complete information of the next order was required for theoretical calculations in PDF fits. This provides a significant advantage moving forward in precision phenomenology, since as we move to higher orders, this information takes increasingly longer to calculate. We have analysed the resulting set of PDFs, denoted MSHT20aN$^{3}$LO, and made two sets available as described in Section~\ref{sec: availability}. The aN$^{3}$LO PDF fits have been performed to the same set of global hard scattering data and PDF parameterisations included for the MSHT20 NNLO PDF fits.

The NNLO theoretical framework for MSHT20 PDFs has been extended in Section~\ref{sec: theo_framework} to include the addition of general N$^{3}$LO theory parameters into the fit. Subsequently, we have outlined how these N$^{3}$LO theory parameters can be included into the Hessian procedure as controllable nuisance parameters where they are not yet known. Two methods of handling subsets of the N$^{3}$LO theory parameters in the Hessian matrix have then been discussed; i.e. including or ignoring correlations with aN$^{3}$LO $K$-factors across distinct processes. 

In Section~\ref{sec: n3lo_split} to Section~\ref{sec: n3lo_K} we have presented the N$^{3}$LO additions to the relevant splitting functions, transition matrix elements, heavy coefficient functions and $K$-factors. We present usable and computationally efficient approximations to N$^{3}$LO based on known information in the small and large-$x$ regimes and the available Mellin moments (and make these available as described in Section~\ref{sec: availability}). In all cases the best fit prediction for each N$^{3}$LO function is in good agreement with the prior expected behaviour. Also in Section~\ref{sec: n3lo_K}, we find good agreement with recent progress towards N$^{3}$LO DY and top production $K$-factors~\cite{Gehrmann:DYN3LO,Kidonakis:tt}. As more information becomes available surrounding each of these functions, the framework we present here can be easily adapted, aiding in the reduction in sources of MHOUs from N$^{3}$LO. As we have stressed, we interpret our theoretical uncertainty as being mainly due to the remaining uncertainty at N$^3$LO, but with some small, but significant contribution from even higher orders, particularly at small-$x$. Our results seem consistent with this interpretation. However, in the future we expect the N$^{3}$LO description to become more exact. Hence, at some point the remaining N$^3$LO uncertainty will become comparable to, or smaller than effects beyond N$^3$LO. We would then have to modify our procedure. However, we expect that once the N$^3$LO theory becomes very largely known, there will at this point also be more information known about even higher orders (i.e. N$^{4}$LO), which could then be incorporated in a similar manner to maintain an estimate for MHOUs. Alternatively, in the event that the available information is not suitable to provide approximations (or indeed to complement these approximations), a treatment similar in principle, but more sophisticated in practice, to that of the $K$-factors may be adopted for DIS quantities. On this note, we acknowledge that the method of constructing aN$^{3}$LO $K$-factors for non-(inclusive) DIS processes presented here is a first step towards a more robust and flexible procedure, which is left for future work.

Combining together all N$^{3}$LO information, in Section~\ref{sec: results} the results of an approximate N$^{3}$LO global PDF fit are presented. The new MSHT20 approximate N$^{3}$LO PDFs show a significant reduction in $\chi^{2}$ from the MSHT20 NNLO PDF set, with the leading NNLO tensions between HERA and non-HERA datasets heavily reduced at aN$^{3}$LO (most notably with the ATLAS 8 TeV $Z\ p_{T}$ dataset~\cite{ATLASZpT}). With this being said, the aN$^{3}$LO set does fit selected Jets datasets worse in an aN$^3$LO global fit than at NNLO, although these are an exception to the behaviour seen for the other datasets. In performing a fit not including ATLAS 8 TeV $Z\ p_{T}$ data we provide evidence that similar tensions seen at NNLO (see \cite{Thorne:MSHT20}) remain between this dataset and jet production data at aN$^{3}$LO. Further to this, we show that since HERA and ATLAS 8 TeV $Z\ p_{T}$ data are more in agreement in the form of the high-$x$ gluon at aN$^{3}$LO, one can observe that the tension with the jet production data is shared between HERA and ATLAS 8 TeV $Z\ p_{T}$ data. Finally, as discussed, we highlight that in future work it will be interesting to observe if this increased tension may be alleviated when considering these jet datasets instead as dijet cross sections. 

Investigating the correlations present within an aN$^{3}$LO PDF fit, a natural separation between process independent and process dependent parameters can be observed. With this motivation, a PDF set with decorrelated aN$^{3}$LO $K$-factor eigenvectors is constructed. The validity of this is then also verified by comparison with a second PDF set which includes correlations between all parameters. Each of these sets exhibits similarly well behaved eigenvectors and levels of dynamical tolerance.

Considering the form of the individual PDFs, the aN$^{3}$LO PDFs include a much harder gluon at small-$x$ due to contributions from the splitting functions as discussed in Section~\ref{subsec: n3lo_contrib}. This enhancement then translates into an increase in the charm and bottom PDFs due to the gluon input into the heavy flavour sector via the transition matrix elements. At very low-$Q^{2}$ the result of the N$^{3}$LO additions is a non-negative charm and gluon PDF at small-$x$. As a consistency check, the fit dependence on $\alpha_{s}$ and $m_{c}$ has been investigated. In both of these cases we show a preference for values which suppress the heavy flavour contributions (slightly lower $\alpha_{s}$ and slightly higher $m_{c}$ than NNLO). Considering the predicted aN$^{3}$LO $\alpha_{s}$, we observe a slightly lower than $1\sigma$ effect when comparing with the NNLO world average. While an extensive analysis of the aN$^{3}$LO $\alpha_{s}$ value is left for further study, since the world average is determined by NNLO results, one could expect a small systematic effect from moving to N$^{3}$LO.

Taking this analysis further and using the approximate N$^{3}$LO PDFs as input to N$^{3}$LO cross section calculations, we consider the cases of gluon and vector boson fusion in Higgs production. We present the first aN$^{3}$LO calculation for these cross sections and show how the aN$^{3}$LO prediction differs from the case with NNLO PDFs including scale variations, highlighting the importance of matching orders in calculations. 
In VBF we provide an example where cancellation is not realised between orders. However in this case the quark sector is much more constrained and due to the smaller variation between orders, there is naturally less scope for cancellation.

In summary, we have presented a set of approximate N$^{3}$LO PDFs that are able to more accurately predict physical quantities involving PDFs (given that all ingredients in these calculations are included at N$^{3}$LO or aN$^{3}$LO). In producing these PDFs, we have provided a more controllable method for estimating theoretical uncertainties from MHOs in a PDF fit than scale variations. While some ambiguity remains in this method in how the prior variations are chosen, we argue that the current knowledge and intuition surrounding each source of uncertainty can be utilised as and when available. This is therefore  much more in line with what one can expect a theoretical uncertainty to encompass. Another potential shortcoming is the possibility of fitting to sources of uncertainty other than higher orders (or higher order corrections elsewhere in theory calculations included in a PDF fit). Although this is a possibility, the position of the considered sources of uncertainty in the underlying theory combined with the prior variations and penalties should act to minimise this effect. In any case, if a separate source of uncertainty is significantly affecting the fit, this will present itself as a source of tension with the N$^{3}$LO penalties and the $\chi^{2}$ (and PDF uncertainty) will be adapted accordingly. 

In future work it will be interesting to investigate the effects in the high-$x$ gluon, which is a region of phenomenological importance and where the interpretation of LHC constraints is not always straightforward. We also note that there are N$^{3}$LO results available from di-lepton rapidity in DY processes~\cite{Gehrmann:DYN3LO}. Considering the results in Section~\ref{sec: n3lo_K} which display an agreement with these recent results, we hope that these approximate N$^{3}$LO PDFs may be of interest in this analysis. Similarly for recent results considering top production~\cite{Kidonakis:tt}. Furthermore, any approximate information from these results could be included in the N$^{3}$LO $K$-factor priors, which was not done for this iteration of the aN$^{3}$LO PDFs. Finally, in order to continually improve the description of aN$^{3}$LO PDFs, the inclusion of more sub-leading sources of MHOUs could be addressed. With the upcoming wealth of experimental data from future colliders such as the HL-LHC and the EIC, it will be of interest to gain a better understanding of the transition matrix elements and also describe better the charged current and longitudinal structure functions, where currently theoretical uncertainties are much smaller than the experimental uncertainties.

\section*{Acknowledgements}

J.M. thanks the Science and Technology Facilities Council (STFC) part of U.K. Research and Innovation for support via Ph.D. funding. T.C. and R.S.T. thank STFC for support via grant awards ST/P000274/1 and ST/T000856/1. L.H.L. thanks STFC for support via grant awards ST/L000377/1 and ST/T000864/1. We would like to thank Xuan Chen, Thomas Gehrmann, Nigel Glover and Alex Huss for providing details of N$^3$LO calculations. We would like to thank members of the PDF4LHC working group for numerous discussions on PDFs and theoretical uncertainties. We would also like to thank Alan Martin for long collaboration on the MSHT series of PDFs.  

\pagebreak

\setcounter{table}{0}
\renewcommand\thetable{\Alph{section}.\arabic{table}}
\appendix
\section{List of \texorpdfstring{N$^{3}$LO}{N3LO} Ingredients}\label{app: n3lo_known}

\begin{table}[H]
\centerline{
\begin{tabular}{|c|c|c|c|c|}
\hline
N$^{3}$LO & No. of & Moments & \multirow{2}{*}{Small-$x$} & \multirow{2}{*}{Large-$x$}\\
Function & Moments & (Even only) &  & \\
\hline
$P_{qq}^{\mathrm{NS}}$ & 8 & $N=2 - 16$ \cite{4loopNS} & \cite{4loopNS} & \cite{4loopNS} \\
$P_{qq}^{\mathrm{PS}}$ & 4 & $N=2 - 8$ \cite{S4loopMoments,S4loopMomentsNew} & LL \cite{Catani:1994sq} & N/A \\
$P_{qg}$ & 4 & $N=2 - 8$ \cite{S4loopMoments,S4loopMomentsNew} & LL \cite{Catani:1994sq} & N/A \\
$P_{gq}$ & 4 & $N=2 - 8$ \cite{S4loopMoments,S4loopMomentsNew} & LL \cite{Lipatov:1976zz,Kuraev:1977fs,Balitsky:1978ic} & N/A \\
$P_{gg}$ & 4 & $N=2 - 8$ \cite{S4loopMoments,S4loopMomentsNew} & LL \& NLL \cite{Lipatov:1976zz,Kuraev:1977fs,Balitsky:1978ic,Fadin:1998py,Ciafaloni:1998gs} & N/A \\
\hline
$A_{qq,H}^{\mathrm{NS}}$ & 7 & $N=2 - 14$ \cite{bierenbaum:OMEmellin} & N/A & N/A \\
$A_{Hq}^{\mathrm{PS}}$ & 6 & $N=2 - 12$ \cite{bierenbaum:OMEmellin} & \cite{ablinger:3loopPS} & \cite{ablinger:3loopPS} \\
$A_{Hg}$ & 5 & $N=2 - 10$ \cite{bierenbaum:OMEmellin} & LL \cite{Vogt:FFN3LO3} & N/A \\
$A_{gq, H}$ & 7 & $N=2 - 14$ \cite{bierenbaum:OMEmellin} & \cite{ablinger:agq}  & \cite{ablinger:agq} \\
$A_{gg, H}$ & 5 & $N=2 - 10$ \cite{bierenbaum:OMEmellin} & N/A & N/A \\
\hline
\end{tabular}
}
\caption{\label{tab: known_splitOME}List of all the N$^{3}$LO ingredients used to construct the approximate N$^{3}$LO splitting functions and transition matrix elements. Where only a citation is provided, extensive knowledge i.e. beyond NLL is used. This table is a non-exhaustive list of the current knowledge about these functions, however information beyond that which is provided here is not currently in a usable format for phenomological studies.}
\end{table}

\begin{table}[H]
\centerline{
\begin{tabular}{|c|c|c|c|}
\hline
GM-VFNS N$^{3}$LO & \multirow{2}{*}{Known N$^{3}$LO Components}\\
Function & \\
\hline
$C_{H,q}$ & $C_{H,q}^{(3),\ \mathrm{FF}}\left(Q^{2} \leq m_{h}^{2}\right)$ LL \cite{Catani:FFN3LO1,Laenen:FFN3LO2,Vogt:FFN3LO3}, $C_{H, q}^{\mathrm{VF},\ (3)}$ \cite{Vermaseren:2005qc} \\
$C_{H,g}$ & $C_{H,g}^{(3),\ \mathrm{FF}}\left(Q^{2} \leq m_{h}^{2}\right)$ LL\cite{Catani:FFN3LO1,Laenen:FFN3LO2,Vogt:FFN3LO3}, $C_{H, q}^{\mathrm{ZM},\ (3)}$ \cite{Vermaseren:2005qc} \\
\hline
$C_{q,q}^{\mathrm{NS}}$ & $C_{q, q,\ \mathrm{NS}}^{\mathrm{ZM},\ (3)}$ \cite{Vermaseren:2005qc} \\
$C_{q,q}^{\mathrm{PS}}$ & $C_{q, q,\ \mathrm{PS}}^{\mathrm{ZM},\ (3)}$ \cite{Vermaseren:2005qc} \\
$C_{q,g}$ & $C_{q, g}^{\mathrm{ZM},\ (3)}$ \cite{Vermaseren:2005qc} \\
\hline
\end{tabular}
}
\caption{\label{tab: known_coeff}List of all N$^{3}$LO ingredients used to construct the approximate N$^{3}$LO GM-VFNS coefficient functions. Note that lower order components that contribute to these functions are also known and are cited in the text. This table only considers contributing 3-loop functions.}
\end{table}
Table's~\ref{tab: known_splitOME} and \ref{tab: known_coeff} summarise the available (at the time of writing) and used information regarding the N$^{3}$LO splitting functions and coefficient functions respectively. The formalism presented in Section~\ref{sec: theo_framework} currently makes use of all this information and is able to be adapted as and when more information becomes available.
\setcounter{table}{0}
\section{\texorpdfstring{$\chi^{2}$}{Chi-squared} Results without HERA}\label{app: noHERA}

\subsection{NNLO}

Table \ref{tab: no_HERA_fullNNLO} shows the differences in $\chi^{2}$ found when omitting HERA data from a PDF fit using the MSHT NNLO PDFs. This table is copied here from \cite{Thorne:MSHT20} for the ease of the reader. We see similarities between these results and the $\Delta \chi^{2}$'s seen in the case of N$^{3}$LO PDFs. Specifically the ATLAS $8\ \text{TeV}\ Z\ p_{T}$ displaying a substantial reduction from the global NNLO fit. This therefore provides evidence that the inclusion of the N$^{3}$LO contributions is aiding in reducing tensions between the HERA and non-HERA datasets.
\begin{table}[H]
\centerline{
\begin{tabular}{|P{6.5cm}|P{1cm}|P{2cm}|P{2cm}|}
\hline
Dataset & $N_{\mathrm{pts}}$ & $\chi^{2}$ & $\Delta \chi^{2}$ \\
\hline
BCDMS $\mu p$ $F_{2}$ \cite{BCDMS} &   163  &  174.7  &  $-5.5$ \\
BCDMS $\mu d$ $F_{2}$ \cite{BCDMS} &   151  &  143.9  &  $-2.1$ \\
NMC $\mu p$ $F_{2}$ \cite{NMC} &   123  &  119.6  &  $-4.5$ \\
NMC $\mu d$ $F_{2}$ \cite{NMC} &   123  &  96.6  &  $-16.1$ \\
SLAC $ep$ $F_{2}$ \cite{SLAC,SLAC1990} &   37  &  33.0  &  $+0.9$ \\
SLAC $ed$ $F_{2}$ \cite{SLAC,SLAC1990} &   38  &  24.1  &  $+1.1$ \\
E665 $\mu d$ $F_{2}$ \cite{E665} &   53  &  63.5  &  $+3.9$ \\
E665 $\mu p$ $F_{2}$ \cite{E665} &   53  &  68.9  &  $+4.3$ \\
NuTeV $\nu N$ $F_{2}$ \cite{NuTev} &   53  &  38.0  &  $-0.3$ \\
NuTeV $\nu N$ $xF_{3}$ \cite{NuTev} &   42  &  27.5  &  $-3.2$ \\
NMC $\mu n / \mu p$ \cite{NMCn/p} &   148  &  132.7  &  $+1.9$ \\
E866 / NuSea $pp$ DY \cite{E866DY} &   184  &  228.0  &  $+2.9$ \\
E866 / NuSea $pd/pp$ DY \cite{E866DYrat} &   15  &  9.1  &  $-1.3$ \\
CCFR $\nu N \rightarrow \mu\mu X$ \cite{Dimuon} &   86  &  66.2  &  $-1.5$ \\
NuTeV $\nu N \rightarrow \mu\mu X$ \cite{Dimuon} &   84  &  49.0  &  $-9.5$ \\
CHORUS $\nu N$ $F_{2}$ \cite{CHORUS} &   42  &  29.6  &  $-0.6$ \\
CHORUS $\nu N$ $xF_{3}$ \cite{CHORUS} &   28  &  18.2  &  $-0.3$ \\
CDF II $p\bar{p}$ incl. jets \cite{CDFjet} &   76  &  60.9  &  $+0.5$ \\
D{\O} II $Z$ rap. \cite{D0Zrap} &   28  &  16.6  &  $+0.3$ \\
CDF II $Z$ rap. \cite{CDFZrap} &   28  &  38.7  &  $+1.5$ \\
D{\O} II $W \rightarrow \nu \mu$ asym. \cite{D0Wnumu} &   10  &  17.4  &  $+0.1$ \\
CDF II $W$ asym. \cite{CDF-Wasym} &   13  &  19.0  &  $+0.0$ \\
D{\O} II $W \rightarrow \nu e$ asym. \cite{D0Wnue} &   12  &  30.0  &   $-3.9$ \\
D{\O} II $p\bar{p}$ incl. jets \cite{D0jet} &   110  &  119.3  &   $-0.9$ \\
ATLAS $W^{+},\ W^{-},\ Z$ \cite{ATLASWZ} &   30  &  29.5  &   $-0.4$ \\
\hline
\end{tabular}
}
\caption{\label{tab: no_HERA_fullNNLO}The change in $\chi^{2}$ for a NNLO fit(with negative indicating an improvement in the fit quality) when the combined HERA data sets including $F_{L}$ and heavy flavour data are removed, illustrating the tensions of these data sets with several of the other data sets in the global fit. $\Delta \chi^{2}$ represents the change from a full global fit at the same order in $\alpha_{s}$.}
\end{table}
\begin{table}\ContinuedFloat
\centerline{
\begin{tabular}{|P{6.5cm}|P{1cm}|P{2cm}|P{2cm}|}
\hline
Dataset & $N_{\mathrm{pts}}$ & $\chi^{2}$ & $\Delta \chi^{2}$ \\
\hline
CMS W asym. $p_{T} > 35\ \text{GeV}$ \cite{CMS-easym} &   11  &  6.6  &   $-1.2$ \\
CMS W asym. $p_{T} > 25, 30\ \text{GeV}$ \cite{CMS-Wasymm} &   24  &  7.5  &   $+0.1$ \\
LHCb $Z \rightarrow e^{+}e^{-}$ \cite{LHCb-Zee} &   9  &  24.2  &   $+1.5$ \\
LHCb W asym. $p_{T} > 20\ \text{GeV}$ \cite{LHCb-WZ} &   10  &  12.1  &   $-0.3$ \\
CMS $Z \rightarrow e^{+}e^{-}$ \cite{CMS-Zee} &   35  &  17.3  &   $-0.6$ \\
ATLAS High-mass Drell-Yan \cite{ATLAShighmass} &   13  &  16.9  &   $-2.0$ \\
Tevatron, ATLAS, CMS $\sigma_{t\bar{t}}$ \cite{Tevatron-top,ATLAS-top7-1,ATLAS-top7-2,ATLAS-top7-3,ATLAS-top7-4,ATLAS-top7-5,ATLAS-top7-6,CMS-top7-1,CMS-top7-2,CMS-top7-3,CMS-top7-4,CMS-top7-5,CMS-top8} &   17  &  14.2  &   $-0.4$ \\
CMS double diff. Drell-Yan \cite{CMS-ddDY} &   132  &  134.2  &   $-10.3$ \\
LHCb 2015 $W, Z$ \cite{LHCbZ7,LHCbWZ8} &   67  &  97.4  &   $-1.9$ \\
LHCb $8\ \text{TeV}$ $Z \rightarrow ee$ \cite{LHCbZ8} &   17  &  24.4  &   $-1.8$ \\
CMS $8\ \text{TeV}\ W$ \cite{CMSW8} &   22  &  13.7  &   $+0.9$ \\
ATLAS $7\ \text{TeV}$ jets \cite{ATLAS7jets} &   140  &  228.0  &   $+6.5$ \\
CMS $7\ \text{TeV}\ W + c$ \cite{CMS7Wpc} &   10  &  9.2  &   $+0.6$ \\
ATLAS $7\ \text{TeV}$ high prec. $W, Z$ \cite{ATLASWZ7f} &   61  &  116.8  &   $+0.2$ \\
CMS $7\ \text{TeV}$ jets \cite{CMS7jetsfinal} &   158  &  179.5  &   $+3.8$ \\
D{\O} $W$ asym. \cite{D0Wasym} &   14  &  11.3  &   $-0.8$ \\
ATLAS $8\ \text{TeV}\ Z\ p_{T}$ \cite{ATLASZpT} &   104  &  149.3  &   $-39.2$ \\
CMS $8\ \text{TeV}$ jets \cite{CMS8jets} &   174  &  259.5  &   $-1.8$ \\
ATLAS $8\ \text{TeV}$ sing. diff. $t\bar{t}$ \cite{ATLASsdtop} &   25  &  24.5  &   $-1.1$ \\
ATLAS $8\ \text{TeV}$ sing. diff. $t\bar{t}$ dilep. \cite{ATLASttbarDilep08_ytt} &   5  &  2.3  &   $-1.1$ \\
ATLAS $8\ \text{TeV}$ High-mass DY \cite{ATLASHMDY8} &   48  &  60.9  &   $+3.7$ \\
ATLAS $8\ \text{TeV}\ W + \text{jets}$ \cite{ATLASWjet} &   30  &  16.4  &   $-1.7$ \\
CMS $8\ \text{TeV}$ double diff. $t\bar{t}$ \cite{CMS8ttDD} &   15  &  23.3  &   $+0.8$ \\
ATLAS $8\ \text{TeV}\ W$ \cite{ATLASW8} &   22  &  54.4  &   $-3.0$ \\
CMS $2.76\ \text{TeV}$ jet \cite{CMS276jets} &   81  &  102.9  &   $+0.0$ \\
CMS $8\ \text{TeV}$ sing. diff. $t\bar{t}$ \cite{CMSttbar08_ytt} &   9  &  10.6  &   $-2.6$ \\
ATLAS $8\ \text{TeV}$ double diff. $Z$ \cite{ATLAS8Z3D} &   59  &  108.3  &   $+22.7$ \\
\hline
Total & 3042 & 3379.6 & $-61.6$ \\
\hline
\end{tabular}
}
\caption{\textit{(Continued)} The change in $\chi^{2}$ (with negative indicating an improvement in the fit quality) when the combined HERA data sets including $F_{L}$ and heavy flavour data are removed, illustrating the tensions of these data sets with several of the other data sets in the global fit. $\Delta \chi^{2}$ represents the change from a full global fit at the same order in $\alpha_{s}$.}
\end{table}

\subsection{\texorpdfstring{aN$^{3}$LO}{aN3LO}}

Table~\ref{tab: no_HERA_fullN3LO} shows the differences in $\chi^{2}$ found when omitting HERA data from a PDF fit using the MSHT aN$^{3}$LO PDFs. These results show that at aN$^{3}$LO the fit no longer experiences large tensions between HERA and ATLAS $8\ \text{TeV}\ Z\ p_{T}$~\cite{ATLASZpT} datasets. The main tensions at N$^{3}$LO are now concerning the Jets data with HERA (and most likely some non-HERA datasets). This result is not unexpected due to the known issues surrounding jets especially as we move to higher precision~\cite{AbdulKhalek:2020jut}.
\begin{table}[H]
\centerline{
\begin{tabular}{|P{6.5cm}|P{1cm}|P{2cm}|P{2cm}|}
\hline
Dataset & $N_{\mathrm{pts}}$ & $\chi^{2}$ & $\Delta \chi^{2}$ \\
\hline
BCDMS $\mu p$ $F_{2}$ \cite{BCDMS} &   163  &  175.8  &  $+1.4$ \\
BCDMS $\mu d$ $F_{2}$ \cite{BCDMS} &   151  &  144.2  &  $-0.0$ \\
NMC $\mu p$ $F_{2}$ \cite{NMC} &   123  &  113.6  &  $-7.8$ \\
NMC $\mu d$ $F_{2}$ \cite{NMC} &   123  &  87.6  &  $-16.6$ \\
SLAC $ep$ $F_{2}$ \cite{SLAC,SLAC1990} &   37  &  30.7  &  $-0.9$ \\
SLAC $ed$ $F_{2}$ \cite{SLAC,SLAC1990} &   38  &  23.2  &  $+0.4$ \\
E665 $\mu d$ $F_{2}$ \cite{E665} &   53  &  65.2  &  $+1.3$ \\
E665 $\mu p$ $F_{2}$ \cite{E665} &   53  &  69.0  &  $+1.5$ \\
NuTeV $\nu N$ $F_{2}$ \cite{NuTev} &   53  &  35.4  &  $-0.4$ \\
NuTeV $\nu N$ $xF_{3}$ \cite{NuTev} &   42  &  29.2  &  $-5.6$ \\
NMC $\mu n / \mu p$ \cite{NMCn/p} &   148  &  131.1  &  $-0.5$ \\
E866 / NuSea $pp$ DY \cite{E866DY} &   184  &  225.4  &  $+2.1$ \\
E866 / NuSea $pd/pp$ DY \cite{E866DYrat} &   15  &  8.2  &  $-0.2$ \\
CCFR $\nu N \rightarrow \mu\mu X$ \cite{Dimuon} &   86  &  67.0  &  $-1.3$ \\
NuTeV $\nu N \rightarrow \mu\mu X$ \cite{Dimuon} &   84  &  47.6  &  $-9.1$ \\
CHORUS $\nu N$ $F_{2}$ \cite{CHORUS} &   42  &  29.0  &  $-0.2$ \\
CHORUS $\nu N$ $xF_{3}$ \cite{CHORUS} &   28  &  18.5  &  $+0.4$ \\
CDF II $p\bar{p}$ incl. jets \cite{CDFjet} &   76  &  65.9  &  $-0.6$ \\
D{\O} II $Z$ rap. \cite{D0Zrap} &   28  &  17.6  &  $+0.3$ \\
CDF II $Z$ rap. \cite{CDFZrap} &   28  &  42.0  &  $+1.5$ \\
D{\O} II $W \rightarrow \nu \mu$ asym. \cite{D0Wnumu} &   10  &  18.9  &  $+2.4$ \\
CDF II $W$ asym. \cite{CDF-Wasym} &   13  &  19.2  &  $+1.0$ \\
D{\O} II $W \rightarrow \nu e$ asym. \cite{D0Wnue} &   12  &  31.0  &   $+0.3$ \\
D{\O} II $p\bar{p}$ incl. jets \cite{D0jet} &   110  &  114.2  &   $+0.9$ \\
ATLAS $W^{+},\ W^{-},\ Z$ \cite{ATLASWZ} &   30  &  29.4  &   $-0.6$ \\
\hline
\end{tabular}
}
\caption{\label{tab: no_HERA_fullN3LO}The change in $\chi^{2}$ for an N$^{3}$LO fit (with negative indicating an improvement in the fit quality) when the combined HERA data sets including $F_{L}$ and heavy flavour data are removed, illustrating the tensions of these data sets with several of the other data sets in the global fit. $\Delta \chi^{2}$ represents the change from a full global fit at the same order in $\alpha_{s}$.}
\end{table}
\begin{table}\ContinuedFloat
\centerline{
\begin{tabular}{|P{6.5cm}|P{1cm}|P{2cm}|P{2cm}|}
\hline
Dataset & $N_{\mathrm{pts}}$ & $\chi^{2}$ & $\Delta \chi^{2}$ \\
\hline
CMS W asym. $p_{T} > 35\ \text{GeV}$ \cite{CMS-easym} &   11  &  7.0  &   $+0.3$ \\
CMS W asym. $p_{T} > 25, 30\ \text{GeV}$ \cite{CMS-Wasymm} &   24  &  7.6  &   $-0.1$ \\
LHCb $Z \rightarrow e^{+}e^{-}$ \cite{LHCb-Zee} &   9  &  22.4  &   $-1.7$ \\
LHCb W asym. $p_{T} > 20\ \text{GeV}$ \cite{LHCb-WZ} &   10  &  12.6  &   $+0.0$ \\
CMS $Z \rightarrow e^{+}e^{-}$ \cite{CMS-Zee} &   35  &  17.1  &   $-0.4$ \\
ATLAS High-mass Drell-Yan \cite{ATLAShighmass} &   13  &  17.2  &   $-0.9$ \\
Tevatron, ATLAS, CMS $\sigma_{t\bar{t}}$ \cite{Tevatron-top,ATLAS-top7-1,ATLAS-top7-2,ATLAS-top7-3,ATLAS-top7-4,ATLAS-top7-5,ATLAS-top7-6,CMS-top7-1,CMS-top7-2,CMS-top7-3,CMS-top7-4,CMS-top7-5,CMS-top8} &   17  &  14.4  &   $+0.1$ \\
CMS double diff. Drell-Yan \cite{CMS-ddDY} &   132  &  126.0  &   $-3.5$ \\
LHCb 2015 $W, Z$ \cite{LHCbZ7,LHCbWZ8} &   67  &  97.6  &   $-6.3$ \\
LHCb $8 \text{TeV}$ $Z \rightarrow ee$ \cite{LHCbZ8} &   17  &  26.6  &   $-2.2$ \\
CMS $8\ \text{TeV}\ W$ \cite{CMSW8} &   22  &  11.9  &   $+0.1$ \\
ATLAS $7\ \text{TeV}$ jets \cite{ATLAS7jets} &   140  &  217.7  &   $+1.8$ \\
CMS $7\ \text{TeV}\ W + c$ \cite{CMS7Wpc} &   10  &  10.8  &   $+0.0$ \\
ATLAS $7\ \text{TeV}$ high prec. $W, Z$ \cite{ATLASWZ7f} &   61  &  94.3  &   $+0.2$ \\
CMS $7\ \text{TeV}$ jets \cite{CMS7jetsfinal} &   158  &  187.8  &   $+1.0$ \\
D{\O} $W$ asym. \cite{D0Wasym} &   14  &  10.1  &   $-2.1$ \\
ATLAS $8\ \text{TeV}\ Z\ p_{T}$ \cite{ATLASZpT} &   104  &  121.2  &   $+12.8$ \\
CMS $8\ \text{TeV}$ jets \cite{CMS8jets} &   174  &  259.8  &   $-11.5$ \\
ATLAS $8\ \text{TeV}$ sing. diff. $t\bar{t}$ \cite{ATLASsdtop} &   25  &  24.1  &   $-0.2$ \\
ATLAS $8\ \text{TeV}$ sing. diff. $t\bar{t}$ dilep. \cite{ATLASttbarDilep08_ytt} &   5  &  3.0  &   $+0.3$ \\
ATLAS $8\ \text{TeV}$ High-mass DY \cite{ATLASHMDY8} &   48  &  65.0  &   $+2.0$ \\
ATLAS $8\ \text{TeV}\ W + \text{jets}$ \cite{ATLASWjet} &   30  &  18.0  &   $-0.8$ \\
CMS $8\ \text{TeV}$ double diff. $t\bar{t}$ \cite{CMS8ttDD} &   15  &  22.8  &   $-0.8$ \\
ATLAS $8\ \text{TeV}\ W$ \cite{ATLASW8} &   22  &  52.5  &   $-5.5$ \\
CMS $2.76\ \text{TeV}$ jet \cite{CMS276jets} &   81  &  103.0  &   $-6.8$ \\
CMS $8\ \text{TeV}$ sing. diff. $t\bar{t}$ \cite{CMSttbar08_ytt} &   9  &  12.3  &   $+2.0$ \\
ATLAS $8\ \text{TeV}$ double diff. $Z$ \cite{ATLAS8Z3D} &   59  &  95.5  &   $+3.9$ \\
\hline
\end{tabular}
}
\caption{\textit{(Continued)} The change in $\chi^{2}$ for an N$^{3}$LO fit (with negative indicating an improvement in the fit quality) when the combined HERA data sets including $F_{L}$ and heavy flavour data are removed, illustrating the tensions of these data sets with several of the other data sets in the global fit. $\Delta \chi^{2}$ represents the change from a full global fit at the same order in $\alpha_{s}$.}
\end{table}
\begin{table}[t]\ContinuedFloat
\centerline{
\begin{tabular}{|c|c|c|c|}
\hline
Low-$Q^{2}$ Coefficient & & & \\
\hline
$c_{q}^{\mathrm{NLL}}$ $ = -3.844$  &   0.006  &  $c_{g}^{\mathrm{NLL}}$ $ = -3.875$  &   0.004 \\
\hline
Transition Matrix Elements & & & \\
\hline
$a_{Hg}$ $ = 17788.000$  &   5.607  &  $a_{qq,H}^{\mathrm{NS}}$ $ = -63.950$  &   0.000 \\
$a_{gg,H}$ $ = -1334.500$  &   0.001  & & \\
\hline
Splitting Functions & & & \\
\hline
$\rho_{qq}^{NS}$ $ = 0.007$  &   0.000  &  $\rho_{gq}$ $ = -1.647$  &   0.001 \\
$\rho_{qq}^{PS}$ $ = -0.579$  &   0.429  &  $\rho_{gg}$ $ = 9.237$  &   0.023 \\
$\rho_{qg}$ $ = -1.343$  &   0.131  & & \\
\hline
K-factors & & & \\
\hline
$\mathrm{DY}_{\mathrm{NLO}}$ $ = -0.279$  &   0.080  &  $\mathrm{DY}_{\mathrm{NNLO}}$ $ = 0.072$  &   0.005 \\
$\mathrm{Top}_{\mathrm{NLO}}$ $ = -0.204$  &   0.042  &  $\mathrm{Top}_{\mathrm{NNLO}}$ $ = 0.412$  &   0.170 \\
$\mathrm{Jet}_{\mathrm{NLO}}$ $ = -0.254$  &   0.065  &  $\mathrm{Jet}_{\mathrm{NNLO}}$ $ = -0.861$  &   0.741 \\
$p_{T}\mathrm{Jets}_{\mathrm{NLO}}$ $ = 0.461$  &   0.213  &  $p_{T}\mathrm{Jets}_{\mathrm{NNLO}}$ $ = 0.016$  &   0.000 \\
$\mathrm{Dimuon}_{\mathrm{NLO}}$ $ = -0.329$  &   0.109  &  $\mathrm{Dimuon}_{\mathrm{NNLO}}$ $ = 0.587$  &   0.345 \\
\hline
\multicolumn{2}{c|}{} & Total & 3306.8 / 3042\\
\multicolumn{2}{c|}{} & $\Delta \chi^{2}$ from N$^{3}$LO & $-49.0$ \\
\cline{3-4}
\end{tabular}
}
\caption{\textit{(Continued)} The change in $\chi^{2}$ for an N$^{3}$LO fit (with negative indicating an improvement in the fit quality) when the combined HERA data sets including $F_{L}$ and heavy flavour data are removed, illustrating the tensions of these data sets with several of the other data sets in the global fit. $\Delta \chi^{2}$ represents the change from a full global fit at the same order in $\alpha_{s}$.}
\end{table}

\setcounter{table}{0}
\clearpage
\section{Dynamic Tolerances}

In this section we provide an exhaustive breakdown of the $\Delta\chi^{2}_{\mathrm{global}}$ behaviour for all eigenvectors found where N$^{3}$LO $K$-factor parameters are considered completely decorrelated ($H_{ij} + K_{ij}$) or correlated ($H_{ij}^{\prime}$) with all other parameters.

\subsection{Case 1: Decorrelated \texorpdfstring{$K$}{K}-factor Parameters}\label{app: tolerance_decorr}

Fig.~\ref{fig: full_tol_decorr} displays the tolerance landscape for each eigenvector found from the decorrelated ($H_{ij} + K_{ij}$) Hessian described in Section~\ref{sec: theo_framework}. Across all 52 eigenvectors (42 PDF + N$^{3}$LO DIS theory and 10 N$^{3}$LO $K$-factor) we show an overall general agreement with the quadratic assumption similar to that found at NNLO.
\begin{figure}[H]
    \begin{center}
    \includegraphics[width=0.3\textwidth]{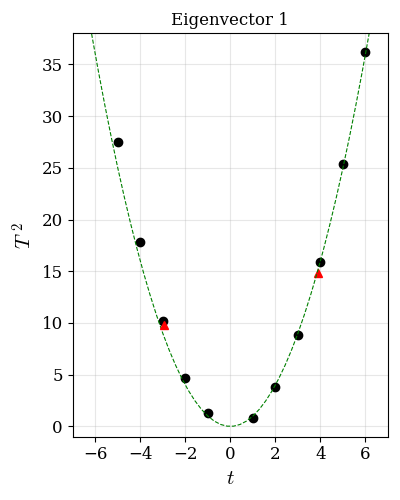}
    \includegraphics[width=0.3\textwidth]{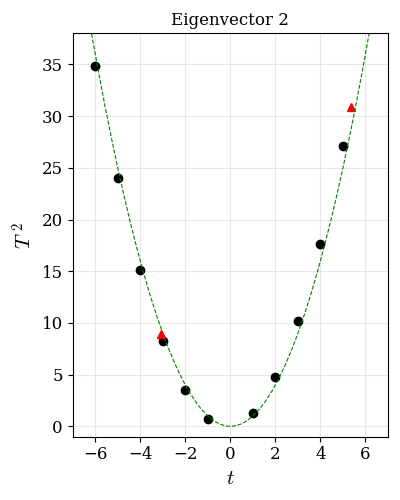}
    \includegraphics[width=0.3\textwidth]{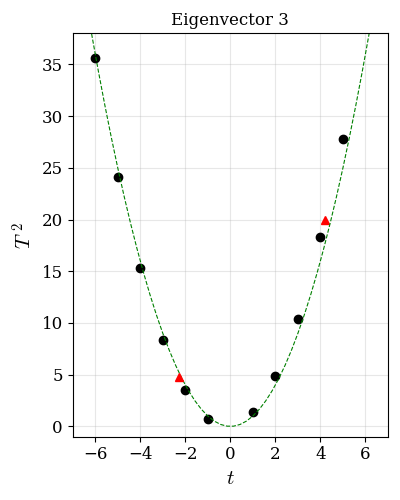}
    \includegraphics[width=0.3\textwidth]{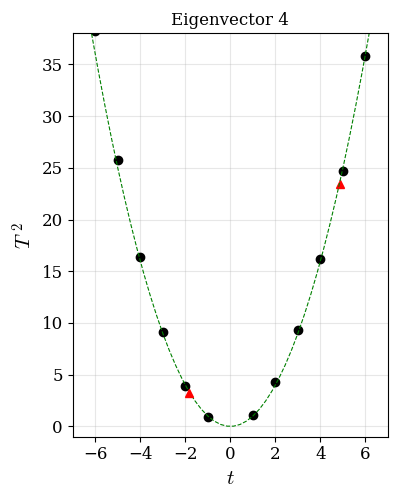}
    \includegraphics[width=0.3\textwidth]{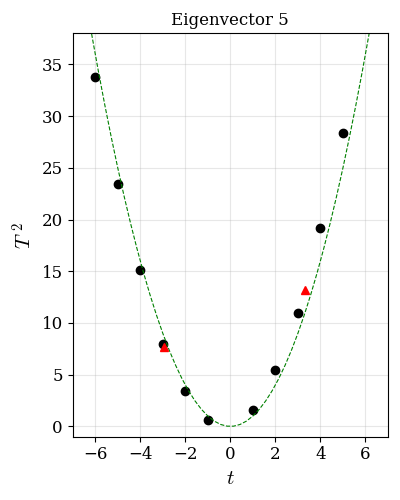}
    \includegraphics[width=0.3\textwidth]{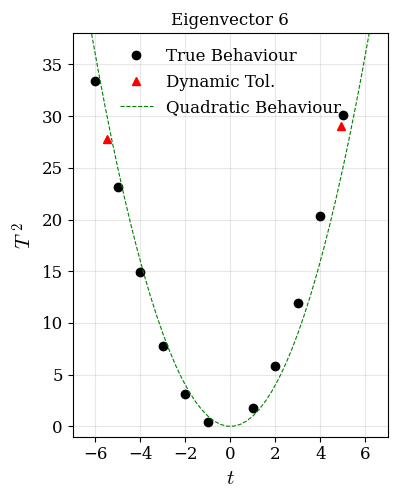}
    \end{center}
    \caption{\label{fig: full_tol_decorr}Dynamic tolerances for each eigenvector direction in the case of complete decorrelation between the theory and PDF parameters, and the $K$-factor parameters included in the PDF fit.}
\end{figure}
\begin{figure}[H]\ContinuedFloat
    \begin{center}
    \includegraphics[width=0.3\textwidth]{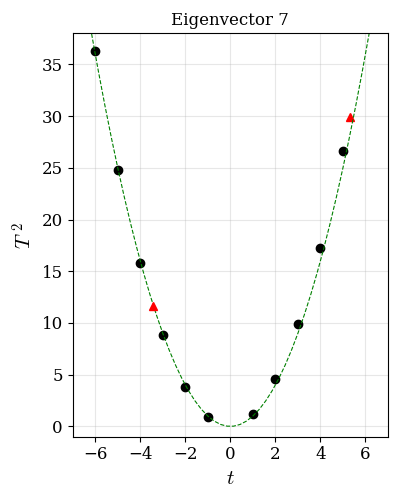}
    \includegraphics[width=0.3\textwidth]{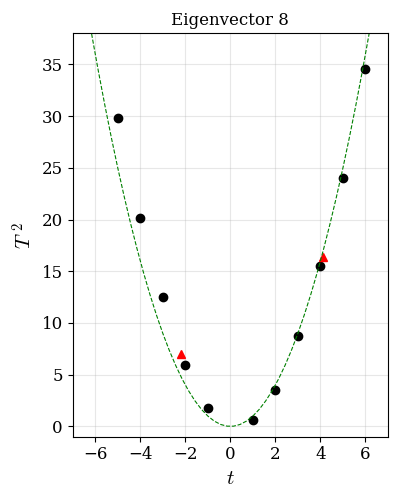}
    \includegraphics[width=0.3\textwidth]{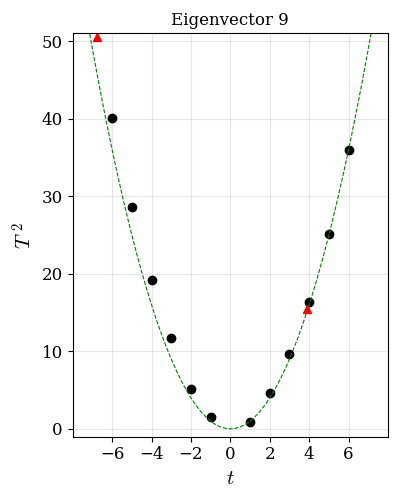}
    \includegraphics[width=0.3\textwidth]{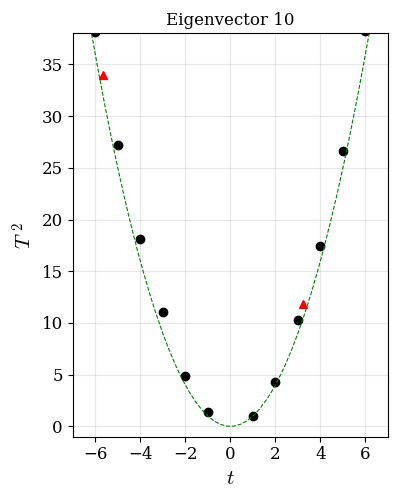}
    \includegraphics[width=0.3\textwidth]{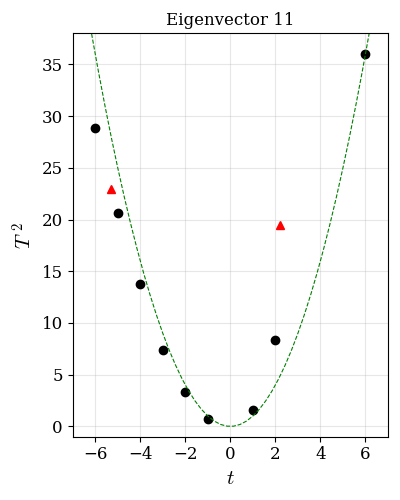}
    \includegraphics[width=0.3\textwidth]{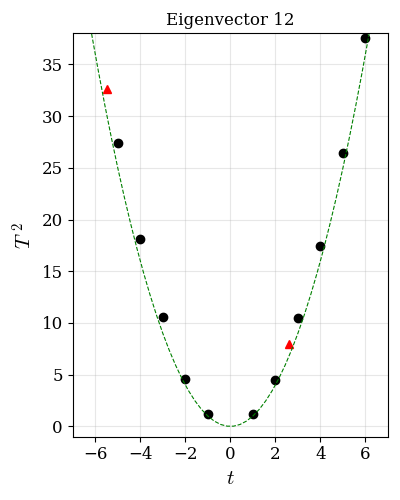}
    \includegraphics[width=0.3\textwidth]{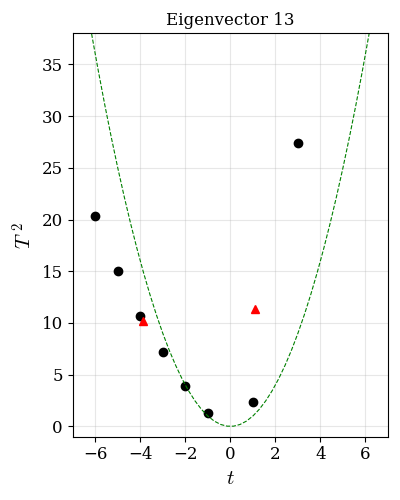}
    \includegraphics[width=0.3\textwidth]{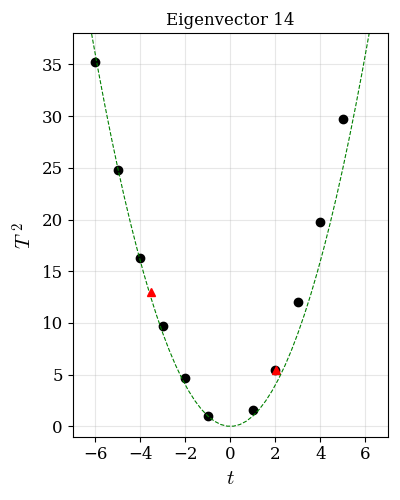}
    \includegraphics[width=0.3\textwidth]{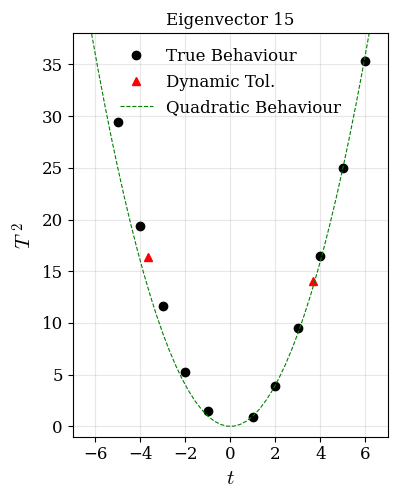}
    \end{center}
    \caption{\textit{(Continued)} Dynamic tolerances for each eigenvector direction in the case of complete decorrelation between the theory and PDF parameters, and the $K$-factor parameters included in the PDF fit.}
\end{figure}
\begin{figure}\ContinuedFloat
    \begin{center}
    \includegraphics[width=0.3\textwidth]{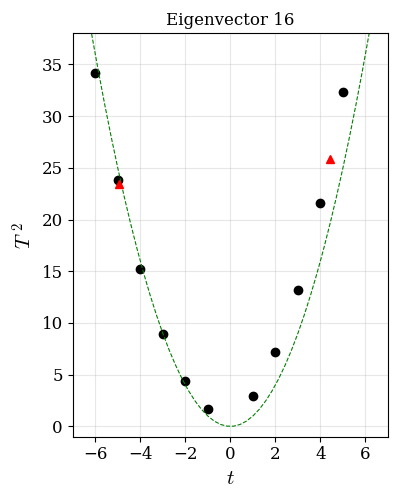}
    \includegraphics[width=0.3\textwidth]{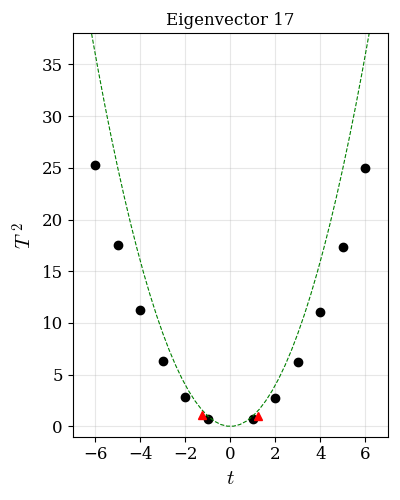}
    \includegraphics[width=0.3\textwidth]{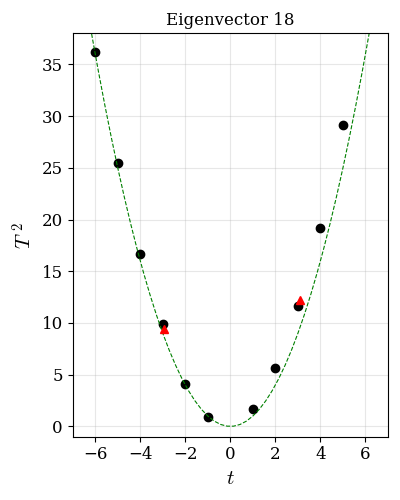}
    \includegraphics[width=0.3\textwidth]{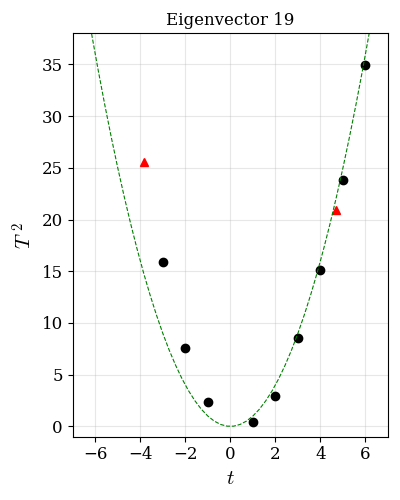}
    \includegraphics[width=0.3\textwidth]{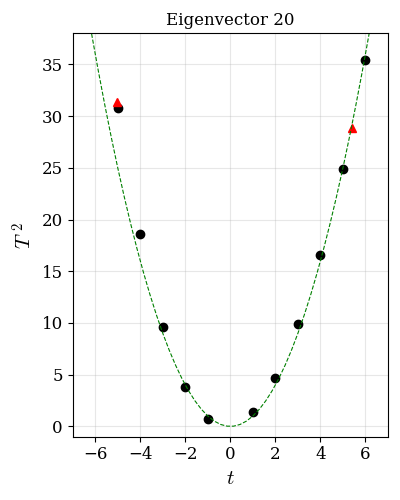}
    \includegraphics[width=0.3\textwidth]{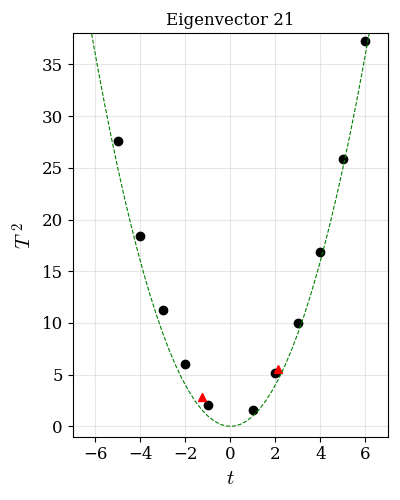}
    \includegraphics[width=0.3\textwidth]{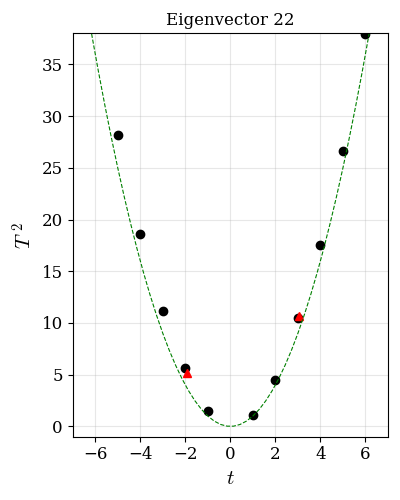}
    \includegraphics[width=0.3\textwidth]{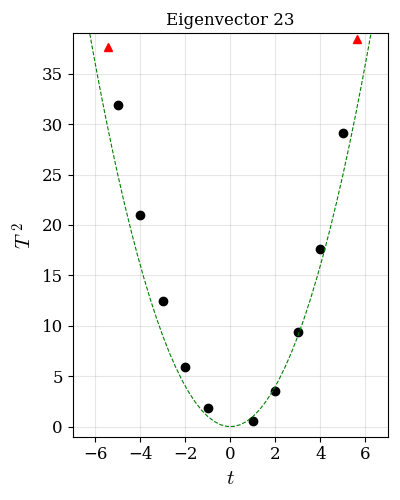}
    \includegraphics[width=0.3\textwidth]{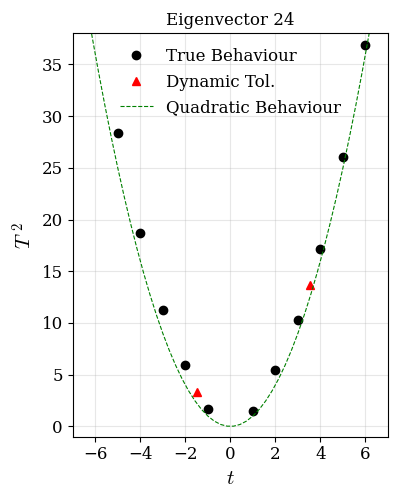}
    \end{center}
    \caption{\textit{(Continued)} Dynamic tolerances for each eigenvector direction in the case of complete decorrelation between the theory and PDF parameters, and the $K$-factor parameters included in the PDF fit.}
\end{figure}
\begin{figure}\ContinuedFloat
    \begin{center}
    \includegraphics[width=0.3\textwidth]{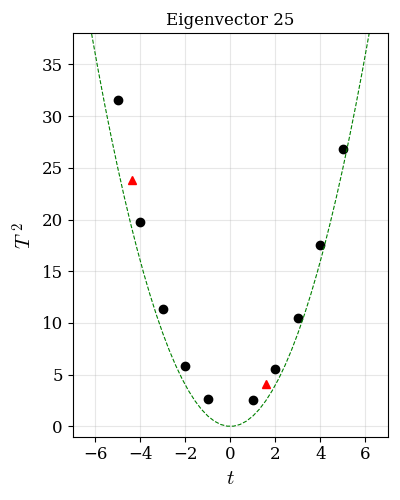}
    \includegraphics[width=0.3\textwidth]{figures/section8/section8-3/N3LO/eigenvector_analysis_26.png}
    \includegraphics[width=0.3\textwidth]{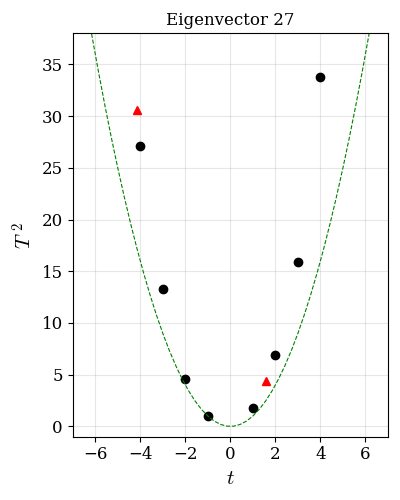}
    \includegraphics[width=0.3\textwidth]{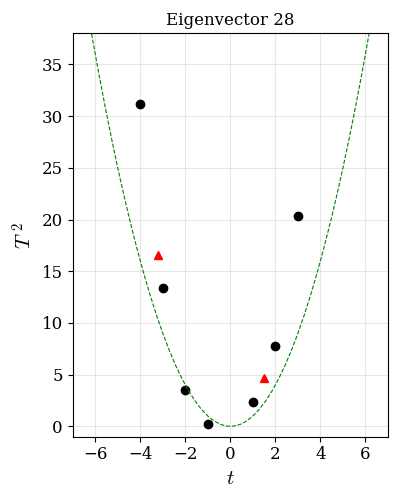}
    \includegraphics[width=0.3\textwidth]{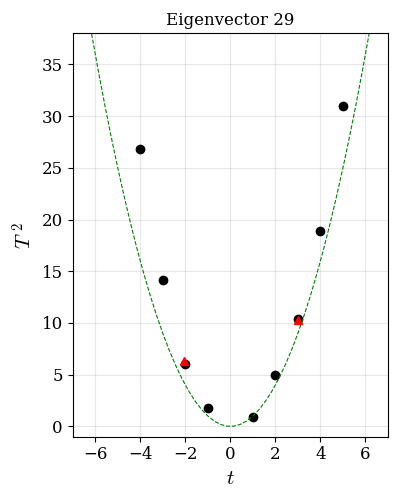}
    \includegraphics[width=0.3\textwidth]{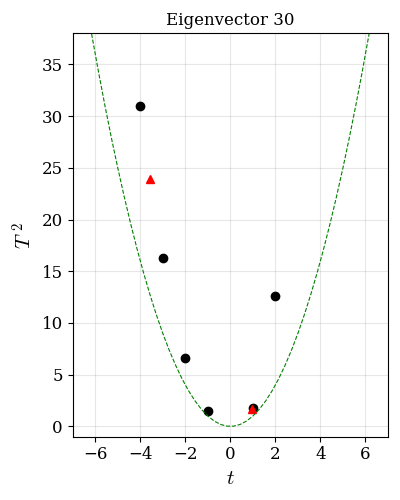}
    \includegraphics[width=0.3\textwidth]{figures/section8/section8-3/N3LO/eigenvector_analysis_31.png}
    \includegraphics[width=0.3\textwidth]{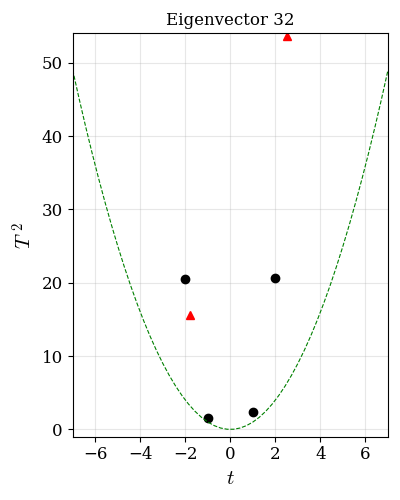}
    \includegraphics[width=0.3\textwidth]{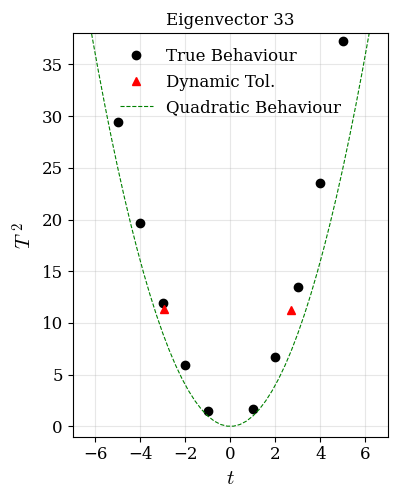}
    \end{center}
    \caption{\textit{(Continued)} Dynamic tolerances for each eigenvector direction in the case of complete decorrelation between the theory and PDF parameters, and the $K$-factor parameters included in the PDF fit.}
\end{figure}
\begin{figure}\ContinuedFloat
    \begin{center}
    \includegraphics[width=0.3\textwidth]{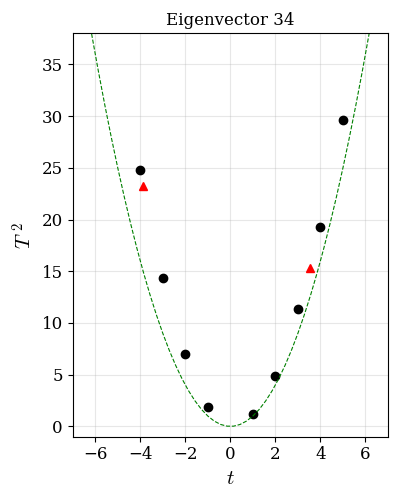}
    \includegraphics[width=0.3\textwidth]{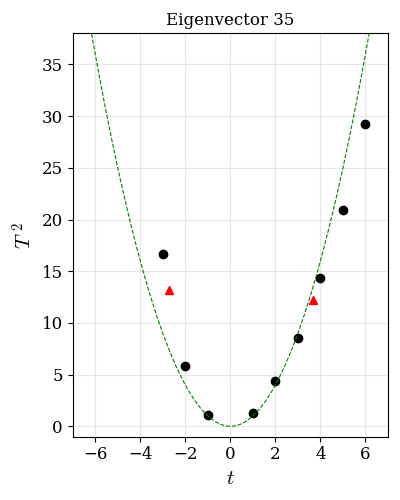}
    \includegraphics[width=0.3\textwidth]{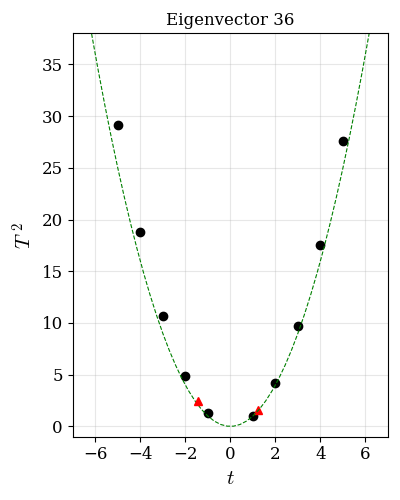}
    \includegraphics[width=0.3\textwidth]{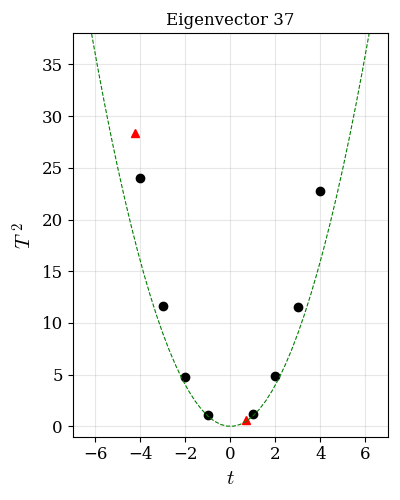}
    \includegraphics[width=0.3\textwidth]{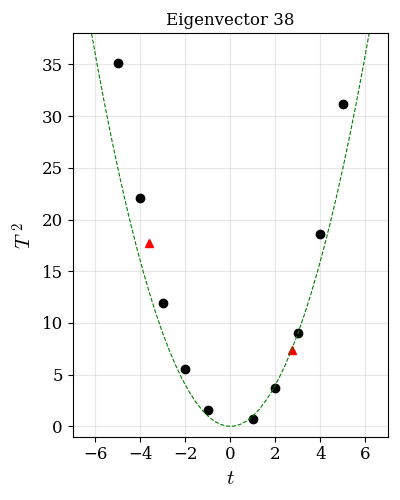}
    \includegraphics[width=0.3\textwidth]{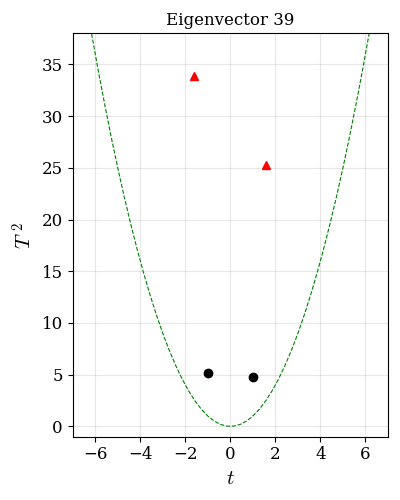}
    \includegraphics[width=0.3\textwidth]{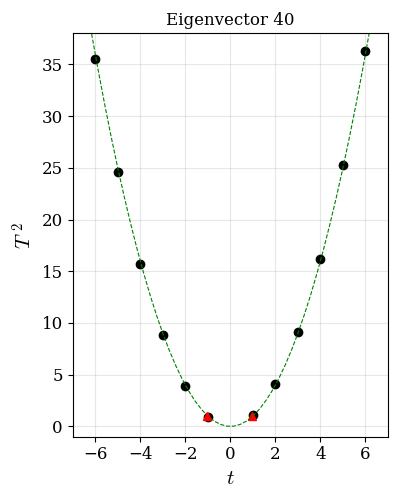}
    \includegraphics[width=0.3\textwidth]{figures/section8/section8-3/N3LO/eigenvector_analysis_41.png}
    \includegraphics[width=0.3\textwidth]{figures/section8/section8-3/N3LO/eigenvector_analysis_42.png}
    \end{center}
    \caption{\textit{(Continued)} Dynamic tolerances for each eigenvector direction in the case of complete decorrelation between the theory and PDF parameters, and the $K$-factor parameters included in the PDF fit.}
\end{figure}
\begin{figure}\ContinuedFloat
    \begin{center}
    \includegraphics[width=0.3\textwidth]{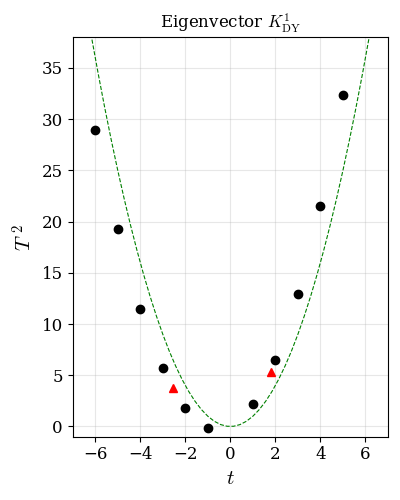}
    \includegraphics[width=0.3\textwidth]{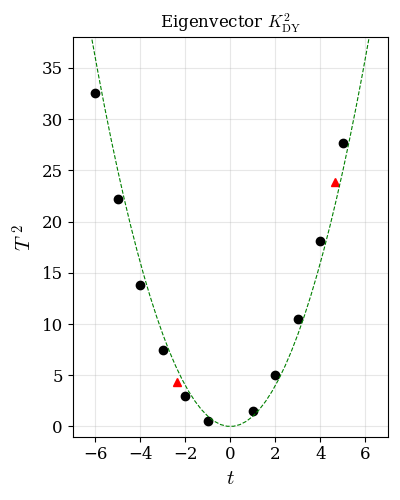}
    \includegraphics[width=0.3\textwidth]{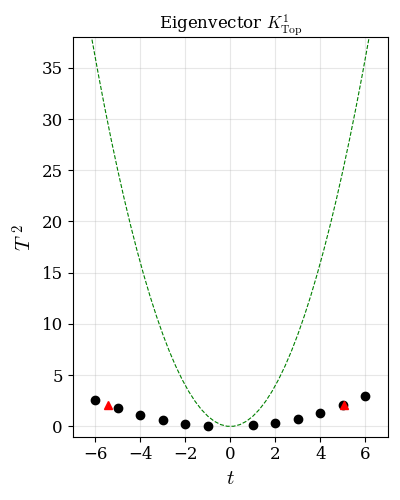}
    \includegraphics[width=0.3\textwidth]{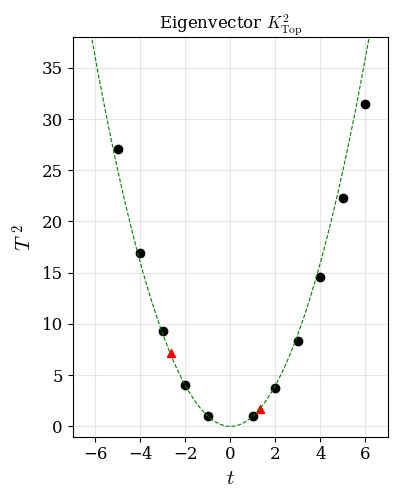}
    \includegraphics[width=0.3\textwidth]{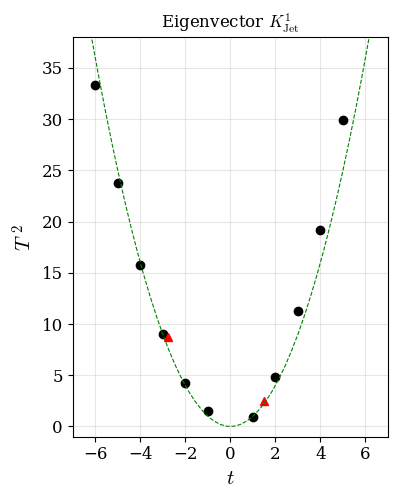}
    \includegraphics[width=0.3\textwidth]{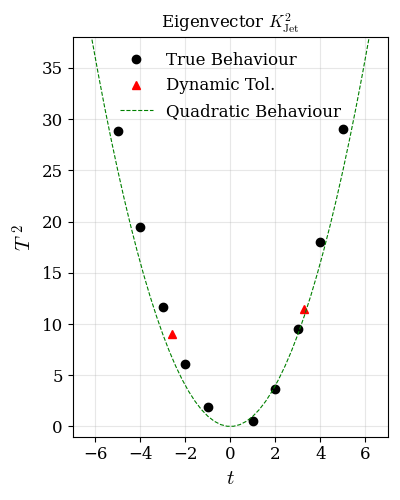}
    \end{center}
    \caption{\textit{(Continued)} Dynamic tolerances for each eigenvector direction in the case of complete decorrelation between the theory and PDF parameters, and the $K$-factor parameters included in the PDF fit.}
\end{figure}
\begin{figure}\ContinuedFloat
    \begin{center}
    \includegraphics[width=0.3\textwidth]{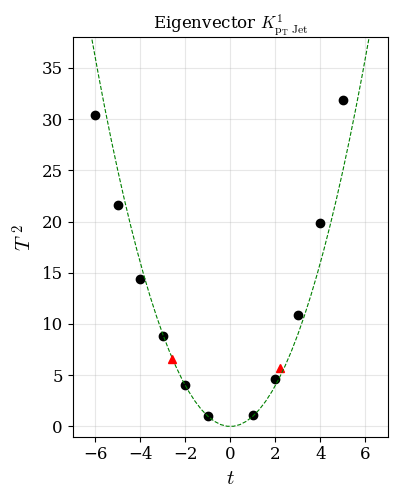}
    \includegraphics[width=0.3\textwidth]{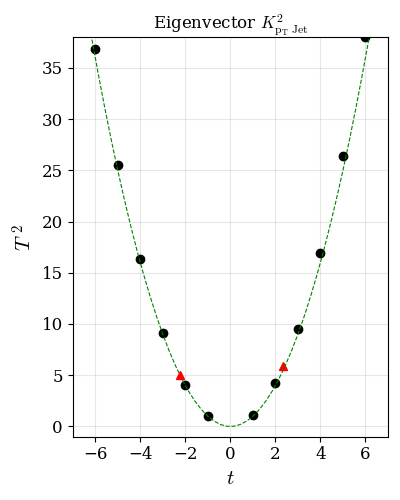}
    \\
    \includegraphics[width=0.3\textwidth]{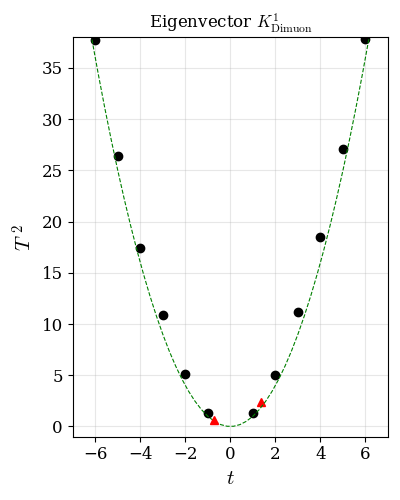}
    \includegraphics[width=0.3\textwidth]{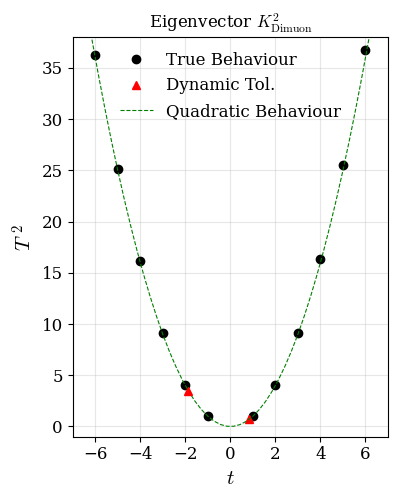}
    \end{center}
    \caption{\textit{(Continued)} Dynamic tolerances for each eigenvector direction in the case of complete decorrelation between the theory and PDF parameters, and the $K$-factor parameters included in the PDF fit.}
\end{figure}

\subsection{Case 2: Correlated \texorpdfstring{$K$}{K}-factor Parameters}\label{app: tolerance_corr}

Fig.~\ref{fig: full_tol_corr} displays the tolerance landscape for each eigenvector found from the correlated ($H_{ij}^{\prime}$) Hessian described in Section~\ref{sec: theo_framework}. Across all 52 eigenvectors we show an overall general agreement with the quadratic assumption similar to that found at NNLO.

\begin{figure}[H]
    \begin{center}
    \includegraphics[width=0.3\textwidth]{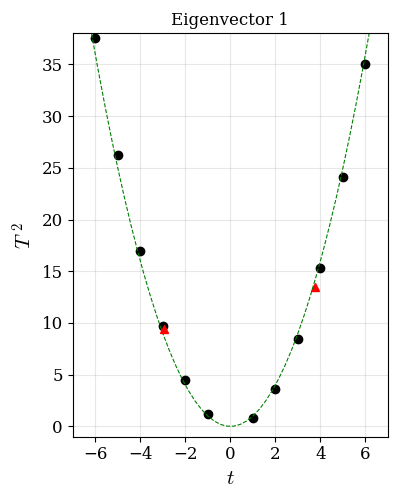}
    \includegraphics[width=0.3\textwidth]{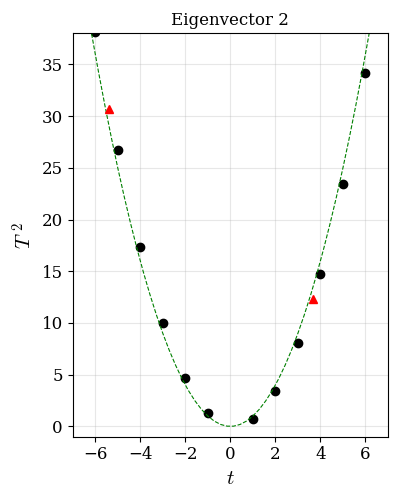}
    \includegraphics[width=0.3\textwidth]{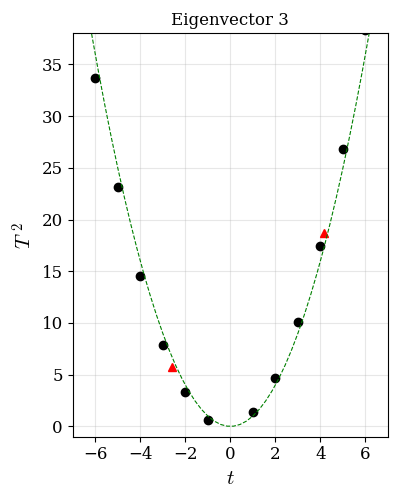}
    \includegraphics[width=0.3\textwidth]{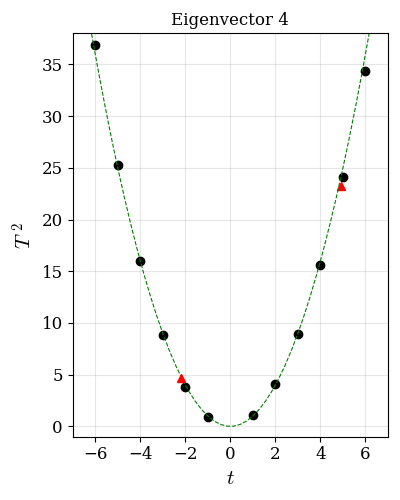}
    \includegraphics[width=0.3\textwidth]{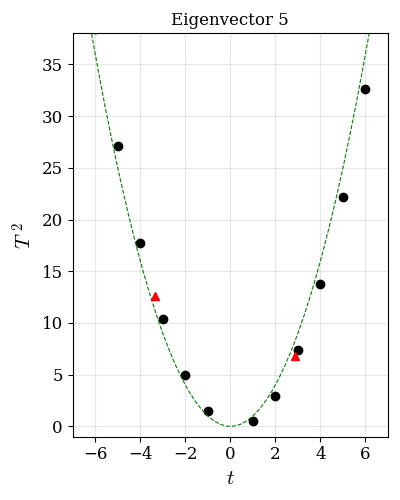}
    \includegraphics[width=0.3\textwidth]{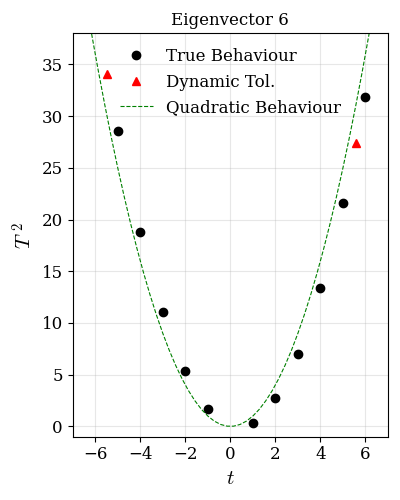}
    \end{center}
    \caption{\label{fig: full_tol_corr}Dynamic tolerances for each eigenvector direction in the case of complete correlation between all theory, PDF and $K$-factor parameters included in the PDF fit.}
\end{figure}
\begin{figure}\ContinuedFloat
    \begin{center}
    \includegraphics[width=0.3\textwidth]{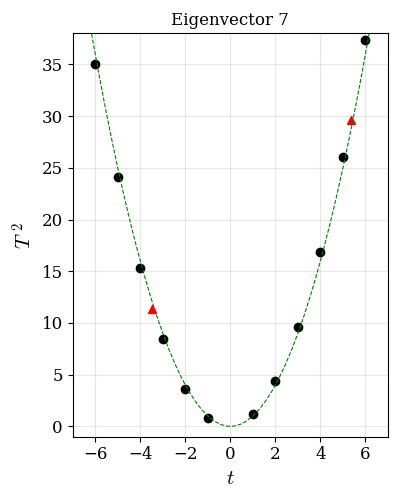}
    \includegraphics[width=0.3\textwidth]{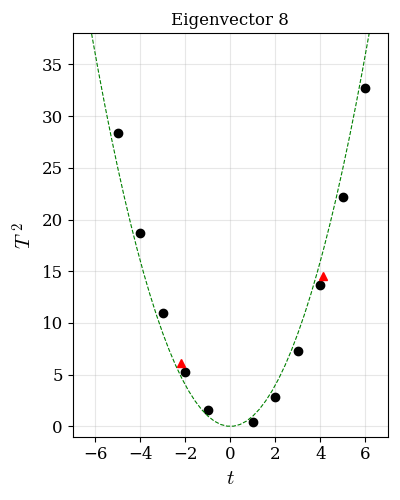}
    \includegraphics[width=0.3\textwidth]{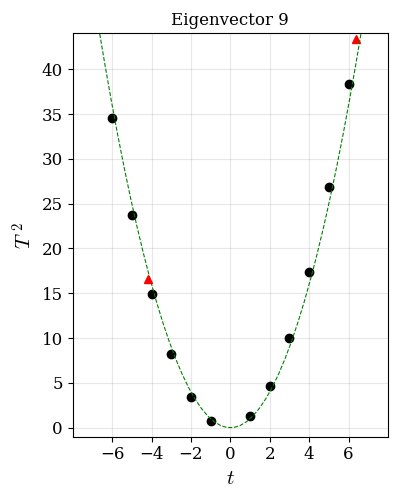}
    \includegraphics[width=0.3\textwidth]{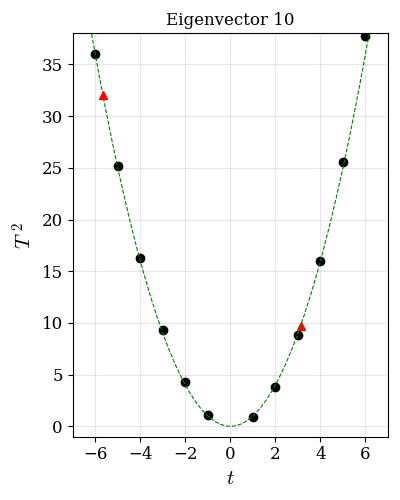}
    \includegraphics[width=0.3\textwidth]{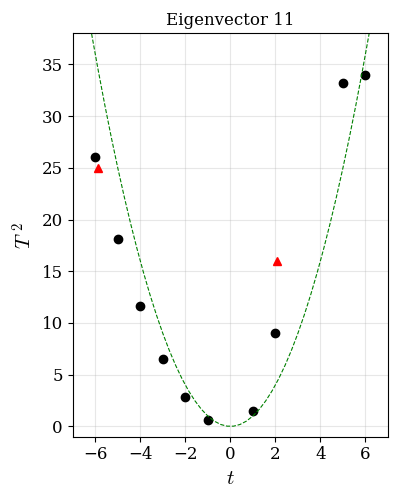}
    \includegraphics[width=0.3\textwidth]{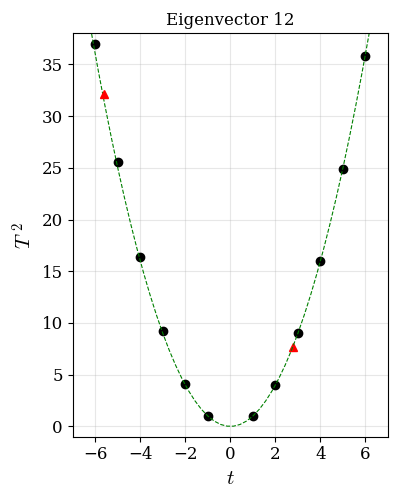}
    \includegraphics[width=0.3\textwidth]{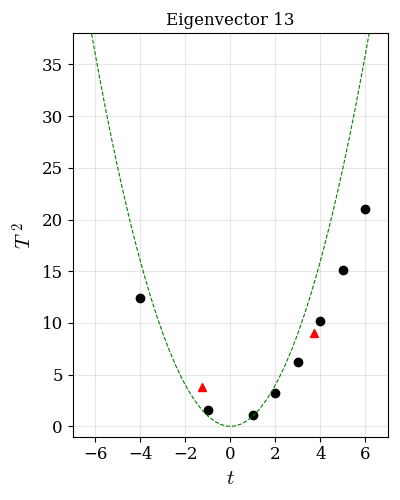}
    \includegraphics[width=0.3\textwidth]{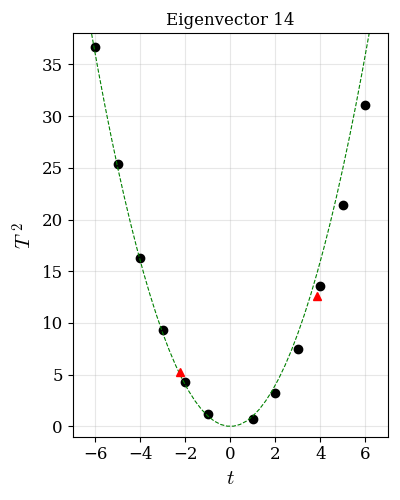}
    \includegraphics[width=0.3\textwidth]{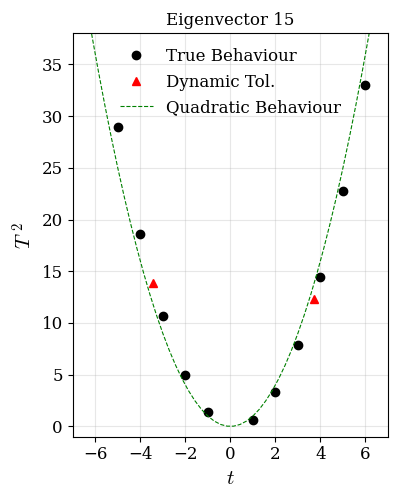}
    \end{center}
    \caption{\textit{(Continued)} Dynamic tolerances for each eigenvector direction in the case of complete correlation between all theory, PDF and $K$-factor parameters included in the PDF fit.}
\end{figure}
\begin{figure}\ContinuedFloat
    \begin{center}
    \includegraphics[width=0.3\textwidth]{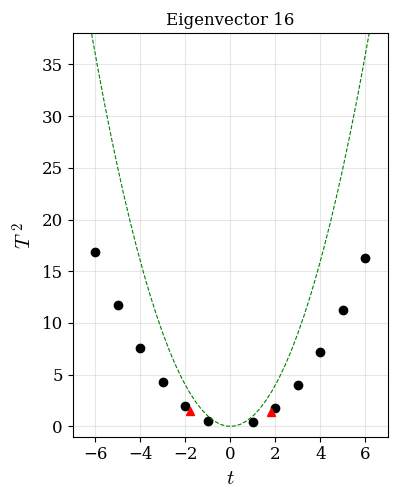}
    \includegraphics[width=0.3\textwidth]{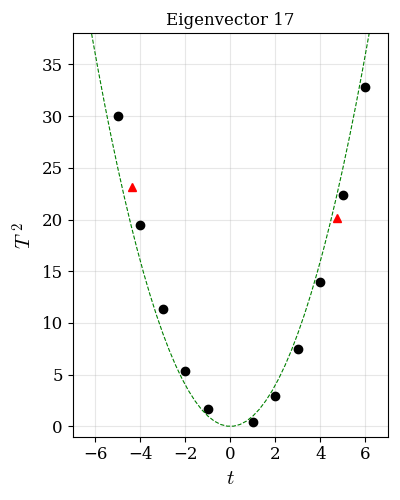}
    \includegraphics[width=0.3\textwidth]{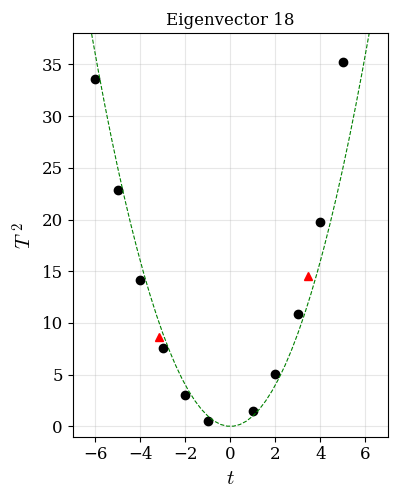}
    \includegraphics[width=0.3\textwidth]{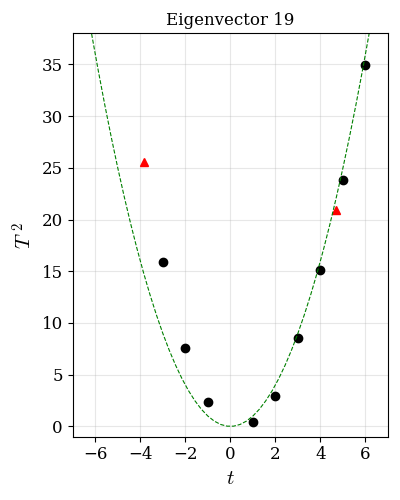}
    \includegraphics[width=0.3\textwidth]{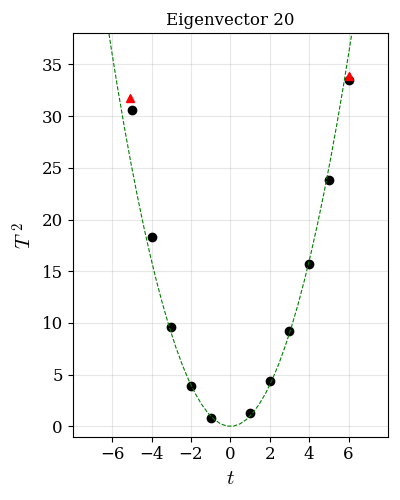}
    \includegraphics[width=0.3\textwidth]{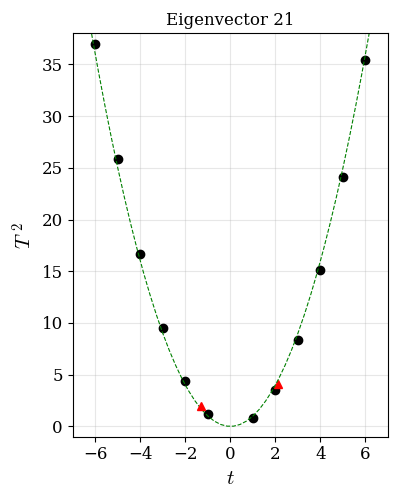}
    \includegraphics[width=0.3\textwidth]{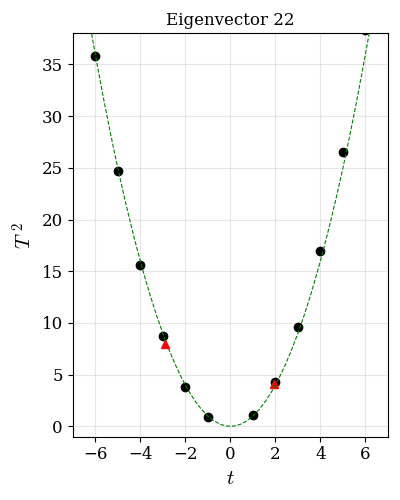}
    \includegraphics[width=0.3\textwidth]{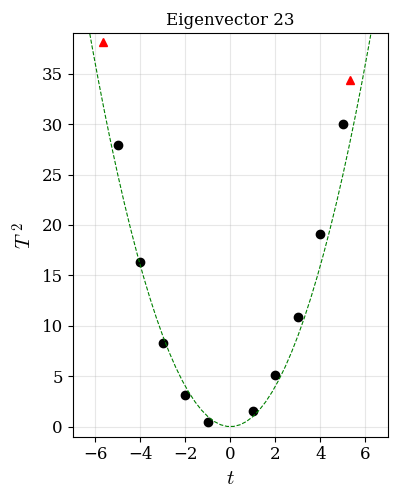}
    \includegraphics[width=0.3\textwidth]{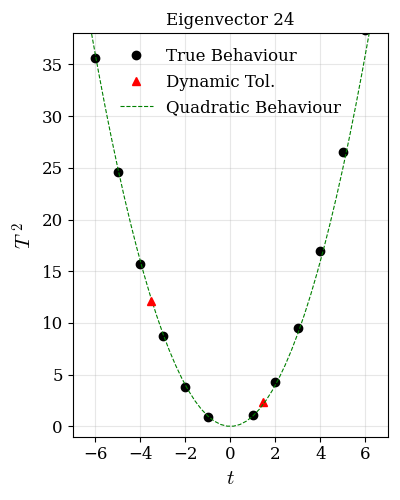}
    \end{center}
    \caption{\textit{(Continued)} Dynamic tolerances for each eigenvector direction in the case of complete correlation between all theory, PDF and $K$-factor parameters included in the PDF fit.}
\end{figure}
\begin{figure}\ContinuedFloat
    \begin{center}
    \includegraphics[width=0.3\textwidth]{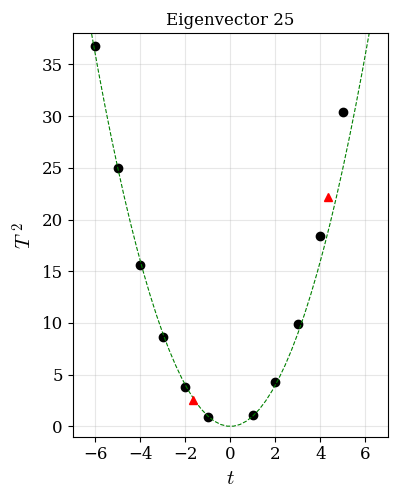}
    \includegraphics[width=0.3\textwidth]{figures/section8/section8-3/N3LO_Kcorr/eigenvector_analysis_26.png}
    \includegraphics[width=0.3\textwidth]{figures/section8/section8-3/N3LO_Kcorr/eigenvector_analysis_27.png}
    \includegraphics[width=0.3\textwidth]{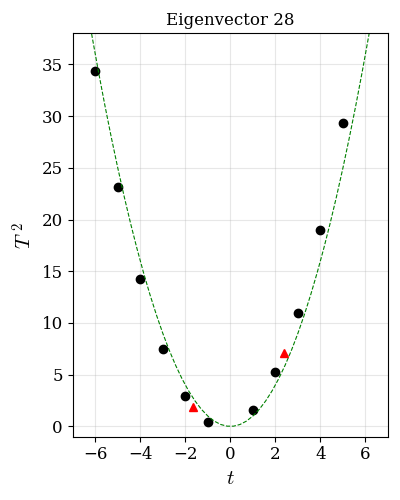}
    \includegraphics[width=0.3\textwidth]{figures/section8/section8-3/N3LO_Kcorr/eigenvector_analysis_29.png}
    \includegraphics[width=0.3\textwidth]{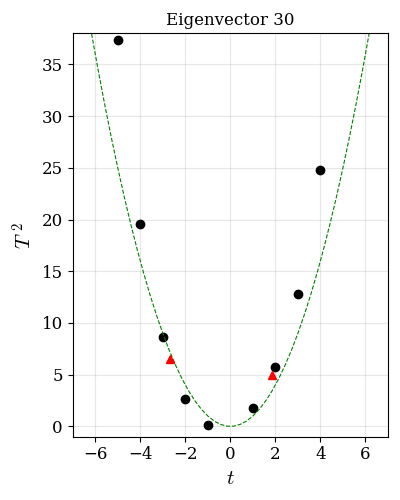}
    \includegraphics[width=0.3\textwidth]{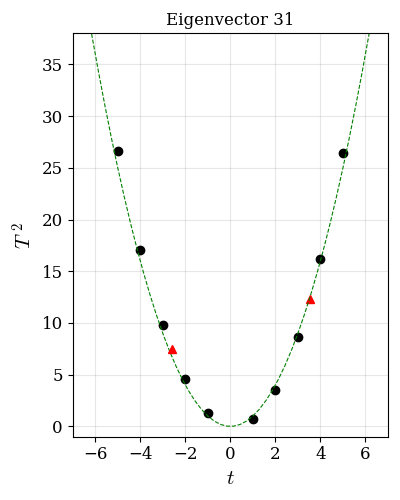}
    \includegraphics[width=0.3\textwidth]{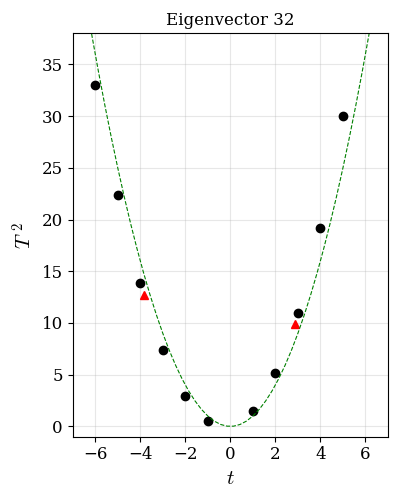}
    \includegraphics[width=0.3\textwidth]{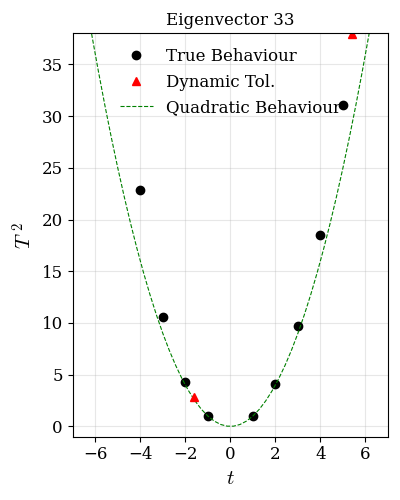}
    \end{center}
    \caption{\textit{(Continued)} Dynamic tolerances for each eigenvector direction in the case of complete correlation between all theory, PDF and $K$-factor parameters included in the PDF fit.}
\end{figure}
\begin{figure}\ContinuedFloat
    \begin{center}
    \includegraphics[width=0.3\textwidth]{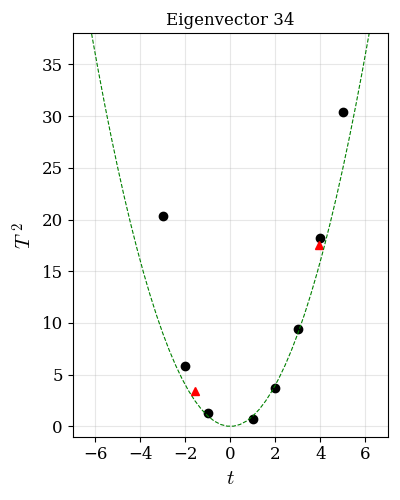}
    \includegraphics[width=0.3\textwidth]{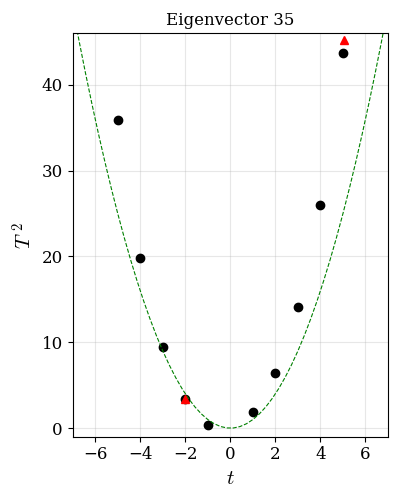}
    \includegraphics[width=0.3\textwidth]{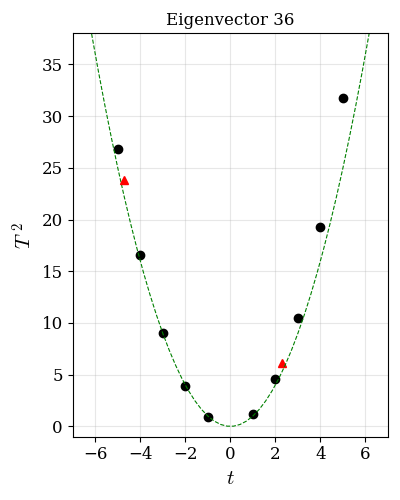}
    \includegraphics[width=0.3\textwidth]{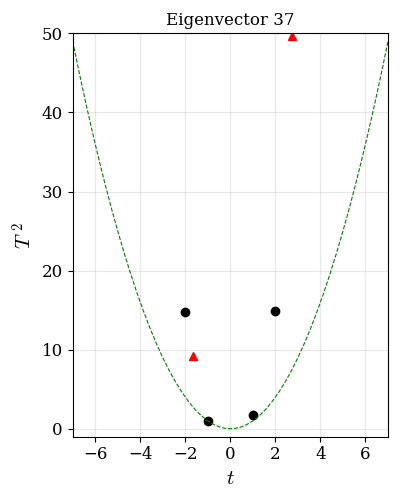}
    \includegraphics[width=0.3\textwidth]{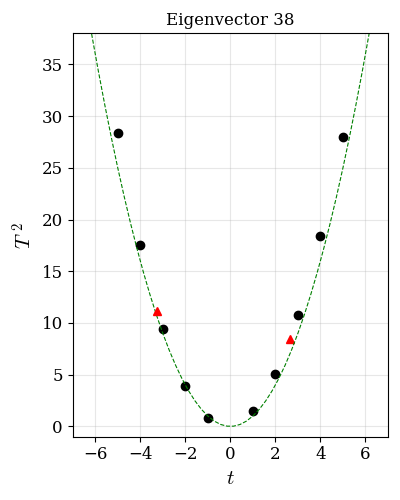}
    \includegraphics[width=0.3\textwidth]{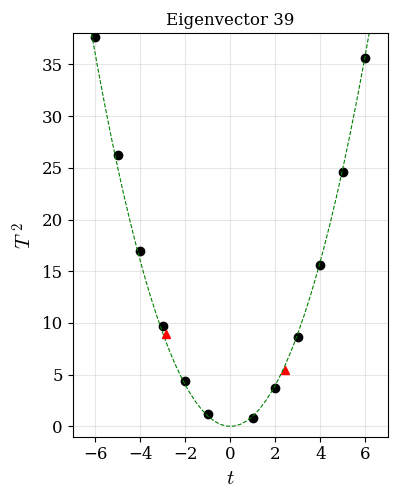}
    \includegraphics[width=0.3\textwidth]{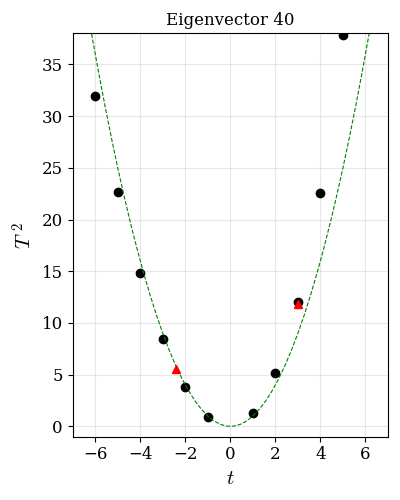}
    \includegraphics[width=0.3\textwidth]{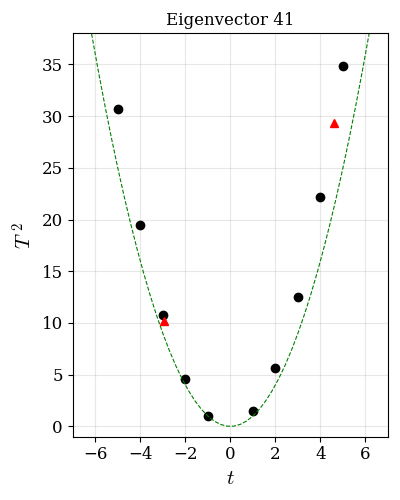}
    \includegraphics[width=0.3\textwidth]{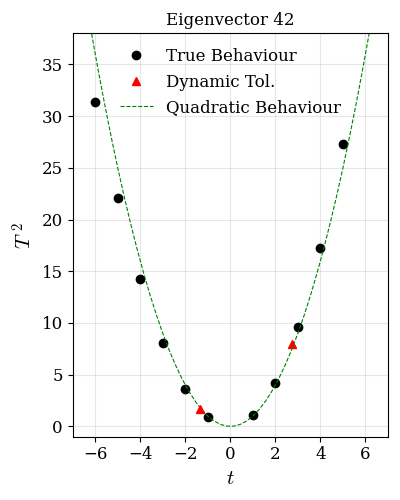}
    \end{center}
    \caption{\textit{(Continued)} Dynamic tolerances for each eigenvector direction in the case of complete correlation between all theory, PDF and $K$-factor parameters included in the PDF fit.}
\end{figure}
\begin{figure}\ContinuedFloat
    \begin{center}
    \includegraphics[width=0.3\textwidth]{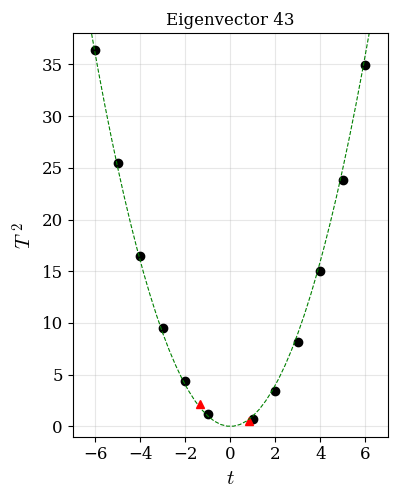}
    \includegraphics[width=0.3\textwidth]{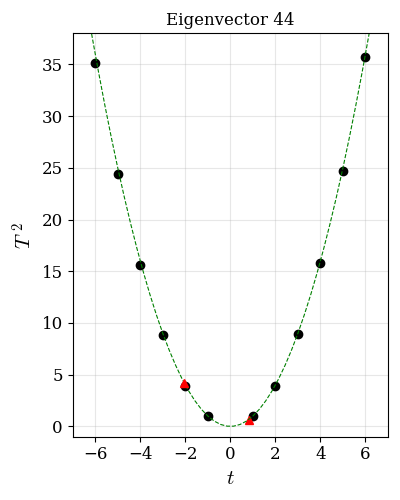}
    \includegraphics[width=0.3\textwidth]{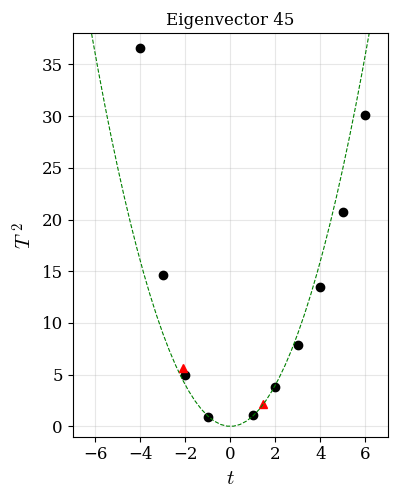}
    \includegraphics[width=0.3\textwidth]{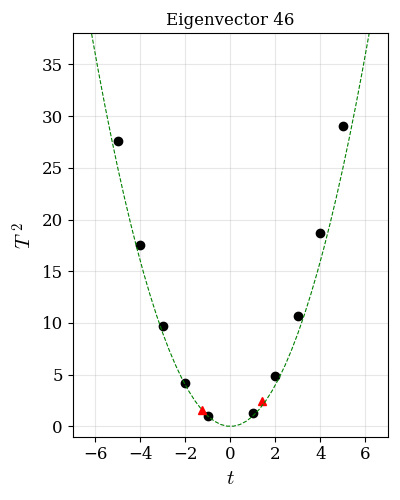}
    \includegraphics[width=0.3\textwidth]{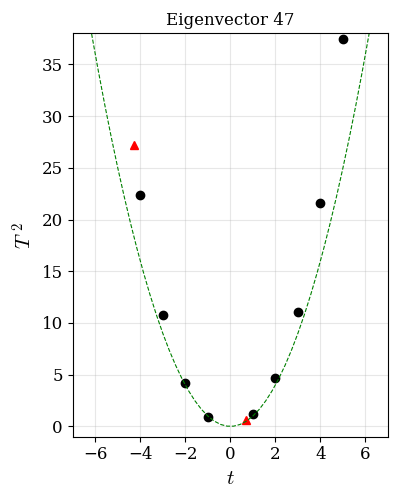}
    \includegraphics[width=0.3\textwidth]{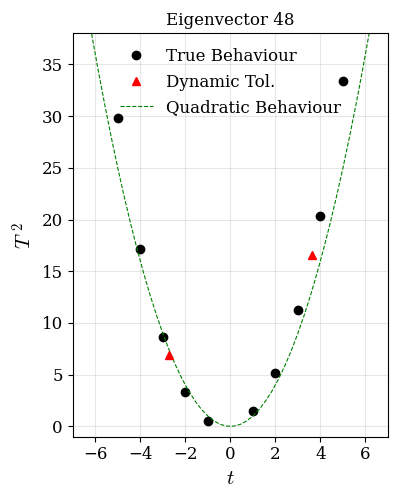}
    \end{center}
    \caption{\textit{(Continued)} Dynamic tolerances for each eigenvector direction in the case of complete correlation between all theory, PDF and $K$-factor parameters included in the PDF fit.}
\end{figure}
\begin{figure}\ContinuedFloat
    \begin{center}
    \includegraphics[width=0.3\textwidth]{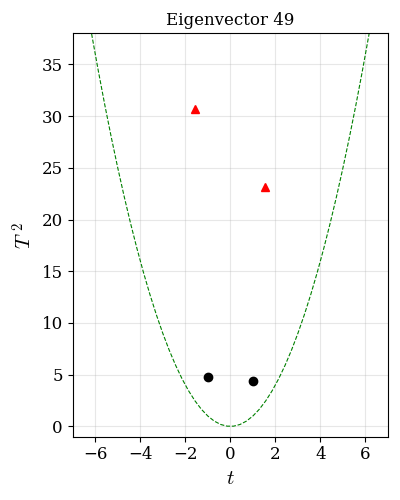}
    \includegraphics[width=0.3\textwidth]{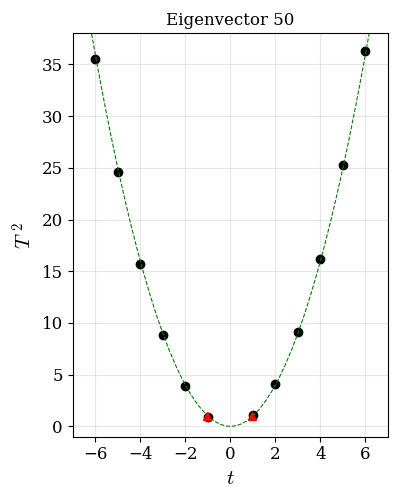}
    \includegraphics[width=0.3\textwidth]{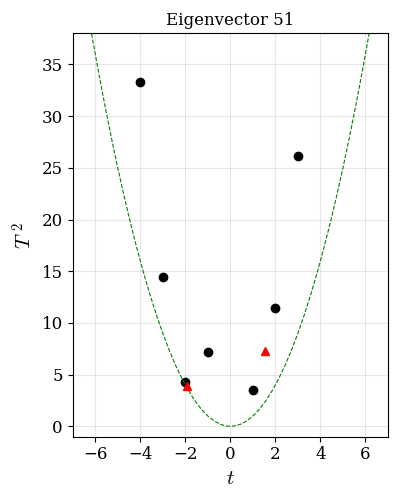}
    \includegraphics[width=0.3\textwidth]{figures/section8/section8-3/N3LO_Kcorr/eigenvector_analysis_52.png}
    \end{center}
    \caption{\textit{(Continued)} Dynamic tolerances for each eigenvector direction in the case of complete correlation between all theory, PDF and $K$-factor parameters included in the PDF fit.}
\end{figure}

\setcounter{table}{0}
\section{Higgs Gluon Fusion \texorpdfstring{$\mu = m_{H}$}{Higgs mass scale} Results}

\begin{table}[H]
\centerline{
\begin{tabular}{c|ccc}
\hline
$\sigma$ order & PDF order & $\sigma + \Delta \sigma_{+} - \Delta \sigma_{-}$ (pb) & $\sigma$ (pb) $+\ \Delta \sigma_{+} - \Delta \sigma_{-}$ (\%) \\
\hline
\multicolumn{4}{c}{PDF uncertainties} \\
\hline
\multirow{4}{*}{N$^{3}$LO} 
 & aN$^{3}$LO (no theory unc.) & $43.803 + 0.685 - 0.526$ & $42.709 + 2.81\% - 3.14\%$ \\
 & aN$^{3}$LO ($H_{ij} + K_{ij}$) & $43.803 + 0.795 - 0.732$ & $42.709 + 3.30\% - 3.17\%$ \\
 & aN$^{3}$LO ($H_{ij}^{\prime}$) & $43.803 + 0.785 - 0.737$ & $42.709 + 3.39\% - 3.08\%$ \\
 & NNLO & $46.243 + 0.524 - 0.563$  & $46.243 + 1.13\% - 1.22\%$ \\
\hline
NNLO & NNLO & $42.129 + 0.472 - 0.510$  & $42.129 + 1.12\% - 1.21\%$ \\
\hline
\multicolumn{4}{c}{PDF + Scale uncertainties} \\
\hline
\multirow{4}{*}{N$^{3}$LO}  & aN$^{3}$LO (no theory unc.) & $43.803 + 1.723 - 2.519$ & $42.709 + 4.68\% - 6.44\%$ \\
 & aN$^{3}$LO ($H_{ij} + K_{ij}$) & $43.803 + 1.770 - 2.570$ & $42.709 + 4.89\% - 6.45\%$ \\
 & aN$^{3}$LO ($H_{ij}^{\prime}$) & $43.803 + 1.766 - 2.571$ & $42.709 + 4.95\% - 6.41\%$ \\
 & NNLO & $46.243 + 1.845 - 3.078$  & $46.243 + 3.99\% - 6.66\%$ \\
\hline
NNLO & NNLO & $42.129 + 4.989 - 5.106$  & $42.129 + 11.84\% - 12.12\%$ \\
\end{tabular}
}
\caption{\label{tab: ggH_results_mh}Higgs production cross section results via gluon fusion using N$^{3}$LO and NNLO hard cross sections combined with NNLO and aN$^{3}$LO PDFs. All PDFs are at the standard choice $\alpha_{s} = 0.118$. These results are found with $\mu = m_{H}$ unless stated otherwise, with the values for $\mu = m_{H}/2$ supplied in Table~\ref{tab: ggH_results_mh2}.}
\end{table}
Provided in Table~\ref{tab: ggH_results_mh} are the results analogous to those in Table~\ref{tab: ggH_results_mh2} but with the central scale set to $\mu = \mu_{f} = \mu_{r} = m_{H}$. These results show a higher level of stability for aN$^{3}$LO PDFs with the chosen central scale. By the renormalisation group arguments, this scale dependence should disappear at all orders in perturbation theory. Therefore the results here suggest that the aN$^{3}$LO PDFs are following this trend.

\clearpage

\bibliographystyle{h-physrev}
\bibliography{MSHT_references}

\end{document}